\pdfoutput=1
\documentclass[10pt]{article}

\usepackage{fullpage}
\usepackage{setspace}
\usepackage{parskip}
\usepackage{titlesec}
\usepackage[section]{placeins}
\usepackage{xcolor}
\usepackage{breakcites}
\usepackage{lineno}
\usepackage{hyphenat}
\usepackage[colorlinks = true,
            linkcolor = blue,
            urlcolor  = blue,
            citecolor = blue,
            anchorcolor = blue]{hyperref}
\usepackage{etoolbox}


\usepackage{natbib}

\titlespacing{\section}{0pt}{*3}{*1}
\titlespacing{\subsection}{0pt}{*2}{*0.5}
\titlespacing{\subsubsection}{0pt}{*1.5}{0pt}

\usepackage{authblk}
\usepackage{graphicx}
\usepackage[space]{grffile}
\usepackage{latexsym}
\usepackage{textcomp}
\usepackage{longtable}
\usepackage{tabulary}
\usepackage{booktabs,array,multirow}
\usepackage{amsfonts,amsmath,amssymb}
\providecommand\citet{\cite}
\providecommand\citep{\cite}

\newif\iflatexml\latexmlfalse
\providecommand{\tightlist}{\setlength{\itemsep}{0pt}\setlength{\parskip}{0pt}}

\AtBeginDocument{\DeclareGraphicsExtensions{.pdf,.PDF,.eps,.EPS,.png,.PNG,.tif,.TIF,.jpg,.JPG,.jpeg,.JPEG}}

\usepackage[utf8]{inputenc}
\usepackage[ngerman,greek,english]{babel}
\usepackage{float}

\begin{document}

\title{Roadmap to Neuromorphic Computing with Emerging Technologies}

\author[1]{Adnan Mehonic}
\author[2]{Daniele Ielmini}
\author[3]{Kaushik Roy}
\author[4]{Onur Mutlu}
\author[5]{Shahar Kvatinsky}
\author[6]{Teresa Serrano-Gotarredona}
\author[6]{Bernabe Linares-Barranco}
\author[7]{Sabina Spiga}
\author[8]{Sergey Savel'ev}
\author[8]{Alexander G. Balanov}
\author[9]{Nitin Chawla}
\author[9]{Giuseppe Desoli}
\author[2]{Gerardo Malavena}
\author[2]{Christian Monzio Compagnoni}
\author[10]{Zhongrui Wang}
\author[11]{J. Joshua Yang}
\author[12]{Ghazi Sarwat Syed}
\author[12]{Abu Sebastian}
\author[13]{Thomas Mikolajick}
\author[14]{Beatriz Noheda}
\author[13]{Stefan Slesazeck}
\author[15]{Bernard Dieny}
\author[16]{Tuo-Hung (Alex) Hou}
\author[17]{Akhil Varri}
\author[17]{Frank Brückenhoff-Plickelmann}
\author[18]{Wolfram Pernice}
\author[19]{Xixiang Zhang}
\author[19]{Sebastian Pazos}
\author[19]{Mario Lanza}
\author[20]{Stefan Wiefels}
\author[20]{Regina Dittmann}
\author[1]{Wing H Ng}
\author[1]{Mark Buckwell}
\author[1]{Horatio RJ Cox}
\author[1]{Daniel J Mannion}
\author[1]{Anthony J Kenyon}
\author[21]{Yingming Lu}
\author[21]{Yuchao Yang}
\author[22]{Damien Querlioz}
\author[22]{Louis Hutin}
\author[23]{Elisa Vianello}
\author[3]{Sayeed Shafayet Chowdhury}
\author[2]{Piergiulio Mannocci}
\author[21]{Yimao Cai}
\author[21]{Zhong Sun}
\author[24]{Giacomo Pedretti}
\author[28]{John Paul Strachan}
\author[25]{Dmitri Strukov}
\author[12]{Manuel Le Gallo}
\author[26]{Stefano Ambrogio}
\author[27]{Ilia Valov}
\author[27]{Rainer Waser}

\affil[1]{University College London}
\affil[2]{Politecnico di Milano}
\affil[3]{Purdue University}
\affil[4]{ETH Zurich}
\affil[5]{Technion - Israel Institute of Technology}
\affil[6]{Institute of Microelectronics of Seville}
\affil[7]{Consiglio Nazionale delle Ricerche}
\affil[8]{Loughborough University}
\affil[9]{STMicroelectronics}
\affil[10]{University of Hong Kong}
\affil[11]{University of Southern California}
\affil[12]{IBM Research Europe - Zurich}
\affil[13]{NaMLab gGmbH}
\affil[14]{University of Groningen}
\affil[15]{SPINTEC}
\affil[16]{National Yang Ming Chiao Tung University}
\affil[17]{University of Muenster}
\affil[18]{Heidelberg University}
\affil[19]{King Abdullah University of Science and Technology}
\affil[20]{Forschungszentrum Jülich Peter Grünberg Institut}
\affil[21]{Peking University}
\affil[22]{Université Paris-Sud}
\affil[23]{CEA - LETI}
\affil[24]{Hewlett Packard Labs}
\affil[25]{UC Santa Barbara}
\affil[26]{IBM Research - Almaden}
\affil[27]{Aachen University}

\date{} 

\begingroup
\let\center\flushleft
\let\endcenter\endflushleft
\maketitle
\endgroup

\newpage

\subsection*{1. Introduction to the roadmap}

{\label{675037}}

\begin{quote}
\emph{Adnan Mehonic,~Daniele Ielmini,~Kaushik Roy}
\end{quote}

\subsection*{1.1. Taxonomy \& Motivation}

{\label{541649}}

The growing adoption of data-driven applications, such as artificial
intelligence (AI), is transforming the way we interact with technology.
Currently, the deployment of AI and machine learning tools in previously
uncharted domains generate considerable enthusiasm for~further research,
development, and utilisation. These innovative applications often
provide effective solutions to complex, longstanding challenges that
have remained unresolved for years. By expanding the reach of AI and
machine learning, we unlock new possibilities and facilitate
advancements in various sectors.~These include but are not limited to,
scientific research, education, transportation, smart city planning,
eHealth, and the metaverse.

However, our predominant focus on performance can sometimes lead to
critical oversights. For instance, our constant dependence on immediate
access to information might cause us to ignore the energy consumption
and environmental consequences associated with the computing systems
that enable such access. Balancing performance with sustainability is
crucial for the technology's continued growth.

~From this standpoint, the environmental impact of AI is a cause for
growing concern. Additionally, applications such as the Internet of
Things (IoT) and autonomous robotic agents may not always rely on
resource-intensive deep learning algorithms but still need to minimize
energy consumption. Realizing the vision of IoT is contingent upon
reducing the energy requirements of numerous connected devices. Demand
for computing power is growing at a rate that far exceeds improvements
achieved through Moore's law scaling. Figure 1a shows the computing
power demands, quantified in Peta floating-point operations (PetaFLOPS,
one peta = 10\textsuperscript{15}) per day, as a function of time,
indicating an increase of a factor 2 every two months in recent
years~\textsuperscript{\hyperref[csl:1]{1}}. In addition to Moore's law, significant
advancements have been made through the combination of intelligent
architecture and hardware-software co-design. For instance, NVIDIA GPUs'
performance has improved by a factor of 317 from 2012 to 2021,
surpassing expectations based on Moore's law alone. Research and
development efforts have demonstrated further impressive performance
improvements~\textsuperscript{\hyperref[csl:2]{2}},~\textsuperscript{\hyperref[csl:3]{3}},~\textsuperscript{\hyperref[csl:4]{4}}
suggesting that more can be achieved. However, conventional computing
solutions alone are unlikely to meet demand in the long term,
particularly when considering the high costs of training associated with
the most complex deep learning models (Figure 1b). It is essential to
explore alternative approaches to tackle these challenges and ensure the
long-term sustainability of AI's rapid advancements.~While global energy
consumption is crucial and important, there is a relevant issue which is
perhaps just as significant: the ability of low-power systems to execute
complex AI algorithms without relying on cloud-based computing. It is
important to keep in mind that the challenge of global AI power
consumption and the ability to implement complex AI on low-power systems
are two somewhat separate challenges.~It might be the case that these
two challenges need to be addressed with somewhat different strategies
(e.g., the power consumption in data centers for the most complex,
largest AI models, such as large language models, might be addressed
differently than implementing mid-sized AI models, such as voice
recognition, on low-power, self-contained systems that might need to run
at a few milliwatts of power).~The latter strategy might not be scalable
for the largest models, or the optimization of the largest models might
not be applicable for simpler models running on much lower power
budgets. However, undeniably, for both, we need to improve the overall
energy efficiency of our computing systems that are designed to execute
AI workloads.\selectlanguage{english}
\begin{figure}[H]
\begin{center}
\includegraphics[width=0.70\columnwidth]{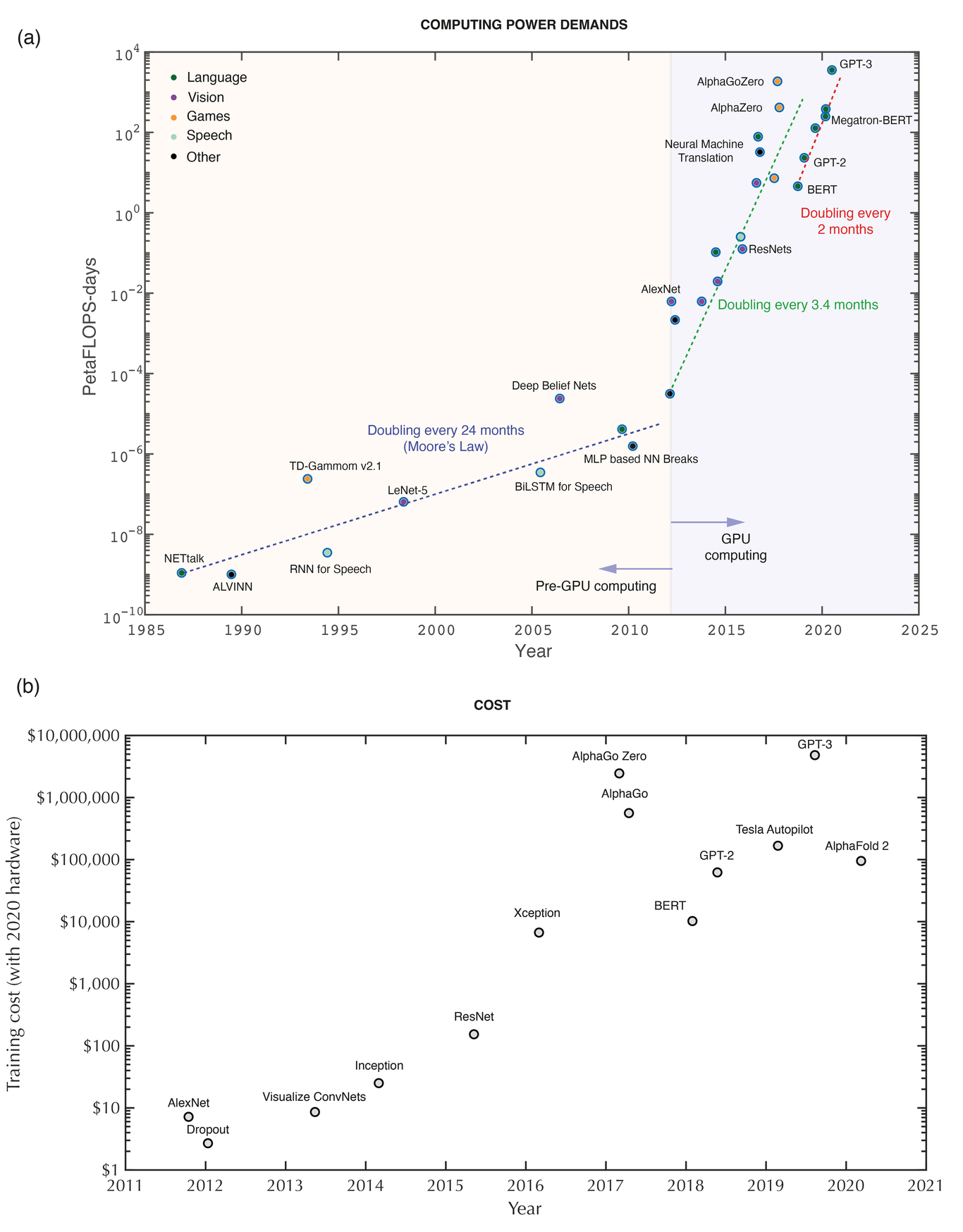}
\caption{{a) Increase in computing power demands to run state-of-the-art AI
models. b) The cost associated with training AI models. ~Adapted and
reproduced from~\protect\textsuperscript{\hyperref[csl:1]{1}}~.
{\label{263337}}%
}}
\end{center}
\end{figure}

The energy efficiency and performance of computing can largely benefit
from new paradigms that aim at replicating or being inspired by specific
characteristics of the brain's biological mechanisms.~It is important to
note that biological systems might be highly specialized and
heterogenious, and therefore different tasks are addressed by different
computational schemes. However, we can still aim to take inspiration
from general features when they are advantageous for specific
applications. It is unlikely that a single architecture or broader
approach will be best applicable for all targeted applications.~

Adopting an interdisciplinary methodology, experts in materials science,
device and circuit engineering, system design, and algorithm and
software development are brought together to collectively contribute to
the progressive field of neuromorphic engineering and computing. This
collaborative approach is instrumental in fueling innovation and
promoting advancements in a domain that seeks to bridge the gap between
biological systems and artificial intelligence.~Coined by Carver Mead in
the late 1980s~\textsuperscript{\hyperref[csl:5]{5}}, the term `neuromorphic' ~originally
referred to systems and devices replicating certain aspects of
biological neural systems, but now it varies across different research
communities. While the term's meaning continues to evolve, it generally
refers to a system embodying brain-inspired properties such as in-memory
computing, hardware learning, spike-based processing, fine-grained
parallelism, and reduced precision computing, among others.~One can also
draw analogies and identify more complex phenomenological similarities
between biological units (e.g., neurons) and electronic components
(e.g., memristors). For example, phenomenological similarities between
models of the redox-based nanoionic resistive memory cell and common
neuronal models, such as Hodgkin-Huxley conductance model and the leaky
integrate-and-fire model, have been
demonstrated~\textsuperscript{\hyperref[csl:6]{6}}.~Even more complex biological
functionalities have been demonstrated using a single third-order
nanocircuit elements \textsuperscript{\hyperref[csl:7]{7}}. It should be noted that many
paradigms related to the neuromorphic approach have also been
independently investigated. For instance, in-memory computing
\textsuperscript{\hyperref[csl:8]{8}}, while being a cornerstone of the neuromorphic
paradigm, is also examined separately. It represents one of the most
promising avenues to enhance the energy efficiency of AI hardware or
more general computing, offering a break from the traditional Von
Neumann architecture paradigm.

Neuromorphic research can be divided into three areas. Firstly,
``neuromorphic engineering'' employs either CMOS technology (e.g.
transistors working in a sub-threshold regime) or cutting-edge post-CMOS
technologies to reproduce the brain's computational units and
mechanisms. Secondly, ``neuromorphic computing'' explores new data
processing methods, frequently drawing inspiration from biological
systems and considering alternative algorithms, such as spike-based
computing. Lastly, the development of ``neuromorphic devices'' marks the
third field. Taking advantage of advancements in electronic and photonic
technologies, it develops innovative nano-devices that frequently
emulate biological components like neurons and synapses or efficiently
implement desired properties, such as in-memory computing.{}

Furthermore, various approaches to neuromorphic research can be
identified based on their primary objectives. Some systems focus on
delivering efficient hardware platforms to enhance our understanding of
biological nervous systems, while others employ brain-inspired
principles to create innovative, efficient computing applications.~This
roadmap primarily focuses on the latter. While there are already
outstanding roadmaps~\textsuperscript{\hyperref[csl:9]{9}}, reviews~\textsuperscript{\hyperref[csl:10]{10}},
~\textsuperscript{\hyperref[csl:11]{11}}, \textsuperscript{\hyperref[csl:12]{12}} and the special issues
\textsuperscript{\hyperref[csl:13]{13}} that offer comprehensive overviews of neuromorphic
technologies, encompassing the integration of hardware and software
solutions as well as the exploration of new learning paradigms, this
particular roadmap focuses on emphasizing the significance of materials
engineering in advancing cutting-edge complementary
metal-oxide-semiconductor (CMOS) and post-CMOS
technologies.~Simultaneously, it offers a holistic perspective on the
general challenges of computing systems, the reasoning behind adopting
the neuromorphic approach, and concise summaries of current technologies
to better contextualize the role of materials engineering within the
broader neuromorphic landscape.~Of course, there are other critical
aspects in the development of neuromorphic technologies that need to be
taken into account. For example, an excellent recent review on thermal
management materials, devices, and networks is one such
example~\textsuperscript{\hyperref[csl:14]{14}}.

The roadmap is organized into several thematic sections, outlining
current computing challenges, discussing the neuromorphic computing
approach, analyzing mature and currently utilized technologies,
providing an overview of emerging technologies, addressing material
challenges, exploring novel computing concepts, and finally examining
the maturity level of emerging technologies while determining the next
essential steps for their advancement.

The roadmap starts with a concise introduction to the current digital
computing landscape, primarily characterized by Moore's law scaling and
the Von Neumann architecture. It then explores the challenges in
sustaining Moore's law and examines the significance and potential
advantages of post-CMOS technologies and architectures aiming to
integrate computing and memory. Following this, the roadmap presents a
historical perspective on the neuromorphic approach, emphasizing its
potential benefits and applications. It provides a thorough review of
cutting-edge developments in various emerging technologies, comparing
them critically. The discussion addresses how these technologies can be
utilized to develop computational building blocks for future computing
systems. The roles of two mature technologies, static random access
memory (SRAM) and Flash, are also explored. The overview of emerging
technologies includes resistive switching and memristors, phase change
materials, ferroelectric materials, magnetic materials, spintronic
materials, optoelectronic and photonic materials, and 2D devices and
systems. Material challenges are discussed in detail, covering types of
challenges, possible solutions, and experimental techniques to study
these. Novel computing concepts are examined, focusing on embracing
device and system variability, spiking-based computing systems, analog
computing for linear algebra, and the use of analog content addressable
memory (CAM) for in-memory computing and optimization solvers. The final
section discusses technological maturity and potential future
directions.

\vspace{0.5cm}
\subsection*{2. Computing challenges}

{\label{535011}}

\begin{quote}
\emph{Onur Mutlu,~Shahar Kvatinsky}
\end{quote}

\subsection*{2.1. Digital computing}

{\label{692296}}

\subsubsection*{2.1.1. Status~}

{\label{799103}}

Digital computing has a long and complex history that stretches back
over a century. The earliest electronic computers were developed in the
1930s and 1940s, and they were large, expensive, and difficult to use.
However, these early computers laid the foundation for the development
of the modern computers that we use today and their principles are still
in widespread use.

One of the key figures in the early history of digital computing was
John von Neumann, a mathematician and computer scientist known for his
contributions to the field of computer science. Von Neumann advocated
the stored program concept and sequential instruction processing, two
vital features of the von Neumann architecture \textsuperscript{\hyperref[csl:15]{15}}~that
are still used in most computers today. Another key feature of the von
Neumann architecture is the separation of the CPU (control unit) and the
main memory. This separation allows the CPU to access the instructions
and data it needs from the main memory while executing a program, and
assigns the computation and control responsibilities specifically to the
CPU.~

Throughout the years, the rapid scaling of semiconductor logic
technology, known as Moore's law \textsuperscript{\hyperref[csl:16]{16}},~ has led to
tremendous improvements in computer performance and energy efficiency.
With the exponential increase in the number of transistors placed on a
single chip provided by technology scaling, engineers have explored many
ways to increase the speed and performance of computers. One way they
did this was by exploiting parallelism, which is the ability of a
computer to perform multiple tasks simultaneously. There are several
different types of parallelism, including SISD (single instruction,
single data), SIMD (single instruction, multiple data), MIMD (multiple
instruction, multiple data), and MISD (multiple instruction, single
data) \textsuperscript{\hyperref[csl:17]{17}}, all of which are exploited in modern computing
systems ranging from general-purpose single-core and multi-core
processors, GPUs, and specialized accelerators.

Technology scaling has also allowed for the development of more
processing units, starting from duplicating the processing cores and,
more recently, adding accelerators. These accelerators can offload
specific tasks, e.g., video processing, compression/decompression,
vision processing, graphics, and machine learning, from the central
processor, further improving performance and energy efficiency (the
required energy to perform a certain task) by specializing the
computation units to the task at hand. As such, modern systems are
heterogeneous, with many different types of logic-based computation
units integrated into the same processor die.

\subsubsection*{2.1.2. Challenges}

{\label{131804}}

While the performance and energy of logic-based computation units have
scaled very well via technology scaling, those of interconnect and
memory systems have not scaled as well. As a result, communication
(e.g., data movement) between computation units and memory units has
emerged as a major bottleneck, partly due to the disparity in scaling
and partly due to the separation and disparity between processing and
memory offered in von Neumann architecture, which both have limited the
ability of computers to take full advantage of the improvements in logic
technology. This bottleneck is broadly referred to as the ``von Neumann
bottleneck'' or the ``memory wall,'' as it can greatly limit the speed
and energy at which the computer can execute instructions.

For decades, the transistor size has scaled down while the power density
has remained constant. This phenomenon, first observed in the 1970s by
Robert Dennard \textsuperscript{\hyperref[csl:18]{18}}, means that as transistors become
smaller and more densely packed onto a chip, the overall performance and
capabilities of the chip improve. However, since the early 2000s, it has
become increasingly challenging to maintain Dennard scaling as voltage
(and thus frequency) scaling has greatly slowed down. The end of Dennard
scaling has increased the importance of energy efficiency of different
processing units and led to phenomena such as ``dark silicon,''
\textsuperscript{\hyperref[csl:19]{19}} where large parts of the chip are powered off. The
rapid move towards more specialized processing units, powered on for
specific tasks, exemplifies the influence of the end of Dennard scaling.

Furthermore, in recent years, it has become increasingly challenging to
maintain the pace of Moore's law due to the physical limitations of
transistors and the challenges of manufacturing smaller and more densely
packed chips. As a result, the looming end of Moore's law has been a
topic of discussion in the tech industry, as this could potentially
limit the future performance improvements of computer chips. New
semiconductor technologies and novel architectural solutions are
required to continue computing systems' performance and energy
efficiency improvements at a similar pace as in the past.

\subsubsection*{2.1.3. Potential Solutions and
Conclusion}

{\label{335174}}

In recent years, different semiconductor and manufacturing technologies
have emerged to overcome the slowdown of Moore's law. These devices
include new transistor structures and materials, advanced packaging
techniques, and new (e.g., nonvolatile) memory devices. Some of those
technologies have similar functionality as standard CMOS technology but
with improved properties. Other technologies also offer radically new
properties, different from CMOS. For example, memristive technologies,
such as resistive RAM \textsuperscript{\hyperref[csl:20]{20}}, have varying resistance and
provide analog data storage that also supports computation. Such novel
technologies with their unique properties may serve as enablers for new
architectures and computing paradigms, which could be different from and
complementary to the von Neumann architecture.

The combination of Moore's law slowdown and von Neumann's bottleneck
requires fresh thinking on computing paradigms. Data movement between
the memory and the processing units is the primary impediment against
high performance and high energy efficiency in modern computing systems
\textsuperscript{\hyperref[csl:21]{21},\hyperref[csl:22]{22},\hyperref[csl:23]{23},\hyperref[csl:24]{24}}. And, this impediment only worsens with the improved
processing abilities and the increased need for data. All modern
computers employ a variety of methods to mitigate the memory bottleneck,
all of which increase the complexity and power requirements of the
system with limited (and sometimes little) success in mitigating the
bottleneck. For example, modern computers have several levels of cache
memories to reduce the latency and power of memory accesses by
exploiting data locality. Cache memories, however, have limited capacity
and are effective only when significant spatial and temporal locality
exists in the program. Cache memories are not always (completely)
effective due to low locality in many modern workloads, which can worsen
the performance and energy efficiency of computers~\textsuperscript{\hyperref[csl:25]{25},\hyperref[csl:26]{26}}.
Similarly, modern computers employ prefetching techniques across the
memory hierarchy to anticipate future memory accesses and load data into
caches before it is needed by the processor. While partially effective
for relatively simple memory access patterns, prefetching is not
effective for complicated memory access~patterns and it increases system
complexity and memory bandwidth consumption \textsuperscript{\hyperref[csl:27]{27}}. Thus,
memory bottleneck remains a tough challenge and hundreds of research
papers and patents are written every year to mitigate
it~\textsuperscript{\hyperref[csl:28]{28}}.~

Overcoming the performance and energy costs of off-chip memory accesses
is an increasingly difficult task as the disparity between the
efficiency of computation and efficiency of memory access continues to
grow. There is therefore a need to examine more disruptive technologies
and architectures that much more tightly integrate logic and memory at a
large scale, avoiding the large costs of data movement across system
components.

Many efforts to move computation closer to and inside the memory units
have been made \textsuperscript{\hyperref[csl:29]{29}}, including adding processing units in
the same package as DRAM chips~\textsuperscript{\hyperref[csl:30]{30},\hyperref[csl:31]{31}}, performing digital
processing using memory cells 10.1038/nature08940 \textsuperscript{\hyperref[csl:32]{32},\hyperref[csl:33]{33}},
and using analog computation capabilities of both DRAM and NVM devices
\textsuperscript{\hyperref[csl:34]{34},\hyperref[csl:35]{35},\hyperref[csl:36]{36},\hyperref[csl:37]{37}}. One exciting novel computing paradigm to eliminate
the von Neumann bottleneck is to reconsider the way computation, and
memory tasks are performed by getting inspiration from the brain, where,
unlike von Neumann architecture, processing and storage are not
separated. Many recent works demonstrate orders of magnitude performance
and energy improvements using various kinds of processing-in-memory
architectures~\textsuperscript{\hyperref[csl:29]{29}}. Processing-in-memory and, more
broadly, neuromorphic (or brain-inspired) computing thus offers a
promising way to overcome the major performance and energy bottleneck in
modern memory systems. However, it also introduces significant
challenges for adoption as it is a disruptive technology that affects
all levels of the system stack, from hardware devices to software
algorithms. 
\vspace{0.5cm}

\subsection*{3.~Neuromorphic computing basics and its
evolution}

{\label{963240}}

\emph{Teresa Serrano-Gotarredona, Bernabe Linares Barranco}

\subsection*{3.1 What is neuromorphic computing/engineering}

{\label{244393}}

Neuromorphic computing can be defined as the underlying computations
performed by neuromorphic physical systems. Neuromorphic physical
systems carry out robust and efficient neural computation using hardware
implementations that operate in physical time. Typically, they are
event- or data-driven, and typically they employ low-power, massively
parallel hybrid analog, digital or mixed VLSI circuits, and they operate
using similar physics of computation used by the nervous system~.~

Spiking neural networks (SNN) are one very good example of a
neuromorphic computing system. Computation is performed whenever a spike
is transmitted and received by destination neurons. Computation can be
performed at the dendritic tree, while spikes travel to their
destinations, as well as at the destination neurons where they are
collected to update the internal states of the neurons. Neurons collect
pre-weighted and pre-filtered spikes coming from different source
neurons or sensors, perform some basic computation on them and generate
an output spike whenever their internal state reaches some threshold. A
neuron firing typically means the ``feature'' this neuron represents has
been identified in place and time. The collective computation of
populations of neurons can give rise to powerful system level behavior,
such as pattern recognition, decision making, sensory fusion, knowledge
abstraction, and so on. Additionally, neuromorphic computing systems can
also be enabled to acquire new knowledge through both supervised and
unsupervised learning, either off-line or while they perform, which is
typically known as on-line learning and which can be life-long.
Neuromorphic computing covers typically from sensing to processing to
learning.

\subsubsection*{3.1.1. Neuromorphic Sensing}

{\label{631188}}

Probably the most clarifying example of what neuromorphic computation is
about, is the paradigm of neuromorphic visual computation. Neuromorphic
visual computation exploits the data encoding provided by neuromorphic
visual sensors. Today, the most wide-spread neuromorphic vision sensor
is the Dynamic Vision Sensor (DVS) {[}\textsuperscript{\hyperref[csl:38]{38}}{]}. In a DVS
each pixel sends out its (x,y) coordinate whenever its photodiode
perceives a relative change of light beyond some pre-set
thresholds~\(\theta^->\frac{I_{n+1}}{I_n}>\theta^{_+}\), with~\(\theta^+\)slightly greater
than 1 and~\(\theta^-\)slightly less than 1. This is typically
referred to as an ``Address Event''. If ~\(I_{n+1}>I_n\)then light
has increased. If~\(I_{n+1}<\ I_n\)then light has decreased. To
differentiate both situations, the address event can also be a signed
event, by adding a sign bit `s', (x,y,s). If events are recorded using
some event-recording hardware, then a
timestamp~\emph{t}\textsubscript{\emph{n}}~is added to each event
(\emph{x}\textsubscript{\emph{n}}\emph{,y}\textsubscript{\emph{n}}\emph{,s}\textsubscript{\emph{n}}\emph{,t}\textsubscript{\emph{n}}).
The full recording consists then of a list of time-stamped address
events.~Fig. 2 illustrates this. In~Fig. 2(a) a DVS camera is observing
a 7KHz spiral on a classic phosphor oscilloscope (without any extra
illumination source).~Fig. 2(b) plots in \{x,y,t\} space the recorded
events. The camera was a 128x128 pixel high-contrast sensitivity DVS
camera~{[}\textsuperscript{\hyperref[csl:39]{39}}{]}. Therefore, x-y coordinates in~~Fig.
2(b) spawn from 0 to 127. The vertical axis is time, which spawns over
about 400us, slightly less than 100us per spiral turn. Each dot in~Fig.
2(b) is an address event, and we can count several hundreds of them
within the 400us. This DVS camera is capable of generating over 10
million events per second (about one every 100ns). This produces a very
fine timing resolution when sensing dynamic scenes.

The information (events) produced by this type of sensors can be sent
directly to event-driven neuromorphic computing hardware, which would
process this quasi-instantaneous dynamic visual information event by
event.

\par\null\selectlanguage{english}
\begin{figure}[H]
\begin{center}
\includegraphics[width=0.80\columnwidth]{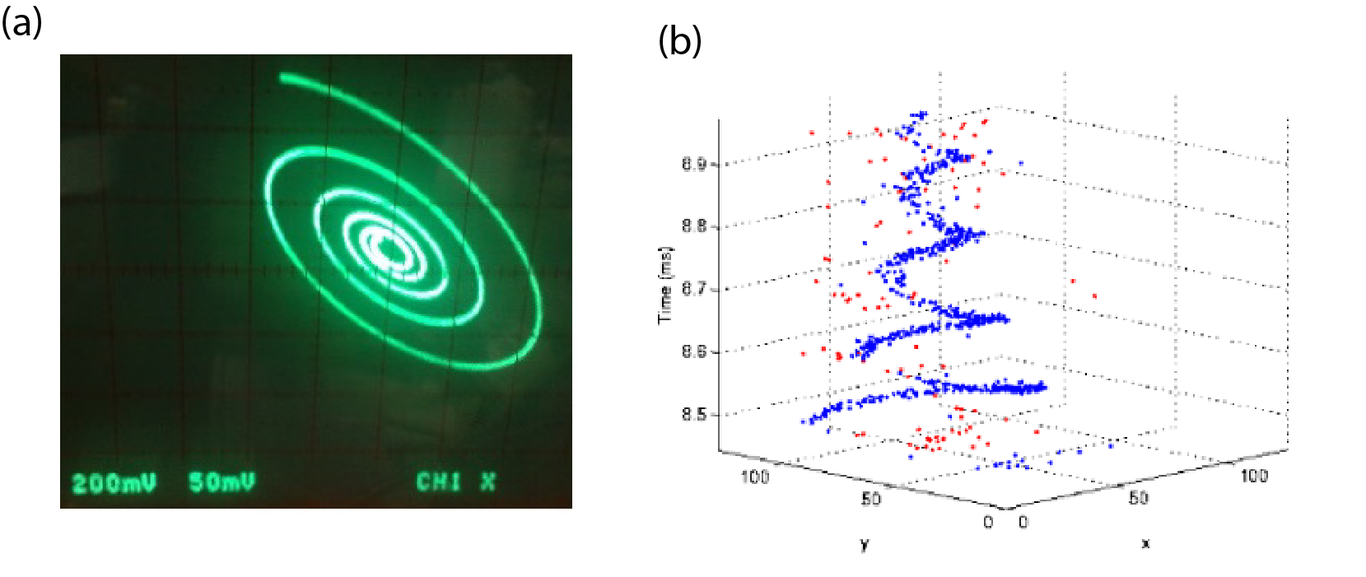}
\caption{{(a) 7KHz spiral observed in a classic phosphor oscilloscope set in X/Y
mode. (b) DVS output event stream when observing the oscilloscope in
(a).
{\label{273121}}%
}}
\end{center}
\end{figure}

DVS cameras have evolved over the past 20 years, since they first
appeared~{[}\textsuperscript{\hyperref[csl:40]{40}}{]}, combined with frames, sensitive to
color~{[}\textsuperscript{\hyperref[csl:41]{41}}{]}, and of resolutions up to
1-Mega-pixel~{[}\textsuperscript{\hyperref[csl:42]{42}}{]}.

Other sensory modality event-driven neuromorphic devices have been
reported, such as auditory cochleae~{[}\textsuperscript{\hyperref[csl:43]{43}}{]}, olfactory
noses~{[}\textsuperscript{\hyperref[csl:44]{44}}{]}, or tactile
sensing~{[}\textsuperscript{\hyperref[csl:45]{45}}{]}.

\subsubsection*{3.1.2. Neuromorphic Processing}

{\label{829241}}

Neuromorphic signal information encoding in the form of sequences of
events reduces information so that only meaningful data, such as
changes, are transmitted and processed. This follows the underlying
principle in biological nervous systems, as information transmission (in
the form of nervous spikes) and their consequent processing affects
energy consumption. Thus, biological systems tend to minimize the number
of spikes (events) to be transmitted and processed for a given
computational task. This principle is what neuromorphic computing
intends to pursue.~Fig. 3 shows an illustrative example of this
efficient frame-free event-driven information
encoding~{[}\textsuperscript{\hyperref[csl:46]{46}}{]}. In Fig. 3(a) we see a poker card deck
being browsed at natural speed, recorded with a DVS, and played back at
real-time seed with a reconstructed frame time of about 20ms. In Fig.
3(b) the same recorded list of events is played back at 77us frame time.
In Fig. 3(c) we show the tracked symbol input fed to a spiking
convolutional neural network for object recognition, displaying the
recognized output symbol. In Fig. 3(d) we show the 4-layer spiking
convent structure, and in Fig. 3~(e) we show the \{x,y, time\}
representation of 20ms input and output events occurring during a change
of card so that the recognition switches from one symbol to the next in
less than 2ms. Note that here the system is composed of both, the sensor
and the network executing the recognition. Both working together need
less than 2ms. This contrasts dramatically with conventional artificial
systems, in which the sensor first needs to acquire two consecutive
images (typically 25ms per image) and then process both to capture the
change.

\par\null\selectlanguage{english}
\begin{figure}[H]
\begin{center}
\includegraphics[width=0.80\columnwidth]{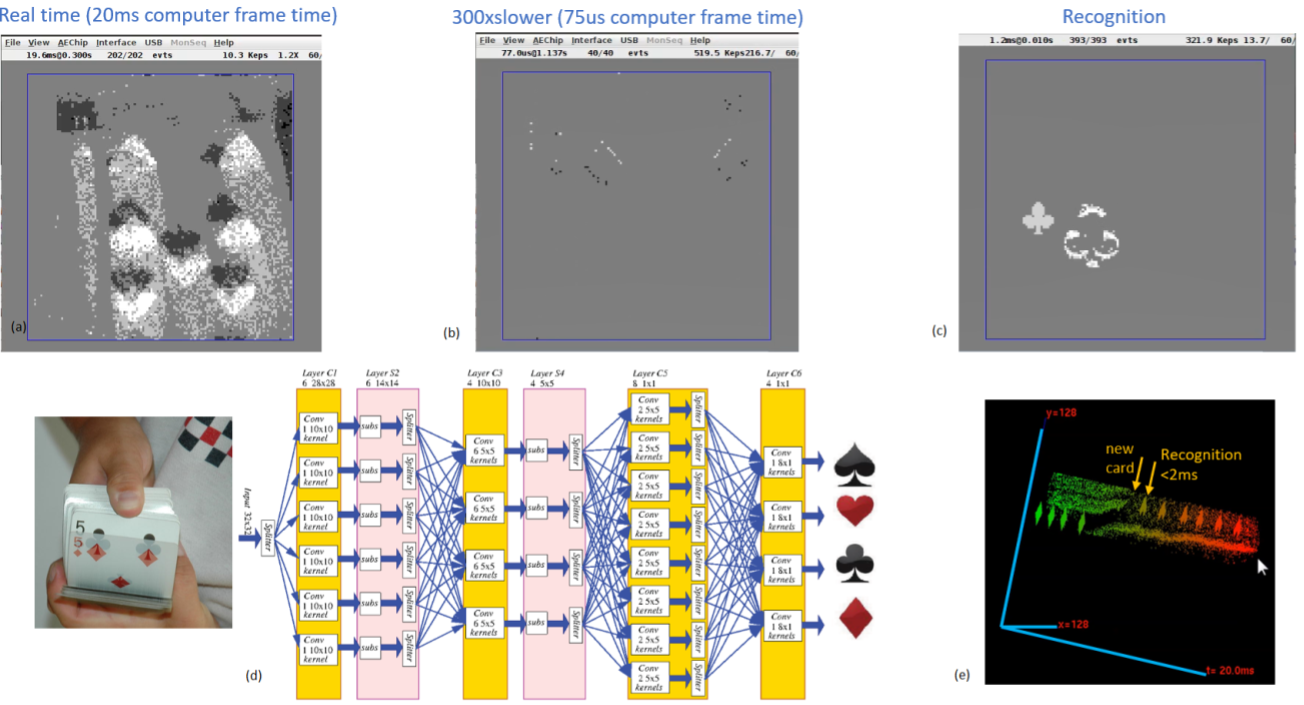}
\caption{{{(a) Fast speed poker deck browsing: events are collected about every
20ms to build a frame to display on a computer screen. (b) Slow speed
playback at 77us per reconstructed frame. (c) Poker symbol tracked and
displayed on the right, and recognition output on the left. (d)
Event-driven CNN to classify four poker symbols. (e) \{x,y,t\} display
during 20ms showing camera events together with recognition events
during a change of card with a recognition of less than 2ms.}
{\label{415251}}%
}}
\end{center}
\end{figure}

Fig. 3 illustrates a simple version of a neuromorphic sensing and
processing system. By today, much larger neuromorphic systems, inspired
in the same information encoding scheme, have been developed and
demonstrated. Some powerful example systems are:

-~~~~~~~~The SpiNNaker Platform {[}\textsuperscript{\hyperref[csl:47]{47}}{]}, developed
partly within the human Brain Project~{[}\textsuperscript{\hyperref[csl:48]{48}}{]}, is based
on an 18 ARM core SpiNNaker chip, 48 of which are assembled into a node
PCB, and about 1200 of which are assembled in a set of furnitures each
with racks, hosting all together about 1 million ARM cores. This system
is capable of emulating 1 billion neurons in real time. An updated
SpiNNaker chip has already been developed performing about 10x in
efficiency, neuron emulation capability, and event traffic handling,
while keeping similar power consumption.

-~~~~~~~~The BrainScales Platform~{[}\textsuperscript{\hyperref[csl:49]{49}}{]}, also
developed during the the Human Brain Project~{[}\textsuperscript{\hyperref[csl:47]{47}}{]},
implements physical silicon neurons fabricated on full silicon 8-inch
wafers, and interconnecting 20 of these wafers in a cabinet, together
with 48 FPGA based communication modules. It implements accelerated time
computations with respect to real time (about 10,000x), with
Spike-Timing-Dependent plastic synapses. Each wafer can host about 200k
neurons and 44 million synapses.

-~~~~~~~~The IBM TrueNorth chip~{[}\textsuperscript{\hyperref[csl:50]{50}}{]}~could host 1
million very simple neurons, or be reconfigured to trade-off number of
neurons versus neuron model complexity. They were structured into 4096
identical cores, consuming about 63 mW each.

-~~~~~~~~Loihi from Intel is probably by today the most advanced
neuromorphic chip. In its first version~{[}\textsuperscript{\hyperref[csl:51]{51}}{]},
fabricated in 14nm, it contains 128 cores, each capable of implementing
1k spiking neuronal units (compartments), and including plastic
synapses. More recently, Loihi 2 chip was introduced, with up to 1
million neurons per chip, manufactured in Intel 4 technology (7nm). Up
to 768 of Loihi chips have been assembled into the Pohoiki Springs
system, while operating at less than 500 watts~{[}\textsuperscript{\hyperref[csl:52]{52}}{]}.

\subsubsection*{3.1.3. Challenges and Conclusion}

{\label{858515}}

Neuromorphic computing algorithms should be optimum when run on
neuromorphic hardware, where events travel and are processing in a fully
parallel manner. One of the main challenges in present day neuromorphic
computing is to train and execute powerful computing systems directly on
neuromorphic hardware. Traditionally, neuromorphic computing problems
were mapped to more traditional deep neural networks to obtain their
parameters through back-propagation based
training~{[}\textsuperscript{\hyperref[csl:53]{53}}{]}, which would then be mapped to their
neuromorphic/spiking counterpart\textsuperscript{\hyperref[csl:46]{46}}{]}. However, these
transformations always resulted in a loss of performance. By today there
are many proposals of training directly in the spiking domain, combining
variants of Spike-Timing-Dependent plasticity rules, with surrogate
training techniques that adapt backpropagation to spiking systems, and
tested on either on fully connected or convolution based deep spiking
neural networks. For an updated review readers are referred
to~{[}\textsuperscript{\hyperref[csl:54]{54}}{]}.

On the other hand, it remains to see whether novel nanomaterial devices,
such as memristors, can provide truly giga-scale compact chips with
billions of neurons on a single chip and self-learning algorithms. Some
initial demonstrations of single~{[}\textsuperscript{\hyperref[csl:55]{55}}{]}~or multi-core
systems~{[}\textsuperscript{\hyperref[csl:56]{56}}{]}~exploiting a nano-scale memristor
combined with a selector transistor as synaptic element have been
reported, with highly promising outlooks once synapse elements could be
provided as pure nanoscale devices while stacking multiple layers of
synapse fabrics together with other nano-scale
neurons~{[}\textsuperscript{\hyperref[csl:57]{57}}{]}. In the end, the success of
neuromorphic computing will rely on the optimum combined progress in
neuromorphic hardware, most probably exploiting emerging nano-scale
devices massively, in an event- and data-driven information and
energy-efficient processing methodologies, and finally in providing
efficient, resilient, and quick learning methodologies for mapping
real-world applications into the available hardware and computational
neuromorphic substrates.

\vspace{0.5cm}

\subsection*{3.2 Different neuromorphic technologies and state of the
art}

{\label{595902}}

\begin{quote}
\emph{Sabina Spiga}
\end{quote}

\subsubsection*{3.2.1. Status}

{\label{813536}}

The research field of neuromorphic computing has been growing
significantly over the last three decades, following the pioneering
research at Caltech (USA) by Carved Mead \&
co-workers\textsuperscript{\hyperref[csl:5]{5}}, and it is currently attracting the
interest of a wide and interdisciplinary community from device, circuits
and systems to neuroscience, biology, computer science, materials and
physics. Within this framework, the developed
neuromorphic~\emph{hardware technologies} span from fully CMOS-based
systems\textsuperscript{\hyperref[csl:58]{58}}\textsuperscript{\hyperref[csl:59]{59}} to solutions exploiting the
use of charge-based or resistive non-volatile memory
technologies\textsuperscript{\hyperref[csl:60]{60}}\textsuperscript{\hyperref[csl:61]{61}}\textsuperscript{\hyperref[csl:62]{62}}, and
to emerging memristive device concepts and novel
materials\textsuperscript{\hyperref[csl:63]{63}}\textsuperscript{\hyperref[csl:64]{64}}\textsuperscript{\hyperref[csl:65]{65}}\textsuperscript{\hyperref[csl:66]{66}}.
Fig. 4 reports a schematic (and non-exhaustive) evolution of the main
technologies of interest. A common feature of these approaches is to
take inspiration from the brain computation, by co-locating memory and
processing (in-memory computing-IMC approach), and then overcoming the
von-Neumann architecture. Hardware artificial neural networks (ANN) can
implement IMC computing and provides an efficient physical substrate for
machine learning algorithms and artificial intelligence (AI). On the
other side, spiking neural networks (SNN), encoding and processing
~information using spikes, hold great promise for applications requiring
always-on real-time processing of sensory signals, for example in edge
computing, personalized medicine and Internet of Things.~\selectlanguage{english}
\begin{figure}[H]
\begin{center}
\includegraphics[width=0.98\columnwidth]{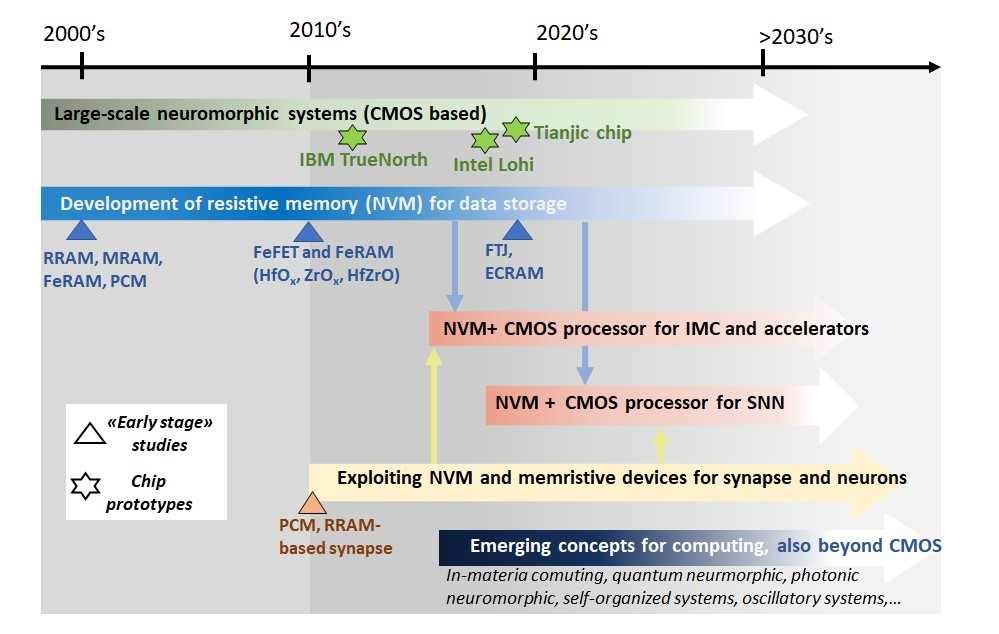}
\caption{{Schematic~ evolution of the main hardware technologies of interest for
neuromorphic computing (the indicates decades represent only a time
frame). Triangular symbols mark the refernce period for early stage
studies or starting interest in the technology development. From bottom
to top of the figure, the listed technologies are today at higher
maturity level and more advanced at system integration level.~
{\label{622193}}%
}}
\end{center}
\end{figure}

In terms of the maturity of neuromorphic technologies, we can discuss
three main bocks. ~

(i) \emph{} Current \emph{large-scale hardware neuromorphic computing
systems} are fully CMOS-based and exploit digital or
analogue/mixed-signal technologies. Examples of fabricated chips are the
IBM TrueNorth, Intel Loihi, Tianjic, ODIN, and others as discussed in
these review papers \textsuperscript{\hyperref[csl:58]{58}}\textsuperscript{\hyperref[csl:59]{59}}. In these
systems, the neuron and synapse functionalities are emulated by using
circuit blocks based on CMOS transistors, capacitors and volatile SRAM
memory. The scientific community is now exploiting these chips to
implement novel algorithms for AI applications.

(ii\emph{) Non-volatile memory technologies}. In the last decade,
resistive non-volatile memory (NVM) technologies, such as ~Resistive
Random Access Memory (RRAM), Phase change memory (PCM), Ferroelectric
memory (FeRAM) and ferroelectric transistor (FeFET), and
magnetoresistive random access memories (MRAM), have been proposed as
possible compact, low power and dynamical elements to implement in
hardware the synaptic nodes, replacing SRAMs, or as key element of
neuronal block\textsuperscript{\hyperref[csl:67]{67},\hyperref[csl:61]{61},\hyperref[csl:68]{68}}. While these NVMs have been developed
over the last twenty years mainly for data storage applications, and
introduced in the market, they can be considered emerging technologies
in the field of neuromorphic computing and their great potential is
still not fully exploited. Over the last 10 years, novel concepts for
computing, based on hybrid CMOS/non-volatile resistive memory circuits
and chips\textsuperscript{\hyperref[csl:56]{56}},~ have been proposed in the literature. In
parallel, also more conventional charge-based non-volatile memories such
as FLASH and NRAM are currently being investigated for IMC since they
are mature technologies. Finally, it is worth mentioning the emerging
memory technologies that are attracting increasing interest in the field
of IMC and neuromorphic computing, namely the ferroelectric tunnel
junction (FTJ)\textsuperscript{\hyperref[csl:69]{69}} and the 3-terminal electrochemical
random access memory (ECRAM)\textsuperscript{\hyperref[csl:70]{70}}.

ii)~\emph{Advanced memristive materials, devices and novel computation
concepts~}that are currently investigated include 2D materials, organic
material, perovskite, nanotubes, self-assembled nano-objects and
nanowire networks, advanced device concepts in the field of spintronics
(domain wall, race-trace memory, skyrmions), devices based on
metal-insulator transition (for instance VO\textsubscript{2}-based
devices), and volatile
memristors.\textsuperscript{\hyperref[csl:66]{66}}\textsuperscript{\hyperref[csl:71]{71}}\textsuperscript{\hyperref[csl:65]{65}}~\textsuperscript{\hyperref[csl:72]{72}}\textsuperscript{\hyperref[csl:73]{73}}.
These technologies are currently proof of concepts at a single device
level and circuit blocks connecting a reduced number of devices. The
computing system is sometimes demonstrated with a mixed
hardware/software approach, where the measured device characteristics
are used to simulate large systems. Finally, it is worth mentioning the
increasing interest in architectures that can exploit photonics
components for computing, towards the building of neuromorphic photonics
processors taking advantage of the silicon photonic platforms and
co-integration with novel optical memory devices and advanced materials
such as phase-change materials~~\textsuperscript{\hyperref[csl:74]{74}}\textsuperscript{\hyperref[csl:75]{75}}.~

Fig.5 shows schematically examples of the material systems currently
most investigated in various approaches and technologies for
neuromorphic computing.\selectlanguage{english}
\begin{figure}[H]
\begin{center}
\includegraphics[width=0.98\columnwidth]{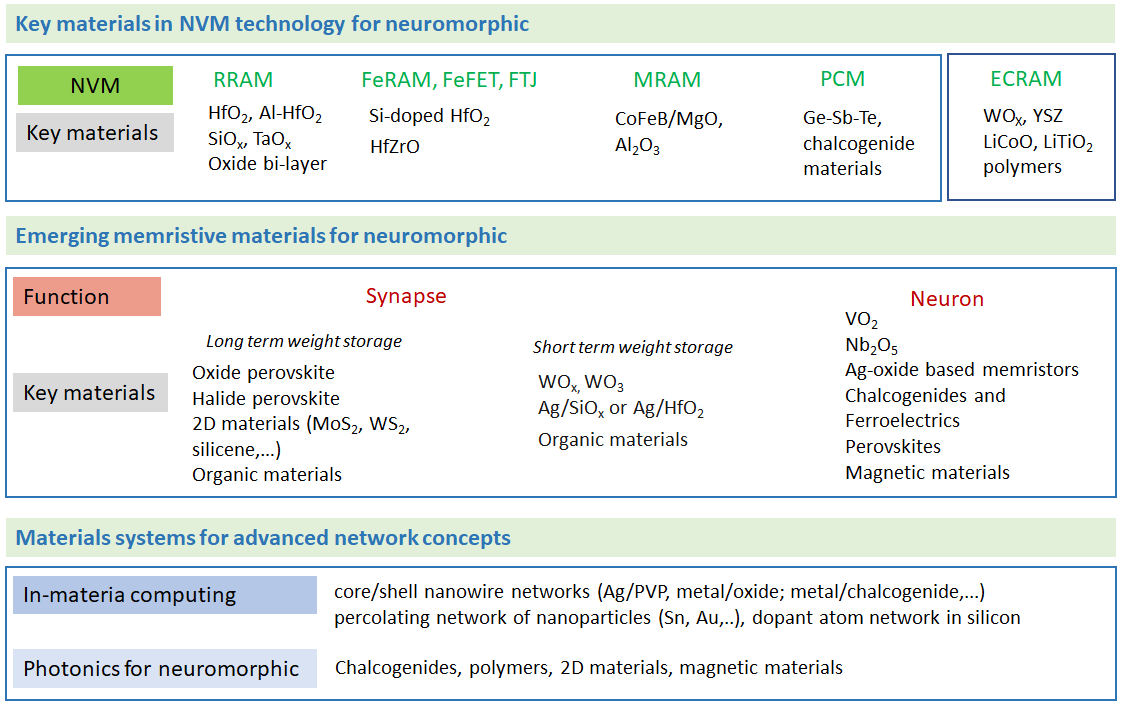}
\caption{{Examples of materials systems currently employed in memristive
technologies. The list of materials is not exhastive andinclude only
some of the most used ones. For the NVM devices(top line), the main
active material is indicated, but each device includes also various
types of material electrodes depending on the technology.~
{\label{112066}}%
}}
\end{center}
\end{figure}

{\label{792140}}

\subsubsection*{3.2.2. Challenges}

{\label{792140}}

The current and future challenges can be considered at various levels.

(i) For~\emph{large-scale neuromorphic processors}, the progress of
CMOS-based technologies and their scaling still provide room to advance
the research field. The main challenges are at the architecture and
algorithms level. On the other side, most~\emph{NVM memories} (RRAM,
PCM, FeRAM, FeFET and MRAM) have been already integrated with CMOS at
scaled technological nodes and large integration density, and hold
interesting properties (depending on the specific technology) such as
small size, scalability, possible easy integration also in 3D array
stacking, low programming energy, multilevel programming capability.~~
Therefore, it is expected that NVM technologies will play an increasing
role in future IMC chips or neuromorphic processors, by enabling
energy-efficient computation. Prototype IMC chips have been reported in
the literature\textsuperscript{\hyperref[csl:56]{56}}\textsuperscript{\hyperref[csl:76]{76}}, as well as
innovative circuits for SNN implementing advanced learning rules to
compute with dynamics\textsuperscript{\hyperref[csl:77]{77}}\textsuperscript{\hyperref[csl:78]{78}}.

On the other side, It is worth mentioning that the NVM technologies
exhibit several device-level non-idealities., as discussed in more
detail in the following sections of this roadmap. As examples and
non-exhaustive list, we can mention: nonlinearity and stochasticity in
conductance update vs a number of pulses at a fixed voltage (PCM, RRAM,
FeRAM), asymmetry (RRAM) in the bidirectional tuning of conductance,
conductance drift (PCM) or broadening of the resistance distribution
(RRAM) after programming, device-to-device and cycle-to-cycle
variability of the programmed states, low resolution due to the limited
number of programmable levels (up to 8 or 16 are demonstrated for RRAM
and PCM at array level), restricted memory window (MRAM) or limited
endurance (general issue except than for MRAM), relative high conduction
also in the OFF state. ~~All these aspects can impact the neural network
accuracy and reliability, although proper algorithms/architecture can
take advantage from stochasticity or asymmetry of conductance
tuning~\textsuperscript{\hyperref[csl:79]{79}}. Therefore, a careful co-design of hardware
and algorithms is required together with an improvement of circuit
design and/or programming device strategies to fully exploit NVMs in
combination with CMOS and in large systems. Further discussion on
specific challenges and~ possible specific applications of the listed
technologies will be discussed in the sections 5.1 - 5.4 of this
roadmap, while a more deep view on application scenario is reported in
section 7.

ii) Regarding the plethora of~\emph{emerging materials/devices and novel
concepts} proposed for neuromorphic computing (beyond the ones discussed
in the previous point, see some examples in section 5.4-5.6 of the
roadmap), the main challenge is that they are mostly demonstrated at the
single device level or in early stage proof of concepts in small
array/large device size, then extending their implementation in ANN or
SNN only at simulation level. To leaverage these concepts at higher TRL,
it is necessary to prove that the device characteristics are
reproducible and scalable, the working principle well understood, and to
provide more advanced characterizations on several down-scaled devices,
and fianlly to close the current gap between laboratory exploration of
single material/devices and integration in array or circuits. Another
challenge is to address more into details how to exploit the nanodevcies
peculiarities, such as dynamic or stochastic behavior, to implement in
hardware more complex bio-inspired functionalities or even to perform
radically new computation paradigms. Indeed, while the more standard
technologies (CMOS, Flash,SRAM) can also be used in hardware neural
network to implement complex functions,~ this is possible only at the
high cost of increased circuit complexity. To give an example, the
required dynamic to reproduce the synaptic or neuronal fucntionality in
SNN is implemented at circuit level and/or using large area capacitors
which are not easy scalable in view of large systems. One possible
approach is to exploit the emerging memristve technologies and their
properties (variability, stochasticity, non-idealities) to implement
complex functions with more compact and low-power devices. One example
is the use of resistivity drift in PCM (usually an unwanted
characteristics for IMC or storage applications) to implement advanced
learning rules in hardware SNN\textsuperscript{\hyperref[csl:80]{80}}. Another example
(discussed in section 7.1) is to use the inherent variability and
stochasticity of some nanodevices to build efficient random number
generators ( for data security applications) and stochastic computing
models.~Overall, this scenario point out a long-term development
research, likely up to ten years or more, to close the gap between these
novel concepts and real industrial applications.~

\subsubsection*{3.2.3. Potential Solutions}

{\label{539217}}

To pursue advances in the development of neuromorphic hardware chips, it
is necessary to develop a common framework to compare and benchmark
different approaches, also in view of some metrics such as computing
density, energy efficiency, computing accuracy, learning algorithms,
theoretical framework as well as target possible killer applications
that might significantly benefit from neuro-inspired chips. Within this
framework, materials strategies can be still relevant to address some of
the outlined challenges for NVMs, but materials need to be co-developed
together with a demonstration of a device at the scaled node and array
level. An important strategy for future is also the possibility to
substitute current mainstream materials with green materials or to
identify fabrication processes more sustainable in term of cost and
environmental impact, without compromising the hardware functionality.
Moreover, other important aspects include the development of hardware
architecture that can lead to the integration of several devices, and
exploiting a large connectivity among them; the implementation of
efficient algorithms supporting online learning, also on different time
scales as in biological systems; and to address the low power analysis
of large amount of data also for internet of things and edge devices.
Overall, it is necessary and holistic view which includes the
materials/device/architecture/algorithms co-design to develop
large-scaled neuromorphic chip.

\subsubsection*{3.2.4 Concluding remarks}

{\label{542359}}

The development of advanced neuromorphic hardware than can support
efficiently AI applications is becoming more and more important. Despite
the several prototypes and results presented in literature,
neuro-inspired chips are still only an early stage of development and
there is plenty of room for further development. Many mature NVM devices
are definitely candidates to became a future mainstream technology for
large scale neuromorphic processor that can outperform the current
platform based only on CMOS circuits. In the long term, it is also
necessary to close the gap between emerging materials and concepts,
currently demonstrated only at proofs of concepts, and their possible
integration in functional systems.~ Materials research and understading
of physical principles enabling novel functioalities are an important
parts fo this scenario.

\vspace{0.5cm}

\subsection*{3.3. Possible future computational primitives for
neuromorphic
computing}

{\label{383742}}

\begin{quote}
\emph{Sergey Savel'ev, Alexander Balanov}
\end{quote}

The core idea of neuromorphic computing to develop and design
computational systems mimicking electrochemical activities in brain
cortex is currently booming, embracing areas of deep physical neural
networks~\textsuperscript{\hyperref[csl:81]{81}}, classical and quantum reservoir
computing~\textsuperscript{\hyperref[csl:82]{82}},\textsuperscript{\hyperref[csl:83]{83}}, oscillator-based
computing~~\textsuperscript{\hyperref[csl:84]{84}}, and spiking networks~\textsuperscript{\hyperref[csl:85]{85}}~
among many other concepts~\textsuperscript{\hyperref[csl:86]{86}}. These computational
paradigms imply new ways for information processing and storage
different from conventional computing and therefore require elementary
base and primitives which often involve unusual novel physical
principles~\textsuperscript{\hyperref[csl:87]{87}}.

Presently, memristors - electronic switchers with memory - and their
circuits demonstrate great potential for application in the primitives
for future neuromorphic computing systems. In particular, different
types of volatile and non-volatile memristors can serve as artificial
neurons and synapses, respectively, which facilitate the transfer,
storage, and processing of information~\textsuperscript{\hyperref[csl:88]{88}}. For example,
volatile Mott memristors~~\textsuperscript{\hyperref[csl:89]{89}} can work as an electric
oscillator with either regular or chaotic dynamics~\textsuperscript{\hyperref[csl:7]{7}},
while memristors with filament-formation~\textsuperscript{\hyperref[csl:90]{90}} demonstrate
tunable stochasticity~\textsuperscript{\hyperref[csl:91]{91}},~\textsuperscript{\hyperref[csl:92]{92}} allows
designing neuromorphic circuits with different degree of plasticity,
chaoticity, and stochasticity to address diverse computational aims in
mimicking dynamics of different neuron populations. Furthermore, a
crossbar of non-volatile memristors (servicing to memorise training)
attached to volatile memristors (working as readouts) enables the design
of AI hardware with unsupervised learning
capability~\textsuperscript{\hyperref[csl:93]{93}}. Thus, combining memristive circuits with
different functionalities paves the way to building a wide range of
in-memory computational blocks for a broad spectrum of artificial neural
networks (ANNs) starting from deep learning accelerators to spiking
neuron networks~\textsuperscript{\hyperref[csl:94]{94}}.

A rapidly developing class of volatile memristive
elements~\textsuperscript{\hyperref[csl:95]{95}} has been shown to demonstrate a rich
spectrum of versatile dynamical
patterns~\textsuperscript{\hyperref[csl:96]{96}},\textsuperscript{\hyperref[csl:7]{7}}~\textsuperscript{\hyperref[csl:97]{97}},
which makes them suitable for the realisation of a range of
neuroscience-motivated AI
concepts~\textsuperscript{\hyperref[csl:98]{98}},~\textsuperscript{\hyperref[csl:99]{99}},~\textsuperscript{\hyperref[csl:100]{100}}.
For instance, the ANNs based on volatile memristors can go well beyond
usual oscillator-based computing~\textsuperscript{\hyperref[csl:84]{84}} or spiking neural
networks~\textsuperscript{\hyperref[csl:85]{85}}. They rely on manipulating information by
utilising complexity in dynamical regimes that offer a novel
computational framework~\textsuperscript{\hyperref[csl:98]{98}},~\textsuperscript{\hyperref[csl:99]{99}} with
cognitive abilities closer to biological brains. There is a specific
emphasis on using dynamical behaviours of memristors, instead of only
static behaviours.~\textsuperscript{\hyperref[csl:101]{101}}

Remarkably memristive elements can be realised not only in electronic
devices but also within spintronics or photonics frameworks, which have
their own advantages compared to electronics. Therefore, hybridised
design promises great benefits in the further development of
neuromorphic primitives. For example, a combination of memristive
chipsets with spintronic or/and photonic components can potentially
create AI hardware with enhanced parallelism offered by optical devices
operating simultaneously at many frequencies (e.g., optical cavity
eigenfrequencies)~\textsuperscript{\hyperref[csl:102]{102}}, energy-efficient magnetic
non-volatile memory~and flexible memristive spiking network
architecture. An important step in the realisation of this approach is
the development of interface technologies for bringing electronic,
photonic and spintronic technologies together. A possible example is the
spintronic
memristor~\textsuperscript{\hyperref[csl:103]{103}},~\textsuperscript{\hyperref[csl:104]{104}},~\textsuperscript{\hyperref[csl:65]{65}},
where the transformation of magnetic structure influences the resistance
of the system. Control of resistance through magnetic structures by
either current or voltage and vice versa offers a possibility to combine
{the} same chipset spintronic both components~\textsuperscript{\hyperref[csl:105]{105}}
affecting the magnetic structure of~the spintronic memristor and
memristive components influencing current through the spintronic
memristor. Therefore, this promotes crosstalk between the electric and
magnetic subsystems of the same device. An interface between
neuromorphic optical and electronic sub-systems of a hybrid device could
be realised using optically controlled electronic memristive
systems~\textsuperscript{\hyperref[csl:106]{106}}, thus, paving a path towards neuromorphic
opto-electronic systems~\textsuperscript{\hyperref[csl:107]{107}}.~

The conventional ANNs with a large number of connections require
training to is is less efficient in the task requiring frequent
retraining for `'moving target'' problems, for example in recognition of
characteristics changing in time. A potential solution for such tasks is
to implement filtering or pre-processing data by a
``reservoir''~\textsuperscript{\hyperref[csl:82]{82}}, usually consisting of neuron units
connected by fixed weights. The reservoir is assisted by a small readout
ANN, which requires much less data for training thus removing
significant retraining burden. Recently, an important evolution has
taken place in the development reservoir computing systems, where the
function of the reservoir is realised by photon, phonon or/and magnon
mode mixing in spintronic~\textsuperscript{\hyperref[csl:108]{108}},{ }\textsuperscript{\hyperref[csl:109]{109}},
and photonics~\textsuperscript{\hyperref[csl:110]{110}} devices. Substitution of the
interaction of many artificial neurons by wave processes resembles
neural wave computation in the visual cortex~\textsuperscript{\hyperref[csl:99]{99}} and
promotes miniaturisation, robustness, and energy efficiency of the
reservoirs (neuromorphic accelerators) which in future could become an
additional class of primitive, especially in neuromorphic computational
systems dealing with temporal or sequential data
processing~\textsuperscript{\hyperref[csl:111]{111}}. In AI training, it has also been shown
that memristive matrix multiplication hardware can enable noisy local
learning algorithms, which perform training at the edge with significant
energy efficiencies compared to graphics processing
units.~\textsuperscript{\hyperref[csl:112]{112}}

Finally, we briefly outline another exciting perspective constituted by
a combination of quantum and neuromorphic
technologies~\textsuperscript{\hyperref[csl:113]{113}}. Currently, quantum
AI~\textsuperscript{\hyperref[csl:114]{114}} attracts significant attention by increasingly
competing with more traditional quantum computing. One of the most
promising quantum AI paradigms is quantum reservoir
computing,~\textsuperscript{\hyperref[csl:115]{115}} which offers not only much larger state
space than classical reservoir computing but also
essentially-nonclassical quantum feedback on the reservoir via
measurements. A quantum reservoir built from quantum
memristors~\textsuperscript{\hyperref[csl:116]{116}},~\textsuperscript{\hyperref[csl:117]{117}}, could significantly
gain quantum AI efficiency as it can readily be integrated with existing
quantum and classical AI devices and also lead to an `'exponential
growth''~\textsuperscript{\hyperref[csl:118]{118}} in the performance of ``reservoirs'' with
the possibility of relaxing requirements on decoherence comparing to
traditional quantum computing. ~~

The above trends and directions in the development of the primitives for
neuromorphic computing are obviously only a slice of exciting future AI
hardware technology. Even though we recognise that our choice is
subjective, we hope that the outlined systems should provide a flavour
of future computational hardware, which should be based on
reconfigurable life-mimicking devices utilising different physical
principles in combination with novel mathematical cognitive
paradigms~\textsuperscript{\hyperref[csl:119]{119}},~\textsuperscript{\hyperref[csl:120]{120}},~\textsuperscript{\hyperref[csl:121]{121}},~\textsuperscript{\hyperref[csl:122]{122}}.
\vspace{0.5cm}

\subsection*{4. Mature Technologies}

{\label{348857}}

\subsection*{4.1 SRAM}

{\label{348857}}

\begin{quote}
\emph{Nitin Chawla, Giuseppe Desoli}
\end{quote}

\subsubsection*{4.1.1 Status}

{\label{448876}}

\begin{figure}[H]
\begin{center}
\includegraphics[width=0.70\columnwidth]{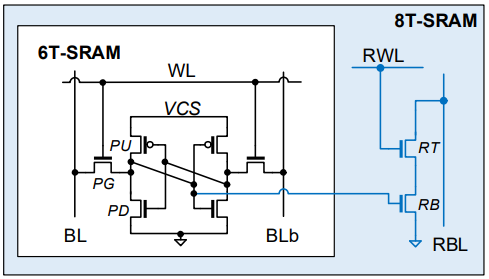}
\caption{{Standard SRAMs bitcells are usually designed with 6 or 8 transistors
{\label{911848}}%
}}
\end{center}
\end{figure}

SRAM-based computing in memory (CIM) or in-memory computing, is seen as
a mature and widely available technology for accelerating matrix and
vector calculations in deep learning applications, yet many technology
driven optimizations are still possible. To make CIM more compatible,
researchers have been exploring ways to improve the design of the
bit-cell (Fig.~{\ref{911848}}), which is the basic unit
of memory. This has led to the development of high-end SRAM chips with
large capacities, such as 107, 128 Mb, and 256 Mb SRAM chips at 10 nm, 7
nm, 5 nm, and 4
nm~\textsuperscript{\hyperref[csl:123]{123}}\textsuperscript{\hyperref[csl:124]{124}}\textsuperscript{\hyperref[csl:125]{125}}\textsuperscript{\hyperref[csl:126]{126}}.
These large SRAM capacities help to reduce the need for off-chip DRAM
access. However, in more cost-sensitive applications, such as embedded
systems and consumer products, modifying the bit-cell design can be too
costly and may limit the ability to easily transfer the technology to
different manufacturing nodes.~

A key difference exists between analog and digital SRAM CIM. Analog CIM
has been heavily studied using capacitive or resistive sharing
techniques to maximize row parallelism \textsuperscript{\hyperref[csl:127]{127}}, but this
comes at the cost of inaccuracies and loss of resolution due to
variations in devices across PVT and the limitations of SNR and dynamic
range in ADC/readout circuits. Impacts of device variations for
different kinds of devices:

\begin{itemize}
\tightlist
\item
  Resistive devices like PCM or RRAM experience a variation in the
  resistive values across the nominal behavior which can vary based on
  process and for a case of +/- 10-20\% change in resistance value there
  will be corresponding change of current values which are then input to
  the read out circuits and hence this will impact the quantization step
  of read out circuits hence impacting the SNR which will then need a
  higher dynamic range to compensate for the same. Temperature behavior
  for resistors also needs to be taken care in the noise margin.
\item
  MOS devices:~ These devices can vary in their performance (threshold
  voltage) due to:
\end{itemize}

\begin{quote}
\begin{quote}
\begin{enumerate}
\tightlist
\item
  Global lot positioning like slow, typical, fast which can again vary
  around +/- 20\% which can be less or more based on technology and
  voltage of operation. This is a deterministic shift.
\item
  Local variation: within the same lot there are device to device
  variations which are random in nature and need statistical analysis
  based on capacity in use to analyze the impact of variations. These
  impact the SNR and Quantization like in case of resistive devices and
  will need higher dynamic range to compensate for the loss in accuracy.
\end{enumerate}
\end{quote}
\end{quote}

Analog SRAM CIM solutions often use large logic bit cells and aggressive
reduction in ADC/readout bit width, resulting in low memory density and
computing inaccuracies, making it difficult to use in situations where
functional safety, low-cost testing, and system scalability are
required. On the other hand, digital CIM offers a fast path for the next
generation of Neural Processing systems due to its deterministic
behavior and compatibility with technology scaling rules.

Researchers have improved the SRAM-based CIM's performance by modifying
the SRAM bitcell structure and developing auxiliary peripheral circuits.
They proposed read-write isolation cells to prevent storage damage and
transposable cells to overcome storage arrangement limitations.
Peripheral circuits, such as DAC, redundant reference columns, and
multiplexed ADC, were proposed to convert between analog and digital
signals.~ The memory cell takes up most of the SRAM area in the core
module of a standard SRAM cut. However the complexity of the additional
operations performed in the memory unit, poses additional problems to
utilize the memory cells to their full potential. Researchers have
explored various trade-offs to implement the necessary computational
functionality while preserving density, power, and, last but not least,
minimizing the additional cost associated with bitcell modifications
required for requalification when deployed in standard design flows.~
Most system-on-chips (SoCs) use standard 6T structures due to their high
robustness and access speed and to minimize area overhead. The 6T
storage cell is made up of two PMOSs and four NMOSs to store data
stably. To perform CIM using the conventional 6T SRAM cell, operands are
represented by the WL voltage and storage node data, and processing
results are reflected by the voltage difference between BL and BLB.~\selectlanguage{english}

\begin{quote}
\end{quote}

{\label{258265}}

The figure above shows the conventional 6T and 8T bitcells which form
the basic building block of SRAM design. The 8T bitcell is made out of a
conventional 6T and read port which allows read and write in parallel.
These bitcells were never designed for parallel access across rows and
this poses one of the main challenges for enabling analog SRAM CIM.~

Dual-split 6T cell with double separation have been
proposed~\textsuperscript{\hyperref[csl:128]{128},\hyperref[csl:129]{129}}, allowing for more sophisticated functions
due to the separated WL and GND, which can use different voltages to
represent various types of information.~\textsuperscript{\hyperref[csl:130]{130}}~ proposed a
4+2T SRAM cell to decouple data from the read path. The read is akin to
that of the standard 6T SRAM, writes instead, use the N-well as the WWL
and two PMOS source as the WBL and WBLB. In computational mode,
different voltages on the WL and storage node encode the operands. ~

In general,~CIM adopting the 6T bitcell structure are unable to
efficiently perform computing operations and may not fully meet the
requirements of future CIM architectures. Hence, many studies on CIM
have modified the 6T structure because using the 6T standard cell
directly poses a reliability challenge as the contents of the bitcells
get effectively shorted if accessed in parallel on the same bitline.
This means special handling on the wordline voltage is required which
adds lot of complexity and limits the dynamic range. Further the
variability and linearity of devices become very difficult to control if
when limiting the device operation to reduced voltage levels due to
these reliability constraints, impacting the overall energy efficiency
of the solution~\textsuperscript{\hyperref[csl:131]{131},\hyperref[csl:132]{132}}.

For practical applications, and specifically for AI ones, it's important
to evaluate the end to end algoritmic accuracy vs. the key metrics, to
this end, recent
research~\textsuperscript{\hyperref[csl:133]{133}}\textsuperscript{\hyperref[csl:134]{134}}\textsuperscript{\hyperref[csl:135]{135}}\textsuperscript{\hyperref[csl:136]{136}}~\textsuperscript{\hyperref[csl:137]{137}}
has suggested various analytical models to examine the balance between
the costs (accuracy) and benefits (primarily energy efficiency and
performance) of digital versus analog SRAM CIM. This is based on the
idea that many machine and deep learning algorithms can tolerate some
degree of computational errors, and that there are methods such as
retraining and fine-tuning as well as hardware-aware training to address
these errors.

The implementation of Neural Processing Units incorporating CIM
components for large-scale deep neural networks (DNNs) presents
significant difficulties, CIM macro can incur substantial column current
magnitudes, which can result in power delivery difficulties and sensing
malfunctions. Furthermore, the utilization of analog domain operations
necessitates the incorporation of ADCs and DACs, which consume a
significant amount of area and energy resources. Further to this, the
pitch matching of ADC with SRAM~ bitcell also poses a big challenge for
array and ADC interface. It is clear that the realization of the full
potential of SRAM-CIM necessitates the development of innovative and
sophisticated techniques.

\subsubsection*{4.1.2 Challenges and Potential
Solutions}

{\label{596347}}

\begin{figure}[H]
\begin{center}
\includegraphics[width=0.56\columnwidth]{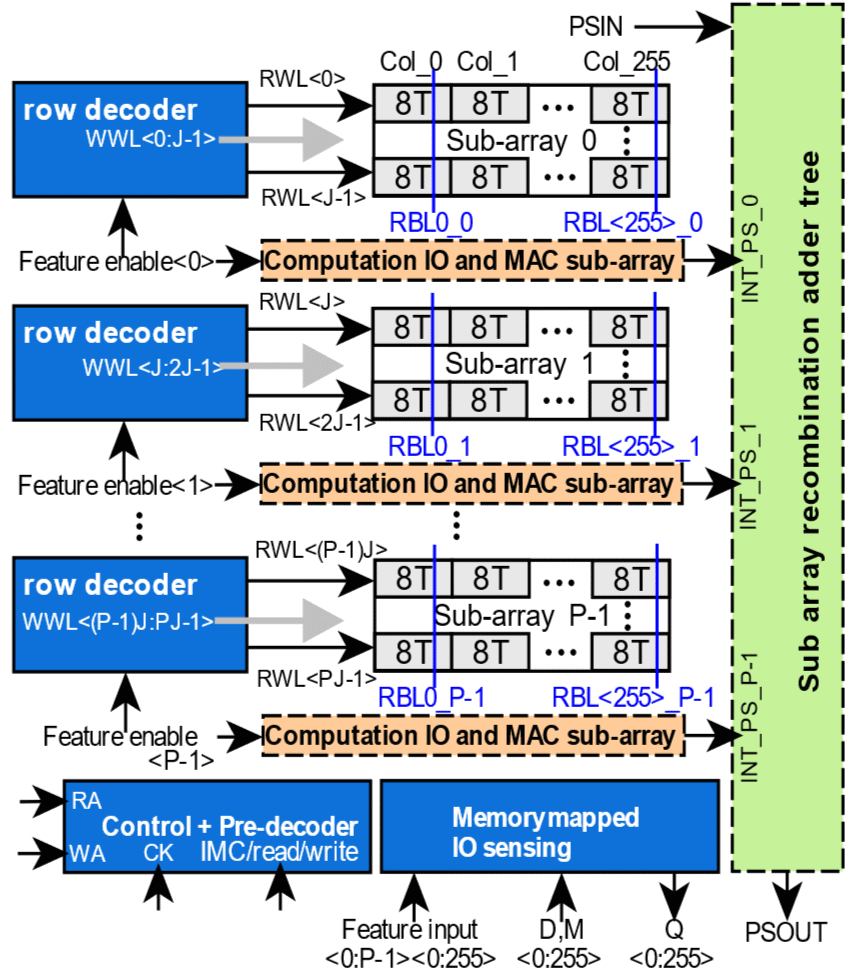}
\caption{{Digital CIM memory macro with 8T bitcells and embedded digital logic
{\label{448861}}%
}}
\end{center}
\end{figure}

In deep learning, convolutional kernels and other types of kernels rely
heavily on matrix/vector and matrix/matrix multiplications (MVM). These
operations are computationally expensive and involve dot product
operations between activation and kernel values. In-memory
multiplication in CIM macro devices can be classified into three primary
categories: current-based, charge-sharing-based for analog computation,
and one for all-digital. All-digital CIMs exhibit the same level of
precision as purely digital ASIC implementations. Various implementation
topologies ranging from bit-serial to all parallel arithmetic
implementations have been proposed for digital CIM solutions. Digital
CIMs like in research work~\textsuperscript{\hyperref[csl:138]{138}} represent a modified logic
bit-cell to support element-wise multiplication followed by a digital
accumulation tree sandwiched within the SRAM array. The solution
improves on energy efficiency by reducing data movement alongside the
efficiency benefits of a custom-built MAC pipeline with improved levels
of parallelism over traditional Digital NPU's for example as in
Fig.~{\ref{448861}} \textsuperscript{\hyperref[csl:139]{139}}. The Digital CIM
implementations also have a wide voltage and frequency dynamic range
allowing runtime reconfigurability between the competing TOPS/W and
TOPS/mm2 performance criteria. The operating range and mission profiles
of these architectures can also be extended by leveraging read-and-write
assist schemes as is commonly done for ultra-low voltage SRAM design.
The digital CIM solution's energy efficiency depends on the operand
precision and due to the deterministic precision and bit true
computation nature, it begins to decline as we increase the operand
precision. ~

\par\null\selectlanguage{english}

Current-based CIMs as represented in one of the early research
works~\textsuperscript{\hyperref[csl:140]{140}}, implements a WL DAC driving a multi-level
feature input with multiple rows active in parallel. The results of the
element-wise multiplications of all the parallel rows are accumulated as
current on the bit lines of the CIM macro that terminates in a
current-based readout/ADC The current accumulation on the bit line
essentially implements a reduction operation limited by the SNR of the
readout circuit. Current-based CIMs as presented in this work suffers
from significant degradation in accuracy due to bit cell variabilities
and nonlinearities of the WL DAC while the throughput is limited by the
readout circuits.~\textsuperscript{\hyperref[csl:141]{141}} implements a variation with a CIM
using a 6 T-derived bit cell with a PWM WL modulation and a focus on
storing and computing multibit weights per column. The modulation scheme
uses binary weighted pulse duration based on the index of the bit-cell
in the column effectively encoding the multi-bit weight in the column to
impact the value of the global bit line. The multiplication is
effectively done in the periphery of this CIM using a switched capacitor
circuit.Bit-cell variation and non-linearity like in the previous case
significantly limit the accuracy of this implementation thus restricting
the industrialization potential of these current-based CIM solutions.
The work in~\textsuperscript{\hyperref[csl:142]{142}} tries to address the limitations of the
above analog CIM techniques and implements a charge-based CIM by using a
modified SRAM logic bit-cell that performs an element-wise binary
multiplication(XNOR) and transfers the results to a small capacitor.
Multiple rows operating in parallel as key to the energy efficiency of
these CIM topologies. In this work, the element-wise multiplication
result is transferred as a charge to the global bit line followed by a
voltage-based readout. The inherent implementation benefits from the
fact that capacitors suffer less from process variability and present
fairly linear transfer characteristics. This architecture however like
other Analog CIM is impacted by dynamic range compression due to the
limited SNR regime of the readout at the end of the
column.~\textsuperscript{\hyperref[csl:143]{143}} extends this approach to support multi-bit
implementations using a bit-sliced architecture. The multi-bit weights
are mapped to different columns while the feature data is essentially
transferred as 1-bit serial data on parallel word lines and each column
performs a binary multiplication followed by accumulation on the
respective bit lines. The near-memory all-digital recombination unit in
this approach performs the shift and scale operations based on the
column index to recreate the results of the multi-bit MAC operation. The
approach is flexible to support asymmetric features and weight precision
and can be made reconfigurable to support different features and weight
precisions on the fly. This however still suffers from the same SNR
constraints as each column operation is compressing the dynamic range
and is limited by the peak dynamic range of the readout ADC. The ADC in
most of these schemes is mostly shared across multiple columns thus
making it a critical design component in determining the throughput of
such CIM architectures. The specific bit-sliced approach has impressive
TOPS/W numbers for the lower weights and activation precision regime but
starts to taper off due to the quadratic increase in the computation
energy with increasing weight and activation
precision.~\textsuperscript{\hyperref[csl:144]{144}} instantiates multiple of these CIM macros
to demonstrate a system-level approach connecting these CIM macros with
a flexible interconnect and adding digital SIMD and scalar arithmetic
units to support real-world Neural Network execution. This specific work
due to the limited readout speed of the CIM macros and the overhead of
the other digital units suffers from a moderate TOPS/mm2 number for the
full solution but presents impressive TOPS/W numbers, especially at the
lower precision regime. The research work~\textsuperscript{\hyperref[csl:145]{145}} represents
another effort with a system-level solution of a hybrid NPU comprising
analog CIM units and traditional digital accelerator blocks. The work
leverages a low-precision (2-bit) Analog CIM macro coupled with a
traditional 8-bit digital MAC accelerator. The two orders of magnitude
difference in energy efficiency between the 8-bit digital MAC engine and
the 2-bit Analog CIM macro can be leveraged by mapping different layers
to the appropriate computation engine but needs careful articulation of
mapping algorithms with the precision constraints of the Analog CIM's
while keeping the overhead of the write refresh and other digital
vector/scalar operators low. This to some expect is a tradeoff between a
very specialized use case as opposed to a general-purpose NN
accelerator.

\subsubsection*{4.1.3 Conclusions}

{\label{617326}}

Analog CIM solutions based on charge-based CIMs display a lower degree
of variability when compared to current-based CIMs, due to variability
in the technologies employed for capacitors and threshold voltage
effects. Additionally, charge-based CIMs are able to activate a greater
number of word lines per cycle and thus achieving higher amounts of row
parallelism. However, both current-based and charge-based CIMs are
limited in terms of accuracy and the equivalent bit precision of the
dynamic range of the accumulation. Selecting an appropriate ADC
bit-precision and MVM parallelism is a challenging task that requires
balancing accuracy and power consumption. Measurements and empirical
evidence suggest that an increase in the accumulation values is
correlated with a higher degree of variability. However, it is important
to note that such high values are relatively rare in practical neural
network models, as shown by the statistical distribution of activation
data and resultant accumulation outcomes. This characteristic along with
noise-aware training can be leveraged to optimize the precision and
throughput demands of the analog-to-digital converter, thereby improving
the FOM of these Analog CIM techniques. The research on noise-aware
training in the state of the art is limited to academic works on
relatively small neural networks and data sets. This for an industrial
deployment still needs to mature and demonstrate the scalability to
larger models and data sets.~~

~All-digital CIMs provide a deterministic and scalable path to intercept
the implementation of NPUs by bringing an order or more of gain vs
traditional all-digital NPUs. Digital CIM solutions provide excellent
scaling for area and energy efficiency as we move towards more advanced
CMOS nodes with a wide operating voltage and frequency range tunability
while still maintaining a general purpose and application-agnostic view
of embedded neural network acceleration at the edge.~

~On the other hand for applications that can handle approximate
computing, Analog SRAM CIM-based solutions provide a much-increased
level of computation parallelism and energy efficiency while still
operating in an SNR-limited regime. The impact of dynamic range
compression and readout throughput are key algorithmic and design
tradeoffs while designing an Analog CIM solution which tries to operate
in a much more restricted voltage and frequency regime as opposed to a
Digital CIM solution. Given the application choices being more
vertically defined as opposed to general purpose is also a deciding
factor in choosing an Analog SRAM CIM-based solution as opposed to~
Digital CIM solutions. In conclusion, due to the rapid industrialization
potential of SRAM-based CIM solutions and the opportunity of exploiting
the duality of these CIM instances to serve as SRAM capacity to support
the system in other operating modes, there are enough reasons to remain
invested in SRAM CIM. The scope to improve both digital and analog SRAM
CIM remains very high, both at the design and technology level, to
exploit the best gains out of these two solutions which in the future
can also be combined to form a hybrid solution serving multiple
modalities of Neural Network execution at the Edge.~

\vspace{0.5cm}

{\label{258265}}

\subsection*{4.2~ Flash memories}

{\label{788056}}

\begin{quote}
\emph{Gerardo Malavena, Christian Monzio Compagnoni~}
\end{quote}

\subsubsection*{4.2.1 Status~}

{\label{448107}}

Thanks to a relentless expansion in all the application fields of
electronics since their conception in the 80's of the 20th century,
Flash memories became ubiquitous nonvolatile storage media in everyday
life and a source of market revenues exceeding \$60B in 2021. The origin
of this success can be traced back to their capability to solve the
trade-off against cost, performance and reliability in data storage much
better than any other technology. Multiple solutions to that trade-off,
besides, were devised through different design strategies that, in the
end, allowed Flash memories to target a great variety of applications in
the best possible way. Among these different design strategies, the two
leading to the so-called NOR Flash memories~\textsuperscript{\hyperref[csl:146]{146}} and NAND
Flash memories~\textsuperscript{\hyperref[csl:147]{147}} became by far the most important.~

As in all Flash memory designs, NOR and NAND Flash memories store
information in memory transistors arranged in an array whose operation
relies on an initialization, or erase, step performed \emph{in a flash}
on a large number of devices simultaneously. In particular, the erase
step moves the threshold-voltage (V\textsubscript{T}) of all the memory
transistors in a block/sector of the array to a low value. From that
initial condition, data are stored through program steps performed in
parallel on a much smaller subset of memory transistors and raising
their V\textsubscript{T} to one or more predefined levels. This working
scheme of the array allows to minimize the number of service elements
needed for information storage and, in the end, is at the basis of the
high integration density, high performance and high reliability of Flash
memories. Starting from it, the structure of the memory transistors, the
architectural connections among them to form the memory array, the array
segmentation in the memory chip, the physical processes exploited for
the erase and program steps, and many other aspects are markedly
different in NOR and NAND Flash memories.

\begin{figure}[H]
\begin{center}
\includegraphics[width=0.70\columnwidth]{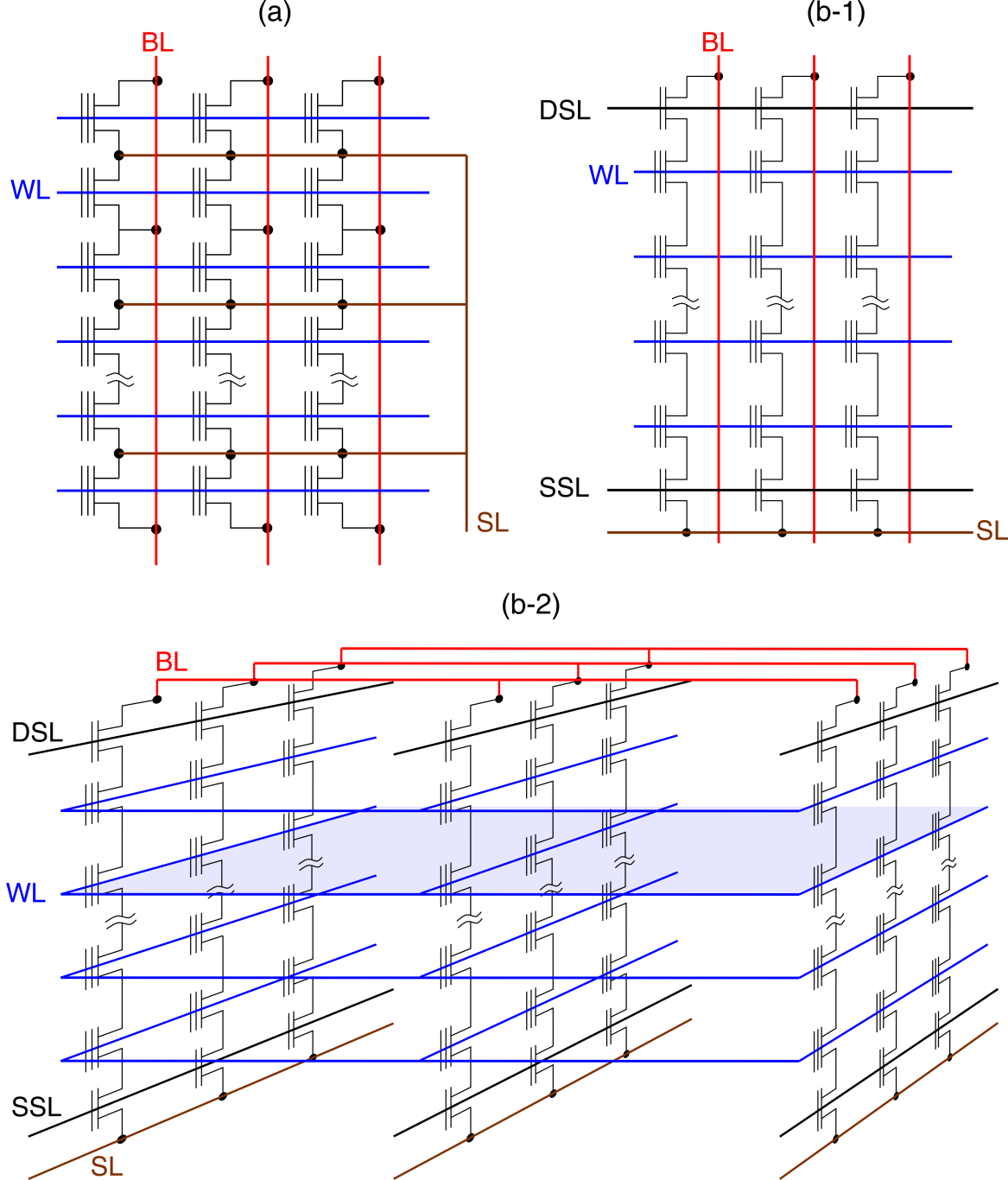}
\caption{{Schematic for the connection of the memory transistors in: (a) a NOR
Flash memory array (a common ground architecture of stacked-gate memory
transistors has been assumed);~ (b-1) a planar and (b-2) a vertical
(3-D) NAND Flash memory arra
{\label{126295}}%
}}
\end{center}
\end{figure}

NOR Flash memories follow a design strategy targeting the minimization
of the random access time to the stored data, reaching latencies as
short as a few tens of nanoseconds. A strong array segmentation is then
adopted to reduce the delay time of the word-lines (WLs) and bit-lines
(BLs) driving the memory transistors. As depicted in
Fig.{\ref{126295}}(a), moreover, the memory transistors
are independently connected to the WLs, BLs and source lines (SLs) of
the array to simplify and speed up the sequence of steps needed to
randomly access the stored data and to allow device operation at
relatively high currents (currents in the microAmpere scale are typical
to sense the data stored in the memory transistors). Fast random access
is also achieved through a very robust raw array reliability, with no or
limited adoption of error correction codes (ECCs). This design strategy,
on the other hand, does not make NOR Flash memories the most convenient
solution from the standpoint of the area and, hence, the cost of the
memory chip and limit the chip storage capacity to low or medium sizes
(up to a few Gbit).

NAND Flash memories rely on a design strategy pointing to the
minimization of the data storage cost. Therefore, limited array
segmentation is adopted and the memory transistors are in series
connection along strings to reduce the area occupancy of the memory
chip. Figs.{\ref{126295}}(b-1) and (b-2) schematically
show the arrangement of the memory transistors in a planar and in a
vertical (or, 3-D) NAND Flash array, respectively. 3-D arrays represent
today the mainstream solution for NAND Flash memories, capable of
pushing their bit storage density up to~ [?]
~15Gbit/mm\textsuperscript{2}~\textsuperscript{\hyperref[csl:148]{148}}, a level unreachable by
any other storage technology. Such an achievement was made possible also
by the use of multi-bit storage per memory transistor and resulted in
memory chips with capacity as high as 1Tbit~\textsuperscript{\hyperref[csl:148]{148}} . The
NAND Flash memory design strategy, on the other hand, makes the random
access time to the stored data relatively long (typically, a few tens of
microseconds). That is the outcome of time delays of the long WLs and
BLs in the microsecond timescale, low sensing currents (tens of
nanoAmpere) during data retrieval due to series resistance limitations
in the strings, and the need of multi-bit detection per memory
transistor. Besides, array reliability relies on powerful ECCs.

Given the successful achievements of Flash memories as nonvolatile
storage media for digital data, exploiting them in the emerging
neuromorphic-computing landscape appears as a natural expansion of their
application fields and is attracting widespread interests. In this
landscape, Flash memories may work not only as storage elements for the
parameters of artificial neural networks (ANNs), but also as active
computing elements to overcome the von Neumann bottleneck of
conventional computing platforms. The latter may represent, of course,
the most innovative and disruptive application of Flash memories in the
years to come. At the same time, the use of Flash memories as active
computing elements may boost the performance, enhance the
power-efficiency and reduce the cost of ANNs, making their bright future
even brighter. In this context, relevant research has been focusing on
employing Flash memory arrays as artificial synaptic arrays in hardware
ANNs and as hardware accelerators for the vector-by-matrix
multiplication (VMM), representing the most common operation in ANN.
Quite promising results have already been reported in the field, through
either NOR~\textsuperscript{\hyperref[csl:149]{149},\hyperref[csl:150]{150}} or NAND~\textsuperscript{\hyperref[csl:151]{151},\hyperref[csl:152]{152},\hyperref[csl:153]{153},\hyperref[csl:154]{154},\hyperref[csl:155]{155},\hyperref[csl:156]{156}} Flash
memories. In these proofs of concept, different encoding schemes for the
inputs (e.g., voltage amplitude or pulse width modulation, with signals
on the BLs or WLs of the memory array) and different working regimes of
the memory transistors have been successfully explored. ~ Interested
readers may go through the references provided in this section for a
detailed description of the most relevant schemes proposed so far to
operate a Flash array as a computing element.

\subsubsection*{4.2.2 Challenges and potential
solutions}

{\label{611756}}

In spite of the encouraging proofs of concept already reported, the path
leading to Flash memory-based ANNs still appears long and full of
challenges. The latter can be classified in the following categories:

\subsubsection*{\texorpdfstring{\emph{Challenges arising from changes in
the design strategy of the
array}}{Challenges arising from changes in the design strategy of the array}}

{\label{697950}}

As previously mentioned, the success of Flash memories as nonvolatile
storage media for digital data arises from precise design strategies.
Modifying those strategies to meet the requests of ANNs may deeply
impact the figures of merit of the technology and should be carefully
done. For instance, ANN topologies requiring to decrease the
segmentation of NOR Flash arrays may worsen their performance in terms
of working speed. Increasing the segmentation of NAND Flash arrays to
meet possible ANN topology constraints or to enhance their working speed
may significantly worsen their cost per memory transistor.

The cost per memory transistor of Flash memories, besides, is strictly
related to the array capacity. Modifying the latter or not exploiting it
all through the ANN topology may reduce the cost effectiveness of the
technology. In this regard, the very different capacities of NOR and
NAND Flash arrays make the former suitable for small/medium size ANNs
(less than~1 Giga parameters) and the latter suitable for large size
ANNs (more than~1 Giga parameters). The organization of the memory
transistors into strings in NAND Flash arrays represents an additional
degree of complexity for the exploitation of their full capacity in
ANNs. In fact, the number of memory transistors per string is the
outcome of technology limitations and cost minimization and, therefore,
cannot be freely modified. Exploiting all the memory transistors per
string, then, necessarily sets some constraints on the ANN topology
(number of hidden layers, number of neurons,\ldots{}), which, of course,
should be compatible with the required ANN performance.

Another important aspect to consider is that the accurate calibration of
the V\textsubscript{T} of the memory transistors needed by
high-performance and reliable ANNs may not be compatible with the
block/sector erase scheme representing a cornerstone of all the design
strategies of Flash memories. Solutions to carry out the erase step on
single memory transistors are then to be devised. These solutions may
require a change of the array design as in~\textsuperscript{\hyperref[csl:150]{150},\hyperref[csl:149]{149}}~or new
physical processes and biasing schemes of the array lines to accomplish
the erase step as in~\textsuperscript{\hyperref[csl:157]{157},\hyperref[csl:158]{158},\hyperref[csl:159]{159}}. All of these approaches,
however, necessarily impact relevant aspects of the technology,
affecting its cost, performance or reliability, and should be carefully
evaluated.

The change of the typical working current of the memory transistors when
exploiting Flash memories for ANN applications is another critical point
to address. In fact, reducing the working current of the memory
transistors may make it more affected by noise and time instabilities.
Increasing it too much, on the other hand, may raise issues related to
the parasitic resistances of the BLs, SLs, and, in the case of NAND
Flash arrays, of the unselected cells in the strings.

\subsubsection*{\texorpdfstring{\emph{Challenges arising from array
reliability}}{Challenges arising from array reliability}}

{\label{868392}}

\emph{~}Flash memories are highly-reliable nonvolatile storage media for
digital data. That, however, does not assure that they can
satisfactorily meet the reliability requirements needed to operate as
computing elements for ANN applications. Especially in the case of NAND
Flash memories, in fact, array reliability in digital applications is
achieved through massive use of ECCs and a variety of smart system-level
stratagems to take under control issues such as electrostatic
interference between neighboring memory transistors, lateral migration
of the stored charge along the charge-trap storage layer of the strings,
degradation of memory transistors after program/erase cycles, and so on.
All of that can hardly be exploited to assure the reliable operation of
Flash arrays as computing elements. Besides, the requirements on the
accuracy of the placement and the stability over time of the
V\textsubscript{T} of the memory transistors when using Flash arrays as
computing elements may be more severe than in the case of digital data
storage. The possibility to satisfy those requirements in the presence
of the well-known constraints to the reliability of all Flash memory
designs~\textsuperscript{\hyperref[csl:147]{147},\hyperref[csl:160]{160},\hyperref[csl:161]{161}} is yet to be fully demonstrated. In this
context, periodic recalibration of the V\textsubscript{T} of the memory
transistors and on-chip learning~\textsuperscript{\hyperref[csl:156]{156}} may mitigate the
array reliability issues. \emph{~}

\subsubsection*{\texorpdfstring{\emph{Challenges arising from the
peripheral circuitry of the
array}}{Challenges arising from the peripheral circuitry of the array}}

{\label{172146}}

As in the case of Flash memory chips for nonvolatile storage of digital
data, the peripheral circuitry of Flash memory arrays used as computing
elements for ANNs should not introduce severe burdens on the chip area,
cost, power efficiency and reliability. In the latter case, this aspect
is particularly critical due to the need to integrate on the chip not
only the circuitry to address the memory transistors in the array and to
carry out operations on them, but also, for instance, to switch between
the digital and the analog domain in VMM accelerators or to implement
artificial neurons in hardware ANNs. Along with effective design
solutions at the circuit level~\textsuperscript{\hyperref[csl:151]{151}}, process solutions
such as CMOS-Under-Array integration~\textsuperscript{\hyperref[csl:148]{148}} ~or
heterogeneous integration schemes~\textsuperscript{\hyperref[csl:153]{153}} ~should be
exploited for successful technology development.

\subsubsection*{4.2.3 Conclusion}

{\label{923386}}

Flash memories may play a key role in the neuromorphic-computing
landscape. Expanding their fields of application, they can be the
elective storage media for ANN parameters. But they can also be active
computing media for high-performance, power-efficient and cost-effective
ANNs. To achieve this intriguing goal, relevant challenges must be faced
from the standpoint of the array design, reliability and peripheral
circuitry. Winning those challenges will be a matter of engineering and
scientific breakthroughs and will pave the way to years of unprecedented
prosperity for both Flash memories and ANNs.

\par\null\selectlanguage{english}

\vspace{0.5cm}

\subsection*{5. Emerging Technologies (computing
approaches)}

{\label{675016}}

Zhongrui Wang, J. Joshua Yang

\subsection*{5.1 Resistive switching and
memristor}

{\label{949762}}

\subsubsection*{5.1.1. Status}

{\label{853744}}

Resistive switches (often called memristors when device nonlinear
dynamics are emphasized) are electrically tunable resistors, of a simple
metal-insulator-metal structure. Typically, their resistance changes as
a result of redox reactions and ion migrations, driven by electric
fields, chemical potentials and temperature ~\textsuperscript{\hyperref[csl:64]{64}}. There
are two types of resistive switches according to the mobile ion species.
In many dielectrics, especially transition metal oxides and perovskites,
anions such as oxygen ions (or equivalently oxygen vacancies) are
relatively mobile and can form a conduction percolation path, leading to
the so-called valence change switching. For example, a conical
pillar-shaped nanocrystalline filament of
Ti\textsubscript{4}O\textsubscript{7} Magn\selectlanguage{ngerman}éli phase filament was
visualized by a transmission electron microscope (TEM) in a
Pt/TiO\textsubscript{2}/Pt resistive switch\textsuperscript{\hyperref[csl:162]{162}}. On the
other hand, the conduction channels can also be created by the redox and
migration of cations, which involves the oxidation of an
electrochemically active metal such as Ag and Cu, followed by the drift
of mobile cations in the solid electrolyte and the nucleation of cations
to establish a conducting channel upon reduction. The dynamic switching
process of a planar Au/SiO\textsubscript{x}:Ag/Au diffusive resistive
memory cell was captured by~\emph{in situ} TEM\textsuperscript{\hyperref[csl:90]{90}}.

Resistive switches provide a hardware solution to address both the von
Neumann bottleneck and the slowdown of Moore's law faced by conventional
digital computers. When these resistive switches are grouped into a
crossbar array, they can naturally perform vector-matrix multiplication,
one of the most expensive and frequent operations in machine learning.
The matrix is stored as the conductance of the resistive memory array,
where Ohm's law and Kirchhoff's current law physically govern the
multiplication and summation, respectively\textsuperscript{\hyperref[csl:64]{64}}. As a
result, the data is both stored and processed in the same location. This
in-memory computing concept can largely obviate the energy and time
overheads incurred by expensive off-chip memory access on conventional
digital hardware. In addition, the resistive memory cells are of simple
capacitor-like structures, equipping them with excellent scalability and
3D stackability. So far resistive in-memory computing has been used for
hardware implementation of deep learning models to handle both
unstructured (\emph{e.g.} general graphs, images, audio and texts) and
structured data, as discussed in the following:

General graph: Graph-type data consists of a set of nodes together with
a set of edges. The theoretical formulation has been made for graph
learning using resistive memory on datasets such as
WikiVote\textsuperscript{\hyperref[csl:163]{163},\hyperref[csl:164]{164}}. Experimentally, a resistive memory-based
echo state graph neural network has been used to classify graphs in
MUTAG and COLLAB datasets as well as nodes in the CORA
dataset\textsuperscript{\hyperref[csl:165]{165}}, including few-shot learning of the
latter\textsuperscript{\hyperref[csl:166]{166}}.

Images: Images are special graph-type data. Both supervised and
unsupervised learning of ordinary images have been experimentally
implemented on resistive memory. For supervised learning, offline
trained resistive memory, where optimal conductance of memory cells is
calculated by digital computers and transferred to resistive memory, is
used to classify simple patterns\textsuperscript{\hyperref[csl:167]{167},\hyperref[csl:168]{168}}, MNIST handwritten
digits\textsuperscript{\hyperref[csl:169]{169},\hyperref[csl:170]{170},\hyperref[csl:171]{171},\hyperref[csl:172]{172}}, CIFAR-10/100 datasets\textsuperscript{\hyperref[csl:173]{173},\hyperref[csl:174]{174},\hyperref[csl:175]{175}},
ImageNet\textsuperscript{\hyperref[csl:176]{176}}; as well as Omniglot one-shot learning
dataset\textsuperscript{\hyperref[csl:177]{177}}. In addition to offline training, online
training adjusts the conductance of resistive memory in the course of
learning, which is more resilident to hardware nonidealities in
classifying simple patterns\textsuperscript{\hyperref[csl:167]{167},\hyperref[csl:37]{37}}, Yale face and MNIST
datasets\textsuperscript{\hyperref[csl:178]{178},\hyperref[csl:179]{179}}, CIFAR-10 dataset\textsuperscript{\hyperref[csl:180]{180}}, and
meta-learning of Omniglot dataset\textsuperscript{\hyperref[csl:181]{181}}. Besides supervised
learning, unsupervised offline learning with resistive memory is used
for sparse coding of images\textsuperscript{\hyperref[csl:182]{182}} and MNIST image
restoration\textsuperscript{\hyperref[csl:183]{183}}.

Audios and texts: Learning sequence data such as audios and texts have
been implemented on resistive memory. Supervised online learning using
recurrent nodes has been done on the Johann Sebastian Bach chorales
dataset\textsuperscript{\hyperref[csl:184]{184}}. In addition, delayed-feedback systems based
on dynamic switching of resistive memory are used for temporal sequence
learning, such as spoken number recognition and chaotic series
prediction\textsuperscript{\hyperref[csl:185]{185},\hyperref[csl:186]{186},\hyperref[csl:187]{187}}. For offline learning, resistive memory is
used for modeling the Penn Treebank dataset\textsuperscript{\hyperref[csl:188]{188}},
Wortschatz Corpora language dataset and Reuters-21578 news
dataset\textsuperscript{\hyperref[csl:189]{189}}, as well as Bonn epilepsy
electroencephalogram dataset and NIST TI-46 spoken digit
dataset\textsuperscript{\hyperref[csl:190]{190},\hyperref[csl:191]{191}}.

Structured data: Despite unstructured data, structured data such as
those of a tabular format has been tackled by resistive memory,
including supervised classification of the Boston housing dataset on an
extreme learning machine\textsuperscript{\hyperref[csl:192]{192}}, K-means clustering of the
IRIS dataset and principal component analysis of the breast cancer
Wisconsin (diagnostic) dataset\textsuperscript{\hyperref[csl:193]{193},\hyperref[csl:194]{194}} and correlation
detection of quality controlled local climatological
database\textsuperscript{\hyperref[csl:195]{195}}.

\par\null\selectlanguage{english}
\begin{figure}[H]
\begin{center}
\includegraphics[width=0.80\columnwidth]{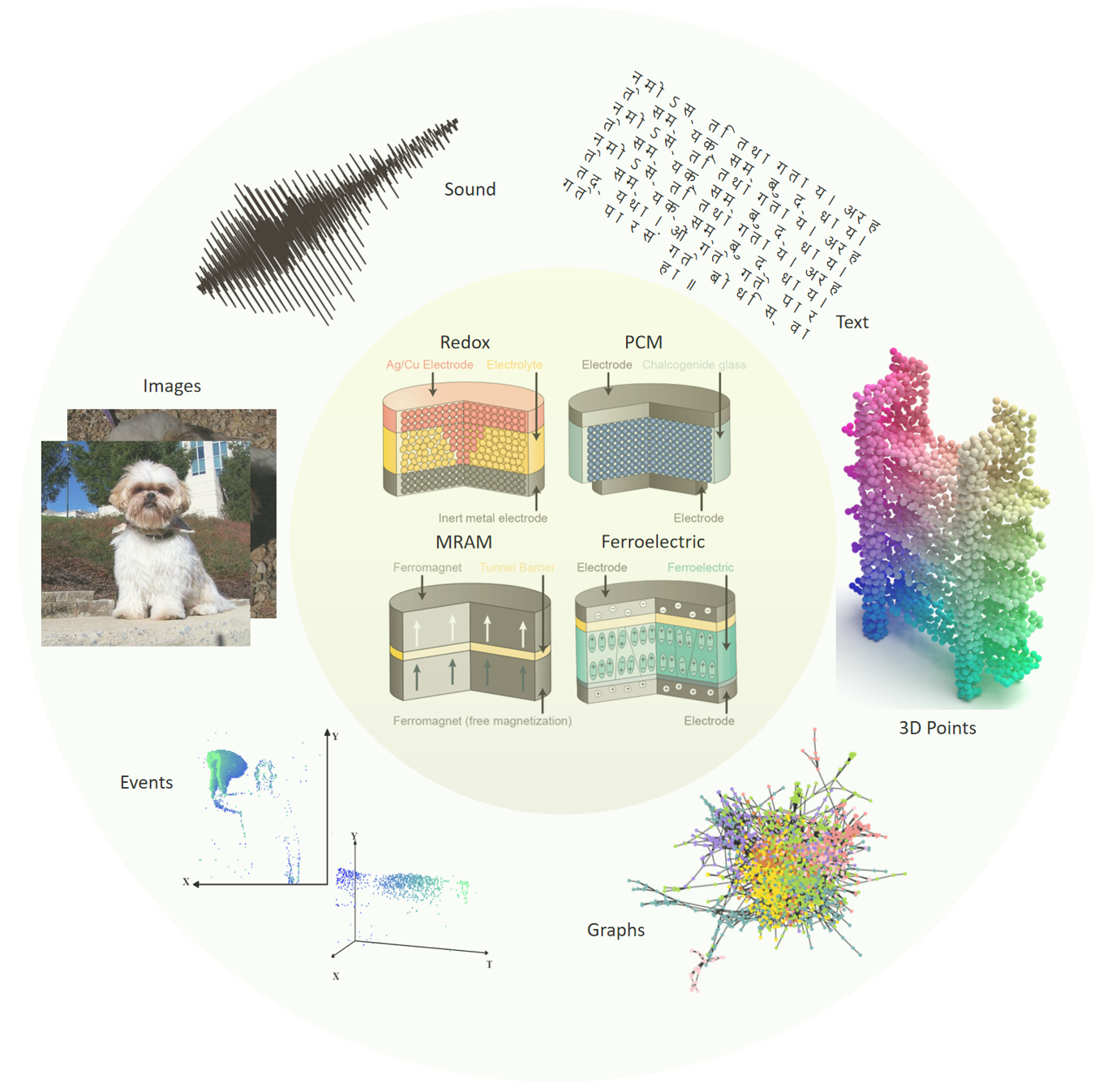}
\caption{{Summary of emerging memory such as memristors and their capabilities in
processing various data types such as images, audios, text, 3D points,
graphs, and events. (Image samples are taken from ImageNet dataset.
Audio waveform visualizes a sample from TIDIGITS dataset. 3D points
visualize a sample from ModelNet10 dataset. Graph sample is from CORA
dataset.
{\label{492910}}%
}}
\end{center}
\end{figure}

\subsubsection*{5.1.2. Challenges}

{\label{440833}}

Major challenges can be categorized at different levels.

Device level: The ionic nature of resistive switching, although benefits
data retention, imposes challenges to programming precision, energy and
speed. The programming precision limits the representation capability of
the resistive switch, or equivalently how many bits a device can encode.
In addition, the programming energy and speed impact online learning
performance. In addition, the degradation of the representation
capability is further intensified by the read noise, manifestation by
the current fluctuation under a constant voltage bias.

Circuit level: Analog resistive memory arrays are mostly interfaced with
up- and downstream digital modules in a computing pipeline. As such,
there is inevitable signal acquisition and conversion cost, which leads
to the question of how to trade off between signal acquisition rate,
precision, and power consumption. In addition, the parasitic resistance
and capacitance, like the non-zero wire resistance, incur the so-called
IR drop in the resistive memory crossbar array.

Algorithm level: So far many applications of resistive memory suffer
from significant performance loss in the presence of resistive memory
nonlinearities (e.g. noises), thus defeating their efficiency advantage
over alternative digital hardware.

\subsubsection*{5.1.3. Potential Solutions}

{\label{622911}}

Device level: Various approaches are used to address the programming
stochasticity, such as the local confinement of conducting
filament~\textsuperscript{\hyperref[csl:196]{196}}.~ A denoising protocol using sub-threshold
voltages has recently been developed to suppress the fluctuation of the
device state and achieve up to 2048 conductance
levels\textsuperscript{\hyperref[csl:197]{197}}. In addition, homogenous switching may suppress
stochasticity at the cost of larger program energy and time
overheads\textsuperscript{\hyperref[csl:64]{64}}. In terms of programming energy, small redox
barriers and large ion mobilities may reduce switching energy and
accelerate switching speed, at the expense of retention and thermal
stability though.Circuit level: Typically, resistive in-memory computing
relies on Ohm's law and Kirchoff's current law, resulting in current
summation. However, there is a recent surge of interest in replacing
current summation by voltage summation, which lowers down the static
power consumption by eliminating current summation incurred Joule
heating. In addition, fully analog neural networks have been proposed to
get rid of the frequent analog-to-digital and digital-to-analog
conversions~\textsuperscript{\hyperref[csl:198]{198}}. To combat with the parasitic wire
resistance, a simple solution is to increase device resistance in both
ON and OFF states, such as that demonstrated in a 256\selectlanguage{ngerman}×256 in-memory
computing macros~\textsuperscript{\hyperref[csl:197]{197}}.

Algorithm: A recent trend is hardware-software codesign to leverage
resistive memory nonlinearities and turn them into advantages. For
example, the programming stochasticity can be exploited by neural
networks of random features (e.g. echo state networks~\textsuperscript{\hyperref[csl:165]{165},\hyperref[csl:166]{166}}
and extreme learning machines\textsuperscript{\hyperref[csl:192]{192}}) and Bayesian inference
using Markov Chain Monte Carlo (MCMC) such as Metropolis--Hastings
algorithm\textsuperscript{\hyperref[csl:199]{199}}. Also, such programming noise is a natural
regularization to suppress overfitting in online
learning\textsuperscript{\hyperref[csl:200]{200}}. Moreover, hyperdimensional
computing\textsuperscript{\hyperref[csl:189]{189}} and mixed-precision design such as
high-precision iterative refinement algorithm paired with low-precision
conjugate gradient\textsuperscript{\hyperref[csl:201]{201}} can withstand resistive memory
programming noise. The reading noise can also be exploited for solving
combinatorial optimization problems using simulated annealing, serving
as a natural noise source to prevent the system from falling into the
local minimum\textsuperscript{\hyperref[csl:202]{202},\hyperref[csl:203]{203}}.

\subsubsection*{5.1.4. Conclusion}

{\label{523719}}

The fast advent of resistive switch-based in-memory computing in the
last decade has demonstrated a wide spectrum of applications in machine
learning and neuromorphic computing, reflected by its handling of
different types of data.

However, there are still plenty of room, at device, circuit, and
algorithm levels, to improve, which will help to fully unleash the power
of in-memory computing with resistive switches and potentially yield a
transformative impact on future computing.~

\vspace{0.5cm}

\subsection*{5.2 Phase change materials}

{\label{223352}}

\begin{quote}
\emph{Abu Sebastian, Ghazi Sarwat Syed}
\end{quote}

\subsubsection*{5.2.1.Introduction}
Phase-change memory (PCM) is arguably the most advanced memristive technology. Similar to conventional metal-oxide based memristive devices, information is stored in terms of changes in atomic configurations in a nanometric volume of material and the resulting change in resistance of the device \textsuperscript{\hyperref[csl:204]{204}}. However, unlike the vast majority of memristive devices, PCM exhibits volumetric switching as opposed to filamentary switching. The volumetric switching is facilitated by certain material compositions along the GeTe-Sb$_2$Te$_3$ pseudo-binary tie line, such as Ge$_2$Sb$_2$Te$_5$ that can be switched reversibly between amorphous and crystalline phases of different electrical resistivity \textsuperscript{\hyperref[csl:205]{205}}. Both transitions are Joule-heating assisted. The crystalline to amorphous phase transition relies on a melt-quench process whereas the reverse transition relies mostly on crystal growth. \\\selectlanguage{english}
\begin{figure}[H]
\begin{center}
\includegraphics[width=0.80\columnwidth]{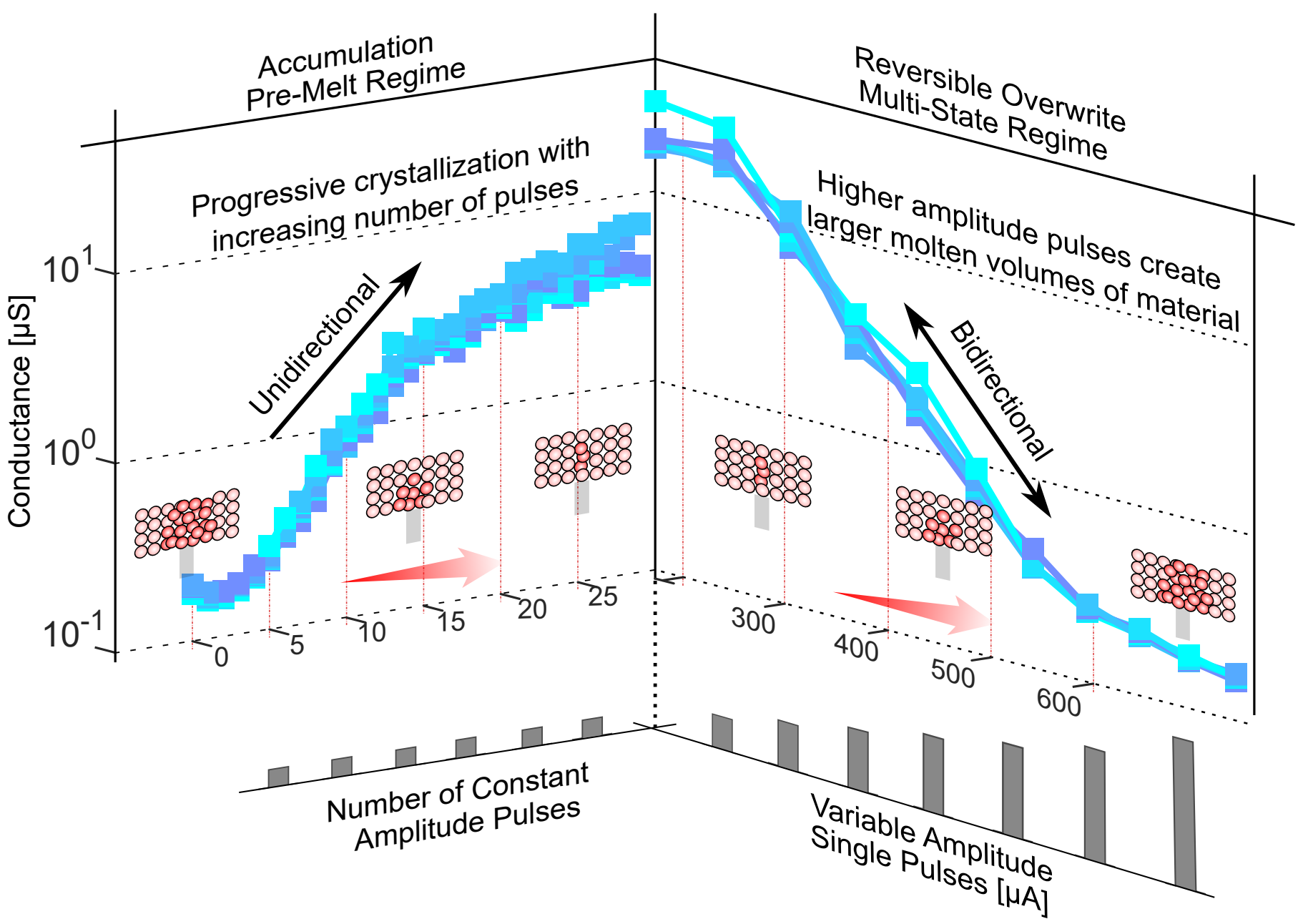}
\caption{{\textbf{Operational regimes of a phase change device when used for
neuromorphic computing}. On the right plot, the direct overwrite regime
utilizing melt-quench dynamics is illustrated. The programming curves
display the achievable conductance values in response to partial RESET
pulses of varying amplitudes. As the RESET pulse is increased in
amplitude, a larger amorphous volume is created mostly independent of
the phase configuration prior to the application of the pulse. On the
left plot, the characteristic accumulative property is demonstrated. It
showcases the evolution of conductance values over successive
applications of a SET pulse with a constant amplitude. As the amorphous
region reduces in size due to crystallization dynamics, the device
conductance progressively increases. Multiple experimental traces are
overlaid in both plots. ~
{\label{163638}}%
}}
\end{center}
\end{figure}

There are essentially two key properties that make PCM devices particularly well suited for neuromorphic computing (see Figure \ref{163638}) \textsuperscript{\hyperref[csl:206]{206}}. Interestingly this was pointed out by Stanford Ovshinsky, a pioneer of PCM technology, way back in 2003 when PCM was being considered just for memory applications \textsuperscript{\hyperref[csl:207]{207}}. The first property is that PCM devices can store a range of conductance values by modulating the size of the amorphous region typically achieved by partial RESET pulses that melt and quench a certain volume of PCM material. This analogue storage capability, combined with a crossbar topology, allows for matrix-vector multiply (MVM) operations to be carried out in O(1) time complexity by leveraging Kirchhoff's circuit laws. This makes it possible to realize an artificial neural network on crossbar arrays of PCM devices, with each synaptic layer of the DNN mapped to one or more of the crossbar arrays \textsuperscript{\hyperref[csl:67]{67},\hyperref[csl:208]{208}}. The second property referred to as accumulative property results from the progressive crystallization of PCM material upon application of an increasing number of partial SET pulses (see Figure 1). It is used for implementing DNN training \textsuperscript{\hyperref[csl:209]{209}}, temporal correlation detection \textsuperscript{\hyperref[csl:210]{210}}, continual learning \textsuperscript{\hyperref[csl:211]{211}}, local learning rules such as spike-timing-dependent plasticity \textsuperscript{\hyperref[csl:212]{212},\hyperref[csl:213]{213}} and neuronal dynamics \textsuperscript{\hyperref[csl:214]{214}}. \\

PCM is at a very high maturity level of development and has been commercialized as both standalone memory\textsuperscript{\hyperref[csl:215]{215}} and embedded memory \textsuperscript{\hyperref[csl:216]{216}}. This fact, together with the ease of embedding PCM on logic platforms (embedded PCM) \textsuperscript{\hyperref[csl:208]{208}} make this technology of unique interest for neuromorphic computing.
\subsubsection*{5.2.2.Challenges}

PCM devices offer write operations in the tens of nanosecond timescale which is sufficient for most neuromorphic applications in particular those targetting deep learning inference. The cycling endurance could also exceed a billion cycles (dependent on the device geometry) which is several orders of magnitudes higher than commercial Flash memory\textsuperscript{\hyperref[csl:217]{217}}. This is sufficient for deep learning inference applications. The cycling endurance for partial SET pulses is much higher than that for full SET-RESET cycling and hence is widely considered sufficient for other neuromorphic applications as well. The read endurance is almost infinite for PCM when sufficiently low read bias is applied. Another key attribute is retention which is typically tuned through material choice \textsuperscript{\hyperref[csl:218]{218}}.  
 However, the use of analogue conductance states in neuromorphic computing makes the retention time of intermediate phase configurations even more important which could be substantially lower than that of fully RESET states.\\
 
One of the primary challenges for PCM is integration density. For example, for DNN inference, it is desirable to have at least 10-100 Million on-chip weight capacity. The crossbar array for neuromorphic computing comprises metal lines intersected by synaptic elements, which are composed of one or more PCM devices and selector devices. Access devices such as bipolar junction transistors or metal-oxide-semiconductor field effect transistors are preferred for accurate programming, while two-terminal poly-silicon diodes offer scalability. To achieve high memory density, stacking multiple crossbar layers vertically is beneficial. BEOL selectors such as ovonic threshold switches show promise but face challenges in achieving precise current control. Edge effects and thermal cross-talk between neighboring cells become significant at smaller feature sizes \textsuperscript{\hyperref[csl:219]{219},\hyperref[csl:220]{220},\hyperref[csl:221]{221}}.\\\selectlanguage{english}
\begin{figure}[H]
\begin{center}
\includegraphics[width=0.98\columnwidth]{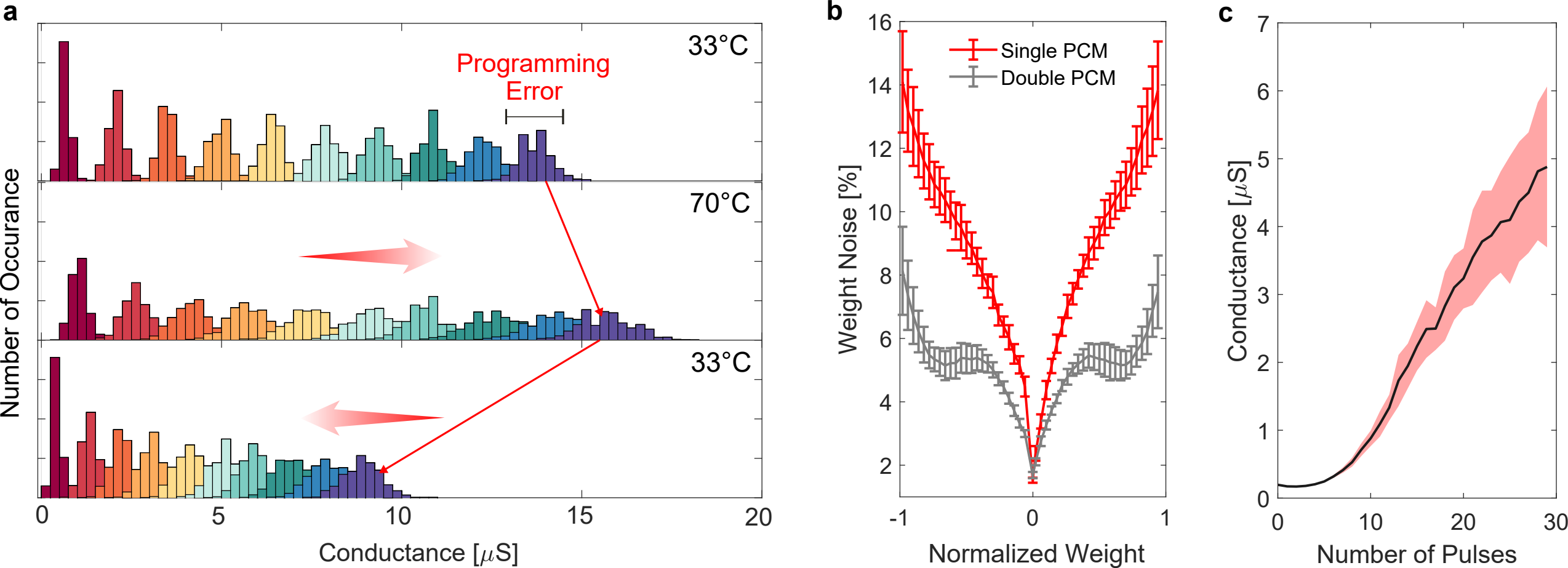}
\caption{{\textbf{PCM non-idealities}. (a) Device data after programming shows
variability, reflected in broad distributions of analogue conductance
values due to programming inaccuracies, read noise and drift variability
(the top panel). Temperature increase raises state conductivity due to
thermal carrier excitation and accelerates structural relaxation. (b)
The conductance fluctuations manifest as synaptic weight noise, here
shown as additive noise in terms of the percentage of the maximum
synaptic weight. Using two PCM devices per synapse reduces this error.
(c) The accumulative behaviour exhibit significant stochasticity mostly
attributed to variations in the crystallization kinetics.
{\label{880438}}%
}}
\end{center}
\end{figure}

\vspace{0.5cm}

\subsection*{5.3 Ferroelectric materials}

{\label{934562}}

\emph{Thomas Mikolajick, Stefan Slesazeck and~\emph{Beatriz Noheda}}

\subsubsection*{5.3.1.Status~}

{\label{791507}}

Ferroelectric materials are, in theory, ideally suited for information
storage tasks since their switching is purely field-driven, holding the
promise of extremely low write energy, and non-volatile at the same
time. Moreover, unlike competing concepts like resistive switching or
magnetic switching, ferroelectric materials offer three different
readout possibilities giving a lot of flexibility in device design
\textsuperscript{\hyperref[csl:222]{222}}. In detail, the following read schemes can be applied
(see also middle part of Fig. {\ref{591962}}):

\begin{itemize}
\tightlist
\item
  Direct sensing of the switched charge during polarization reversal, as
  used in the ferroelectric RAM (FeRAM) concept, results in a cell
  design similar to a dynamic random-access memory
  (DRAM)~\textsuperscript{\hyperref[csl:223]{223}}.
\item
  Coupling of the ferroelectric to the gate of a field effect transistor
  and readout of the resulting drain current, as used in the
  ferroelectric field effect transistor (FeFET). This results in a cell
  that is similar to classical transistor-based charge storage (floating
  gate or a charge trapping) memory cell, which is most prominently used
  in Flash memories~\textsuperscript{\hyperref[csl:224]{224}}.~
\item
  Modulation of the tunnelling barrier in a ferroelectric tunneling
  junction (FTJ). As a result, we can realize a two-terminal device,
  which is essentially a special version of a resistive switching memory
  cell (see Chapter 5.1) \textsuperscript{\hyperref[csl:225]{225}}.
\end{itemize}

Each of the mentioned read-out schemes has advantages and disadvantages
and, therefore, the flexibility to use one of the three is a plus,
especially in applications that go beyond pure memories, like
neuromorphic computing.

However, traditionally ferroelectricity was only experienced in
chemically complex materials, such as lead-zirconium titanate (PZT),
strontium bismuth tantalate (SBT) or bismuth ferrite (BFO) which all are
very difficult to incorporate into the processing flow for integrated
electronic circuits, due to their limited stability in reducing
environments. Another pervasive issue for the integration of
ferroelectrics is their tendency to depolarize upon downscaling, an
issue that is accentuated by their high permittivity. Organic
ferroelectrics, the most prominent example being Polyvinylidenfluorid
(PVDF), can mitigate this problem, as their low permittivities reduce
the depolarization fields; while a rather high coercive field increases
the stability of the polarization state. Such materials are ideally
suited for lab scale demonstrations of new device concepts, due to their
simple fabrication using a solution-based process, and are highly
preferred for flexible and biocompatible electronics \textsuperscript{\hyperref[csl:226]{226}}.
However, their limited thermal stability has taken them out of the game
for devices in integrated circuits. Therefore, although the technology
in form of FeRAM \textsuperscript{\hyperref[csl:227]{227}} is on the market for more than 25
years, it has lacked the ability to scale in a similar manner as
conventional memory elements and, therefore, it is still limited to
niche applications that require a high re-write frequency together with
nonvolatile like in data logging applications.~

\subsubsection*{5.3.2.Challenges}

{\label{356463}}

With the discovery of ferroelectricity in hafnia (HfO\textsubscript{2})
and zirconia (ZrO\textsubscript{2}), the biggest obstacle of the limited
compatibility with integrated circuit fabrication could be
solved~\textsuperscript{\hyperref[csl:228]{228}}. HfO\textsubscript{2} and ZrO\textsubscript{2}
are stable both in reducing ambient and in contact with silicon and
their fabrication using established atomic layer deposition processes is
standard in modern semiconductor process lines.~ However, new
difficulties, especially with respect to reliability~\textsuperscript{\hyperref[csl:229]{229}},
need to be solved. Challenges in this direction are aggravated by the
metastable nature of the ferroelectric phase, which appears mostly at
the nanoscale, making a full understanding of the polar phase quite
demanding. While their high coercive field makes them very stable with
respect to classical retention, the ferroelectric phase typically exists
together with other non-polar phases, which prevents them from reaching
the predicted polarization values (of the order of 50
uC/cm\textsuperscript{2})~\textsuperscript{\hyperref[csl:230]{230}} . Moreover, the most
serious problem of any nonvolatile ferroelectric device, the imprint,
becomes very complex to manage in hafnia/zirconia-based ferroelectrics.
Imprint is a shift of the hysteresis loops due to an internal bias.
While this effect leads to a classical retention of the stored state
that may look perfect, after switching, retention will be degraded and
fixing the so-called opposite-state retention loss needs to be carefully
done by material and interface engineering. Moreover, the high coercive
field in this materials class becomes a problem as HfO\textsubscript{2}
and ZrO\textsubscript{2} often show a pronounced wake-up and fatigue
behavior and the field-cycling endurance is in many cases limited by the
dielectric breakdown of the material.~

While the issues mentioned so far are valid for any nonvolatile device
application, in neuromorphic systems additional challenges arise,
including the linearity of the switching behavior and tuning of the
retention to achieve both short-term and long-term plasticity, as well
as specific effects to mimic neurons, like accumulative
switching~\textsuperscript{\hyperref[csl:231]{231}} \textsuperscript{\hyperref[csl:222]{222}}, which need to be
explored using material and device design measures. Finally, large-scale
neuromorphic systems will require a high integration density that
demands 3-dimensional integration schemes, either realized by the
punch-and-plug technology well-known from NAND Flash or by integrating
devices into the back-end of the line.

\subsubsection*{5.3.3. Potential Solutions}

{\label{444700}}

Since the original report on ferroelectricity in hafnium oxide
\textsuperscript{\hyperref[csl:228]{228}} , the boundary conditions for stabilizing the
ferroelectric phase have been much better understood, although there are
still a number of open questions. The goal is to achieve a high fraction
of the ferroelectric phase without dead layers of non-ferroelectric
phases at the interface to the electrodes or in the bulk of the film.
This needs to be done under the boundary conditions of a realistic
fabrication process, which means that sophisticated methods to control
the crystal structure based on epitaxial growth are not possible.
Epitaxial growth can help clarify scientific questions but the achieved
results need to be transferred to chemical vapor deposition (CVD),
including most prominently atomic layer deposition (ALD), or physical
vapor deposition (PVD) processes using electrodes like TiN or TaN that
can be integrated into electronic processes.

\begin{figure}[H]
\begin{center}
\includegraphics[width=0.91\columnwidth]{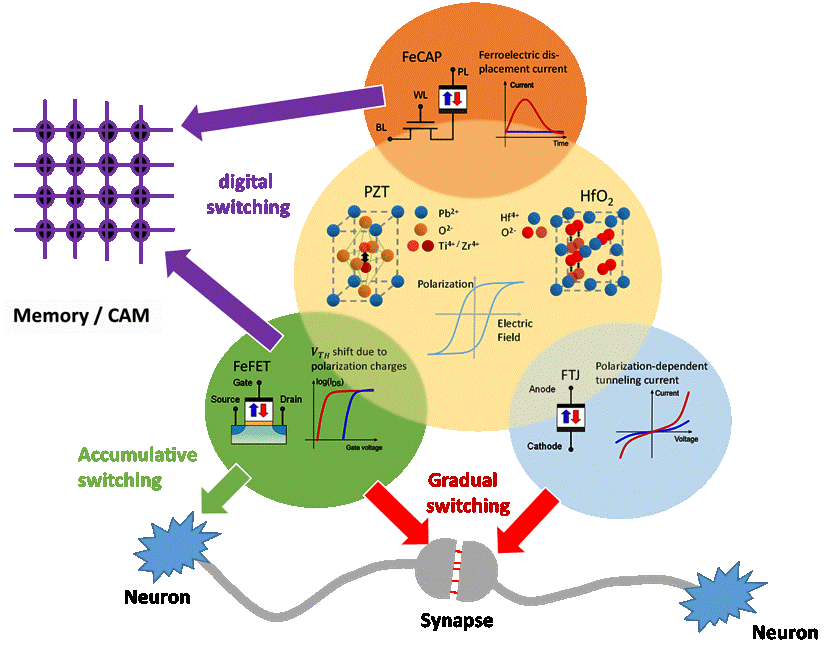}
\caption{{Ferroelectric materials (center) enable three different basic memory
cells (middle ring). These can be used in various ways in neuromorphic
circuits (examples see outer part of the figure). The rich switching
dynamics of ferroelectrics allow to tailor new devices mimicking neurons
and synapses in a much more flexible and area efficient way as compared
to their pure CMOS counterparts.
{\label{591962}}%
}}
\end{center}
\end{figure}

In the last years, it became obvious, that oxygen vacancies are, on the
one hand, required to stabilize the ferroelectric
phase~~\textsuperscript{\hyperref[csl:232]{232}} and, on the other hand, ~detrimental to both
the imprint and the field cycling behavior~\textsuperscript{\hyperref[csl:233]{233}}.
Therefore, many proposals to integrate ~the ferroelectric layer with
additional thin layers in the film stack have been made and currently a
lot of work is going in that direction. Moreover, it is clear, that the
interface to the electrodes needs careful consideration. In this
direction~\textsuperscript{\hyperref[csl:234]{234}}, facilitating the transport of oxygen, not
only in the ferroelectric layer but also across the electrode
interfaces, by minimizing the strain effects, may be the key to
improving device performance~ ~\textsuperscript{\hyperref[csl:235]{235}} . When it comes to
structures that are in direct contact with silicon, a recent observation
of a quasi-epitaxial growth of extremely thin hafnium-zirconium oxide
films on silicon could be an interesting direction~\textsuperscript{\hyperref[csl:236]{236}}.
~For concrete neuromorphic applications, the~rich switching dynamics can
be very helpful (see
Fig.~{\ref{591962}})~\textsuperscript{\hyperref[csl:237]{237}}. While in
large devices a continuous switching between different polarization
states is possible, devices scaled in the 10nm regime show abrupt and
accumulative switching ~\textsuperscript{\hyperref[csl:231]{231}} . The former can be used for
mimicking synaptic functions while the latter is helpful to mimic
neurons. In classical nonvolatile memories, the depolarization fields
created by non-ferroelectric layers or portions of the layer in series
to the ferroelectric are a concern for the retention of the device.
However, when creating short and long-term plasticity in synaptic
devices, this can be turned into an advantage such that the device
retention can be tailored.~\selectlanguage{english}

\vspace{0.5cm}

\subsection*{5.4 Spintronic materials for neuromorphic
computing}

{\label{850040}}

\begin{quote}
\emph{Bernard Dieny, Tuo-Hung (Alex) Hou}
\end{quote}

\subsubsection*{5.4.1. Status}

{\label{885100}}

Spintronics is a merging of magnetism and electronics in which the spin
of electrons is used to reveal new phenomena and implement them in
devices with improved performances and/or new functionalities.
Spintronics already found many applications for magnetic field sensors
in particular in hard disk drives and more recently as non-volatile
memory (MRAM) in replacement of e-FLASH and last-level CACHE memory.
Spintronics can also bring very valuable solutions in the field of
neuromorphic computing both as artificial synapses or neurons.

\begin{figure}[H]
\begin{center}
\includegraphics[width=0.70\columnwidth]{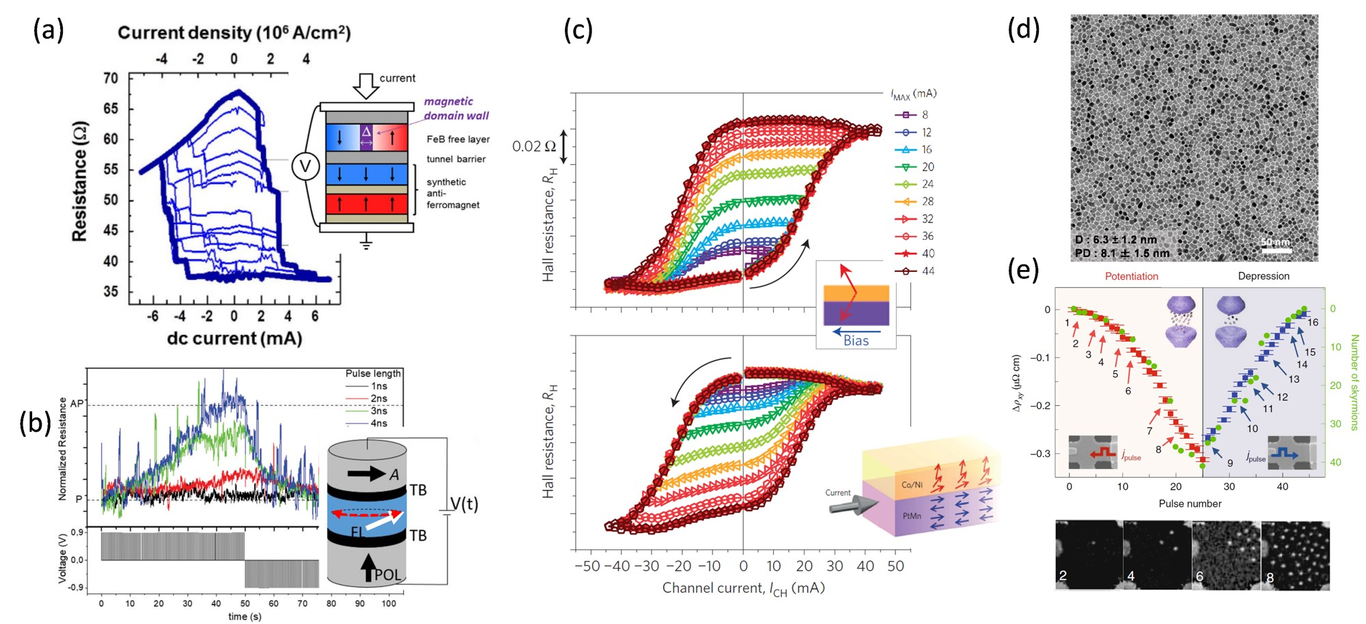}
\caption{{Various realizations of spintronic memristors: (a) based on domain wall
propagation in the storage layer; (b) based on variation of angle
between storage layer and reference layer magnetization; (c) based on
SOT in ferromagnetic storage layer exchange coupled to an
antiferromagnetic SOT line; (d) implementing a storage layer made of a
granular layer similar to the one used in recording technology; (e)
based on a controlled number of skyrmions nucleated in the storage
layer.
{\label{430593}}%
}}
\end{center}
\end{figure}

Artificial synapses are devices supposed to store the potential weight
of the bounds linking two neurons. Various types of spintronic synapses
have been proposed and demonstrated~\textsuperscript{\hyperref[csl:65]{65}}. They are
magnetoresistive non-volatile memory cells working either as binary
memory, multilevel memory, or even in an analogue fashion. Their
resistance depends on the history of the current that has flown through
the device (memristor). Most of these devices are based on magnetic
tunnel junctions (MTJ) which basically consist of two magnetic layers
separated by a tunnel barrier. One of the magnetic layers has a fixed
magnetization (the reference layer) whereas the magnetization of the
other (the storage layer) can be changed by either pulse of magnetic
field or current using phenomena such as spin transfer torque (STT) or
spin-orbit torque (SOT)\textsuperscript{\hyperref[csl:238]{238}}. The resistance of the device
depends on the amplitude and orientation of the magnetic moment of the
storage layer relative to that of the reference layer (tunnel
magnetoresistance effect: TMR). For a binary memory as in STT-MRAM or
SOT-MRAM, only the parallel and antiparallel magnetic configurations are
used~\textsuperscript{\hyperref[csl:239]{239}}. For multilevel or analogue memory, several
options are possible as illustrated in
~Fig.{{\ref{430593}}}. One consists of varying the
proportion of the storage layer area which is in parallel or
antiparallel magnetic alignment with the reference layer magnetization.
This can be achieved by step-by-step propagating a domain wall within
the storage layer using the STT produced by successive current pulses
~{(Fig.{\ref{430593}}a)}~\textsuperscript{\hyperref[csl:240]{240}}, or by
gradually switching the magnetization of the storage layer exchange
coupled to an antiferromagnet using the SOT produced by the pulsed
current flow in the
antiferromagnet~{(Fig.{\ref{430593}}c)}\textsuperscript{\hyperref[csl:241]{241}}
or by gradually switching the grains of a granular storage media similar
to the ones used in hard disk drives
~{(Fig.{\ref{430593}}d)}~\textsuperscript{\hyperref[csl:242]{242}}, or by
nucleating a controlled number of magnetic spin nanotextures in the
storage layer such as skyrmions{
(Fig.{\ref{430593}}e)}~\textsuperscript{\hyperref[csl:243]{243}}.
Alternatively, the memristor resistance can also be varied by changing
the relative angle between the magnetization of the reference and
storage layers using all intermediate angles between 0\selectlanguage{ngerman}° and 180° instead
of only parallel and antiparallel configurations~
(Fig.{\ref{430593}}b) ~\textsuperscript{\hyperref[csl:104]{104}}. Chains of
binary magnetic tunnel junctions can also be used to achieve spintronic
memristors but at the expense of larger footprint~\textsuperscript{\hyperref[csl:244]{244}}.

~\selectlanguage{english}

Concerning artificial neurons, the conventional CMOS neuron circuit is
limited by its large area because a large number of transistors and a
large-area membrane capacitor is required for implementing
Integrate-and-Fire I\&F functions~\textsuperscript{\hyperref[csl:245]{245}}. Recently, several
spintronic neuron devices were reported to generate spike signals by
leveraging nonlinear and stochastic magnetic dynamics without the need
for additional capacitors and complex peripheral circuitry. Spintronic
neurons potentially show a great advantage for compact neuron
implementation~\textsuperscript{\hyperref[csl:246]{246}}.~

Assembly of interacting spin-torque nano-oscillators (STNO) based on the
structure of magnetic tunnel junctions (MTJs) was proposed to achieve
neuron functionality. Unstable conductance oscillation that mimics spike
generation is induced at hundreds of MHz to several tens of GHz by
flowing a current through the device. The frequency and amplitude of
oscillation vary with the applied current and magnetic field.
Torrejon~\emph{et al.} demonstrated spoken-digit or vowel recognition
using such an array of nanoscale oscillators~\textsuperscript{\hyperref[csl:247]{247}}.~

Superparamagnetic tunnel junctions can also be used to mimic stochastic
neurons. They have much lower thermal stability compared to the MTJ used
for memory so they stochastically switch between antiparallel (AP) and
parallel (P) states due to thermal fluctuations, which is referred to as
telegraphic switching~\textsuperscript{\hyperref[csl:248]{248}}. This switching mode can be
used to generate Poisson spike trains in spiking neural networks (SNNs)
as well as for probabilistic computing~\textsuperscript{\hyperref[csl:249]{249}}. The MTJ
device with high thermal stability, which can implement not only
synapses but also neurons in an all-spin neural network, was proposed by
Wu~\emph{et al.}~\textsuperscript{\hyperref[csl:250]{250}}. The reduction of the thermal
stability factor is induced by self-heating at high bias voltage for
neuron operations~\textsuperscript{\hyperref[csl:251]{251}}. At low bias, it stably stores
weight information as synapses.

Many other new spintronic materials and mechanisms were also
investigated for the feasibility of neuron devices, in particular based
on magneto-electric effects. For instance, by playing with magneto-ionic
effects influencing the anisotropy at magnetic metal/oxide interfaces,
the density of skyrmions \textsuperscript{\hyperref[csl:252]{252}} and even their chirality
could be controled electrically~\textsuperscript{\hyperref[csl:253]{253}}. ~Jaiswal~\emph{et
al.} designed a magnetoelectric neuron device for SNNs
~\textsuperscript{\hyperref[csl:254]{254}} . Zahedinejad et al demonstrated that electrically
manipulated spintronic memristors can be used to control the
synchronization of spin Hall nano-oscillators for neuromorphic
computing~\textsuperscript{\hyperref[csl:255]{255}}.

\subsubsection*{5.4.2. Challenges}

{\label{935584}}

Building useful fully functional neuronal circuits requires large-scale
integration of layers of artificial neurons interconnected with
spintronic synapses. Crossbar architectures can achieve cumulate and
multiply functions very efficiently in an analogue manner. An advantage
of magnetic tunnel junctions over other technologies based on materials
such as resistive oxide or phase change is their write endurance
associated with the fact that their resistance change does not involve
ionic migration. However, they exhibit a lower
R\textsubscript{OFF}/R\textsubscript{ON} ratio (\textasciitilde{}4 for
MRAM versus 10 to 100 for RRAM or PCRAM) but also narrower cell-to-cell
distribution of resistance in R\textsubscript{OFF} and
R\textsubscript{ON} states. In crossbar architecture, MTJ should have
high resistance to minimize power consumption. Therefore efforts should
be pursued to further increase the TMR amplitude of MgO-based MTJ and
bring it closer to the expected theoretical values of several
1000\%~\textsuperscript{\hyperref[csl:256]{256}}. In high-resistance MTJ, other approaches
such as SOT or voltage control of anisotropy (VCMA) could be used to
change the MTJ resistance. In all cases, the control of the resistance
change induced by current or voltage pulses must be improved. The
operating temperature has often also a significant impact on the
magnetic properties which imposes challenges on system design.

Concerning artificial neurons, the DC power required to trigger the
magnetization dynamics of STNO neurons is still relatively high (mW
range)~\textsuperscript{\hyperref[csl:257]{257}}. Ways must be found to reduce it by using
different materials or new designs. The switching speed and endurance in
superparamagnetic tunnel junctions and self-heating-assisted MTJ neurons
could be further enhanced to improve processing speed and system
reliability~\textsuperscript{\hyperref[csl:258]{258}}\textsuperscript{\hyperref[csl:259]{259}}. How to continue
improving variability across millions of synapses and thousands of
neurons to ensure high accuracy in future neuromorphic systems remains
an actively research topic. I{nterconnecting all these devices is also a
challenge and innovative approaches beyond classical interconnects must
be found notably by taking advantage of 3D integration.}

\subsubsection*{5.4.3. Potential Solutions}

{\label{235218}}

STT-MRAM entered volume production in 2019 at major microelectronic
companies\textsuperscript{\hyperref[csl:260]{260}}. This marked the adoption of this hybrid
CMOS/magnetic technology by the microelectronic industry. Thanks to the
combined efforts of the chip industry, equipment suppliers, and academic
laboratories, spintronics is progressing very fast. Material research is
very important to increase magnetoresistance amplitude, switching
currents, STT and SOT efficiency, VCMA efficiency, reduce dependence on
operating temperature, reduce current to trigger oscillations in STNO,
and reduce disturbance due to parasitic field. Investigations are in
progress involving antiferromagnetic materials for reduced sensitivity
to the field and access to THz frequency operation~\textsuperscript{\hyperref[csl:261]{261}} ,
half-metallic materials such as Heusler alloys for enhanced TMR
amplitude and reduced write current \textsuperscript{\hyperref[csl:105]{105}} , topological
insulators for very efficient spin/charge current interconversion
possibly combined with ferroelectric materials~\textsuperscript{\hyperref[csl:262]{262}} .

Concerning interconnects, fortunately, magnetic materials are grown in
back-end technology and can be stacked but at the expense of complexity
and cost. Long-range information transmission can be carried out via
spin-current or magnons or by propagating magnetic textures such as
domain walls~\textsuperscript{\hyperref[csl:263]{263}} or skyrmions~\textsuperscript{\hyperref[csl:264]{264}}. Light
could also be used to transmit information in conjunction with recent
developments related to all-optical switching of
magnetization~\textsuperscript{\hyperref[csl:265]{265}}. Besides, a great advantage of
spintronic stacks is that they can be grown on almost any kind of
substrates provided the roughness of the substrate is low enough
compared to the thickness of the layers comprised in the stack. This
enables the use of the third dimension by stacking several spintronic
structures thereby gaining in interconnectivity~\selectlanguage{ngerman}\textsuperscript{\hyperref[csl:266]{266}}.

\subsubsection*{5.4.4. Concluding remarks}

{\label{874441}}

Spintronics can offer valuable solutions for neuromorphic computing.
Considering that STT-MRAM is already in commercial production, it is
very likely that the first generation of spintronic neuromorphic
circuits will integrate this technology. Next, crossbar arrays
implementing analogue MTJ may be developed as well as neuronal circuits
based on the dynamic properties of interacting STNO for learning and
inference.~ ~Still, many challenges are on the way towards practical
applications, including speed, reliability, scalability, and variation
tolerance which need to be addressed in future research. ~

\vspace{0.5cm}

\subsection*{5.5 Optoelectronic and photonic
implementations}

{\label{434238}}

\begin{quote}
\emph{Akhil Varri}, \emph{Frank Brückerhoff-Plückelmann}, \emph{Wolfram Pernice}
\end{quote}

\subsubsection*{5.5.1. Status}

{\label{587475}}

Computing using light offers significant advantages in highly parallel
operation exploiting concepts like wavelength and time multiplexing.
Moreover, optical data transfer enables low power consumption, better
interconnectivity, and ultra-low latency. Already in the 1980s, first
prototypes were developed, however, the bulky tabletop experiments could
not keep pace with the flourishing CMOS industry. Nowadays, novel
fabrication processes and materials enable the (mass) production of
photonic integrated circuits, allowing photonic systems to compete with
their electronic counterparts. Especially in the area of data-heavy
neuromorphic computing, the key advantages of photonic computing can be
exploited.~

Scientific efforts in neuromorphic photonic computing can be segregated
in two major directions: (i) one approach is building hardware
accelerators that excel at specific tasks, e.g., computing matrix-vector
multiplications, by partially mimicking the working principles of the
human brain; and (ii) designs which aim to emulate the functionality of
biological neural networks. Such devices are able to replicate the
behavior of a neuron, synapse, learning mechanisms and ultimately
implement a spiking neural network.~

There has been considerable progress in the (i) direction leading back
to 2017 when Shen et al.~\textsuperscript{\hyperref[csl:267]{267}} demonstrated vowel
recognition where every node of the artificial neural network is
physically represented in the hardware using a cascaded array of
interferometers. This scheme has also been scaled to implement a 3-layer
deep neural network with in-situ training capability
\textsuperscript{\hyperref[csl:268]{268}}. In addition, Feldmann et al. \textsuperscript{\hyperref[csl:269]{269}}
have demonstrated neurosynaptic networks on-chip and used them to
perform image recognition. The photonic circuit deploys non-volatile
phase change material (PCM) to emulate the synapses and exploits the
switching dynamics as a non-linear activation function. As highlighted
in section 5.2, integration of PCMs also lead to in-memory computing
functionality owing to their nonvolatile nature.~

For the (ii) direction, significant work has been done on a device level
to mimic individual components of the brain. Excitable lasers combining
different material platforms such as III-V compounds, and graphene have
been shown to demonstrate leaky integrate and fire-type characteristics
of a neuron \textsuperscript{\hyperref[csl:74]{74}}. Also, neurons based on optoelectronic
modulators have been shown in the literature. For synapses, photonic
devices combined with PCMs, amorphous oxide semiconductors, and 2D
materials have been used to demonstrate synaptic behavior such as
spike-time-dependent plasticity, memory, etc \textsuperscript{\hyperref[csl:74]{74},\hyperref[csl:270]{270},\hyperref[csl:271]{271}}.~~~~

In the following, we break down the challenge of building neuromorphic
photonic hardware \textbf{} to various subtopics, ranging from
increasing the fabrication tolerance of the photonic circuit to
co-packaging the optics and electronics. Then, we review the current
advances in those areas and provide an outlook on the future development
of neuromorphic photonic hardware.~~~

\subsubsection*{5.5.2. Challenges}

{\label{939574}}\par\null\selectlanguage{english}
\begin{figure}[H]
\begin{center}
\includegraphics[width=0.70\columnwidth]{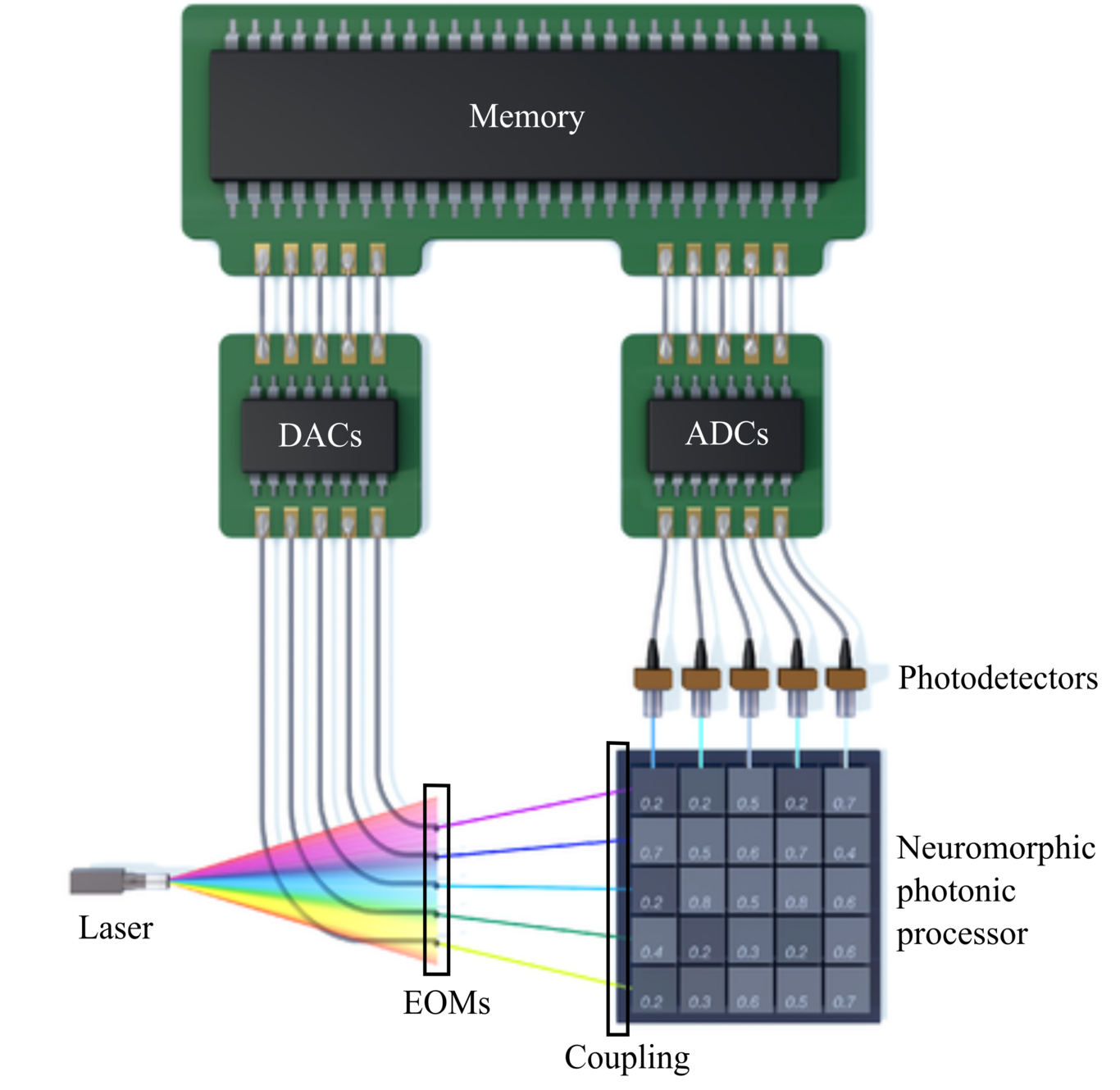}
\caption{{An illustration of an opto-electronic system capable of performing
computing operations. The electro-optic modulator (EOM) encodes the
input data from the memory or a real-time sensor into the light fields.
The encoded light fields are then coupled to a photonic processor which
emulates the neurosynaptic behavior. The results from the processor is
passed on to detectors which transform the signals back into the
electronic domain and are finally stored in a memory. ~
{\label{608518}}%
}}
\end{center}
\end{figure}

A major challenge is combining the various building blocks shown in
Figure 14. Silicon on insulator is the platform of choice for building
large circuits owing to the matured CMOS process flow and the high
refractive index contrast between the silicon waveguide and oxide
cladding. However, silicon has no second-order nonlinearity as it is
centrosymmetric. Further, silicon being an indirect bandgap material
cannot emit light. This strongly limits the options for implementing
nonlinear functions and spiking dynamics crucial for an all-optical
neural network. Therefore, most of the research on mimicking neurons is
focused on novel material platforms that support gain. A key challenge
is integrating those different material platforms. For example, a
circuit may deploy neurons based on III-V semiconductor heterojunction
and synapses that are built with PCMs on silicon. Therefore compact and
fabrication error tolerant optical interconnects are crucial for the
performance of the whole system.~

Apart from packaging various optical components, the electro-optic
interface imposes an additional challenge. Typically the input data is
provided by digital electrical signals whereas optical data processing
is analog. This requires analog-to-digital converters (ADCs) and
digital-to-analog converters (DACs) for digital systems to interface
with the chip as shown in Figure 14. For large circuits, co-packaging
electronics and photonics increases the footprint and cost
significantly. This negatively affects the throughput.

In addition to the above, fabrication imperfections will also impact the
performance of a photonic circuit. Components such as ring resonators,
cavities, interferometers, etc. employed in many photonic circuits are
designed to operate at a certain wavelength. But, due to factors such as
etching rate, sidewall angle, and surface roughness, the wavelength of
operation many times does not match with the design. Hence, in many
cases, active methods such as thermo-optic phase shifters are employed
to post-fabricate trim the wavelength. This results in unnecessarily
increased electronic circuitry adversely affecting the scalability of
the system.~~\textbf{~}

Lastly, a challenge that may be critical in the future is the
electro-optic modulator (EOM) efficiencies that depend on the material
properties and configuration. An important figure of merit for EOM
efficiency is VpiL. This merit shows the voltage that needs to be
applied and the length of the modulator required to obtain a pi phase
shift to the input. A smaller merit figure suggests increased power
efficiency and a compact footprint. As photonic neuromorphic circuits
are supposed to scale up in the future, the power budget and space
available on-chip will play an essential role in influencing the
designs.~

\subsubsection*{5.5.3. Potential~Solutions}

{\label{566176}}

Solutions addressing the challenges mentioned above lie on multiple
fronts. First, we discuss how the scalability can be improved from a
device-level perspective. The compact footprint, power efficiency, and
cascadability of the neurons are essential characteristics for improving
the scalability. In this regard, modulator-based neurons can be improved
by integrating with materials such as electro-optic polymers. These
materials have an order of magnitude higher r33 electro-optic
coefficient compared to bulk lithium niobate which has been
conventionally the material of choice for modulators. As a result,
electro-optic polymers integrated with silicon waveguides show very low
VpiL among fast modulators \textsuperscript{\hyperref[csl:272]{272}}. In addition, novel
materials such as epsilon-near-zero (ENZ) which are promising for
optical nonlinearity can also be explored \textsuperscript{\hyperref[csl:273]{273}}.
Nevertheless, for the widespread use of these devices, a better
understanding of the material properties and engineering efforts to
integrate them into the existing manufacturing process flow is
required.~

Particularly, integration techniques such as micro-transfer printing,
flip-chip bonding, and photonic wire bonding will play a key role. To
solve the problem of packaging with electronics, strategies such as
monolithic fabrication where the photonics and electronics are on the
same die need to be investigated. Foundries are now offering
multi-project wafer runs with these state-of-the-art packaging
techniques.~~~~

For improving the scalability of spike-based processing systems, another
class of neurons that is very promising is the vertical cavity surface
emitting lasers (VCSELs). VCSELs can integrate 100 picosecond-long
pulses and fire an excitable spike when the sum crosses a certain
threshold emulating biological neurons. Recently, it was shown that the
output of one layer of VCSEL neurons combined with a
software-implemented spiking neural network can perform 4-bit image
recognition \textsuperscript{\hyperref[csl:274]{274}}. In order to build the entire system on
hardware and perform larger experiments, 2D VCSEL arrays flip-chip
bonded on a silicon die can be examined.\textbf{~~}

Finally, to address the challenge of fabrication imperfections, passive
tuning approaches can be of interest which need no additional circuitry
and is non-volatile. One direction could be the use of phase change
materials such as GaS, and Sb2Se3 to correct for the variability in
photonic circuits \textsuperscript{\hyperref[csl:275]{275}}. These materials are very
interesting since their real part of the refractive index can be tuned
while keeping low absorption at telecom wavelengths. Another approach
for post-fabrication passive trimming could be to use an electron beam
or ion beam to change the material properties of the waveguide. This
method is also scalable as these tools are widely used in the
semiconductor industry.~

\subsubsection*{5.5.4. Concluding remarks}

{\label{492911}}

Applications like neuromorphic computing are particularly promising for
optics where its unique advantages (i.e high throughput, low latency,
and high power efficiency) can be utilized. Presently, there have been
instances in literature where different devices have been proposed to
emulate the individual characteristics of a neurosynaptic model.
However, there is a lot of scope for research in materials science to
pave the way toward more compact, cascadable, and fabrication-friendly
implementations. Further, large-scale networks are expected to scale in
the near future by integrating state-of-the-art packaging techniques
that are now available to research groups and startups.~

To summarize, the growth of integrated photonics has led to a resurgence
of optical computing not only as a research direction but also
commercially. It is exciting to see how the field of neuromorphic
photonics will shape as advancements in science and technology continue
to happen. \textbf{~ ~}

\vspace{0.5cm}

\subsection*{5.6 2D materials}

{\label{218997}}

\begin{quote}
\emph{Mario Lanza, Xixiang Zhang, Sebastian Pazos}
\end{quote}

\subsubsection*{5.6.1. Status}

{\label{281119}}

Multiple studies have claimed the observation of resistive switching
(RS) in two-dimensional layered materials (2D-LMs), but very few of them
reported excellent performance (i.e., high endurance and retention plus
low switching energy, time and voltages) in a reliable and trustable
manner, and in a device small enough to be attractive for
high-integration-density applications (e.g., memory, computation).

The best RS performance observed in 2D-LMs is based on out-of-plane
ionic movements. In such types of devices, the presence and quality of
the RS phenomenon mainly depend on three factors: the density of native
defects, the type of electrode used, and the volume of the dielectric
(thickness and area). In general, 2D-LMs with excellent crystallographic
structure (i.e., without native defects, such as those produced by
mechanical exfoliation) do not exhibit stable resistive switching.
Reference~\textsuperscript{\hyperref[csl:276]{276}} reported that mechanically exfoliated
multilayer MoS\textsubscript{2} does not show RS; only after oxidizing
it (i.e., introducing defects) it shows RS based on the migration of
oxygen ions. Along these lines, reference~\textsuperscript{\hyperref[csl:277]{277}} showed that
mechanically exfoliated multilayer hexagonal boron nitride (h-BN) does
not exhibit RS; instead, the application of voltage produces a violent
dielectric breakdown (DB) followed by material removal. The more violent
DB phenomenon in h-BN compared to MoS\textsubscript{2} is related to the
higher energy for intrinsic vacancies formation: \textgreater{}10 eV for
boron vacancies in h-BN versus \textless{}3 eV for sulfur vacancies in
MoS\textsubscript{2}. Some articles claimed RS in mechanically
exfoliated 2D-LMs, but very few cycles and poor performance were
demonstrated; those observations are more typical of unstable DB than
stable RS. Reference~\textsuperscript{\hyperref[csl:278]{278}} reported good RS in a crossbar
array of Au/h-BN/graphene/h-BN/Ag cells produced by mechanical
exfoliation, but in that study, the graphene film shows amorphous
structure in the cross-sectional transmission electron microscope
images. Hence, stable and high-quality RS based on ionic movement has
never been demonstrated in as-prepared mechanically exfoliated 2D-LMs.
This is something expected because ionic-movement-based RS is only
observed in materials with high density of defects (e.g., high-k
materials, sputtered SiO\textsubscript{2}), but not in materials with
low density of defects (e.g., thermal SiO\textsubscript{2}), as the
higher energy-to-breakdown forms an irreversible DB event.

On the contrary, 2D-LMs prepared by chemical vapor deposition (CVD) and
liquid phase exfoliation (LPE) have exhibited stable RS in two-terminal
memristors~\textsuperscript{\hyperref[csl:279]{279}} and three-terminal (memtransistors)
configurations~\textsuperscript{\hyperref[csl:280]{280}}, though in the latter the switching
mechanism is largely different. In two-terminal devices . In such cases,
the RS is enabled by the migration of ions across the 2D-LM. In
transition metal dichalcogenides (TMDs), the movement of chalcogenide
ions can be enough to leave behind a metallic path (often referred to as
conductive nanofilament or CNF) that produces the switching (similar to
oxygen movement in metal-oxides)~\textsuperscript{\hyperref[csl:276]{276}}. However, in h-BN
metal penetration from the adjacent electrodes is needed, as this
material contains no metallic atoms~\textsuperscript{\hyperref[csl:277]{277}}. In 2D-LMs
prepared by CVD and LPE methods ionic movement takes place at lower
energies (than in mechanically exfoliated ones) due to the presence of
native defects (mainly lattice distortions and impurities). The best
performance so far has been observed in CVD-grown
\textasciitilde{}6-nm-thick h-BN, as it is the only material enough
insulating and thick to keep low the current in the high resistive
state~{\textsuperscript{\hyperref[csl:73]{73}}}. This includes the coexistence of bipolar
and threshold regimes (the second one with highly-controllable
potentiation and relaxation), bipolar RS with endurances \textgreater{}5
million cycles (similar to commercial RRAM memories and phase-change
memories)~\textsuperscript{\hyperref[csl:281]{281}}, and ultra-low switching energies of
\textasciitilde{}8.8 zJ in threshold regime~\textsuperscript{\hyperref[csl:282]{282}}.
Moreover, high yield (\textasciitilde{}98\%) and low variability has
been demonstrated~\textsuperscript{\hyperref[csl:282]{282}}. In 2D-LMs produced by LPE or
other solution-processing methods~\textsuperscript{\hyperref[csl:283]{283}}, the junctions
between the flakes and their size play a very important role, and while
there is evidence of potentially good endurance, synaptic behaviour and
variability, sub-\selectlanguage{greek}μ\selectlanguage{english}m downscaling still hasn't shown equivalent
performance~\textsuperscript{\hyperref[csl:284]{284}}.

Apart from ionic movement, 2D-LMs can also exhibit RS based on
ferroelectric effect~\textsuperscript{\hyperref[csl:285]{285}}. A remarkable example is
-In\textsubscript{2}Se\textsubscript{3}, which has electrically
switchable out-of-plane and in-plane electric dipoles. Recent works have
demonstrated that RS in ferroelectric -In\textsubscript{2}Se is ensured
by three independent variables (polarization, initial Schottky barrier
and barrier change), and that it delivers multidirectional switching and
photon storage \textsuperscript{\hyperref[csl:286]{286}}. However, the endurance and retention
time are still limited to hundreds of cycles, and stable ferroelectric
RS at the single-layer limit remain unexplored.~ Finally, tunable
optoelectronic properties and unique electronic structure attainable
through 2D-LM heterostructures present enormous potential for near-/
in-sensor computing in neuromorphic systems. The responsiveness to
physical variables (light, humidity, temperature, pressure, torsion) of
2D-LM memristor and memtransistor devices allows to mimic biological
neurosynaptic cells (visual cortex, tactile receptors)
\textsuperscript{\hyperref[csl:287]{287}}.

\subsubsection*{5.6.2. Challenges and Potential
Solutions}

{\label{175997}}

The main challenge of RS devices (of any type) is to exhibit high
endurance in small devices. Many studies have reported RS in large
devices with sizes \textgreater{}10 \selectlanguage{greek}µ\selectlanguage{english}m\textsuperscript{2}, and claimed
that their devices are ``\emph{promising}'' for memory and computing
applications. This is a huge and unreasonable exaggeration; these two
applications require high integration density, as commercial devices for
those applications have sizes down to tens or hundreds of nanometers. It
should be noted that in ionic-movement-based RS devices the CNF always
forms at the weakest location of the sample; when the device size is
reduced the density and size of defects is (statistically) reduced,
which produces an increase of the forming voltage~\textsuperscript{\hyperref[csl:288]{288}}.
Hence, the CNF of smaller devices is wider due to the larger amount of
energy delivered during the forming. This has a huge effect on state
resistances, switching voltages, time and energy, as well as endurance,
retention time and device-to-device variability. In other words, the
fact that a large device (\textgreater{}10 \selectlanguage{greek}µ\selectlanguage{english}m\textsuperscript{2})
exhibits good RS does not mean that a small device (\textless{}1
\selectlanguage{greek}µ\selectlanguage{english}m\textsuperscript{2}) made with the same materials will also exhibit
it; hence, RS ``\emph{promising for memory and computatio}n'' is only
the one that is observed in devices with sizes of tens/hundreds of
square nanometers. ~

Taking this into account, the main challenge in 2D-LMs based devices is
to observe RS in small devices, and the most difficult figure-of-merit
to obtain is (by far) the endurance. Reference~\textsuperscript{\hyperref[csl:289]{289}}
demonstrated good RS in 5 \selectlanguage{greek}µ\selectlanguage{english}m \selectlanguage{ngerman}× 5 \selectlanguage{greek}µ\selectlanguage{english}m Au/h-BN/Au devices, in which the
h-BN was \textasciitilde{}6-nm-thick and grown by CVD method; however,
when the size of the devices was reduced to 320 nm \selectlanguage{ngerman}× 420 nm the yield
and the number of devices observed was very limited. The main issue was
the current overshoot during the switching, which takes place randomly
and produced irreversible DB in most devices. Similarly,
solution-processed Pt/MoS\textsubscript{2}/Ti
devices~\textsuperscript{\hyperref[csl:290]{290}} showed excellent performance across all
figures-of-merit observed in 25 \selectlanguage{greek}µ\selectlanguage{english}m\textsuperscript{2} devices, but such
performance has not been reported for 500 nm \selectlanguage{ngerman}× 500 nm devices patterned
via electron beam lithography. In this case, the large size of the
nanoflakes (slightly below 1 \selectlanguage{greek}µ\selectlanguage{english}m minimum) may be imposing an intrinsic
scaling limitation. Meanwhile, the scaling and overshoot problem was
problem was solved in reference~\textsuperscript{\hyperref[csl:73]{73}}
integrating~CVD-grown h-BN right on top (via wet-transfer method) of a
silicon complementary metal-oxide-semiconductor (CMOS) transistor, which
acted as instantaneous current limitation. Moreover, this approach
brings the advantage of a very small device size (in
reference~\textsuperscript{\hyperref[csl:73]{73}} it was only 0.053 \selectlanguage{greek}µ\selectlanguage{english}m\textsuperscript{2},
as the bottom electrode of the RS device is the via from one of the
metallization levels). The heterogeneous integration of the 2D-LMs at
the back-end-of-line (BEOL) wiring of silicon microchips could be a good
way of testing materials for RS applications and directly integrating
selector devices with each memristor (into one transistor-one memristor
1T1M cells), which is fundamental for the realization of large
memeristive synapse arrays --- all state-of-the-art demonstrations of
memristive neural accelerators based on mature memristor devices use
1T1M cells or differential implementations of such (2T2M, 4T4M). So far
these CMOS testing vehicles for RS materials are mainly employed by the
industry; in the future, wide spreading this type of testing vehicles
among academics working in the field of RS could improve the quality of
the knowledge generated.~~In addition, these devices may benefit from
common practices in the field of silicon microchip manufacturing, such
as surface planarization, plug deposition and high-quality, thick
interconnect techniques.

Next steps in the field of 2D-LMs for RS applications consist on
improving the materials quality to achieve better reproducibility of the
experiments (from one batch to another) and adjust the thickness and
density of defects to achieve better figures-of-merit in nanosized RS
devices while growing 2D-LM at the wafer-scale \textsuperscript{\hyperref[csl:291]{291}}.
Recent studies successfully synthesized large-area single-crystal 2D-LMs
via CVD~\textsuperscript{\hyperref[csl:292]{292}}, although in most cases it is only
monolayer. However, monolayer 2D-LMs are less than 1-nm-thick, and when
they are exposed to an out-of-plane electrical field very high leakage
current is generated even if no defects are present, which increases a
lot the current in HRS and the energy consumption of the device.
Reference~\textsuperscript{\hyperref[csl:293]{293}} recently presented the synthesis of
single-crystal multilayer h-BN using scalable methods, but controlling
the number of layers is still difficult. Electrical studies in such
types of single-crystal multilayer samples should be conducted.
Improving manipulating methods to prevent the formation of cracks during
transfer is also necessary, although it is worth mentioning that
multilayer h-BN materials are more mechanically stable than monolayers.

Recent demonstration of vector-matrix multiplication using
MoS\textsubscript{2~}memtransistors~\textsuperscript{\hyperref[csl:294]{294}} is a promising
advance in terms of a higher-level functional demonstration, though the
fundamental phenomenon exploited is the well-known floating gate memory
effect, not unique to 2D-LM themselves. Meanwhile, understanding the
role of flake size in the functionality of solution-processed 2D-LM
two-terminal synaptic devices is critical to address the true scaling
limitations of such approach, key aspect to define potential realistic
applications in neuromorphic systems. On the other hand, sensing
capabilities emerge with great potential for biological synaptic
mimicking. The full potential of different 2D-LM material
heterostructures and memtransistors opens a huge design-space worth of
exploration. In that sense, the complex physical characteristics offered
by different 2D-LM hold the potential not only for basic neuromorphic
functionality but also for higher-order complexity. This could be
exploited to achieve high-complexity neural and synaptic
functions~\textsuperscript{\hyperref[csl:101]{101}}, more closely mimicking actual biological
systems. However, in parallel to elucidating the physical properties and
capabilities of these material systems, efforts should be put into
strengthening the quality of the reported results, focusing on proper
characterization methods, reliable practices, and statistical
validation. 

\subsubsection*{5.6.3. Concluding remarks}

{\label{402521}}

Leading companies like TSMC, Samsung, IBM and Imec have started to work
with 2D-LMs, but mainly for sensors and transistors. In the field of
2D-LMs based neuromorphic devices most work is being carried out by
academics. In this regard, unfortunately, many studies make a simple
proof-of-concept using a novel nanomaterial without measuring essential
figures-of-merit like endurance, retention and switching time. What is
even worse, in many cases the studies employ unsuitable characterization
protocols that heavily overestimate the performance (the most popular
case is the erroneous measurement of endurance~\textsuperscript{\hyperref[csl:295]{295}}),
withholding information regarding the failure mechanisms that lead to
certain performance metrics not being achieved on some devices . This
working style often result in articles with striking numbers (i.e.,
performance), but those are unreliable, and it is really bad for the
field because it creates a hype of expectations and disillusion among
investors and companies. The most important is that the scientists
working in this field follow a few considerations: i) always aim to show
high performance in small (\textless{}1 \selectlanguage{greek}µ\selectlanguage{english}m\textsuperscript{2}) devices
fabricated using scalable methods (even better if they are integrated on
a functional CMOS microchip, not on an unfunctional SiO\textsubscript{2}
substrate), ii) measure all the figures-of-merit of several
(\textgreater{}100) memristive devices for the targeted application
(this may vary depending on the application)~\textsuperscript{\hyperref[csl:296]{296}}, iii)
define clearly the yield-pass criteria and the yield achieved, as well
as the device-to-device variability observed, and iv) whenever a failure
mode is observed preventing of reaching a desired figure-of-merit,
clearly convey it to maximize the probabilities of finding a solution .

\vspace{0.5cm}

\subsection*{6. Materials challenges and
perspectives}

{\label{762792}}

\begin{quote}
\emph{Stefan Wiefels, Regina Dittmann}
\end{quote}

\subsection*{6.1 Materials challenges}

{\label{380858}}

For the neuromorphic computing approaches addressed in chapter 3, the
use of emerging memories based on novel materials will be key in order
to improve their performance and energy-efficiency.~ ~This chapter
discusses the most relevant properties and challenges for different use
cases and how they relate to the respective materials properties.
However, it is important to note that a dedicated co-development of
material with the read- out and write algorithm and circuitry will be
required ~in order to advance the field.

\par\null\selectlanguage{english}
\begin{figure}[H]
\begin{center}
\includegraphics[width=0.70\columnwidth]{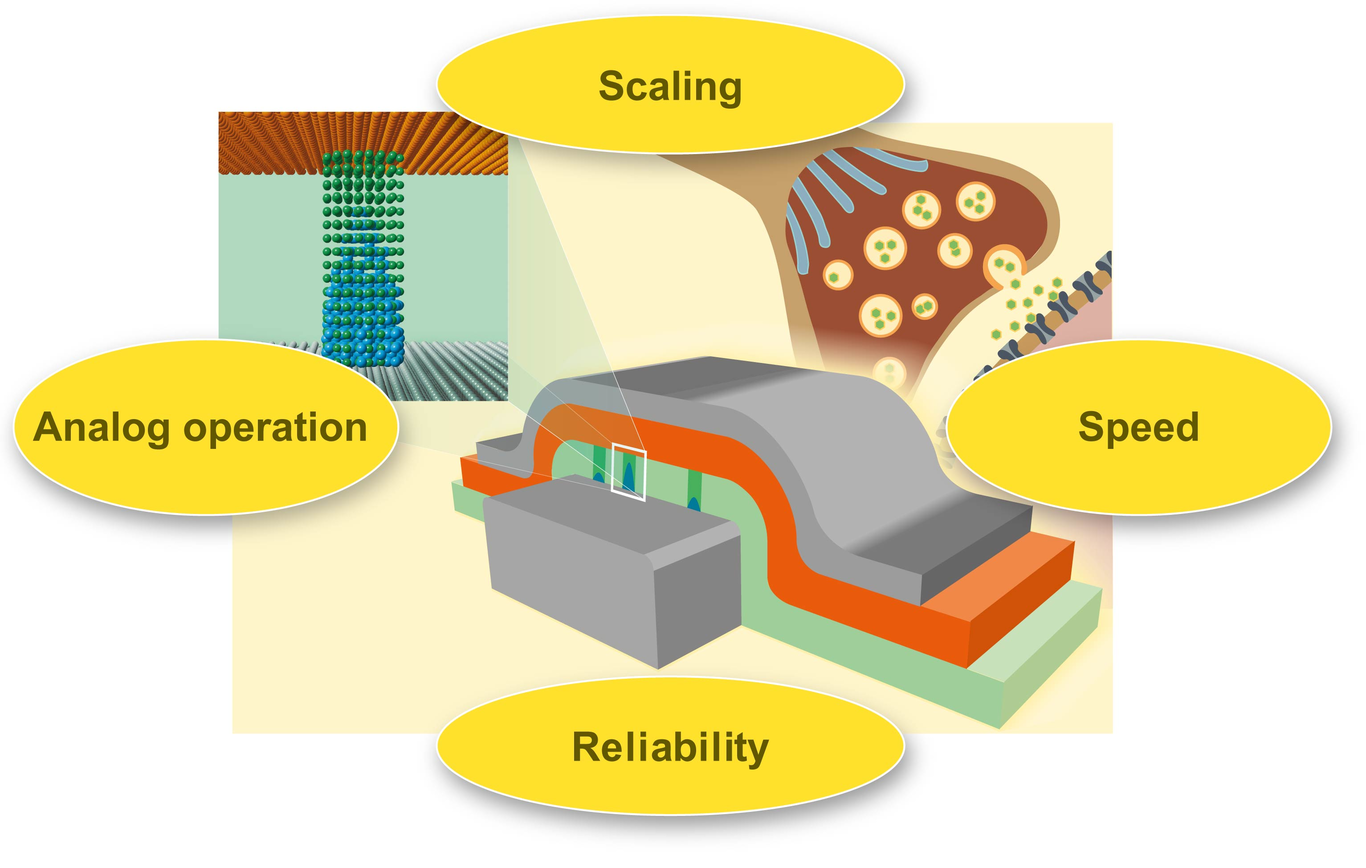}
\caption{{Materials challenges for neuromorphic computing. Novel NVMs need to be
scalable, fast, reliable and allow for analog operation.
{\label{176216}}%
}}
\end{center}
\end{figure}

\subsubsection*{6.1.1. ~Scaling~}

{\label{838199}}

One main driving force to use emerging materials and devices is to gain
space and energy efficiency by the fabrication of highly dense cross-bar
arrays. For STT-MRAM, scaling down to 11 nm cells has been demonstrated
as well as the realization of 2 Mb embedded MRAM in 14 nm FinFET
CMOS~\textsuperscript{\hyperref[csl:297]{297}}. However, due to the small resistance ratio of
2-3, the read-out of magnetic tunnel junctions (MTJs) is more complex
than for other technologies. Nevertheless, a 64 x 64 MTJ array,
integrated into 28 nm CMOS, has recently been
realised~\textsuperscript{\hyperref[csl:76]{76}}. Advancement from the material side will be
needed in order to increase the resistance ratio of MTJs in the future.

For ferroelectric HfO\textsubscript{2}-based devices, the main challenge
with respect to scaling is to decrease the thickness reliably in order
to enable 3D capacitors with 10 nm node and to obtain a uniform
polarisation at the nanoscale of a material that currently still
contains a mixture of different phases. Therefore, ultrathin films with
the pure ferroelectric orthorhombic phase and without any dead-layers at
the interfaces will be key to approach the sub-20 nm regime of
hafnia-based ferroelectric devices \textsuperscript{\hyperref[csl:9]{9}}.

PCM devices can be fabricated on the sub-10 nm scale \textsuperscript{\hyperref[csl:298]{298}}.
The limiting factor for CMOS integrated PCM devices is the high RESET
current which is required to implement larger access transistors
\textsuperscript{\hyperref[csl:299]{299}}. Commercially available ReRAM cells with conventional
geometries have been co-integrated on 28 nm CMOS technology. By
employing~ a sidewall technique and nanofin Pt electrodes, small arrays
with 1~nm~\selectlanguage{ngerman}×~3~nm HfO\textsubscript{2} \textsuperscript{\hyperref[csl:299]{299}} cells ~and 3 ×
3 arrays of Pt/HfO\textsubscript{2}/TiO\textsubscript{x}/Pt cells~ with
a 2~nm feature size and a 6~nm half-pitch ~have been fabricated,
respectively \textsuperscript{\hyperref[csl:300]{300}}.

With respect to ultimate scaling, the loss of oxygen to the environment
might pose limitations to the retention times for ReRAM devices scaled
in the sub-10 nm regime \textsuperscript{\hyperref[csl:301]{301}}. However, filaments in the
size of 1-2 nm can be stable if they are stabilized by structural
defects such as grain boundaries or dislocations. Therefore, finding
materials solution for confining oxygen vacancies to the nanoscale might
retain the required retention for devices in the few nm scale.

\subsubsection*{6.1.2. Speed~}

{\label{413689}}

Although the extensive parallelism leads to high demands for scaling, it
is considered an advantage as it makes the race for ever increasing
clock frequencies obsolete~\textsuperscript{\hyperref[csl:302]{302}}. In contrast, the
operation speed is closely linked to the respective application, i.e.
the timing is based on real physical time \textsuperscript{\hyperref[csl:302]{302}}. As
signals processed by humans are typically on a time scale of
milliseconds or longer, the expected speed benchmark is well below the
reported speed limits of emerging NVMs. Nevertheless, it is reasonable
to understand the ultimate speed limits of NVM concepts in order to
estimate maximum learning rates, to explore the impact of short spiking
stimulation. Furthermore, novel computing concepts, as discussed in
chapter 7, might still benefit from higher clock frequencies. For MRAM,
reliable 250 ps switching has been demonstrated by using double
spin-torque MTJs which consist of two reference layers, a tunnel
barrier, and a non-magnetic spacer ~\textsuperscript{\hyperref[csl:303]{303}}. FeRAM arrays
have successfully been switched with 14 ns at 2.5 V. Ferroelectric field
effect transistors (FeFETs) have been shown to switch with \textless{}
50 ns pulses in 1 Mbit memory arrays \textsuperscript{\hyperref[csl:9]{9}}. PCM devices
can be switched with pulses \textless{} 10 ns \textsuperscript{\hyperref[csl:298]{298}}. In
general, their speed is limited by the crystallization time of the
material. It has been shown exemplarily on
Ge\textsubscript{x}Sn\textsubscript{y}Te samples that this time can be
tuned in a broad range of 25 ns up to 10 ms by adjusting the material
composition \textsuperscript{\hyperref[csl:304]{304}}. Thus, it has a high potential to match
the operation time of an NC system to the respective application. For
VCM ReRAM SET and RESET switching with 50 ps and 400 ps has been
demonstrated \textsuperscript{\hyperref[csl:305]{305}}. Both are so far limited by extrinsic
effects and device failure modes rather than by intrinsic physical rate
limiting steps.~~

\subsubsection*{6.1.3. Reliability~}

{\label{195789}}

Independent of the application, the reliability of the memory technology
has to be taken into account. In the case of implementing NVMs as
artificial synapses, the requirements of learning and inference phase
have to be distinguished. Whereas the endurance is more relevant for
learning schemes, the stability of the programmed state, i.e. the
retention and robustness against read disturb have to be sufficient for
reliable inference operations.

\subsubsection*{6.1.4. Endurance~}

{\label{690949}}

While MRAM has in principle unlimited endurance, all memristive devices
that are based on the motion or displacement of atoms such as ReRAM, PCM
and ferroelectric systems have limited endurance. For silicon-based
FeFETs the endurance is typically in the order of 10\textsuperscript{5},
which is mainly limited by a dielectric breakdown in the
SiO\textsubscript{2} at the Si-HfO\textsubscript{2} interface
\textsuperscript{\hyperref[csl:9]{9}}. Regarding the endurance of VCM ReRAM, it has been
demonstrated with convincing statistics that \textgreater{} 10\^{}6
cycles are realistic. Some reports suggest maximum cycle numbers of more
than 10\^{}10 \textsuperscript{\hyperref[csl:295]{295}}. Depending on the material system,
various failure mechanisms for the endurance are discussed. The
microstructure of the switching material might degrade or be
irreversibly penetrated by metallic atoms \textsuperscript{\hyperref[csl:9]{9}}. In VCM
ReRAM an excessive generation of oxygen vacancies was discussed as
endurance limiting factor \textsuperscript{\hyperref[csl:306]{306}}. Novel material solutions
which confine ions to the intended radius of action might be a pathway
to increase the endurance of ReRAM devices. For PCM, it was suggested to
implement multiPCM synapses. Arbitration over multiple memory elements
might circumvent endurance as well as variability issues
\textsuperscript{\hyperref[csl:9]{9}}.

Typical limitations with respect to a reliable operation of
ferroelectric memories is the so called wake-up effect which causes an
increasing polarization after a few cycles and the fatigue resulting in
a decrease of the polarisation for high cycle numbers. Both are induced
by the motion of defects such as oxygen vacancies and will have to be
tackled in the future by intense material research in this field.

\subsubsection*{6.1.5. Retention~}

{\label{782167}}

After training, the state of the non-volatile memory synapse is required
to be stable for 10 years at an operating temperature of 85 °C. However,
for many applications in the field of neuromorphic applications the
requirement is much more relaxed in particular for the training phase.
From a thermodynamical point of view, the states in ferroelectric or
ferromagnetic memories might both be stable. In contrast, ReRAM and PCM
devices store information in the configuration of atoms where both LRS
and HRS are metastable states and the retention is determined by
material parameters such as the diffusion coefficient of the respective
species \textsuperscript{\hyperref[csl:9]{9}}. Here, the degradation is not a digital
flipping of states but a gradual process. For PCM, the drift of the
resistance state is caused by the structural relaxation of the
melt-quenched amorphous phase \textsuperscript{\hyperref[csl:299]{299}}. Apart from a drifting
of the state, a broadening of the programmed state distribution (e.g.
resistance) is typically observed for ReRAM \textsuperscript{\hyperref[csl:307]{307}}.
~Further, since analogue or multi-level programming are highly relevant
for NC, it should be considered that intermediate resistance states
might have a reduced retention compared to the edge cases of high and
low resistive states as demonstrated for PCM devices
\textsuperscript{\hyperref[csl:9]{9}}.

\subsubsection*{6.1.6.~Read disturb~}

{\label{289455}}

During inference, frequent reading of the memory elements is required
which should not change the learned state. For a bipolar ReRAM memory, a
read disturb in the HRS/LRS occurs mainly when reading with a SET/RESET
polarity since the read-disturb can be considered as an extrapolation of
the SET/RESET kinetics to lower voltages. Nevertheless, the HRS state in
bipolar filamentary VCM has been demonstrated by extrapolation to be
stable for years at read voltages up to 350 mV \textsuperscript{\hyperref[csl:308]{308}}.

\subsubsection*{6.1.7. Variability~}

{\label{134582}}

Variability is most pronounced for systems which rely on the stochastic
motion and redistribution of atoms such as ReRAM and PCM.~ Here, the
variability from device to device (D2D), from cycle to cycle (C2C) as
well as even from one read to the next (R2R) have to be distinguished.
By optimizing fabrication processes, the D2D variability can be kept
comparatively low. In contrast, the C2C variability for filamentary
resistive ReAM and PCM can be significant due to the randomness of
filament~\textsuperscript{\hyperref[csl:308]{308}} or crystal~\textsuperscript{\hyperref[csl:299]{299}} growth,
respectively. However, using smart programming algorithms, the C2C
variability can be very well reduced to a minimum ~\textsuperscript{\hyperref[csl:308]{308}}.
But, R2R variations remain in form of read noise in filamentary VCM. It
is typically attributed to the activation and deactivation of traps or
the random redistribution of defects ~\textsuperscript{\hyperref[csl:308]{308}} and strongly
depends on the material~\textsuperscript{\hyperref[csl:309]{309}}. For PCM, R2R variations are
caused by 1/f noise and temperature induced resistance variations. One
approach to address these issues as well as the drift is to use the so
called projected phase change memory with a non-insulating projection
segment in parallel to the PCM segment.~

Although the variability is a challenge for storage applications, it
might be possible to design NC systems to exploit it
\textsuperscript{\hyperref[csl:302]{302}}. In the end, a thorough understanding of the
intrinsic variability might enable to match NC application and material
\textsuperscript{\hyperref[csl:309]{309}} .

\subsubsection*{\texorpdfstring{6.1.8 Analog
operation\emph{~}}{6.1.8 Analog operation~}}

{\label{950563}}

For most computing concepts described in Chapter 3, their operation with
binary memory devices is strongly limited and the possibility to adjust
multiple states is of crucial importance. For devices with
thermodynamically stable states such as ferroelectric or magnetic
memory, intermediate states rely on the presence of domains. As a
result, the performance strongly depends on the specific domain
structure and scaling might be limited by the size of the domains.
Nevertheless, multilevel switching has been demonstrated by fine tuning
of programming voltages for both FTJ and FeRAM~\textsuperscript{\hyperref[csl:310]{310}}.

For ReRAM and PCM, the metastable intermediate states have to be
programmed in a reliable manner. Since these states are kinetically
stabilized during programming, the success depends strongly on the
switching kinetics of the specific system, the operation regime and the
intrinsic R2R variability of the material. For PCM devices, intermediate
states can be addressed by partial reset pulses which result in partial
amorphisation. As a result of the crystallization kinetics, a gradual
crystallization can be obtained by consecutive pulses.

Filamentary ReRAM devices usually undergo an abrupt SET which is caused
by the self-accelerating, thermally-driven filament formation. ~However,
in a limited operation regime intermediate states can be obtained with a
good control of the SET current, fast pulses or if the kinetics are
slowed down. This is the case for non-filamentary systems which show a
very pronounced gradual behavior for both SET and RESET
\textsuperscript{\hyperref[csl:300]{300}}.

Furthermore, resistive switching devices with purely electronic
switching mechanisms like trapping and de-trapping of electrons at
defects states might be promising for analog operation
\textsuperscript{\hyperref[csl:311]{311}}.

\vspace{0.5cm}

\subsection*{6.2. Characterisation
techniques}

{\label{803061}}

\begin{quote}
\emph{Adnan Mehonic, Wing H Ng, Mark Buckwell, Horatio RJ Cox, Daniel J
Mannion, Anthony J Kenyon}
\end{quote}

\subsubsection*{6.2.1.Status}

{\label{261661}}

Memristive devices pose challenges to the experimentalist both in
investigating the physics underpinning device behaviour and in
optimising functionality. The wide range of physical phenomena involved
in memristance -- from metal diffusion in dielectrics to Mott
metal-insulator transitions, phase changes and the formation of oxygen
vacancy filaments -- requires comprehensive physical, electrical and
chemical characterisation and even the development of novel analytical
techniques~\textsuperscript{\hyperref[csl:312]{312}}. Here, rather than an exhaustive
literature survey we provide examples to illustrate recent progress in
characterisation of memristive materials and devices. While we
concentrate on oxide-based RRAM materials and devices for reasons of
space, most techniques reviewed here are applicable to other memristor
types (PCM, MRAM, FeRAM), which require similar characterisation of
structural, chemical and electrical changes occurring in in devices as a
result of operation. The challenges presented to the experimentalist,
particularly when it comes to structural and chemical analysis, are
largely similar across all memristive devices.

\subsubsection*{6.2.2.Challenges and Potential
Solutions}

{\label{820180}}

Looking first at resistance switching materials and devices, we see that
early work from the 1960s on dielectric breakdown suggested the
formation of conductive filaments in electrically biased
oxide~\textsuperscript{\hyperref[csl:313]{313}}~and, while experimental techniques at the time
were unable to image them, the authors correctly surmised their
existence. The small sizes of these filaments, which can be of the order
of a few nanometres in diameter, makes studying their formation and
disruption difficult. This is particularly true for oxygen vacancy
filaments in oxides; the minimal contrast between the oxide matrix and
the oxygen-deficient filament when imaged using electron beam techniques
such as TEM-EELS means that there are few direct observations of such
filaments. This contrasts with several published TEM studies of metal
filaments in oxides, including seminal work by Yang et al
~\textsuperscript{\hyperref[csl:314]{314}}, which demonstrated field-driven movement of silver
ions through SiO\textsubscript{2},
Al\textsubscript{2}O\textsubscript{3}~and amorphous silicon to form
dendritic conductive filaments. The large contrast between metal ions
and oxide matrix makes it a more tractable problem to image individual
conductive filaments. Subsequent work demonstrated different filament
growth modes that depend on the relative magnitudes of metal ion
mobility and applied field~\textsuperscript{\hyperref[csl:315]{315}}. Work by Waser et
al~\textsuperscript{\hyperref[csl:316]{316}}~details earlier TEM work on electrical and
physical characterisation of resistive switching dielectrics, including
observation dating from as early as 1976 of silver dendrites formed in
AgS under the application of an external field. Here again, the contrast
between the silver filaments and the surrounding matrix provides a
significant advantage. It is worth noting that the use of TEM
measurements to characterise phase change (PCM) materials and devices is
rather easier than in the case of oxide-based RRAM as the contrast
between amorphous and crystalline phases of PCM materials is easier to
detect, though in-situ and in-operando measurements can pose significant
challenges thanks to the fast switching speeds of PCM devices
\textsuperscript{\hyperref[csl:317]{317}}.

The difficulty of imaging oxygen vacancy filaments using electron beam
techniques can be overcome using conductive atomic force microscopy
(CAFM) tomography (``scalpel AFM''), a review of which can be found in
\textsuperscript{\hyperref[csl:318]{318}}. In this technique, sequential CAFM scans of a sample
surface imaged using a conductive diamond tip contacting the sample with
sufficient force to scrape away the surface provide layer-by-layer
conductivity maps that, when stacked, provide three-dimensional images
of conductive regions within the oxide. Such studies reveal that
electroforming generates large-scale changes beneath device top
electrodes, modifying the conductivity of large volumes of material,
while one or more highly localised conductive filaments bridge the
inter-electrode gap. The technique also reveals details of the internal
microstructure of the oxide, showing, for example, the columnar
structure of sputter-deposited oxides, as the edges of columns are more
conductive than their cores \textsuperscript{\hyperref[csl:319]{319}}. It should be noted that
this technique only maps conductive regions that are connected to the
bottom electrode, so may be better thought of as a measure
of~\emph{connectivity}~rather than of regions of high conductivity.

The need to apply multiple analysis techniques to probe the movement of
oxygen within, and emission of oxygen from, oxides under electrical
stress was demonstrated by Mehonic et al~\textsuperscript{\hyperref[csl:320]{320}}. TEM, EELS,
CAFM, XPS and mass spectrometry measurements of oxygen emission from
samples under electrical stress combine with atomistic modelling to give
a fuller picture of the dynamics of oxygen movement and its role in
resistance switching. It is clear from such measurements that
electrically biasing oxides -- particularly those with some structural
inhomogeneity -- can drive large-scale changes in stoichiometry,
reinforcing the CAFM tomography results referred to above. It has been
known for some time that this can cause surface distortions and
localised bubbling of both electrode and oxide
surfaces~\textsuperscript{\hyperref[csl:313]{313}}~\textsuperscript{\hyperref[csl:321]{321}}. While work reported
in~\textsuperscript{\hyperref[csl:320]{320}}~examines such features using AFM and TEM, the
question of how mobile oxygen interacts with electrode materials is
partly addressed in recent work by Cox et al~\textsuperscript{\hyperref[csl:322]{322}}.
Oxide-based RRAM devices rely on the repeated reduction and oxidation of
the switching oxide. To compete with high density Flash, at least
10\textsuperscript{4}~cycles are required, and for many applications
more than 10\textsuperscript{7}~are needed. For such high numbers of
cycles, oxygen should not be lost from the switching region around the
conductive filament, implying the need for an oxygen reservoir that can
both accommodate and release oxygen under appropriate electrical biases.
This may be within the oxide, at an oxide/electrode interface, or within
one or other electrode. The need to measure oxygen movement is critical.
Cox et al~\textsuperscript{\hyperref[csl:322]{322}}~demonstrate that both the electrode
material and the microstructure of the oxide layer influence the
reversibility of oxygen movement. In the case of electrode metals with
high oxygen affinities, oxygen moves easily from the oxide into the
electrode, but its movement back again varies significantly between
metals. Molybdenum, for example, both accepts and releases oxygen
readily when the bias polarity is reversed, while titanium is easily
oxidised when positively biased (and when neutral) but does not release
oxygen back again when negatively biased. Platinum, having a very low
electron affinity, does not accept or release oxygen.

Oxide porosity and sensitivity to moisture is an important factor in
resistance switching both by metal diffusion (extrinsic switching) and
oxygen vacancy formation (intrinsic switching)~\textsuperscript{\hyperref[csl:323]{323}}. The
presence of moisture in the switching oxide can lead to highly variable
resistance switching behaviour. While the origin of such effects remains
somewhat unclear, electrical measurements must be interpreted with care
in the presence of moisture. The difficulty in measuring hydrogen
content in materials and devices reliably and quantitatively is a
particular challenge, not only for RRAM, but also other memristive
devices. Hydrogen, even in relatively low concentration, can affect the
electrical properties of oxides and electrode stacks, as has been
recognised for decades in the CMOS community. However, there have been
very few studies of the role of hydrogen in memristive devices. At the
same time, there is considerable interest in reducing variability and
increasing stability in memristors. One cannot help but suppose that
detailed studies of hydrogen, and control of its presence, could help
these efforts.

Electrical characterisation of memristive devices poses other
challenges. A critical issue in RRAM is electroforming conductive
filaments. While some materials and devices are forming-free, the
majority require an initial conditioning step in which a voltage higher
than the normal programming (set or reset) voltages is applied to form a
conductive filament that will subsequently be partially oxidised (reset)
and reduced (set) by a sequence of voltage sweeps or pulses.
Electroforming causes an abrupt drop in oxide resistance, which can span
several orders of magnitude, over a timescale of nanoseconds or shorter.
Without an appropriate limit on delivered current, this can destroy the
device by irreversible oxide breakdown due to Joule heating. It is
therefore essential to implement fast current limiting during
electroforming. Response times of standard characterisation instruments
are generally too long, so current overshoots can over-stress the oxide,
leading to breakdown or to conductive filaments too large (hence too
strong) to reset. Consequently, full electrical characterisation of
devices requires integrated current limiting devices such as transistors
or series resistors. Care must be taken to avoid parasitic capacitances
or inductances, so the most reliable methods involve on-chip series
resistors or transistors in the 1R1R or 1T1R configurations, where the
initial R or T labels refer to the integrated current limiter (resistor
or transistor), and the second R refers to the resistance switching
element.

On the other hand, electrical characterisation of RRAM devices can
reveal important clues to the physical mechanisms responsible for
resistance switching. Careful analysis of current/voltage curves can
indicate a range of electron transport mechanisms, including various
forms of tunnelling (direct, trap-assisted, Fowler-Nordheim), thermally
assisted transport (Poole-Frenkel) and Ohmic conduction. These, in turn
provide evidence for microscopic processes such as charge
trapping/detrapping, formation of Schottky barriers or various
interface-related electronic states. However, more than one transport
mechanism may contribute as, for example, currents may flow in parallel
both through a conductive filament and through a highly defective
surrounding oxide~\textsuperscript{\hyperref[csl:324]{324}}. In the case of nanoscale filament
formation, the thinnest point of the filament can behave as a quantum
constriction, allowing only currents that are multiples of the
conductance quantum,~\emph{G}\textsubscript{\emph{0}}, to flow. Such
effects can be seen easily at room temperature \textsuperscript{\hyperref[csl:324]{324}}.
Reviews of electrical characterisation techniques for resistance
switching devices can be found in ~~\textsuperscript{\hyperref[csl:295]{295}}and
\textsuperscript{\hyperref[csl:325]{325}}.

\subsubsection*{6.2.3. Concluding Remarks}

{\label{836284}}

The inherently interdisciplinary approach that is needed to fully
characterise memristive devices poses challenges to the experimentalist.
A wide range of techniques are required, which is often beyond the
capabilities of a single laboratory, and in some cases these techniques
are operating close to their limits. While there have been significant
advances in characterisation (CAFM tomography, for example), more work
is needed -- for instance, to better characterise the role of hydrogen
in resistance switching. Where characterisation has been most successful
has been when there is close collaboration, not only between experts in
different experimental techniques, but also with theorists who provide
models to interpret experimental results.

\vspace{0.5cm}

\subsection*{6.3 Comparison between different material
systems}

{\label{264111}}

\begin{quote}
\emph{Yuchao Yang, Yingming Lu}
\end{quote}

\subsubsection*{6.3.1. Status}

{\label{521141}}

Varied types of memristors can be realized based on the abundant
resistive switching mechanisms in different materials. Each type of
memristor has certain characteristics (such as power consumption,
switching speed, etc.) that are suited for specific applications, while
at the same time, different materials also have their own shortcomings
and limitations. A clear understanding of the respective advantages and
shortcomings of each material is key to its development and
applications. \textbf{}

Among various types of resistive switching materials, transition metal
oxides (TMOs) are among the most widely used materials due to their rich
resistive switching mechanisms and characteristics. The retention time
of TMO-based memristive devices, which indicates how long the resistive
state can be maintained after electrical stimulation, is distributed in
a wide range from \selectlanguage{greek}μ\selectlanguage{english}s to years. According to the retention time, TMO can
be generally divided into non-volatile TMO and Mott TMO.

The resistive switching of non-volatile TMO, such as
HfO\textsubscript{x} and TaO\textsubscript{x}, originates from the
migration and redox reactions of oxygen ions or vacancies driven by
external electric field or thermal effects, which in turn create or
destroy conductive filaments between the electrodes. The filaments can
exist stably for a long time, and the continuous electrical modulation
of geometric characteristics of the filaments such as their lengths or
diameters etc., results in multilevel resistive states. Therefore, the
non-volatile TMO can be used to imitate the long-term plasticity (LTP)
of biological synapses \textsuperscript{\hyperref[csl:170]{170}}, and~can accelerate the
computationally intensive matrix-vector multiplication (MVM) in
artificial neural networks. Furthermore, due to the mature and
CMOS-compatible manufacturing process, a variety of in-memory computing
chips based on non-volatile TMO have been demonstrated
\textsuperscript{\hyperref[csl:56]{56}}. One of the challenges of non-volatile TMO in
circuit applications is the existence of the forming process, which
requires a high voltage to initialize the TMO layer and subsequently
increases the requirements for the voltage and robustness of peripheral
circuits. In addition, due to the switching mechanism of non-volatile
TMO, the conductance change during the programming process usually shows
nonlinear characteristics, along with obvious variations and noises,
which increases the difficulty of programming, for example by
necessitating closed-loop write-and-verify programming.

There are many other similar TMO systems, such as WO\textsubscript{x}
and TiO\textsubscript{x}. Depending on the robustness of the formed
filaments, the retention time can be gradually reduced as a result of
filament dissolution, and the conductance of volatile TMO can undergo a
continuous decay process. This can effectively map information in time
series into high-dimensional vectors, which is widely used in
reservoir~\textsuperscript{\hyperref[csl:185]{185}} for image classification or time series
prediction.

Mott TMOs, mainly including VO\textsubscript{2} and
NbO\textsubscript{2}, show high resistance in body-centred tetragonal
(BCT) or monoclinic (M) phases at low temperatures. Once the temperature
of Mott TMO exceeds a threshold, the reversible Mott transition occurs
and results in structural transition into the rutile (R) phase,
accompanied by a significant resistance drop. Based on threshold
switching, Mott TMO is widely used in artificial neurons for
constructing neuromorphic computing or sensory systems
\textsuperscript{\hyperref[csl:326]{326}}. Meanwhile, the nonlinear transport mechanism of
NbO\textsubscript{2} shows high-order complexity during current sweeps,
which can effectively realize the rich dynamics of biological neural
systems~\textsuperscript{\hyperref[csl:7]{7}}. However, as the Mott transition is closely
related to temperature, the operation of Mott TMO can be affected by
ambient temperature. On the other hand, this could serve as the physical
foundation for temperature sensing. Furthermore, the metallic domains
produced by previous switching events have been found to remain in
VO\textsubscript{2} for a long time, which will affect the switching
threshold voltage of the devices \textsuperscript{\hyperref[csl:327]{327}} subsequently. The
transition temperatures of VO\textsubscript{2} and NbO\textsubscript{2}
are around 70 and 800 \selectlanguage{ngerman}°C, respectively, which are too low and too high
from the perspective of circuit applications. A Mott TMO with ideal
transition temperature is yet to be developed.

Phase change materials mainly refer to chalcogenide glass materials with
reversible phase transition processes. Phase change materials show high
resistance in the amorphous state and relatively low resistance in the
crystalline state. The phase change memory (PCM) constructed by these
materials shows multilevel non-volatile conductance states, and hence
PCM can also be exploited to accelerate MVM combined with
cross-bar~\textsuperscript{\hyperref[csl:328]{328}}. Traditional phase change materials are
the ternary GeSbTe compounds along the pseudo-binary tielines of
GeTe--Sb\textsubscript{2}Te\textsubscript{3},
Ge-Sb\textsubscript{2}Te\textsubscript{3} and
GeTe-Sb\textsuperscript{\hyperref[csl:329]{329}}. By adjusting the proportions of the three
elements in the compound or by incorporating other elements into the
compound, these phase change materials can exhibit advantages in various
aspects such as on/off ratio, switching speed, and retention. Among
these materials,
Ge\textsubscript{2}Sb\textsubscript{2}Te\textsubscript{5}, with its
outstanding recyclability, is currently widely used, and has been
applied in commercial products. In addition to compound-type phase
change materials, monatomic Sb exhibits completely different resistances
in the crystalline and amorphous states, making it also suitable as a
phase change material\textsuperscript{\hyperref[csl:330]{330}}. This type of simplest material
can effectively avoid the problem of stoichiometry deviation during the
phase change process, which is important for further reducing the size
of PCM devices. One of the challenges faced by PCM comes from its
conductance drift, caused by the spontaneous structural relaxation in
unstable amorphous material, which leads to gradual conductance decay
over time and seriously affects computing accuracy and reliability.
Therefore, a compensation circuit or strategy for conductance drift is
desired. Another challenge originates from the long heating time
required for the crystallization of PCM, which not only affects the
programming speed~\textsuperscript{\hyperref[csl:331]{331}}, but also results in higher energy
consumption during programming.

Besides, magnetic materials-based magnetic random access memory (MRAM)
is also widely used in neuromorphic computing, which has relatively
mature technology, high endurance, etc. The typical structure of MRAM is
a magnetic tunnel junction (MTJ), which is composed of two layers
(pinned and free) of magnetic material, and an insulator (usually MgO)
sandwiched in between.~ There are two types of magnetic materials mainly
used in MTJ\textsuperscript{\hyperref[csl:332]{332}}. The first one is the multilayers formed
by transition metals (e.g. Co, Fe) and noble metals (e.g. Pt, Pd), such
as (Co/Pd)\textsubscript{n}, (Co/Pt)\textsubscript{n}. These materials
have advantages in terms of thermal stability and scalability. Another
important magnetic material is CoFeB, which shows extremely low
programming currents and good matching with the lattice of the MgO
barrier layer. By tuning the magnetization orientations of the free
magnetic layer to parallel (P) or antiparallel (AP) orientation with the
pinned layer, MTJ can exhibit low or high resistance, respectively.
Because only the magnetization orientation of the material is changed
without large-scale atomic migration or rearrangement, MRAM has low
device variation and high reliability. However, the main shortcoming of
MRAM is reflected in its relatively low resistance even in the high
resistance state~\textsuperscript{\hyperref[csl:76]{76}}, which will increase the power
consumption for MVM computing. Besides, it has a small on/off ratio,
which supports only two states, indicating P and AP, and limits its
applications in analogue computing.

Similar to MTJ, ferroelectric tunneling junction (FTJ) also changes its
resistance by adjusting the polarization orientations of the
ferroelectric material sandwiched between two metal electrodes.~ The
commonly used ferroelectric material systems include perovskite oxides,
fluorite ferroelectric materials and wurtzite ferroelectric materials.
Perovskite oxides are the most widely used ferroelectric materials with
advantages of scalability and switching speed\textsuperscript{\hyperref[csl:333]{333}}. Among
these materials,
Pb{[}Zr\textsubscript{x}Ti\textsubscript{1-x}O\textsubscript{3}{]} (PZT)
and Sr\textsubscript{2}Bi\textsubscript{2}TaO\textsubscript{9} (SBT)
have more mature technology and have been applied in commercial memory
devices. The fluorite ferroelectric materials, such as
HfZrO\textsubscript{2}, are also widely utilized and studied due to its
CMOS-compatible fabrication process and good
scalability\textsuperscript{\hyperref[csl:334]{334}}. The wurtzite materials, such as
Al\textsubscript{1-x}Sc\textsubscript{x}N, belong to a new type of
ferroelectric material system, characterized by their high on/off ratio,
thermal stability, and retention\textsuperscript{\hyperref[csl:335]{335}}. Since the
polarization orientations of the ferroelectric layer can be tuned by the
electric field with a very low tunneling current, the programming of FTJ
consumes low power. Meanwhile, the FTJ shows multilevel
conductance~\textsuperscript{\hyperref[csl:336]{336}} for MVM acceleration in the edge AI
platforms. However, ferroelectric materials can have retention
degradation caused by intrinsic depolarization fields in the
ferroelectric layer, while recent HZO based ferroelectric materials
exhibit relatively high coercive voltages and low endurance.

Ion-gated transistor (IGT) is a type of three-terminal device, where the
electric field from the gate drives small ions (such as
H\textsuperscript{+} and Li\textsuperscript{+}) in electrolytes into the
device channels to continuously tune conductance of the channel
\textsuperscript{\hyperref[csl:337]{337}}.~ Among the reported material systems for IGT, the
most chosen material for channel is TMO, which shows the potential for
mass production and environmental stability. For the material
consideration of electrolyte, the phosphosilicate glass shows clear
advantages over previously used Li-ions electrolyte, which is more
mature and compatible in the COMS process platform. Owing to the low
gate leakage current and separated programming and reading terminals,
the IGT shows significantly lower power consumption during programming
compared with most two-terminal devices. Furthermore, the weight
modulation of IGT can be much more linear, which can greatly reduce the
overhead for weight programming compared with devices based on other
materials. However, the IGT devices shown to date usually have lower
compatibility with standard fabrication processes and have difficulty in
the fabrication of large-scale arrays.

\subsubsection*{6.3.2.Conclusion ~}

{\label{615782}}

In conclusion, memristors based on resistive switching materials with
different mechanisms have demonstrated a variety of encouraging
characteristics, such as low power consumption, high scaling potential,
fast switching speed, long retention, multilevel conductance, high-order
complexity, etc. The MVM engines, artificial neurons, artificial
synapses and sensory systems constructed by memristive devicesvi have
shown great potential in highly efficient and functional neuromorphic
computing. However, depending on the specific resistive switching
mechanism, technological maturity and manufacturing cost, there are
still technical challenges related to the above material systems. It is
important to compensate and correct the adverse effects existing in the
application of the materials through the improvement from materials,
systems and even algorithms, such as reducing the impact of conductance
drift and programming noise on the accuracy of PCM based convolutional
neural networks (CNN) by improving the training
algorithms~\textsuperscript{\hyperref[csl:176]{176}}. On the other hand, we can also design
systems and algorithms to exploit the non-ideal effects in various
memristors as resources for improving computational efficiency, such as
utilizing the variations and noise in memristor arrays for accelerating
the convergence of Hopfield neural networks \textsuperscript{\hyperref[csl:202]{202}}. Once
the physical attributes of the memristive devices are properly utilized,
they can play important roles in efficient neuromorphic computing. ~

\vspace{0.5cm}

\subsection*{7. Novel computing concepts}

{\label{687635}}

\subsection*{7.1 Embracing variability}

{\label{842599}}

\begin{quote}
\emph{Damien Querlioz, Louis Hutin, Elisa Vianello}
\end{quote}

\subsubsection*{7.1.1. Status}

{\label{398288}}

\begin{figure}[H]
\begin{center}
\includegraphics[width=0.70\columnwidth]{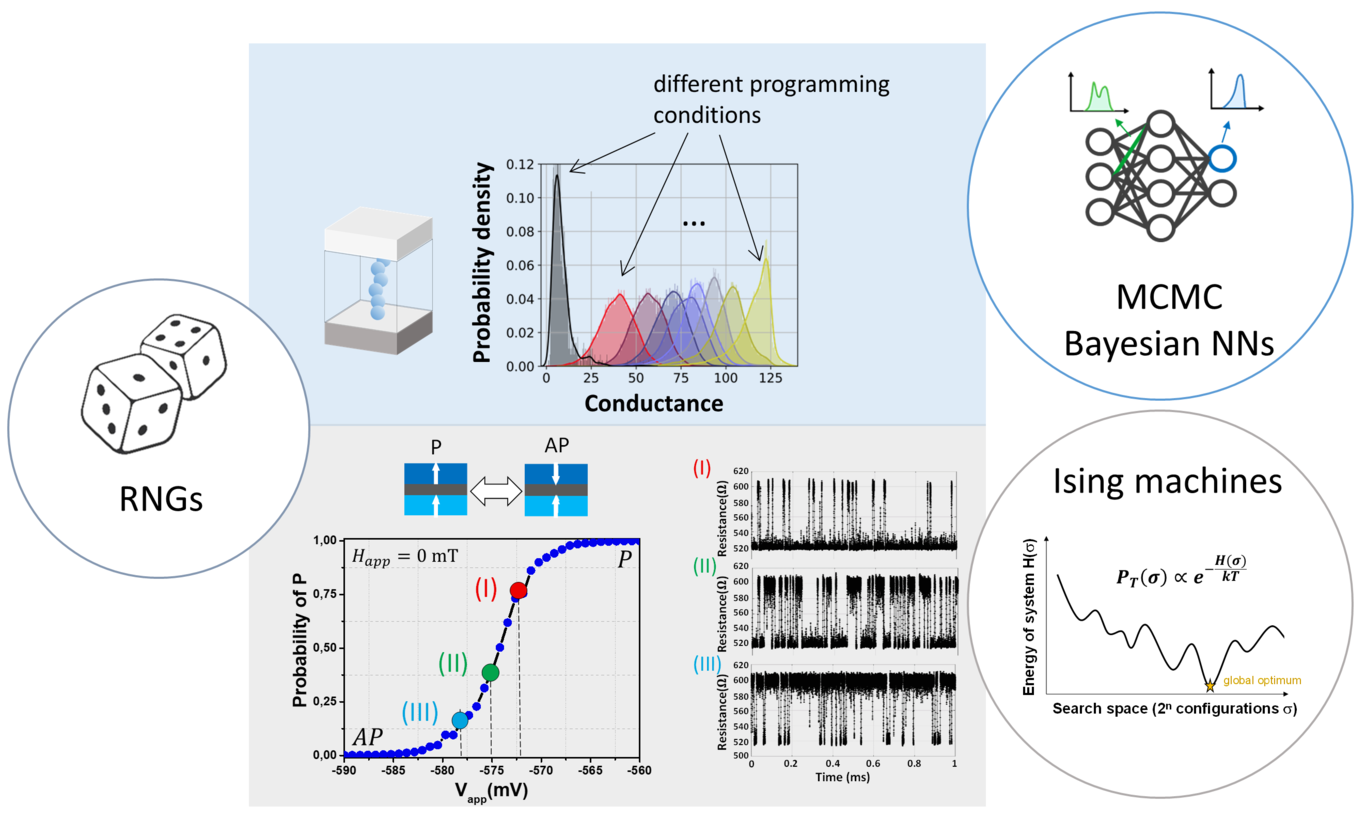}
\caption{{Illustration of some leading approaches exploiting the variability of
nanodevices for computing. Top: measurements of the programming
variability of hafnium-oxide filamentary memristors. The statistical
rule that this variability naturally implements can be used to perform
Markov Chain Monte Carlo training. Bottom: measurement of stochastic
magnetic tunnel junctions, naturally implementing p-bits, used, e.g., in
Ising machines.~ Both devices can also be used for random number
genration.
{\label{769858}}%
}}
\end{center}
\end{figure}

Emerging nanoelectronic components, such as memristors and spintronic
devices, offer exceptional features. However, these devices also exhibit
a high degree of variability in their behavior due to their atomic-level
features and reliance on sophisticated, sometimes
incompletely-understood, physics. This variability has made it necessary
to model these devices using statistical tools, effectively treating
them as random variables. Interestingly, multiple applications,
particularly in machine learning and security, require random variables,
which are expensive to generate using traditional CMOS technology.
Therefore, exploiting the inherent variability of these nanodevices
presents a unique opportunity to develop efficient random number
generators and stochastic computing models (see
Figure~{\ref{769858}}).\selectlanguage{english}

The first idea stems from utilizing the cycle-to-cycle read or
programming variability to create random number generators (RNGs) that
consume less power than traditional pseudo-random or truly random number
generators. By harnessing the inherent variability in these devices, we
can develop RNGs that are not only more energy-efficient but also offer
improved security and robustness for various applications. Experimental
realizations using filamentary memristors~\textsuperscript{\hyperref[csl:91]{91}}, phase
change memories~~\textsuperscript{\hyperref[csl:338]{338}}, and spintronic
devices~\textsuperscript{\hyperref[csl:339]{339},\hyperref[csl:340]{340}} have validated this concept. An interesting
application is stochastic computing, an alternative approximate low
area/low energy computing scheme that has been held back by the lack of
compact RNGs: It requires large amounts of RNGs, which in conventional
implementations dominate the area of circuits~\textsuperscript{\hyperref[csl:341]{341},\hyperref[csl:342]{342}}.~
~Stochastic computing can therefore strongly benefit from such
stochastic nanodevices~\textsuperscript{\hyperref[csl:343]{343},\hyperref[csl:344]{344}}.

The second idea involves making these RNGs adjustable, i.e., the
probability for an output to be one can be controlled by an input
signal, effectively creating probabilistic bits or
``p-bits''~\textsuperscript{\hyperref[csl:345]{345}}. Such structures are analogous to
stochastic binary neurons~\textsuperscript{\hyperref[csl:346]{346},\hyperref[csl:347]{347}}, and they have been used in
adaptive inference models designed for optimization problems, wherein a
set of variables evolves through local and more-or-less random
transformations towards configurations that are increasingly probable as
they minimize energy~\textsuperscript{\hyperref[csl:249]{249},\hyperref[csl:345]{345}}. A particularly exciting
opportunity is their use within Ising machines, which have shown
potential for solving highly complex tasks using reduced
resources~\textsuperscript{\hyperref[csl:348]{348}}. Some of the most promising p-bit
implementations use spintronics, as low-energy barrier magnetic tunnel
junctions provide a p-bit functionality almost
intrinsically~~\textsuperscript{\hyperref[csl:249]{249},\hyperref[csl:346]{346},\hyperref[csl:345]{345}}. Some recent
works~\textsuperscript{\hyperref[csl:349]{349},\hyperref[csl:350]{350}} have used a limited number of stochastic
devices as~fast high-quality randomness sources for larger FPGA-based
circuits, while stressing that the projected benefits of utilizing
nanodevices at a larger scale intrinsically harnessing randomness from
the thermal bath remain significant and appealing.~Memristors with high
random telegraph
noise~\textsuperscript{\hyperref[csl:351]{351}}\href{https://doi.org/10.1109/LED.2012.2199734}{~}could
also be used in that direction.

The third idea underscores the striking parallels between the behavior
of nanodevices and the principles of Markov Chain Monte Carlo (MCMC)
algorithms, a class of stochastic optimization techniques. This
correspondence is particularly evident in the case of the
Metropolis-Hastings MCMC algorithm. In this context, it is required to
generate and store multiple random values for a single parameter, a
process that can be naturally implemented using the inherent variability
of nanodevices. For instance, the programming current can be employed to
determine the mean value, while the imperfections of the devices provide
the necessary randomness. This family of algorithms has been extensively
used for sampling synaptic weight distributions in training Bayesian
models, which excel at modeling uncertainty in complex situations. The
most important experimental demonstration, using filamentary memristors,
is presented in \textsuperscript{\hyperref[csl:199]{199}}.

~While MCMC algorithms serve as a compelling illustration, they are by
no means the only example of stochastic computing methods that can
benefit from the unique properties of nanodevices. Indeed, a diverse and
rapidly evolving field of research is currently exploring the potential
of other stochastic computing techniques for a wide range of
applications, from neural networks to combinatorial optimization
problems. As our understanding of nanodevice behavior continues to
deepen, the opportunities to leverage their inherent variability in
novel and transformative ways promise to fuel further innovation in
stochastic computing and beyond.

\subsubsection*{7.1.2. Challenges}

{\label{937299}}

The primary challenge in exploiting the imperfection of nanodevices lies
in the imperfect nature of the imperfections themselves. To harness the
inherent variability of these devices for practical applications, it is
necessary to achieve a certain level of ``controlled'' imperfection,
which refers to maintaining the desired degree of variability without
compromising the reliability and stability of the devices.

The details of memristor ~or superparamagnetic tunnel
junctions~imperfections~ are subject to variability. This variability
stems from various factors, such as manufacturing variations,
environmental conditions, and the complex interplay of atomic-level
features and underlying physics. In the case of superparamagnetic tunnel
junctions, additional variability in time can occur due to their
sensitivity to magnetic field, which is higher than in stable magnetic
tunnel junctions. Magnetic shielding can be required to suppress this
sensitivity~\textsuperscript{\hyperref[csl:340]{340}}. Overall, modeling the imperfections
of~these nanodevices accurately becomes a challenging task, as capturing
these variations in a consistent manner is difficult.

~This challenge is further exacerbated by the fact that different types
of imperfections have different impacts on the performance and usability
of the devices. Therefore, identifying and understanding the specific
imperfections that can be harnessed for the development of efficient
RNGs and stochastic computing models is of utmost importance.

\subsubsection*{7.1.3. Potential Solutions}

{\label{296752}}

~To address the challenges associated with exploiting the imperfections
in nanodevices, it is essential to acknowledge that not all
imperfections are equally exploitable. For instance, cycle-to-cycle
(C2C) variability and device-to-device (D2D) variability differ in terms
of their usability and intrinsic nature. C2C variability arises from the
inherent variability in the behavior of a single device across different
operational cycles, making it more directly applicable and intrinsic to
the device. On the other hand, D2D variability occurs due to variations
in the performance of different devices, which may be influenced by
manufacturing inconsistencies or other external factors.

C2C variability in nanodevices can manifest in two distinct forms: the
variability resulting from two consecutive read operations on a device,
or the variability arising from two programming operations on the
device. Distinguishing between these two effects can be challenging,
particularly because both types of variability are highly dependent on
the read and programming conditions of the devices. These two forms of
C2C variability offer unique advantages from an algorithmic perspective,
but they are utilized differently in various applications. For instance,
read variability is especially well-suited for RNGs or probabilistic
bits \textsuperscript{\hyperref[csl:340]{340},\hyperref[csl:249]{249}}, while programming variability is specifically
tailored for MCMC algorithms \textsuperscript{\hyperref[csl:199]{199}}.

Even if C2C variability is, in general, more appealing than D2D
variability from an algorithmic perspective, D2D variability also has
applications: It enables the generation of distinct and irreproducible
signatures for each device. Physically unclonable functions (PUFs) are
cryptographic primitives that derive their security from the unique and
unpredictable physical characteristics of individual devices, and are
therefore an excellent example of harnessing D2D variability for
practical applications ~\textsuperscript{\hyperref[csl:352]{352}}.

Understanding the specific physics underlying different nanodevices can
greatly impact their suitability for exploiting imperfections. For
example, in hafnium-oxide memristors, only the low-resistance state
(LRS) can be readily exploited for our proposed applications, as this
state exhibits a much-more controlled degree of variability compared to
the high-resistance state (HRS)~\textsuperscript{\hyperref[csl:199]{199}}. Spintronic devices
also show promising potential in this regard, as they rely on physical
phenomena that inherently exhibit a degree of randomness, which can be
understood and modeled~\textsuperscript{\hyperref[csl:340]{340},\hyperref[csl:346]{346},\hyperref[csl:249]{249},\hyperref[csl:353]{353},\hyperref[csl:347]{347}}. For example, the switching
time of a magnetic tunnel junction is directly connected to the initial
angle of the free layers' magnetization , a random quantity, which can
be well-modeled~\textsuperscript{\hyperref[csl:354]{354}}, and to some extent controlled.~When
using low-barrier MTJs as artificial spins in Ising machines, D2D
variability may cause an unwanted spread in threshold values or even
effective pseudo-temperature across the network. It was shown that these
non-ideal and non-uniform activations could be compensated to some
degree by relearning the synaptic weights following the initial mapping
\textsuperscript{\hyperref[csl:355]{355}}.~

~Another core idea for dealing with nanodevice imperfections can be to
utilize multiple devices instead of relying on a single one. By
employing an ensemble of devices, we can achieve better statistical
properties, resulting in improved performance and robustness of the RNGs
and stochastic computing models. This approach also helps mitigate the
impact of individual device variations and reduces the reliance on any
single device, thereby enhancing the overall reliability and stability
of the systems (examples of this strategy are seen
in~\textsuperscript{\hyperref[csl:356]{356},\hyperref[csl:357]{357},\hyperref[csl:358]{358}}).

\subsubsection*{7.1.4. Concluding remarks}

{\label{420330}}

\textbf{~}Embracing the inherent variability of emerging nanoelectronic
components, such as memristors and spintronic devices, presents a unique
opportunity to advance stochastic computing, machine learning, and
security applications. By understanding and exploiting the intricate
relationship between the variability of these devices and the
requirements of various algorithms, we can develop efficient random
number generators, p-bits, and Markov Chain Monte Carlo implementations,
among other stochastic computing techniques.

However, not every nanodevice imperfection can be exploited. To fully
harness the potential of these nanodevices, it is essential to address
the challenges associated with their imperfect nature. Achieving a
``controlled'' level of imperfection, understanding the impact of
different types of variability, and leveraging the specific physics
underlying each device are crucial steps in this endeavor. By employing
multiple devices in an ensemble, we can further enhance the reliability,
stability, and performance of the resulting systems. As our
understanding of nanodevice behavior and the opportunities to exploit
their inherent variability continues to deepen, we can anticipate a
surge in innovation in stochastic computing and beyond.

\vspace{0.5cm}

\subsection*{7.2 Spiking-based computing}

{\label{753173}}

~ ~ ~ ~ Sayeed Shafayet Chowdhury, Kaushik Roy~

\subsubsection*{7.2.1. Status}

{\label{986590}}

Spiking neural networks (SNNs) are a promising energy efficient
alternative to traditional artificial neural networks (ANNs). While ANN
based deep learning has achieved tremendous progress in fields such as
computer vision and natural language processing, it comes at a cost of
huge compute requirements. Spike-based neuromorphic
computing~\textsuperscript{\hyperref[csl:359]{359}} provides a potential solution to this issue
using brain-inspired event-driven processing. SNNs use~ binary spikes
for computation contrary to analog values used in ANNs. A schematic of
spiking neurons with their temporal dynamics is shown in Fig. 15. The
spiking neuron receives spike inputs over time which are accumulated in
the membrane potential (V\textsubscript{mem}), which upon crossing a
threshold (V\textsubscript{th}), emits an output spike.

\par\null\par\null\selectlanguage{english}
\begin{figure}[H]
\begin{center}
\includegraphics[width=0.70\columnwidth]{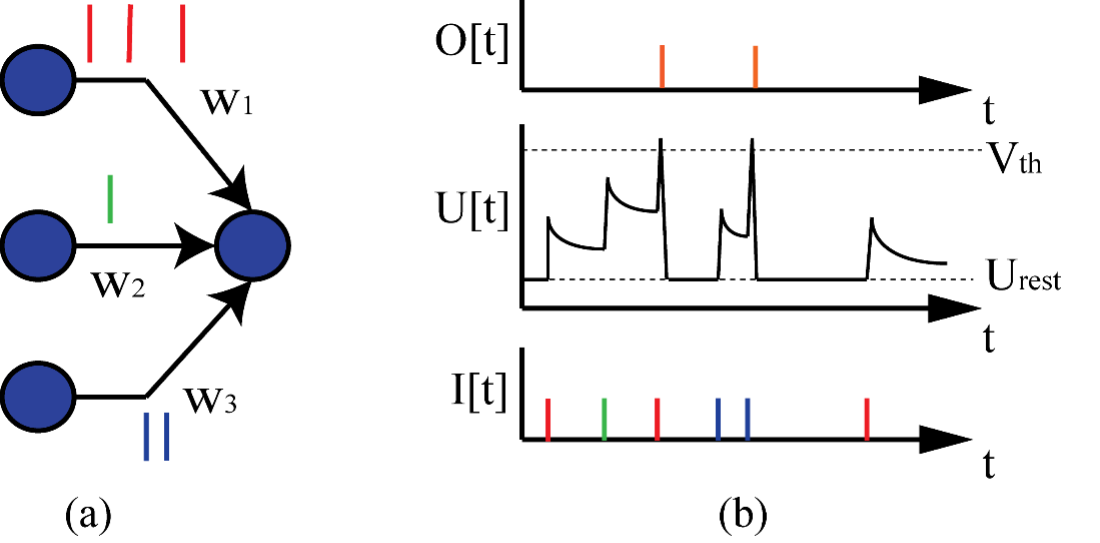}
\caption{{A leaky-integrate-and-fire (LIF) neuron, (a) schematic connection
between three pre-neurons to a post-neuron, (b) temporal dynamics of the
post-neuron; adapted from~\protect\textsuperscript{\hyperref[csl:360]{360}} ~ ~ ~ ~
{\label{447762}}%
}}
\end{center}
\end{figure}

\subsubsection*{}

{\label{539395}}

A key characteristic of SNNs is the notion of time. ~While conventional
feed-forward ANNs are able to map static inputs to outputs with very
impressive performance, learning long-term temporal correlations using
them is challenging. On the other hand, recurrent neural networks (RNNs)
are more suited to process temporal information
~efficiently~\textsuperscript{\hyperref[csl:360]{360}}; although recent developments such as
transformers have shown that ANNs can perform well in temporal
processing too~\textsuperscript{\hyperref[csl:361]{361}}, albeit with a higher training and
memory cost. Different variants of RNNs, such as vanilla
RNNs~\textsuperscript{\hyperref[csl:362]{362}}, long-short term memory networks
(LSTMs)~\textsuperscript{\hyperref[csl:363]{363}},~\textsuperscript{\hyperref[csl:364]{364}} and gated recurrent units
(GRUs)~\textsuperscript{\hyperref[csl:365]{365}},~\textsuperscript{\hyperref[csl:366]{366}} have been proposed which
differ in their degree of complexity and capability to capture temporal
information. However, they all contain explicit feedback connections and
memory elements to handle temporal dependencies. On the contrary, SNNs
can be regarded as simpler form of RNNs, where the recurrent dynamics of
V\textsubscript{mem} acts as an internalized memory~\textsuperscript{\hyperref[csl:367]{367}}.
Interestingly, the leak in SNNs can play the role of a lightweight
gating mechanism, thereby temporally filtering out some irrelevant
information \textsuperscript{\hyperref[csl:368]{368}}. Additionally, SNNs may lead to lower
parameter count and easier training overhead compared to LSTMs
\textsuperscript{\hyperref[csl:369]{369}}.~~~~~~~~~~~~~~~~~~~~~~~~~~~~~~~

The sequential nature of processing in SNNs leads to unique
opportunities in terms of input representation. Traditional feed-forward
ANNs such as CNNs and the more recent Vision
Transformers~\textsuperscript{\hyperref[csl:370]{370}} process several temporal inputs by
merging them into a single large representation. On the other hand, SNNs
can process the sequential inputs in a streaming fashion, using the
inherent recurrence of neuronal membrane potential. As a result, SNNs
are inherently suitable to process~temporal event
camera~\textsuperscript{\hyperref[csl:40]{40}} data. However, for analog inputs, it becomes
a key challenge to efficiently encode the data into a spike train.
Initial works~\textsuperscript{\hyperref[csl:371]{371}},~\textsuperscript{\hyperref[csl:372]{372}} use Poisson
rate-coding where the input is compared to a random number at each
timestep and a spike is generated if the input is higher than the random
number. However, this process suffers from high inference latency.
Temporal coding schemes such as phase~\textsuperscript{\hyperref[csl:373]{373}},
burst~\textsuperscript{\hyperref[csl:374]{374}} coding, DCT-encoding~\textsuperscript{\hyperref[csl:375]{375}},
time-to-first-spike (TTFS) coding~\textsuperscript{\hyperref[csl:376]{376}}, temporal
coding~\textsuperscript{\hyperref[csl:377]{377}} etc. attempt to capture the temporal
correlation in the data. But their accuracy is often lower than ANNs.
More recently, direct encoding
approach~\textsuperscript{\hyperref[csl:378]{378}},~\textsuperscript{\hyperref[csl:379]{379}} has become popular
where analog values are given directly to the SNN and the first layer of
the neural network layer acts as spike generator. Such method has
provided impressive performance on complex tasks with very few
timesteps. Besides the analog input modalities, DVS cameras (such as
DAVIS240 ~\textsuperscript{\hyperref[csl:380]{380}}) provide discrete spikes directly as
inputs which are inherently more amenable to SNNs. Conventional
ANN-based approaches tackle the event streams by accumulating them over
time and subsequently processing the lumped input
altogether~\textsuperscript{\hyperref[csl:381]{381}},~\textsuperscript{\hyperref[csl:382]{382}}. However, the rich
temporal cues present in the event data may not be optimally leveraged
in this process. Therefore, recent works propose to process the event
streams using SNNs~\textsuperscript{\hyperref[csl:383]{383}}, which leads to a synergistic
alliance between the inputs and spike-based processing. We believe this
is a promising direction to pursue as it enables harnessing the inherent
temporal processing capabilities of SNNs with event data.

\subsubsection*{7.2.2. Challenges and Potential
Solutions}

{\label{182694}}

A crucial bottleneck in the advancement of SNNs is the lack of suitable
training methods. Sparsity of activations and the discontinuous
derivative of spike function leads to training complexity in SNNs. To
counter that, initial approaches mostly used ANN-SNN
conversion~\textsuperscript{\hyperref[csl:371]{371}},~\textsuperscript{\hyperref[csl:384]{384}}. Though the
conversion method provides high accuracy, it suffers from significant
inference latency. Following this, surrogate gradient-based
backpropagation (BP) methods have been
proposed~\textsuperscript{\hyperref[csl:385]{385}},~\textsuperscript{\hyperref[csl:372]{372}},~\textsuperscript{\hyperref[csl:386]{386}}
to train SNNs from scratch. Note, due to the sequential nature of
inputs, SNNs are trained with backpropagation-through-time (BPTT), like
RNNs. However, simpler neuron models and lower parameter complexity
makes the optimization of SNNs simpler compared to RNNs. Although these
surrogate gradient-based BPTT methods have advanced the field of SNNs by
obtaining high accuracy, the training workloads are still quite
intensive in addition to considerable inference latency (~100
timesteps). To overcome these, the authors in~\textsuperscript{\hyperref[csl:387]{387}}
propose to merge the conversion and BP-based training methods (termed as
`hybrid' training) where first an ANN is trained to use it as
initialization for subsequent surrogate gradient-based BP. More
recently, advanced training approaches such as temporal
pruning~\textsuperscript{\hyperref[csl:388]{388}}, custom regularizers~\textsuperscript{\hyperref[csl:389]{389}} and
modified neuron models~\textsuperscript{\hyperref[csl:390]{390}} have been proposed which
enable reducing the latency of SNNs to unit timestep. A complementary
research direction proposes to utilize equilibrium propagation to train
SNNs~\textsuperscript{\hyperref[csl:391]{391}},~\textsuperscript{\hyperref[csl:392]{392}}. These approaches provide
a promising more bio-plausible alternative to backpropagation. However,
challenges remain in their large scale implementation.~~

Parallel to algorithmic developments, advancements in neuromorphic
hardware fabrics are equally critical to unearth the true potential of
SNNs. Due to the sequential nature of data processing, SNNs present
unique challenges on the hardware front as current graphics processing
units (GPUs) and tensor processing units (TPUs) are sub-optimal to
exploit the high temporal as well as spatial
sparsity~\textsuperscript{\hyperref[csl:393]{393}}. Furthermore, information processing using
membrane potential over multiple timesteps leads to memory-intensive
operations; an overhead that is non-trivial to mitigate using
off-the-shelf digital accelerators. Taking such issues into account,
several research directions have been pursued in recent years across the
stack from devices and circuits to architectures. Event-driven
neuromorphic chips such as Neurogrid ~\textsuperscript{\hyperref[csl:394]{394}} and
TrueNorth~\textsuperscript{\hyperref[csl:395]{395}} are notable, which are based on mixed
signal analog and~digital circuits, respectively. Two standout features
of these neuromorphic chips are asynchronous address event
representation and networks-on-chip (NOCs). Another promising direction
is investigating various beyond von Neumann computing models to counter
the `memory wall bottleneck'. To this end, near-memory and
in-memory~\textsuperscript{\hyperref[csl:395]{395}},~\textsuperscript{\hyperref[csl:396]{396}},~\textsuperscript{\hyperref[csl:397]{397}}
computing paradigms are being explored to improve throughput and energy
efficiency. To realize these emerging computing platforms, exciting
progress is being achieved in the device domain utilizing non-volatile
technologies~\textsuperscript{\hyperref[csl:398]{398}}. Some noteworthy approaches based on
memristive technologies include resistive random-access memory
(RRAM)~\textsuperscript{\hyperref[csl:399]{399}}, phase-change memory (PCM)
~\textsuperscript{\hyperref[csl:400]{400}} and spin-transfer torque magnetic random-access
memory (STT-MRAM)~\textsuperscript{\hyperref[csl:401]{401}}. RRAMs provide analog programmable
resistance but are prone to process and cycle variations and read/write
endurance. Devices based on PCM can achieve comparable programming
voltages and write speed to RRAMs, however high write-current and
resistance drift over time cause issues. On the other hand, compared to
RRAMs and PCMs, advantages of spin devices~\textsuperscript{\hyperref[csl:402]{402}} are
almost unlimited endurance, lower write energy and faster reversal.
However, their ON/OFF ratio is much smaller than in PCMs and RRAMs,
requiring proper algorithm/hardware co-design~\textsuperscript{\hyperref[csl:403]{403}}. Note,
each of these technologies has its pros and cons and there is no single
winner at the moment. Floating-gate transistors~\textsuperscript{\hyperref[csl:404]{404}} are
another class of non-volatile devices which are being explored for
synaptic storage. While their compatibility with MOS fabrication process
is attractive, challenges persist regarding reduced endurance and high
programming voltage.~ ~~~

{\label{997883}}

\subsubsection*{7.2.3. Conclusion}

{\label{997883}}

To conclude, SNNs are a promising bio-plausible alternative to
conventional deep neural networks. However, it is imperative to
understand the `\emph{why'} and `\emph{where'} of their proper usage.
While SNN algorithms have largely focused on static vision
tasks~\textsuperscript{\hyperref[csl:405]{405}} till now, their true potential lies in
processing sequential information. To that effect, several works are
exploring event-based vision for optical flow, depth estimation,
egomotion etc. We believe immense opportunities lie ahead in further
exploration of SNNs in varied avenues requiring temporal processing such
as video processing, reinforcement learning, speech, control etc. In
order to achieve that, there is a need of concerted synergistic efforts
on algorithms as well as hardware. On the algorithmic aspect, we need to
focus on investigating learning approaches that can leverage the unique
data representation provided by spiking neurons. Additionally,
developing hardware geared towards the SNN specific algorithms is
critical for the whole field to move forward. Overall, while
domain-specific challenges are prevalent in spike-based computing, we
believe the next few years will be exciting as we discover the niche of
SNNs, most likely comprising temporal applications with low power
requirements. ~

\vspace{0.5cm}

\subsection*{7.3 Analog computing for linear
algebra}

{\label{360234}}

\begin{quote}
\emph{Zhong Sun,~ Piergiulio Mannocci, Yimao Cai ~}
\end{quote}

\subsubsection*{7.3.1. Status}

{\label{295439}}

Linear algebra problems are being solved in every corner of the
information world. Solving these problems by running algorithms in
digital computers, however, is generally hard and resource-demanding,
featuring a high computational complexity such
as~\emph{O}(\emph{n}\textsuperscript{3}), where~\emph{n} is the number
of variables. To overcome the inadequacy of digital computers whose
performance is fundamentally limited by the ultimate scaling of Moore's
law and the intrinsic bottleneck of von Neumann architecture, analog
computing arises as a promising solution, thanks to its efficient
information encoding, massive parallelism, fast response, as well as the
emerging resistive memory technology ~\textsuperscript{\hyperref[csl:406]{406}}. Analog matrix
computing (AMC) is conveniently realized with crosspoint resistive
memory array, which forms a physical matrix by storing entries as
crosspoint device conductances and thus can be used for linear algebra
computations. There are several resistive memory device concepts that
rely on distinct underlying physics, including two-terminal devices such
as resistive random-access memory (RRAM), phase change memory (PCM),
magnetoresistive RAM (MRAM), ferroelectric tunnel junction
(FTJ)~\textsuperscript{\hyperref[csl:407]{407}}, and three-terminal devices such as
ferroelectric field-effect transistor (FeFET) and electrochemical RAM
(ECRAM). They are all simply used as programmable resistive devices to
implement AMC
\textsuperscript{\hyperref[csl:408]{408}}\textsuperscript{\hyperref[csl:409]{409}}\textsuperscript{\hyperref[csl:410]{410}}\textsuperscript{\hyperref[csl:411]{411}}\textsuperscript{\hyperref[csl:76]{76}}\textsuperscript{\hyperref[csl:412]{412}}.
In this context, one of their differences lies in the conductance range
that may limit the capacity of mapping matrix elements, say one bit or
multiple bits. It is possible to replace one type of resistive memory
device that has been demonstrated for AMC application by another one.
For simplicity, we limit our discussion to RRAM that we have used
frequently. ~

The most straightforward AMC implementation is to perform the
matrix-vector multiplication (MVM) in one step. By simply applying
simultaneously a set of voltages (representing an input vector) to the
crosspoint columns, the currents through the crosspoint array are
collected at the grounded rows, constituting the output vector which in
turn is converted and read out with transimpedance amplifiers (TIAs)
(Fig. 1a). By adopting the conductance compensation strategy
\textsuperscript{\hyperref[csl:413]{413}}, MVM of mixed matrix that contains negative entries
can be implemented with the same number of TIAs as in Fig. 1a. MVM is
the backbone of many important algorithms, such as neural networks and
discrete transformations. Consequently, RRAM-based AMC has widely been
considered as an accelerator approach, showing more than two orders of
magnitude improvements of throughput and energy efficiency
\textsuperscript{\hyperref[csl:179]{179}}.

In addition to the naive MVM application, more complicated linear
algebra computations have been realized through configuring AMC circuits
with feedback loops. Figs. 1b-1e show closed-loop AMC circuits for other
basic matrix operations, including matrix inversion (INV), generalized
inverse, and eigenvector. The INV circuit in Fig. 1b is constructed
based on the global feedback connections between crosspoint rows and
columns through operational amplifiers (OPAs). It solves a system of
linear equations when an input current vector is provided, with the
output voltages of OPAs representing the solution~\textsuperscript{\hyperref[csl:414]{414}}.
INV is exactly the inverse problem of MVM, both circuits utilize the
same electronic components, whereas the different connection topologies
define the opposite functions. The INV concept can be generalized to
non-square matrices, by configuring AMC circuits with two sets of
crosspoint arrays and amplifiers. Depending on the matrix shape (tall or
broad), the generalized left (Fig. 1c) and right (Fig 1d) inverse
circuits have been designed, both based on the same feedback loops while
showing differences in terms of input terminals and matrix storage
(transpose or not)~\textsuperscript{\hyperref[csl:192]{192}}. INV and generalized inverse find
applications in many scenarios, such as machine learning, wireless
communications, and scientific
computing~\textsuperscript{\hyperref[csl:415]{415}}\textsuperscript{\hyperref[csl:416]{416}}\textsuperscript{\hyperref[csl:417]{417}}.
Unlike MVM, INV suffers from the condition number issue, where the input
error would be amplified for an ill-conditioned matrix (with a large
condition number), resulting in a low computing precision. Consequently,
it is appropriate to use the INV circuit as an analog preconditioner,
and the performance of scientific computing may be improved by more than
three orders of magnitude~\textsuperscript{\hyperref[csl:415]{415}}. The eigenvector circuit
in Fig. 1e uses global feedback as well, but it is a fully
self-sustained system with no external inputs, while working by positive
feedback mechanism~\textsuperscript{\hyperref[csl:414]{414}}. It finds applications in typical
scenarios including quantum simulations, PageRank for Google search or
recommender systems.

\par\null\selectlanguage{english}
\begin{figure}[H]
\begin{center}
\includegraphics[width=0.80\columnwidth]{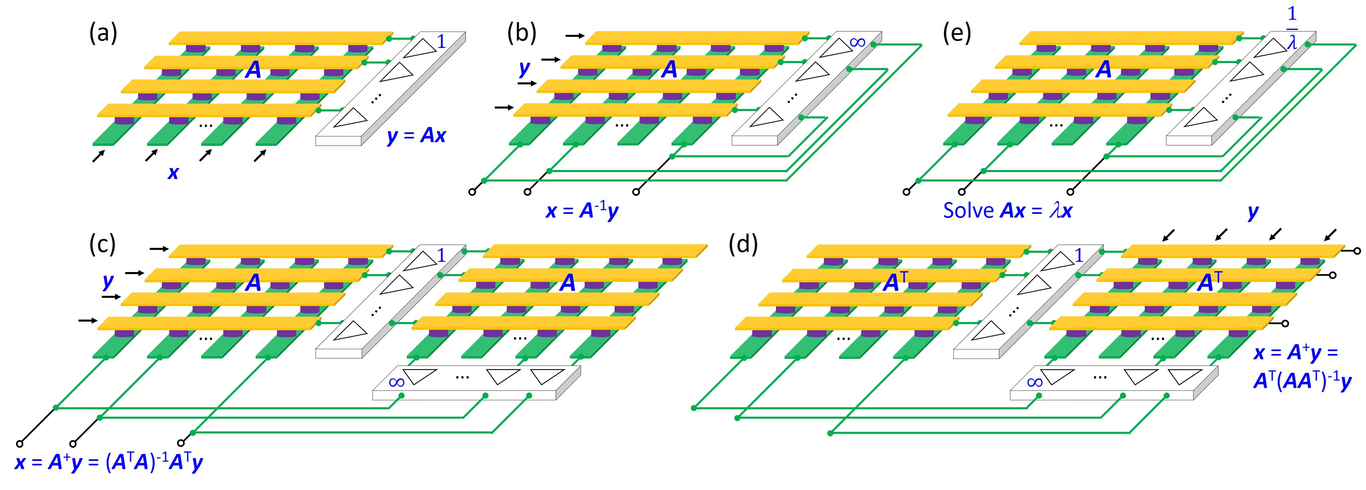}
\caption{{AMC circuits for (a) matrix-vector multiplication, (b) matrix inversion,
(c) generalized left inverse, (d) generalized right inverse, and (e)
eigenvector computations. ~
{\label{576075}}%
}}
\end{center}
\end{figure}

\subsubsection*{}

{\label{463761}}

Recently, more AMC circuits have been developed for solving more
complicated matrix problems. Fig. 2a shows a design for
matrix-matrix-vector multiplication (MMVM), by mapping two matrices (or
two copies of one matrix) in a RRAM array, and assisted by the use of
conductance compensation \textsuperscript{\hyperref[csl:418]{418}}. By connecting this MMVM
circuit with other analog components, and particularly a nonlinear
function module based on operational amplifier, to form a feedback loop,
the resulting circuit solves the sparse approximation problem in Fig. 2b
in one step without discrete iterations. Notably, the nonlinear function
may also be implemented by a volatile resistive switching device, thus
substantially improving the compactness of the AMC circuit
\textsuperscript{\hyperref[csl:419]{419}}. It has been used for compressed sensing recovery pf
sparse signal, natural and medical images, representing a highly
promising solution for the back-end processor to deliver real-time
processing capability in the microsecond regime.\selectlanguage{english}
\begin{figure}[H]
\begin{center}
\includegraphics[width=0.80\columnwidth]{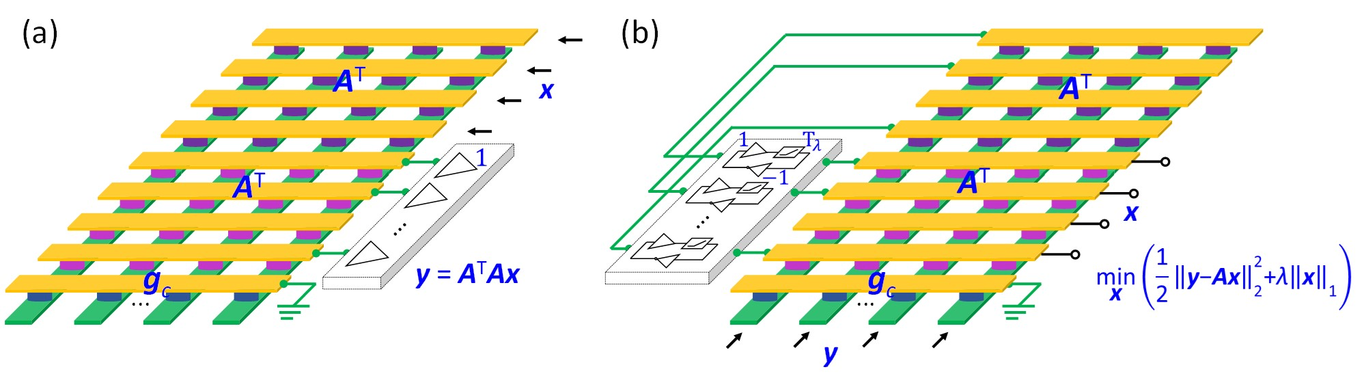}
\caption{{AMC circuits for (a) matrix-matrix-vector multiplication, and (b)
solving sparse approximation problems, \emph{e.g.}, compressed sensing
recovery.
{\label{590687}}%
}}
\end{center}
\end{figure}

\subsubsection*{7.3.2. Challenges}

{\label{463761}}

Thanks to the manufacturability and compatibility of RRAM device in
modern CMOS technology process, large-scale RRAM arrays have been
fabricated for AMC implementations. Particularly, RRAM macros including
peripheral circuitries are usually designed to deliver MVM
accelerations. By contrast, closed-loop AMC has been limited to
small-scale concept demonstration, \emph{e.g.}, for 3\selectlanguage{ngerman}×3 matrices. The
reason behind the developmental stagnation might be ascribed to the
unconventional analog circuitry. Different from MVM that consists of
only local feedback, closed-loop AMC circuits contain complicated,
hard-wired global feedback connections across the entire RRAM array.
Since all elements are involved to provide a collective circuit
response, the operation may become sensitive, risking the damage of RRAM
devices under excessive electric stimulus. In addition, for large-scale
circuits, the non-ideal factors of devices and circuits will jointly
lead to an exaggerated deviation from the correct result. Also, the time
response of the circuit might be influenced by non-idealities such as
parasitic resistances and capacitances, which may even cause an
instability issue.

Despite the lack of large-scale demonstration at this moment, it is
quite promising to build closed-loop AMC circuits based on the
relatively mature RRAM technology in the near term, given that lots of
theoretical and simulation works have intensively examined the potential
issues. In the long term, the following aspects shall be addressed to
support the development of AMC circuits for linear algebra:

~(1) Device/array level: At present, the largest available array size is
512×512 \textsuperscript{\hyperref[csl:174]{174}} . In practice, only a fraction of the array
is turned on for computation, due to the current/power overload and the
accuracy limitation. We believe such an array size is already sufficient
to well support the advantages of AMC over other paradigms. Otherwise,
we should put the stress on the analog conductance tunability of RRAM
devices, which are expected to show as many distinguishable conductance
levels as possible. In turn, fast and accurate programming of the RRAM
conductance is essential to maximizing the AMC efficiency. The linearity
and symmetry of device conductance update has been continuously
emphasized for the online training of neural network weighs
\textsuperscript{\hyperref[csl:420]{420}}, such a characteristic should also be favored for
real-time update of matrix elements in general AMC applications. On the
other hand, there should be a tradeoff consideration on the RRAM
conductance range, which is associated with the power consumption,
alleviation or exacerbation of the impacts of resistive and capacitive
parasitics. The device variations should matter most to affect the
computing precision. In particular, for solving inverse problems with
the closed-loop circuits, the matrix structure related to the condition
number should also be a deterministic factor.

~(2) Circuit/architecture level: The scalability of AMC circuits comes
at the price of the reduced precision, where the accumulated error may
eventually decline the nominal result. As a result, there is a tradeoff
between the desired computing accuracy and the possible array size. For
solving extremely large-scale problems, especially for the inverse
matrix problems, it is imperative to have algorithmic solutions to
recover the result correctly. Therefore, clever design of efficient
algorithms for matrix processing would be very helpful. Particularly, as
RRAM-based AMC uses two physical attributes, namely conductance and
voltage, the cascading of an algorithm may require the transition
between the two attributes, which will inevitably make the operation
complicated. To this end, algorithms featuring as less such transitions
as possible are precious. In addition, since AMC circuits are all based
on the core RRAM array, it is valuable to have a reconfigurable
architecture for performing different matrix operations. As amplifiers
play another critical role in AMC, it is highly beneficial to design
efficient amplifiers that are well suited to the circuits.

~(3) Software \& application: During the AMC operation, multiple
rows/columns of the array are simultaneously activated to carry out
computation, indicating a primary requirement of such instructions in an
AMC program. Additionally, instructions for circuit reconfigurations
should be included to enable a general AMC architecture. Therefore, the
efforts on compiling and programming for AMC will be of significant
importance ~\textsuperscript{\hyperref[csl:421]{421}} . To support the AMC concept, finding a
killer application that well matches the advantages of AMC will be most
convincing. The application of MVM to neural network acceleration has
attracted enormous attention in past years, with a few successful
silicon demonstrations towards real-world applications. For closed-loop
AMC circuits, we believe non-linear matrix problems and related
applications would be most promising, since they are more tolerant to
errors as in the neural network.

\subsubsection*{~7.3.3. Potential Solutions}

{\label{778939}}

RRAM-based AMC is extremely fast in that the computation is basically a
parallel reading process. To prevent the memory writing process being a
bottleneck, parallel writing schemes have been developed. For instance,
the weight matrix in the array can be updated by using the outer product
operations, sometimes assisted by gradient decomposition methods
\textsuperscript{\hyperref[csl:422]{422}}. However, these methods usually overlooked the memory
nature of RRAM, that is, the analog information needs to be reliably
read out. To this end, a verification circuit should be included to
confine the conductance distributions. In particular, fast writing with
less/none iterations will be very beneficial to saving the latency. One
example is the closed-loop write scheme that uses current feedback to
control the resistive switching process \textsuperscript{\hyperref[csl:423]{423}}. Because of
the underlying ionic migration mechanism of RRAM devices, the endurance
capability has constantly been a critical issue for both memory and AMC
applications \textsuperscript{\hyperref[csl:62]{62}}. It is unlikely that this issue can be
overcome solely by test method innovations, new physical mechanisms may
be needed to solve it from the source. Additionally, investigations on
device materials and structures to fundamentally optimize analog RRAM
performance and to empower the array extension are always highly
desired.

The scale-accuracy tradeoff of RRAM-based AMC should be elaborated to
balance the efficiency and reliability. Efficient algorithms for matrix
tiling and result recovery are vital for solving large-scale problems.
Recently, a scalable AMC method, termed BlockAMC, has been proposed for
solving large-scale linear systems. It partitions a large original
matrix into smaller ones on different memory arrays, and performs MVM
and INV operations with the block matrices to recover the original
solution \textsuperscript{\hyperref[csl:424]{424}}. To reduce the impact of non-idealities, AMC
systems with both algorithmic and architectural innovations should be
designed. For MVM, it is convenient to implement the bit slicing method
to extend the computing precision, by using only low-precision memory
devices \textsuperscript{\hyperref[csl:425]{425}}. However, as the INV circuits contain global
feedback loops, it is difficult to apply this method to improve the
precision of INV operations. Eventually, an analog-digital hybrid system
may be required to deliver this capability. With the RRAM array as the
core, the peripheral circuits and the connections can be reconfigured to
perform different AMC operations. In this regard, the basic models shall
include RRAM array, amplifiers, and reconfigurable routing, \emph{etc}.
Different circuit configurations are actually hardware-embedded
instructions of matrix operations.

It is of great practical convenience to have a set of fixed AMC
primitives that can be adopted to realize general matrix computations,
together with vector processing, parameter scaling, and variable
attribute conversion. There is a trend that takes MVM and INV as two
primitives to enable a reconfigurable, general-purpose AMC system, which
may be used for linear regression, generalized regression and
eigen-decomposition
\textsuperscript{\hyperref[csl:426]{426}}\textsuperscript{\hyperref[csl:427]{427}}\textsuperscript{\hyperref[csl:428]{428}}. Given the
intrinsic noises in analog computing, we believe non-linear closed-loop
AMC that are inherent in error tolerance are more promising towards
real-world applications. Typical examples include solving some
optimization problems, such as sparse coding, compressed sensing
recovery, as well as linear/quadratic programming. These problems appear
in many common scenarios such as wireless channel estimation, magnetic
resonance imaging, and signal processing ~\textsuperscript{\hyperref[csl:429]{429}} . Also,
front-end integration with sensors for signal processing would be
encouraging, which helps save data conversions and thus reduces the
accumulation of noise effects ~\textsuperscript{\hyperref[csl:430]{430}} . These problems
typically favor relatively small RRAM arrays, thus alleviating the
rigorous requirements on the device/array performance. Eventually, as
the RRAM technology matures, the application to large-scale and
high-precision problems will be advanced, with the help of innovative
algorithms and architectures.

\subsubsection*{~7.3.4.~Conclusion}

{\label{740229}}

In the modern era, due to the strong demand for linear algebra
acceleration and the rapid development of emerging resistive memory
devices/architecture, analog computing has gained a renewed interest
across academia and the industry. Various AMC circuits have been
successfully demonstrated for fast solutions of matrix problems that
constitute the basic operations for linear algebra computations.
However, most of the AMC concepts still remain in the laboratory
prototype stage, calling for a roadmap covering different aspects to
guide system integration, optimization, and application. Next steps
towards effective AMC shall include developing (1) reliable analog
conductance programming methods, (2) architecture and algorithm designs
for large-scale problems, (3) circuit designs for non-ideality
mitigation, (4) reconfigurable systems, (5) more AMC circuits,
\emph{e.g.}, for non-linear matrix operations, and (6) application to
near-term and long-term typical scenarios.

\vspace{0.5cm}

\subsection*{7.4 Analog Content Addressable Memories (CAMs) for in-memory computing}

{\label{812483}}

\begin{quote}
\emph{Giacomo Pedretti~ ~ ~ ~ ~ ~ ~ ~ ~ ~ ~ ~ ~ ~}
\end{quote}

\subsubsection*{7.4.1. Status}

{\label{142897}}

Content addressable memories (CAMs) are a class of memory structure
which given an input query returns its stored location, or address. A
wildcard, `X', can be added allowing for `fuzzy' searches, resulting in
a ternary CAM (TCAM) ~\textsuperscript{\hyperref[csl:431]{431}}.\selectlanguage{english}
\begin{figure}[H]
\begin{center}
\includegraphics[width=0.80\columnwidth]{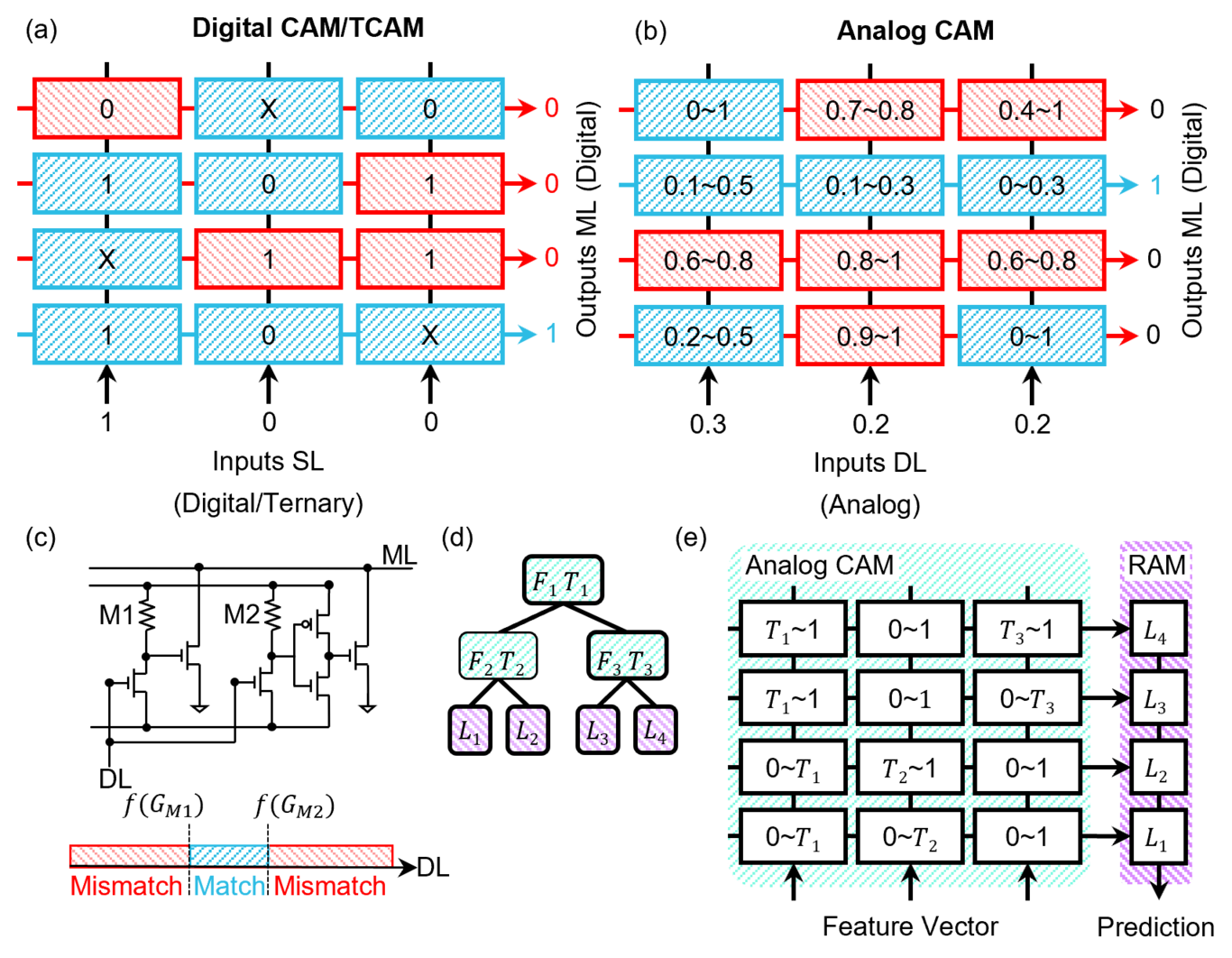}
\caption{{Illustration of a TCAM (a) and an analog CAM (b). (c) Circuit schematic
of a 6T2M analog CAM, able of returning a match if the input on the DL
is within a function of the memristor condutance (inset). (d) Example of
decision tree and (e) its mapping to an in-memory computing circuit for
single/few cycle inference.
{\label{781830}}%
}}
\end{center}
\end{figure}

\begin{quote}
\end{quote}

{\label{926630}}

Figure~{\ref{781830}}(a) shows a TCAM schematic, an
input query~is applied along the columns, or search lines (SLs), and
outputs are returned along the rows, or match lines (MLs). A given row
is matched (returning a `1') if each
value~\(q_{j\ }\)of~\(q\)~is equal to the
corresponding key-value~\(K_{ij}\)~stored in the
row~\(i\)(column~\(j\)), which corresponds to
performing the operation\[ML_i\ =\ \prod_j^{ }\overline{\left(q_j\oplus K_{ij\ }\right)}.\]CMOS/SRAM-based TCAMs are
ubiquitous in networking, but they are usually bulky and power-hungry.
Memristor-based TCAMs have been shown to outperform the CMOS circuits
but given the relatively low speed and inefficient writing of
memristors, they are currently being proposed for in-memory computing
applications.~ In this framework, recently we proposed an analog CAM
~\textsuperscript{\hyperref[csl:432]{432}} which stores~\emph{ranges~}and returns a match on
the MLs if the analog input is within the stored ranges
or\[ML_i=\prod_j^{ }\left(T_{l,ij}\le q_{j\ }<T_{h,lj\ }\right)\]

where~\(T_l\) and~\(T_h\) are the stored upper and
lower bounds thresholds, respectively.~
Figure~{\ref{781830}}(c) shows a conceptual
representation of an analog CAM and~
Figure~{\ref{781830}}(c) the schematic of a
6-transistors-2-memristors (6T2M) analog CAM. Two 1T1R voltage dividers
compare the analog input on their gate with the stored value in the
memristor conductance, and their drain node controls either a pull-down
transistor (left side) or a series of an inverter and a pull down
transistor (right side) to discharge the pre-charged ML if the input is
lower than the lower bound or higher than the upper bound, respectively.
For example, if the input voltage is high enough to turn on the input
transistor resulting in a voltage on the left-most 1T1R drain lower than
the pulldown threshold, the latter won't be activated. Similarly, if~
the same input voltage is low enough such that the input voltage of the
inverter is above its threshold of the right-most 1T1R, its pulldown
transistor would not be activated. This results in a match, given that
the ML would not discharge.

Other implementations of analog CAMs have been proposed based on compact
ferroelectric structures, although given the absence of the inverter, a
proper conditioning of the input should be performed~\textsuperscript{\hyperref[csl:433]{433}}.

While the in-memory computing community in the last decade~ have been
focused on the acceleration of deep neural networks (DNN) (as presented
in
paragraph~{\ref{348857}}~and~{\ref{675016}}),
tree based machine learning (ML) still outperforms DNN when processing
tabular data due mainly to the presence of missing and categorical
features~\textsuperscript{\hyperref[csl:434]{434}}. ~
Figure~{\ref{781830}}(d)~ shows a schematic
representation of a decision tree (DT), which is essentially a
collection of conditional branches (nodes) in a tree structure. Given
the irregular structure, ensembles of DTs are not well suited for being
accelerated in CPU and GPU, due to thread synchronization issues, load
imbalance and uncoalesced memory acces ~\textsuperscript{\hyperref[csl:435]{435}}.~ Recently,
we showed that DTs ensembles can be mapped to analog CAM, providing a
one-cycle and conversion-less inference, given that the output is
already digital and ready to be post processed ~\textsuperscript{\hyperref[csl:436]{436}}.~
Figure~{\ref{781830}}(e)~shows the DT of
Figure~{\ref{781830}}(d) mapped into an analog CAM
array, where each root-to-leaf path is encoded in a row. The digital MLs
output are connected to a conventional RAM storing the leaves values, or
prediction. All branches are executed in parallel, providing a
size-independent inference latency.

\subsubsection*{7.4.2. Challenges}

{\label{740809}}

Most of the challenges for building reliable analog CAM are shared with
crossbar arrays of memristors, although some circuit and device
requirements are different. Similarly, a wide range of programmable
levels in the memristor conductance is desirable, but different from
crossbar array slicing techniques for increasing
precision~\textsuperscript{\hyperref[csl:437]{437}}~can't be performed. Range-based analog CAMs
have a unary encoding thus connecting multiple adjacent analog CAMs on a
given row doubles the number of levels as opposed to bit-slicing in
crossbar arrays which doubles the number of bits, leading to an
exponential overhead. Linear approaches for improved analog CAM
precision have been recently shown~\textsuperscript{\hyperref[csl:438]{438}}, although with
significant circuit complexity overhead. Moreover, given the voltage
divider operation performed directly on the memristor, its conductance
needs to be stable at higher read voltages, e.g. 1 V, to be able to
control efficiently the pulldown transistor.~

While a small 2x2 array was realized~in the 180 nm technology node and
large array simulations at the 16 nm technology node were
performed~\textsuperscript{\hyperref[csl:432]{432}}~experiments on large arrays, i.e. 128x16,
have yet to be realized. There are several concerns to be solved at the
array level, for example, the impact of memristor variation and noise on
the search accuracy and the impact of parasitics capacitance and
resistances on a reliable operation. Moreover, due to the non-linear
input/output relationship, even defining the appropriate patterns for
the program is challenging and ad hoc circuit-aware programming routines
should be performed.~

Finally, a full and general architecture design has yet to be performed,
to feed and pull data with enough bandwidth to take advantage of the
single-cycle memory lookup operation, eventually including multiple
independent cores and performing operations on a large amount of data in
parallel. The architecture should be programmable, not limited to
tree-based ML inference,~ allowing multiple tasks such as associative
searches~\textsuperscript{\hyperref[csl:433]{433}}, and resilient to errors due to memristor
variations for example including error correction codes routines. ~

\subsubsection*{7.4.3. Potential Solutions}

{\label{735339}}

First, large arrays of CMOS-integrated analog CAM with memristor devices
should be designed and fabricated. Techniques for increasing the
memristor conductance stability at relatively high reading voltage have
been presented and should be used to develop analog CAM-specific
memristor stack. For example, a larger oxidation layer or/and a
controlled deposition to limit the oxygen defect in it can be used to
increase the set voltage of memristor devices, which could result in
better stability.

Analog CAM circuits to support a higher number of bits can be designed
by studying the appropriate logic functionalities in the case of a
separate comparison of the least significant and most significant bits,
for example by mapping different logic functions between adjacent
cells~\textsuperscript{\hyperref[csl:439]{439}}. In principle, it is possible to implement any
kind of logic operation between adjacent CAM cells by opportunely
connecting the pulldown transistors. Custom program and verify
algorithms, efficiently including process variation and noise are being
developed~\textsuperscript{\hyperref[csl:440]{440}} , in order to design appropriate target
patterns to program in the memristor conductance while performing
non-linear operations.

Finally, a similar architecture to the crossbar array accelerators with
a custom instruction set architecture (ISA) compiled from popular tools
such as sk-learn, can be implemented~\textsuperscript{\hyperref[csl:441]{441}}. The analog CAM
operation should be abstracted enough in order to efficeintly map
different workloads in a hierarchical way to the programmable
accelerator, with multiple cores handling a different part of the
problems, and a global network on chip accumulating the results to
finalize the computation. In order to make each core operation reliable,
an error detection scheme already designed for TCAMs can be adapted to
the new analog CAM to efficiently re-program a given device once the
state has been drifted out of the desired one.~\textsuperscript{\hyperref[csl:442]{442}}

\subsubsection*{\texorpdfstring{7.4.4. Conclusion
\emph{\emph{~}}}{7.4.4. Conclusion ~}}

{\label{347704}}

While a lot of attention has been spent in the last years by academia
and industry research for developing in memory computing structures
based on linear operators, such as crosspoint arrays, recently a renewed
interest in memristor-based CAMs has arisen. Different circuit for
memristive TCAMs and analog CAM have been proposed, targeting multiple
applications, but a circuit-level integration of CMOS circuitry with a
BEOL-integrated memristor device along with a system-level analysis of
performance at scale has yet to be shown. While engineering such
required milestones, researchers should also focus on exploring new
applications exploiting open-sourced circuit models, given the novelty
of the idea it is in fact likely that we are just scratching the surface
of the potentiality of such computing primitives given the inherited
non-linear operation performed with unprecedented speed and energy
efficiency. 
\vspace{0.5cm}
{\label{926630}}

\subsection*{7.5 Optimization solvers~}

{\label{273267}}

\begin{quote}
\emph{John Paul Strachan, Dmitri Strukov}
\end{quote}

\subsubsection*{7.5.1. Status}

{\label{403971}}

One of the best examples of a high-risk-high-reward area for novel
computing is in the area of optimization solvers.~ Here, the ambition is
to offer~\emph{some~}form of speed-up or reduced resource requirements
(memory, energy, etc.) in solving computationally expensive
combinatorial optimization problems.~ In such problems, the task is
to~minimize a given cost (or energy) function by choosing the best
configuration within a large dimensional space.~ Well-known examples
include the Traveling Salesmen problem (finding a minimal route uniquely
visiting various cities), graph coloring (coloring nodes such that
connected nodes have different colors, while using the fewest colors ),
or training the weights in an artificial neural network (ANN).~ While
some problem classes and instances can be solved approximately and
quickly using greedy algorithms, many remain intractable and no known
algorithm exists that can solve them all with polynomial resources and
in polynomial time.~ Even a modest speed-up could offer immediate
benefits in practical applications across planning, wireless
communications, bioinformatics, routing, finance, and many more.

The above challenge has inspired many research communities to develop
both new algorithms and new physical hardware approaches, often in
conjunction.~ A prominent example of this was the recognition that many
optimization problems can be mapped to problems in
physics\textsuperscript{\hyperref[csl:443]{443}}, such as finding the ground state of an Ising
system of coupled up-or-down spins. This has driven interest in
so-called Ising solvers, which can have many physical realizations
including optical parametric oscillators, coupled electrical
oscillators, magnetic tunnel junctions, analog memristors, and others
~\textsuperscript{\hyperref[csl:348]{348}}.~ From the early neural-inspired communities, John
Hopfield's nonlinear networks of analog neurons turned out to also solve
planted optimization problems\textsuperscript{\hyperref[csl:444]{444}}. Quantum or
quantum-inspired approaches have also been pursued, leveraging the
quantum adiabatic theorem, where a ground state (optimal) solution is
found by evolving slowly from an initial, simple Hamiltonian toward a
final Hamiltonian that encodes the desired problem\textsuperscript{\hyperref[csl:445]{445}}.~
~Dynamical solvers are also being
pursued\textsuperscript{\hyperref[csl:446]{446}}\textsuperscript{\hyperref[csl:447]{447}}, where the large
configuration space is explored rapidly via coupled nonlinear dynamical
equations, with optimal configurations acting as attractor states.~

It is important to realize that the variety of physics and
quantum-inspired approaches discussed above offer primarily
an~algorithmic paradigm for solving optimization problems.~ The
underlying materials, physics, and computational elements remain quite
flexible.~ For example, traditional CPUs, GPUs, and FPGAs can be used to
simulate nonlinear dynamical equations.~ Or electronic CMOS-based ring
oscillators can be built and resistively or capacitively coupled in
order to simulate the magnetically interacting spins in an Ising model.~
Thus, there are many possible combinations of algorithmic or
computational models and the physical substrate where computations are
performed. It is easy to mistake one for the other.~ Yet, despite such
variety in possible physical realizations, some common underlying
challenges emerge that must be handled in order to successfully engineer
an efficient and broadly useful optimization solver.~ The remaining
portion of this document lists these challenges and opportunities.~

\subsubsection*{7.5.2. Challenges}

{\label{933481}}

\subsubsection*{\texorpdfstring{\emph{Challenges of mapping to
hardware~}}{Challenges of mapping to hardware~}}

{\label{587127}}

\begin{figure}[H]
\begin{center}
\includegraphics[width=0.70\columnwidth]{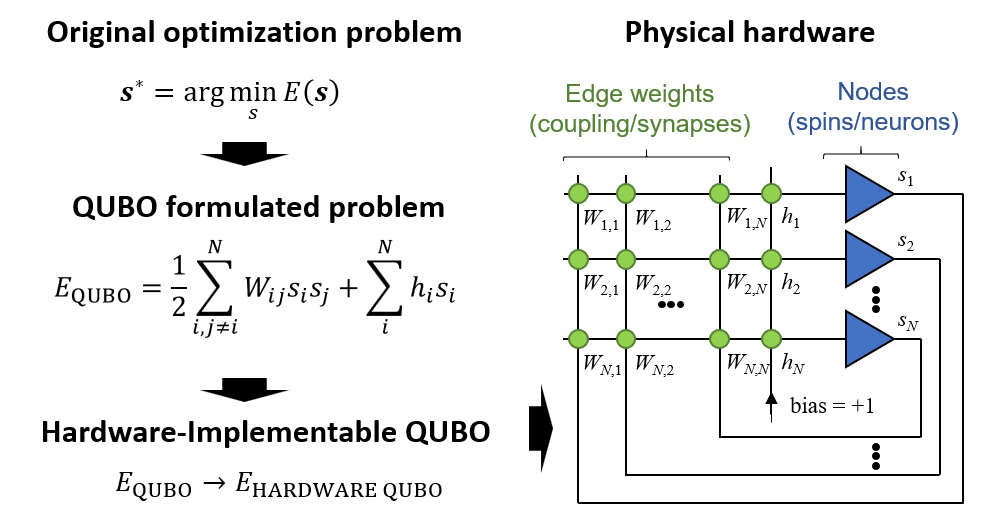}
\caption{{Typical steps for solving combinatorial optimization problems with Ising
machine. The original problem is formulated as quadratic unconstrained
optimization (QUBO) problem. The QUBO problem may need to be modified,
e.g., sparsified or decomposed to smaller subproblems, before it can be
mapped on the targeted Ising machine. Note that the figure shows fully
connected Ising machine and simplest baseline QUBO formulation.
Efficient implementation of different annealing techniques approaches
and solving constrained optimization problems relies on the ability to
adjust coupling weights during runtime.
{\label{928324}}%
}}
\end{center}
\end{figure}

Mapping optimization problems to physical systems, such as a hardware
Ising model (illustrated in Figure {\ref{928324}}), is
an attractive approach, but also highlights many key challenges.~ An
Ising model is based on pair-wise couplings of binary spins, and
therefore, up to quadratic terms appear in an energy function with
binary variables, and there is no explicit mechanism to enforce
additional constraints (such as summing to certain integer values). This
is also known as a Quadratic Unconstrained Binary Optimization (QUBO)
type, having an energy function in the case of an Ising model,

\[E_{\textrm{QUBO}}=\frac{1}{2}\sum_{i,j\ne i}^{ }W_{ij}s_is_j+\sum_i^{ }h_is_i\ .\]

Yet many optimization problems involve higher than quadratic
interactions.~ For example, a~\emph{k}-SAT Boolean satisfiability
problem involves terms of order~\emph{k}. Such higher-order interactions
must be mapped down to quadratic terms, through the introduction of
additional auxiliary variables.~ The result is a new total number of
variables that is polynomially larger than the original number of
variables, leading to, in the worst case, an exponential penalty in
terms of the configuration space that needs to be searched. This has
motivated explorations of algorithms with higher-order
interactions\textsuperscript{\hyperref[csl:448]{448}} and their physical realizations.\selectlanguage{english}

Additionally, QUBO formulations are sometimes inefficient, even for
natively quadratic problems. For example, the typical QUBO approach for
a~\emph{K}-city traveling salesman problem is to encode each route
with~\emph{K} one-hot-encoded~\emph{K}-bit vectors representing the
visitation order of each city~\textsuperscript{\hyperref[csl:443]{443}}.~ The result is a
quadratic scaling for QUBO variables, i.e.,~\emph{N}
=~\emph{K}\textsuperscript{2}, with the number of cities. This can lead
to a worst-case exponential penalty in the configuration space to search
as a function of the number of variables.

\subsubsection*{\texorpdfstring{\emph{Scaling
challenges}}{Scaling challenges}}

{\label{554283}}

Efficient mapping (``embedding'') of an optimization problem to the
underlying Ising hardware may require further modification of the QUBO
formulation.~ Indeed, the naive implementation of \emph{N}-variable QUBO
problems requires~\emph{N}\textsuperscript{2~}coupling weights to allow
programmable coupling with unique weights between any pair of neurons.
\emph{N} can be more than a million for many practical sparsely-coupled
optimization problems (such as already mentioned TSP and SAT problems),
which is clearly unacceptable for most Ising model implementations. To
address this challenge, a general approach is to decompose the original
QUBO problem into smaller subproblems that can be implemented in the
hardware. For example, one can partition a large neuron connectivity
graph into smaller subgraphs, with minimized number of edges between
subgraphs. The corresponding subgraph subproblems are then solved
independently, e.g., by fixing values of variables not participating in
the currently solved subgraph. The downside of such an approach can be
lower solution quality or longer time, even if all subproblems are
individually solved optimally. The scaling problem is further
exacerbated for hardware approaches with limited coupling capabilities.
~For example, if only limited neighbor connectivity is possible -- such
as in the~so-called Chimera graph topology implemented in the DWave
interconnected superconducting bits~ -- then the embedding comes with
even higher overhead. In this case, the original QUBO problem is first
sparsified to meet the connectivity limitations by adding redundant
variables, and then partitioned into subproblems. The result can be an
exponential slowdown\textsuperscript{\hyperref[csl:449]{449}} in solution convergence.

\subsubsection*{\texorpdfstring{\emph{Precision
challenges}}{Precision challenges}}

{\label{138413}}

Many emerging device technologies do not allow for realizing accurate
weight couplings and performing precise computation of node updates. On
the other hand, the weight precision should be sufficient to encode the
weight dynamic range. More generally, the non-idealities in the physical
couplings need to be low enough not to distort the global energy minima.
Detailed studies of non-idealities in, for example, Ising machine
operations are so far very limited; however, an initial insight is
provided by similar studies for training artificial neural networks, a
type of optimization task. Training to the highest accuracy requires at
least 4-bit weight and dot-product computation precision for many deep
learning models\textsuperscript{\hyperref[csl:450]{450}}. The precision requirements are higher
for very compact models such as MobileNet and lower for larger,
redundant models such as VGG. This~is likely inversely correlated with
the number of global minima and their basin volumes. Therefore, higher
precision requirements are expected for solving harder combinatorial
optimization problems and/or when better solution quality is sought.~

\subsubsection*{7.5.3. Potential Solutions}

{\label{986972}}

Advances are needed across algorithmic, device, and architectural
levels.~ On the algorithmic front, it was already stressed above that
improved mappings are needed to ensure efficient conversion from the
original optimization problem to QUBO, as well as~efficient embedding.~

On the device front, we need to identify the best device and material
implementations to overcome challenges in connectivity and interactions.
Are there any material and device concepts that would allow
implementation of the higher order in the hardware? The scaling issues
highlight the importance of efficiently implementing coupling weights.
In this respect, emerging analog memory devices and in-memory computing
with dot-product circuits are especially attractive. Optical computing
approaches are attractive because of the high fan-in interconnect that
can be attained. Quantum computing is attractive because of quantum
annealing, which offers a tunneling mechanism through energy barriers.~
Many tricks on the device and circuit levels can be borrowed from the
work on neuromorphic inference to improve tolerance to non-idealities
and increase precision.

On the architectural side, we need more flexible designs to handle
broadly varying levels of difficulty and size. Many problems are locally
dense but globally sparse. For example, hard~\emph{k-}SAT problems are
sparse, with sparseness increasing with model size, and the architecture
should take advantage of such sparseness by mapping highly
interconnected neuron subgraphs onto fully connected smaller size Ising
machines and interconnecting Ising machines with routing networks.~

\begin{figure}[H]
\begin{center}
\includegraphics[width=0.70\columnwidth]{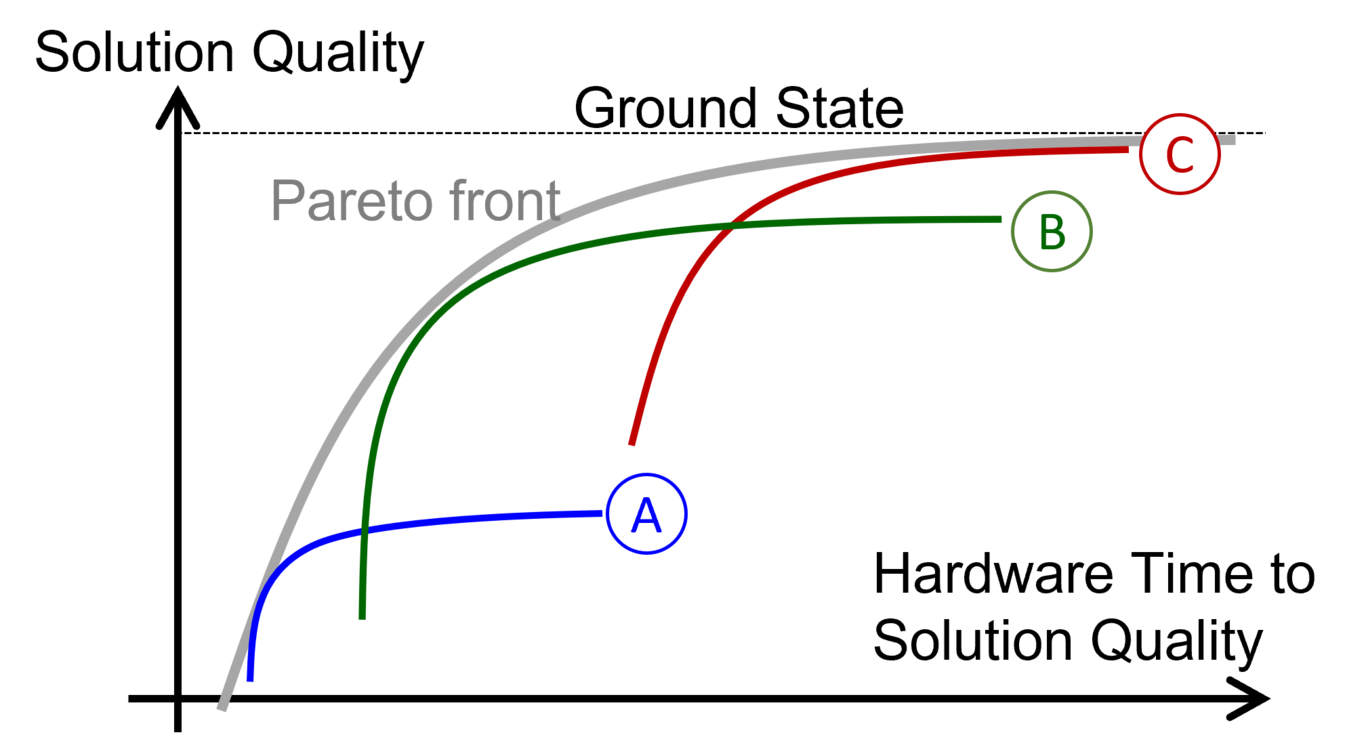}
\caption{{Benchmarking and design space for optimization hardware (shown
schematically). We expect solver implementations of type (A) that are
suitable for quickly finding lower-quality solutions but incapable of
reaching higher solution qualities, such as, for example,~fast but
approximate hardware heavily relying on parallel decomposition
algorithms and/or imprecise circuits. At the other end of the spectrum
are brute-force (complete) approaches (C), e.g., running an algorithm of
checking all possible solutions on the conventional high-precision
computer, that are slow but guaranteed to find the global optimum. Other
approaches (B) may perform better for medium solution quality and time
cost. We expect a Pareto front formed by different hardware approaches
highlighting that linear improvement in solution comes at exponentially
longer hardware times. The figure is motivated by similar dependencies
at the algorithmic level (so-called run-length distribution figures) and
similar hardware tradeoffs in neuromorphic inference and
training~\protect\textsuperscript{\hyperref[csl:451]{451}}. ~
{\label{922461}}%
}}
\end{center}
\end{figure}

A key feature for further advances on every front will be proper
benchmarking. The design space is very broad and is further complicated
by the wide range of applications with different characteristics
(hardness, coupling density, problem size).
Figure~{\ref{922461}}~shows the key metrics -- solution
quality found as a function of hardware time to solution at a fixed
power consumption and hardware resource budgets, highlighting the
important tradeoffs to identify promising solutions and guide further
design. Such a figure can be drawn for a collection of benchmarks or an
individual benchmark and, in the general case, would change with the
scale of the problems.~ Note that the fixed resource budget in
Figure~{\ref{922461}} is essential for a fair
comparison of different designs because~time to solution can be improved
with parallelism, i.e., spending more energy and/or relying on higher
complexity hardware. (Alternatively, Figure
{\ref{922461}} can be extended to show time, energy,
and hardware complexity to the solution quality. Such characterization
would be more complete, though less insightful when sparse data are
available.)~ ~\selectlanguage{english}

For example, we anticipate that low precision and/or restricted
connectivity systems will be much faster and energy-efficient when worse
precision is required. Such systems may never reach the highest quality
solutions, e.g., because inherent error modifies the energy landscape or
the subspace. On the other hand, a brute force algorithm, e.g.,
performed on a conventional computer, would guarantee finding optimal
solutions. The first emerging optimization solvers (such as Ising
machines) can occupy intermediate locations in these trade-off curves,
tuned to the desired priority metrics.

\subsubsection*{7.5.4. Conclusions}

{\label{821792}}

There is a great potential to harness physics- and brain-inspired
approaches to improve today's computing systems for solving optimization
problems.~ Yet, we see challenges that must be overcome at many
materials, circuits, system, and algorithmic levels in order to realize
this potential.~ A variety of approaches are under exploration that use
optical, quantum, magnetic, or electronic components, sometimes in
combinations.~ But the core requirements and issues are similar across
all of them, as outlined here.~ In the end, it will be critical to
engage other, non-hardware communities to become enthusiastic users and
to help tool development for these emerging hardware systems. It will
only be through continuous feedback between the users, developers,
material scientists, and engineers that a reliable and flexible
optimization solver can be developed for large-scale practical problems.

\vspace{0.5cm}
\subsection*{8. Technological Maturity}

{\label{580179}}

\subsection*{8.1. Current status \& next
steps}

{\label{506960}}

\begin{quote}
\emph{Manuel Le Gallo, Stefano Ambrogio}
\end{quote}

\subsubsection*{8.1.1. Status}

{\label{317767}}

The various neuromorphic computing technologies and concepts covered in
this roadmap have shown promising results through various prototype chip
demonstrations realized by academic and industrial institutions.
SRAM-based in-memory computing accelerator chips could demonstrate
energy efficiencies \textgreater{}100 TOPS/W for 4-bit matrix-vector
multiplications and provide support for all the additional CNN and LSTM
inference operations with on-chip digital ALUs~\textsuperscript{\hyperref[csl:452]{452}}. The
recent analog accelerator based on nor-Flash memory from Mythic,
supporting 80M on-chip analog weights, demonstrated 3.3 TOPS/W
system-level energy efficiency for 8-bit calculation precision and could
run a pose-detection application while consuming only
3.73W~\textsuperscript{\hyperref[csl:453]{453}}. While several alternative technologies have
been recently explored, such as ferroelectric and magnetic devices, 2D
transistors and memtransistors (see Section
{\ref{595902}}), the investigations are still at the
level of single device or small array, preventing a proper computation
performance exploration. For this reason, in non-CMOS based
implementations, only the more mature PCM and RRAM emerging technologies
have been integrated into multi-core in-memory computing chips and could
demonstrate various neural network inference tasks, albeit not fully
end-to-end. Near software-equivalent accuracies and energy efficiencies
of 10 TOPS/W or higher for matrix-vector multiplications have been
reported~\textsuperscript{\hyperref[csl:454]{454},\hyperref[csl:174]{174},\hyperref[csl:56]{56},\hyperref[csl:455]{455}}. Less mature technologies such as
ferroelectric and magnetic memories have been successfully integrated
into small arrays with on-chip data converters~\textsuperscript{\hyperref[csl:76]{76}}.~
Memories based on spintronic, 2D and atomistic materials as well as
photonic processors based on resistive memories have been mainly
investigated at the individual device level or integrated into small
arrays without peripheral circuits performing data
conversions~\textsuperscript{\hyperref[csl:269]{269},\hyperref[csl:282]{282}}.~

Besides chips that aim at accelerating matrix-vector multiplications for
deep neural network inference tasks, platforms that can execute more
novel neuromorphic computing concepts via in-memory computing have been
demonstrated as well. A 64k-cell PCM chip from IBM with in-situ learning
capability using STPD and leaky integrate-and-fire neurons performing a
simple associative learning task was demonstrated \textsuperscript{\hyperref[csl:456]{456}}.
Another 1.4M-cell PCM chip implementing a restricted Boltzmann machine
with STDP learning rule could demonstrate low-power on-chip training and
inference on the MNIST dataset \textsuperscript{\hyperref[csl:171]{171}}. Spiking
implementations are also used to efficiently implement ResNET networks
on CIFAR-10 \textsuperscript{\hyperref[csl:457]{457}}. Although those platforms are small
prototypes and do not support end-to-end deployment of a variety of
models, the key computational blocks involved in the execution of the
algorithms have been successfully integrated on-chip together with the
in-memory computing crossbar arrays. Nonetheless, such prototype
demonstrations are still far behind the maturity of digital CMOS
computing platforms, which support full deployment of a wide variety of
models, often with end-to-end software stack. CMOS neuromorphic
computing chips such as IBM TrueNorth~\textsuperscript{\hyperref[csl:50]{50}} and Intel
Loihi~\textsuperscript{\hyperref[csl:51]{51}} have been made available as research platforms
to implement inference and training of spiking neural networks with some
software support being provided. Moreover, several digital
application-specific deep learning accelerators with end-to-end software
stack, such as Google Tensor Processing Unit, Amazon Inferentia,
Facebook's M.2 accelerator, and IBM Artificial Intelligence Unit have
reached a mature state in terms of production and system-level
integration \textsuperscript{\hyperref[csl:458]{458}}.

\subsubsection*{8.1.2.Challenges}

{\label{874428}}

Several challenges at the device- and system-level could hinder the
development of fully end-to-end in-memory computing accelerators. The
growing interest in inference tasks, where the neural network is firstly
trained in software and then deployed on-chip to get high throughput
(number of inferences per second) and low latency (time to process one
input) requires chips with multiple crossbar arrays, to account for
multiple layers in neural networks, and reasonably large size, to map
extended network layers (e.g., first layers in convolutional networks)
\textsuperscript{\hyperref[csl:454]{454},\hyperref[csl:459]{459}}. This leads to several device-level requirements,
such as high HRS/LRS resistance ratio, moderate endurance, high
retention, and low intrinsic variability. It is also beneficial to have
a LRS resistance high enough to limit the impact of the voltage drop in
the lines of the crossbar array during writing and readout. Although a
single device can be designed to easily meet one of those requirements,
the challenge is to build multiple chips each containing multiple arrays
of devices that should meet all of them. Because it is rather
challenging to simultaneously achieve such specifications with emerging
technologies, often tradeoffs have to be made depending on the envisaged
application. As an example, a lower endurance can be acceptable to
tradeoff for a higher HRS/LRS ratio for inference purposes. In addition,
several aspects regarding the operating voltages and currents of these
devices need to be considered when integrating them into crossbar
arrays. Programming voltages in excess of 2V and large programming
currents (\textgreater{}100\selectlanguage{greek}µ\selectlanguage{english}A) necessitate the use of large access
transistors that can block this voltage and drive this current.
Therefore, minimizing the programming voltages and currents is critical
to achieve high device density in the crossbar.

On training chips, where the weights of the neural network are actively
changed on-chip to get high-accuracy processing on several tasks
(classification, language processing, \ldots{}) specifications are
higher. In addition to the inference requirements, data needs to
back-propagate through the network. Such behavior is critical in analog
cores, requiring fully symmetric peripheral circuitry during forward and
back propagation, absence of any polarity dependence of the resistive
devices (since each memory element is generally biased in a different
way between forward and back propagation). Clearly, device challenges
are different: while inference requires high retention, low drift and
low temperature dependence, training requires high endurance and
symmetric conductance update behavior, leading to different material
choices.

Moreover, to achieve highly competitive system-level performance against
existing CMOS-based accelerators, further improvements in device-level
precision and compute density are required \textsuperscript{\hyperref[csl:460]{460}}. While
this poses material challenges on resistive memories, it also influences
the peripheral circuitry by limiting its available area while ensuring
that it does not restrain precision. In addition, real workloads involve
a variety of different operations other than matrix-vector
multiplications that need to be implemented in separate digital
computing units. Therefore, power-hungry analogue-to-digital conversion
is needed at the crossbar outputs, which limits energy efficiency.
Moreover, a fast and flexible communication scheme together with
highly-efficient pipelining of the digital compute units and
intermediate SRAM storage is primordial to ensure that they do not
dominate the latency and power consumption. Finally, a user-friendly
software-stack that tightly integrates such hardware with common machine
learning frameworks (Pytorch/Tensorflow) is key for its widespread
adoption within the community.

\subsubsection*{8.1.3.Potential Solutions}

{\label{268013}}

The advancement in the field requires a joint development of both
material aspects of the memory, algorithms used to train the neural
networks, efficient network mapping techniques over multiple crossbar
arrays and careful design of the peripheral circuitry. In general,
inference/training chips are required to have high-accuracy computation
and high energy (operations per energy) and high area (operations per
millimetre square) efficiency.

To improve the compute precision up to four or five-bit fixed-point
arithmetic, it is essential to minimize the temporal conductance
fluctuations (such as noise, conductance drift and temperature
dependence). To achieve this, additional material research such as
proper incorporation of dopants, memories composed of multi-layer
materials stacks \textsuperscript{\hyperref[csl:461]{461}}, and exploiting material
confinement to change device properties \textsuperscript{\hyperref[csl:330]{330}}, is
essential. To improve the compute density, besides scaling both the
devices and the associated access transistors, high-density arrays need
to be integrated at the back end of a CMOS wafer. To decrease the
computational time required to convert the integrated charge/voltage
after a Multiply-and-Accumulate (MAC) operation, column multiplexing
should be avoided, using a per-column circuitry, shrinking the area to
accommodate Analog-to-Digital Converters (ADCs). Exploration of ADC
designs with low-number of bits enables efficient integration, while
still keeping a reasonably high MAC accuracy \textsuperscript{\hyperref[csl:459]{459}}.
Another approach uses fully analog peripheral circuitry, which is more
power efficient \textsuperscript{\hyperref[csl:454]{454}}, posing however more stringent
limitations on computation precision and available activation functions.
As an example, ReLU has been demonstrated in the analog domain, while
non-linear Sigmoid or Tanh generally require digital processing.

Even in the case of highly-efficient crossbar arrays, the general
performance could still be poor due to Amdahl's Law, since fast analog
cores would generate large amounts of data that the neighbouring digital
cores need to process~\textsuperscript{\hyperref[csl:460]{460}}. In other words, highly
efficient analog circuitry would require highly efficient digital blocks
closely located. Therefore, to improve the hybrid analog/digital chip,
spatial network mapping, analog/digital cores spatial location and data
communication need to be codesigned, leading to a general trade-off
between chip reconfigurability (ability to map any type of network) and
performance (ability to efficiently process specific networks)
\textsuperscript{\hyperref[csl:460]{460}}.

\subsubsection*{8.1.4.Conclusion}

{\label{734420}}

Computing in memory has greatly improved in recent years thanks to
several on-chip demonstrations. However, challenges need to be overcome
to get to product level, such as 4-to-5 bit computing precision together
with performances in the range of 100/200 TOPS/W. Achieving such
specifications is critical to provide a competitive advantage over
existing purely digital processing cores. This will be obtained by
optimizing the full computing stack: devices will require large
uniformity, reasonably high resistance levels~ and overall low noise
behavior. Compact peripheral circuitry will be essential to reduce the
chip area, and careful spatial mapping of analog and digital dedicated
cores will need to improve the performance on a variety of neural
networks.

\vspace{0.5cm}

{\label{262129}}

\subsection*{9. Epilogue and Concluding Remarks for the
Roadmap}

{\label{714363}}

Ilia Valov, Rainer Waser, Adnan Mehonic

After decades of reliance on transistor-based electronics, we are now
delving into an era where the exploration of innovative nanoelectronic
technologies based on functional materials is more crucial than ever.
For instance, nanoscale memristive devices have emerged as key
components for future nanoelectronics. Their straightforward stack
structure, diverse functionalities, and specific benefits such as
scalability, broad temperature stability and operation range, and
resilience to high-energy particles and electromagnetic interference
make them indispensable for numerous applications. Furthermore,
memristors and similar novel technologies now serve as foundational
units for the next generation of brain-inspired computing architectures.
From their introduction in the '60s of the last century as resistive
switching memories, through the relation to Chua's memristor till
nowadays using them as artificial neurons and synapses, memristive
devices have passed decades of intensive research with respect to both
fundamentals and applications by academia and industry.

Examining the foundational aspects of materials science is pivotal for
developing new nanoelectronic technologies. Using memristive technology
as an example: the appeal of memristive devices stems from the multitude
of benefits they provide, which are influenced and modulated by the
materials and processes that dictate their behaviour and
functionalities. Here especially the relation between materials,
materials properties, physicochemical processes, and functionalities
should be highlighted. A huge spectrum of materials has been used for
switching films -- 1D (single molecules or molecular clusters), 2D
(Graphene, hBN, MoS2, MoSe2 etc.) and 3D, including inorganic, organic
and biomaterials. ~Several physical phenomena were reported to lead to
memristive behaviour - phase change, redox reactions and ionic
transport, electronic effects, van der Waals forces, and magnetic and
magneto-resistance changes, all covered in this roadmap.

All different phenomena, of course, depend strongly on the used
materials. The main challenge appears to properly select a combination
of appropriate materials and their dimensions (i.e. thickness and
lateral scale). Apparently, a simple two-electrode cell is composed of a
switching layer or in many cases more switching layers, two electrodes
and capping film(s). The devices are exposed to extreme operating
conditions such as current densities of up to 1 million Amperes per
square centimetre and electric fields of 10\textsuperscript{8} Volts per
meter. Due to these extreme conditions and the nano-dimensions of the
films their thermodynamic and kinetic behaviour typically deviate
strongly from their macroscopic counterparts going much beyond the frame
of the classical knowledge in terms of chemical and mechanical
stability, point defect chemistry and transport properties. The
nano-size effects are especially prominent considering the switching
films. These are typically materials, that in a macroscopic sense are
insulators, some used as well as high-k dielectrics, but turn into mixed
solid electrolytes at thicknesses below \textasciitilde{} 50 nm.
Furthermore, due to enhanced surface energy excess, the layers react
chemically within the stack even if the change in the Gibbs energy for
the reaction is positive. As a result, intermediate films can be formed,
with similar thicknesses as the switching layers, that can either
inhibit or enhance charge and mass transport through interfaces,
strongly influencing all related processes. Other effects caused by
impurities, absorption of moisture and/or oxygen, local changes of
concentrations and nanocavities are known to additionally complicate the
control over the device's behaviour and characteristics. This poses a
challenge, yet also presents an opportunity to uncover unique device
physics and explore alternative device functionalities (e.g.
neuromorphic functionalities).

The way to keep control over the devices and to further expand the
horizon of functionalities is by using a material science-based approach
where materials properties and processes are studied in very detail and
this knowledge should be applied in the design of the devices.

The ``Roadmap to Neuromorphic Computing with Emerging Technologies''
roadmap seeks to tackle these issues and present a contemporary
perspective on the intersections between current materials science,
electronic engineering, and system design. Its ultimate goal is to delve
into alternative computing models, particularly brain-inspired
(neuromorphic) computing. Eminent groups and experts span a wide
spectrum of relevant subjects and technologies, fostering a platform for
idea exchange. Additionally, they highlight next-generation neuromorphic
hardware, emphasizing the foundational role of functional materials and
innovative device technologies.~Although materials have been the main
focus of the roadmap, our aim was to provide a more holistic overview
and highlight a range of emerging and highly active research areas. As
such, there are many details and specific material considerations not
covered here that we strongly recommend the authors explore in the
extensive background literature, much of which is published in special
issues~\textsuperscript{\hyperref[csl:13]{13}},~\textsuperscript{\hyperref[csl:462]{462}},~\textsuperscript{\hyperref[csl:463]{463}},~\textsuperscript{\hyperref[csl:464]{464}}or
excellent reviews~\textsuperscript{\hyperref[csl:12]{12}},~\textsuperscript{\hyperref[csl:8]{8}}. Likewise, in
order to keep the format of the roadmap relatively compact, we have not
elaborated on a number of equally valid and highly promising approaches
in the context of neuromorphic technology development. Some notable
examples include~ adaptive matters and computation based on disorder
~\textsuperscript{\hyperref[csl:287]{287}},~\textsuperscript{\hyperref[csl:465]{465}}, systems based on organic~
perovskites~\textsuperscript{\hyperref[csl:466]{466},\hyperref[csl:467]{467}},~ ionic-liquid based
devices~\textsuperscript{\hyperref[csl:468]{468},\hyperref[csl:469]{469}}, and molecular devices~\textsuperscript{\hyperref[csl:470]{470},\hyperref[csl:471]{471},\hyperref[csl:63]{63}}.
Of course, the list is not exhaustive, and other approaches, physical
systems, and technologies are emerging.

Likewise, in order to keep the format of the roadmap in relatively
conpact form, we have not elaborated on a number of equally valid and
highly promising approahces in context of neuromorphic technology
development.

Finally, recommending the most promising material systems is a complex
task. In general, RRAM technology might have an advantage due to its
simplicity and CMOS-friendliness. However, recent developments in
HfO2-based FeRAMs represent a highly active area of research. MRAM is
likely the most mature technology, with already available products,
while PCM has seen significant interest recently. 2D materials are
expected to integrate with all these technologies, providing further
device improvements. It is important to keep in mind that the
requirements might be dramatically different depending on targeted
applications. For embedded systems, ease of integration and full CMOS
compatibility are likely the most important factors. Conventional NVM
devices need to outperform Flash, while in the context of computing,
higher endurance will likely be needed. Additionally, the requirements
for less conventional analog or neuromorphic functionalities are
somewhat less defined but equally relevant and in development.~One
should also bear in mind the gap that exists between academic research
on proof-of-concept demonstrators and industrial R\&D. Industrial R\&D
must consider not only technical factors but also economic feasibility
\textsuperscript{\hyperref[csl:472]{472}}.

Neuromorphic technologies are undoubtedly poised to be strong contenders
for the future of computing, whether based on conventional digital,
analog, or conceptually different computing and signal processing
paradigms.

\selectlanguage{english}
\FloatBarrier
\section*{References}\sloppy

\phantomsection
\label{csl:1}1 A. Mehonic and A.J. Kenyon, Nature \textbf{604}, 255 (2022).

\phantomsection
\label{csl:2}2 T. Gokmen and Y. Vlasov, Front Neurosci \textbf{10}, 333 (2016).

\phantomsection
\label{csl:3}3 M.J. Marinella, S. Agarwal, A. Hsia, I. Richter, R. Jacobs-Gedrim, J. Niroula, S.J. Plimpton, E. Ipek, and C.D. James, {IEEE} Journal on Emerging and Selected Topics in Circuits and Systems \textbf{8}, 86 (2018).

\phantomsection
\label{csl:4}4 H.-Y. Chang, P. Narayanan, S.C. Lewis, N.C.P. Farinha, K. Hosokawa, C. Mackin, H. Tsai, S. Ambrogio, A. Chen, and G.W. Burr, {IBM} Journal of Research and Development \textbf{63}, 8 (2019).

\phantomsection
\label{csl:5}5 C. Mead, Nature Electronics \textbf{3}, 434 (2020).

\phantomsection
\label{csl:6}6 A. Mehonic and A.J. Kenyon, Frontiers in Neuroscience \textbf{10}, (2016).

\phantomsection
\label{csl:7}7 S. Kumar, R.S. Williams, and Z. Wang, Nature \textbf{585}, (2020).

\phantomsection
\label{csl:8}8 P. Mannocci, M. Farronato, N. Lepri, L. Cattaneo, A. Glukhov, Z. Sun, and D. Ielmini, APL Machine Learning \textbf{1}, (2023).

\phantomsection
\label{csl:9}9 D.V. Christensen, R. Dittmann, B. Linares-Barranco, A. Sebastian, M.L. Gallo, A. Redaelli, S. Slesazeck, T. Mikolajick, S. Spiga, S. Menzel, I. Valov, G. Milano, C. Ricciardi, S.-J. Liang, F. Miao, M. Lanza, T.J. Quill, S.T. Keene, A. Salleo, J. Grollier, D. Markovi{\'{c}}, A. Mizrahi, P. Yao, J.J. Yang, G. Indiveri, J.P. Strachan, S. Datta, E. Vianello, A. Valentian, J. Feldmann, X. Li, W.H.P. Pernice, H. Bhaskaran, S. Furber, E. Neftci, F. Scherr, W. Maass, S. Ramaswamy, J. Tapson, P. Panda, Y. Kim, G. Tanaka, S. Thorpe, C. Bartolozzi, T.A. Cleland, C. Posch, S.C. Liu, G. Panuccio, M. Mahmud, A.N. Mazumder, M. Hosseini, T. Mohsenin, E. Donati, S. Tolu, R. Galeazzi, M.E. Christensen, S. Holm, D. Ielmini, and N. Pryds, Neuromorphic Computing and Engineering \textbf{2}, 022501 (2022).

\phantomsection
\label{csl:10}10 D. Ielmini, Z. Wang, and Y. Liu, {APL} Materials \textbf{9}, 050702 (2021).

\phantomsection
\label{csl:11}11 G. Li, L. Deng, H. Tang, G. Pan, Y. Tian, K. Roy, and W. Maass, (2023).

\phantomsection
\label{csl:12}12 M.-K. Kim, Y. Park, I.-J. Kim, and J.-S. Lee, IScience \textbf{23}, (2020).

\phantomsection
\label{csl:13}13 I.K. Schuller, A. Frano, R.C. Dynes, A. Hoffmann, B. Noheda, C. Schuman, A. Sebastian, and J. Shen, Applied Physics Letters \textbf{120}, (2022).

\phantomsection
\label{csl:14}14 F. Torres, A.C. Basaran, and I.K. Schuller, Advanced Materials \textbf{35}, (2023).

\phantomsection
\label{csl:15}15 A.W. Burks, H.H. Goldstine, and J. Neumann, in \textit{The Origins of Digital Computers} (Springer Berlin Heidelberg, 1982), pp. 399–413.

\phantomsection
\label{csl:16}16 R.R. Schaller, {IEEE} Spectrum \textbf{34}, 52 (1997).

\phantomsection
\label{csl:17}17 M.J. Flynn, Proceedings of the {IEEE} \textbf{54}, 1901 (1966).

\phantomsection
\label{csl:18}18 R.H. Dennard, F.H. Gaensslen, H.-N. Yu, V.L. Rideout, E. Bassous, and A.R. LeBlanc, {IEEE} Journal of Solid-State Circuits \textbf{9}, 256 (1974).

\phantomsection
\label{csl:19}19 H. Esmaeilzadeh, E. Blem, R.S. Amant, K. Sankaralingam, and D. Burger, in \textit{Proceedings of the 38th Annual International Symposium on Computer Architecture} ({ACM}, 2011).

\phantomsection
\label{csl:20}20 H.-S.P. Wong, H.-Y. Lee, S. Yu, Y.-S. Chen, Y. Wu, P.-S. Chen, B. Lee, F.T. Chen, and M.-J. Tsai, Proceedings of the {IEEE} \textbf{100}, 1951 (2012).

\phantomsection
\label{csl:21}21 M. Horowitz, in \textit{2014 {IEEE} International Solid-State Circuits Conference Digest of Technical Papers ({ISSCC})} ({IEEE}, 2014).

\phantomsection
\label{csl:22} A. Pedram, S. Richardson, M. Horowitz, S. Galal, and S. Kvatinsky, \textit{IEEE Design \& Test} \textbf{34}, 39 (2017).

\phantomsection
\label{csl:23}23 A. Boroumand, S. Ghose, Y. Kim, R. Ausavarungnirun, E. Shiu, R. Thakur, D. Kim, A. Kuusela, A. Knies, P. Ranganathan, and O. Mutlu, {ACM} {SIGPLAN} Notices \textbf{53}, 316 (2018).

\phantomsection
\label{csl:24}24 A. Boroumand, S. Ghose, B. Akin, R. Narayanaswami, G.F. Oliveira, X. Ma, E. Shiu, and O. Mutlu, in \textit{2021 30th International Conference on Parallel Architectures and Compilation Techniques ({PACT})} ({IEEE}, 2021).

\phantomsection
\label{csl:25}25 S. Palacharla and R.E. Kessler, in \textit{Proceedings of 21 International Symposium on Computer Architecture} ({IEEE} Comput. Soc. Press, n.d.).

\phantomsection
\label{csl:26}26 R. Bera, K. Kanellopoulos, S. Balachandran, D. Novo, A. Olgun, M. Sadrosadat, and O. Mutlu, in \textit{2022 55th {IEEE}/{ACM} International Symposium on Microarchitecture ({MICRO})} ({IEEE}, 2022).

\phantomsection
\label{csl:27}27 S. Srinath, O. Mutlu, H. Kim, and Y.N. Patt, in \textit{2007 IEEE 13th International Symposium on High Performance Computer Architecture} (IEEE, 2007).

\phantomsection
\label{csl:28}28Supercomputing Frontiers and Innovations \textbf{1}, (2014).

\phantomsection
\label{csl:29}29 O. Mutlu, S. Ghose, J. G{\'{o}}mez-Luna, and R. Ausavarungnirun, in \textit{Emerging Computing: From Devices to Systems} (Springer Nature Singapore, 2022), pp. 171–243.

\phantomsection
\label{csl:30}30 F. Devaux, in \textit{2019 {IEEE} Hot Chips 31 Symposium ({HCS})} ({IEEE}, 2019).

\phantomsection
\label{csl:31}31 Y.-C. Kwon, S.H. Lee, J. Lee, S.-H. Kwon, J.M. Ryu, J.-P. Son, O. Seongil, H.-S. Yu, H. Lee, S.Y. Kim, Y. Cho, J.G. Kim, J. Choi, H.-S. Shin, J. Kim, B.S. Phuah, H.M. Kim, M.J. Song, A. Choi, D. Kim, S.Y. Kim, E.-B. Kim, D. Wang, S. Kang, Y. Ro, S. Seo, J.H. Song, J. Youn, K. Sohn, and N.S. Kim, in \textit{2021 {IEEE} International Solid- State Circuits Conference ({ISSCC})} ({IEEE}, 2021).

\phantomsection
\label{csl:32}32 J. Borghetti, G.S. Snider, P.J. Kuekes, J.J. Yang, D.R. Stewart, and R.S. Williams, Nature \textbf{464}, 873 (2010).

\phantomsection
\label{csl:33}33 S. Kvatinsky, D. Belousov, S. Liman, G. Satat, N. Wald, E.G. Friedman, A. Kolodny, and U.C. Weiser, {IEEE} Transactions on Circuits and Systems {II}: Express Briefs \textbf{61}, 895 (2014).

\phantomsection
\label{csl:34}34 V. Seshadri, D. Lee, T. Mullins, H. Hassan, A. Boroumand, J. Kim, M.A. Kozuch, O. Mutlu, P.B. Gibbons, and T.C. Mowry, in \textit{Proceedings of the 50th Annual {IEEE}/{ACM} International Symposium on Microarchitecture} ({ACM}, 2017).

\phantomsection
\label{csl:35}35 A. Shafiee, A. Nag, N. Muralimanohar, R. Balasubramonian, J.P. Strachan, M. Hu, R.S. Williams, and V. Srikumar, {ACM} {SIGARCH} Computer Architecture News \textbf{44}, 14 (2016).

\phantomsection
\label{csl:36}36 N. Hajinazar, G.F. Oliveira, S. Gregorio, J.D. Ferreira, N.M. Ghiasi, M. Patel, M. Alser, S. Ghose, J. G{\'{o}}mez-Luna, and O. Mutlu, in \textit{Proceedings of the 26th {ACM} International Conference on Architectural Support for Programming Languages and Operating Systems} ({ACM}, 2021).

\phantomsection
\label{csl:37}37 M. Prezioso, F. Merrikh-Bayat, B.D. Hoskins, G.C. Adam, K.K. Likharev, and D.B. Strukov, Nature \textbf{521}, 61 (2015).

\phantomsection
\label{csl:38}38 C. Posch, T. Serrano-Gotarredona, B. Linares-Barranco, and T. Delbruck, Proceedings of the {IEEE} \textbf{102}, 1470 (2014).

\phantomsection
\label{csl:39}39 T. Serrano-Gotarredona and B. Linares-Barranco, {IEEE} Journal of Solid-State Circuits \textbf{48}, 827 (2013).

\phantomsection
\label{csl:40}40 P. Lichtsteiner, C. Posch, and T. Delbruck, {IEEE} Journal of Solid-State Circuits \textbf{43}, 566 (2008).

\phantomsection
\label{csl:41}41 D.P. Moeys, C. Li, J.N.P. Martel, S. Bamford, L. Longinotti, V. Motsnyi, D.S.S. Bello, and T. Delbruck, in \textit{2017 {IEEE} International Symposium on Circuits and Systems ({ISCAS})} ({IEEE}, 2017).

\phantomsection
\label{csl:42}42 T. Finateu, A. Niwa, D. Matolin, K. Tsuchimoto, A. Mascheroni, E. Reynaud, P. Mostafalu, F. Brady, L. Chotard, F. LeGoff, H. Takahashi, H. Wakabayashi, Y. Oike, and C. Posch, in \textit{2020 {IEEE} International Solid- State Circuits Conference - ({ISSCC})} ({IEEE}, 2020).

\phantomsection
\label{csl:43}43 A. van Schaik and S.-C. Liu, in \textit{2005 {IEEE} International Symposium on Circuits and Systems} ({IEEE}, n.d.).

\phantomsection
\label{csl:44}44 S.-W. Chiu and K.-T. Tang, Sensors \textbf{13}, 14214 (2013).

\phantomsection
\label{csl:45}45 S. Caviglia, M. Valle, and C. Bartolozzi, in \textit{2014 {IEEE} International Symposium on Circuits and Systems ({ISCAS})} ({IEEE}, 2014).

\phantomsection
\label{csl:46}46 J.A. Perez-Carrasco, B. Zhao, C. Serrano, B. Acha, T. Serrano-Gotarredona, S. Chen, and B. Linares-Barranco, {IEEE} Transactions on Pattern Analysis and Machine Intelligence \textbf{35}, 2706 (2013).

\phantomsection
\label{csl:47}47 E. Painkras, L.A. Plana, J. Garside, S. Temple, F. Galluppi, C. Patterson, D.R. Lester, A.D. Brown, and S.B. Furber, {IEEE} Journal of Solid-State Circuits \textbf{48}, 1943 (2013).

\phantomsection
\label{csl:48}48 H. Markram, Scientific American \textbf{306}, 50 (2012).

\phantomsection
\label{csl:49}49 J.S. Sebastian Millner Andreas Hartel and K. Meier, .

\phantomsection
\label{csl:50}50 P.A. Merolla, J.V. Arthur, R. Alvarez-Icaza, A.S. Cassidy, J. Sawada, F. Akopyan, B.L. Jackson, N. Imam, C. Guo, Y. Nakamura, B. Brezzo, I. Vo, S.K. Esser, R. Appuswamy, B. Taba, A. Amir, M.D. Flickner, W.P. Risk, R. Manohar, and D.S. Modha, Science \textbf{345}, 668 (2014).

\phantomsection
\label{csl:51}51 M. Davies, N. Srinivasa, T.-H. Lin, G. Chinya, Y. Cao, S.H. Choday, G. Dimou, P. Joshi, N. Imam, S. Jain, Y. Liao, C.-K. Lin, A. Lines, R. Liu, D. Mathaikutty, S. McCoy, A. Paul, J. Tse, G. Venkataramanan, Y.-H. Weng, A. Wild, Y. Yang, and H. Wang, {IEEE} Micro \textbf{38}, 82 (2018).

\phantomsection
\label{csl:52}
52 \href{https://www.intel.com/content/www/us/en/newsroom/news/intel-scales-neuromorphic-research-system-100-million-neurons.html}{\textit{https://www.intel.com/content/www/us/en/newsroom/news/intel-scales-neuromorphic-research-system-100-million-neurons.html}}.

\phantomsection
\label{csl:53}53 Y. LeCun, Y. Bengio, and G. Hinton, Nature \textbf{521}, 436 (2015).

\phantomsection
\label{csl:54}54 A. Tavanaei, M. Ghodrati, S.R. Kheradpisheh, T. Masquelier, and A. Maida, Neural Networks \textbf{111}, 47 (2019).

\phantomsection
\label{csl:55}55 L.A. Camunas-Mesa, E. Vianello, C. Reita, T. Serrano-Gotarredona, and B. Linares-Barranco, {IEEE} Journal on Emerging and Selected Topics in Circuits and Systems \textbf{12}, 898 (2022).

\phantomsection
\label{csl:56}56 W. Wan, R. Kubendran, C. Schaefer, S.B. Eryilmaz, W. Zhang, D. Wu, S. Deiss, P. Raina, H. Qian, B. Gao, S. Joshi, H. Wu, H.-S.P. Wong, and G. Cauwenberghs, Nature \textbf{608}, 504 (2022).

\phantomsection
\label{csl:57}57 B. Linares-Barranco, Nature Electronics \textbf{1}, 100 (2018).

\phantomsection
\label{csl:58}58 S. Furber, Journal of Neural Engineering \textbf{13}, 051001 (2016).

\phantomsection
\label{csl:59}59 H. Hendy and C. Merkel, Journal of Electronic Imaging \textbf{31}, (2022).

\phantomsection
\label{csl:60}60 A. Sebastian, M.L. Gallo, R. Khaddam-Aljameh, and E. Eleftheriou, Nature Nanotechnology \textbf{15}, 529 (2020).

\phantomsection
\label{csl:61}61 S. Brivio, S. Spiga, and D. Ielmini, Neuromorphic Computing and Engineering \textbf{2}, 042001 (2022).

\phantomsection
\label{csl:62}62 G.W. Burr, R.M. Shelby, A. Sebastian, S. Kim, S. Kim, S. Sidler, K. Virwani, M. Ishii, P. Narayanan, A. Fumarola, L.L. Sanches, I. Boybat, M. Le Gallo, K. Moon, J. Woo, H. Hwang, and Y. Leblebici, Advances in Physics: X \textbf{2}, (2016).

\phantomsection
\label{csl:63}63 V.K. Sangwan and M.C. Hersam, Nature Nanotechnology \textbf{15}, (2020).

\phantomsection
\label{csl:64}64 Z. Wang, H. Wu, G.W. Burr, C.S. Hwang, K.L. Wang, Q. Xia, and J.J. Yang, Nature Reviews Materials \textbf{5}, 173 (2020).

\phantomsection
\label{csl:65}65 J. Grollier, D. Querlioz, K.Y. Camsari, K. Everschor-Sitte, S. Fukami, and M.D. Stiles, Nature Electronics \textbf{3}, 360 (2020).

\phantomsection
\label{csl:66}66 G. Milano, G. Pedretti, K. Montano, S. Ricci, S. Hashemkhani, L. Boarino, D. Ielmini, and C. Ricciardi, Nat Mater \textbf{21}, 195 (2022).

\phantomsection
\label{csl:67}67 G.W. Burr, R.M. Shelby, A. Sebastian, S. Kim, S. Kim, S. Sidler, K. Virwani, M. Ishii, P. Narayanan, A. Fumarola, and others, Advances in Physics: X \textbf{2}, 89 (2017).

\phantomsection
\label{csl:68}68 A. Sebastian, M. Le Gallo, R. Khaddam-Aljameh, and E. Eleftheriou, Nature Nanotechnology \textbf{15}, 529 (2020).

\phantomsection
\label{csl:69}69 E. Covi, H. Mulaosmanovic, B. Max, S. Slesazeck, and T. Mikolajick, Neuromorphic Computing and Engineering \textbf{2}, 012002 (2022).

\phantomsection
\label{csl:70}70 A.A. Talin, Y. Li, D.A. Robinson, E.J. Fuller, and S. Kumar, Advanced Materials 2204771 (2022).

\phantomsection
\label{csl:71}71 Y. van de Burgt, A. Melianas, S.T. Keene, G. Malliaras, and A. Salleo, Nature Electronics \textbf{1}, 386 (2018).

\phantomsection
\label{csl:72}72 R. Wang, J.-Q. Yang, J.-Y. Mao, Z.-P. Wang, S. Wu, M. Zhou, T. Chen, Y. Zhou, and S.-T. Han, Advanced Intelligent Systems \textbf{2}, 2000055 (2020).

\phantomsection
\label{csl:73}73 K. Zhu, S. Pazos, F. Aguirre, Y. Shen, Y. Yuan, W. Zheng, O. Alharbi, M.A. Villena, B. Fang, X. Li, A. Milozzi, M. Farronato, M. Mu{\~{n}}oz-Rojo, T. Wang, R. Li, H. Fariborzi, J.B. Roldan, G. Benstetter, X. Zhang, H. Alshareef, T. Grasser, H. Wu, D. Ielmini, and M. Lanza, Nature (2023).

\phantomsection
\label{csl:74}74 B.J. Shastri, A.N. Tait, T.F. de Lima, W.H.P. Pernice, H. Bhaskaran, C.D. Wright, and P.R. Prucnal, Nature Photonics \textbf{15}, 102 (2021).

\phantomsection
\label{csl:75}75 W. Zhou, N. Farmakidis, J. Feldmann, X. Li, J. Tan, Y. He, C.D. Wright, W.H.P. Pernice, and H. Bhaskaran, {MRS} Bulletin \textbf{47}, 502 (2022).

\phantomsection
\label{csl:76}76 S. Jung, H. Lee, S. Myung, H. Kim, S.K. Yoon, S.-W. Kwon, Y. Ju, M. Kim, W. Yi, S. Han, B. Kwon, B. Seo, K. Lee, G.-H. Koh, K. Lee, Y. Song, C. Choi, D. Ham, and S.J. Kim, Nature \textbf{601}, 211 (2022).

\phantomsection
\label{csl:77}77 M. Payvand, F. Moro, K. Nomura, T. Dalgaty, E. Vianello, Y. Nishi, and G. Indiveri, Nature Communications \textbf{13}, (2022).

\phantomsection
\label{csl:78}78 F. Moro, E. Hardy, B. Fain, T. Dalgaty, P. Cl{\'{e}}men{\c{c}}on, A.D. Pr{\`{a}}, E. Esmanhotto, N. Castellani, F. Blard, F. Gardien, T. Mesquida, F. Rummens, D. Esseni, J. Casas, G. Indiveri, M. Payvand, and E. Vianello, Nature Communications \textbf{13}, (2022).

\phantomsection
\label{csl:79}79 S. Brivio, D. Conti, M.V. Nair, J. Frascaroli, E. Covi, C. Ricciardi, G. Indiveri, and S. Spiga, Nanotechnology \textbf{30}, 015102 (2018).

\phantomsection
\label{csl:80}80 Y. Demirag, F. Moro, T. Dalgaty, G. Navarro, C. Frenkel, G. Indiveri, E. Vianello, and M. Payvand, in \textit{2021 IEEE International Symposium on Circuits and Systems (ISCAS)} (IEEE, 2021).

\phantomsection
\label{csl:81}81 L.G. Wright, T. Onodera, M.M. Stein, T. Wang, D.T. Schachter, Z. Hu, and P.L. McMahon, Nature \textbf{601}, 549 (2022).

\phantomsection
\label{csl:82}82 G. Tanaka, T. Yamane, J.B. H{\'{e}}roux, R. Nakane, N. Kanazawa, S. Takeda, H. Numata, D. Nakano, and A. Hirose, Neural Networks \textbf{115}, 100 (2019).

\phantomsection
\label{csl:83}83 S. Ghosh, K. Nakajima, T. Krisnanda, K. Fujii, and T.C.H. Liew, Advanced Quantum Technologies \textbf{4}, 2100053 (2021).

\phantomsection
\label{csl:84}84 G. Csaba and W. Porod, Applied Physics Reviews \textbf{7}, 011302 (2020).

\phantomsection
\label{csl:85}85 K. Yamazaki, V.-K. Vo-Ho, D. Bulsara, and N. Le, Brain Sciences \textbf{12}, 863 (2022).

\phantomsection
\label{csl:86}86 J. Zhu, T. Zhang, Y. Yang, and R. Huang, Applied Physics Reviews \textbf{7}, 011312 (2020).

\phantomsection
\label{csl:87}87 D. Ielmini and S. Ambrogio, Nanotechnology \textbf{31}, 092001 (2019).

\phantomsection
\label{csl:88}88 G. Zhou, Z. Wang, B. Sun, F. Zhou, L. Sun, H. Zhao, X. Hu, X. Peng, J. Yan, H. Wang, W. Wang, J. Li, B. Yan, D. Kuang, Y. Wang, L. Wang, and S. Duan, Advanced Electronic Materials \textbf{8}, 2101127 (2022).

\phantomsection
\label{csl:89}89 M.D. Pickett, G. Medeiros-Ribeiro, and R.S. Williams, Nature Materials \textbf{12}, 114 (2012).

\phantomsection
\label{csl:90}90 Z. Wang, S. Joshi, S.E. Savel'ev, H. Jiang, R. Midya, P. Lin, M. Hu, N. Ge, J.P. Strachan, Z. Li, Q. Wu, M. Barnell, G.-L. Li, H.L. Xin, R.S. Williams, Q. Xia, and J.J. Yang, Nature Materials \textbf{16}, 101 (2016).

\phantomsection
\label{csl:91}91 H. Jiang, D. Belkin, S.E. Savel'ev, S. Lin, Z. Wang, Y. Li, S. Joshi, R. Midya, C. Li, M. Rao, M. Barnell, Q. Wu, J.J. Yang, and Q. Xia, Nature Communications \textbf{8}, (2017).

\phantomsection
\label{csl:92} Y. Ushakov, A. Balanov, and S. Savel'ev, \textit{Chaos, Solitons \& Fractals} \textbf{145}, 110803 (2021).

\phantomsection
\label{csl:93}93 Z. Wang, S. Joshi, S. Savel'ev, W. Song, R. Midya, Y. Li, M. Rao, P. Yan, S. Asapu, Y. Zhuo, H. Jiang, P. Lin, C. Li, J.H. Yoon, N.K. Upadhyay, J. Zhang, M. Hu, J.P. Strachan, M. Barnell, Q. Wu, H. Wu, R.S. Williams, Q. Xia, and J.J. Yang, Nature Electronics \textbf{1}, 137 (2018).

\phantomsection
\label{csl:94}94 A. Mehonic, A. Sebastian, B. Rajendran, O. Simeone, E. Vasilaki, and A.J. Kenyon, Advanced Intelligent Systems \textbf{2}, 2000085 (2020).

\phantomsection
\label{csl:95}95 D. Kim, B. Jeon, Y. Lee, D. Kim, Y. Cho, and S. Kim, Applied Physics Letters \textbf{121}, 010501 (2022).

\phantomsection
\label{csl:96} Y. Ushakov, A. Akther, P. Borisov, D. Pattnaik, S. Savel'ev, and A.G. Balanov, \textit{Chaos, Solitons \& Fractals} \textbf{149}, 110997 (2021).

\phantomsection
\label{csl:97} A.M. Wojtusiak, A.G. Balanov, and S.E. Savel'ev, \textit{Chaos, Solitons \& Fractals} \textbf{142}, 110383 (2021).

\phantomsection
\label{csl:98}98 D. Sussillo, Current Opinion in Neurobiology \textbf{25}, 156 (2014).

\phantomsection
\label{csl:99}99 S. Gepshtein, A.S. Pawar, S. Kwon, S. Savel'ev, and T.D. Albright, Science Advances \textbf{8}, (2022).

\phantomsection
\label{csl:100}100 J.D. Monaco, K. Rajan, and G.M. Hwang, ArXiv \textbf{abs/2105.07284}, (2021).

\phantomsection
\label{csl:101}101 S. Kumar, X. Wang, J.P. Strachan, Y. Yang, and W.D. Lu, Nature Reviews Materials \textbf{7}, (2022).

\phantomsection
\label{csl:102}102 J. Feldmann, N. Youngblood, M. Karpov, H. Gehring, X. Li, M. Stappers, M.L. Gallo, X. Fu, A. Lukashchuk, A.S. Raja, J. Liu, C.D. Wright, A. Sebastian, T.J. Kippenberg, W.H.P. Pernice, and H. Bhaskaran, Nature \textbf{589}, 52 (2021).

\phantomsection
\label{csl:103}103 X. Wang, Y. Chen, H. Xi, H. Li, and D. Dimitrov, {IEEE} Electron Device Letters \textbf{30}, 294 (2009).

\phantomsection
\label{csl:104}104 M. Mansueto, A. Chavent, S. Auffret, I. Joumard, L. Vila, R.C. Sousa, L.D. Buda-Prejbeanu, I.L. Prejbeanu, and B. Dieny, Nanoscale \textbf{13}, 11488 (2021).

\phantomsection
\label{csl:105}105 A. Hirohata, K. Yamada, Y. Nakatani, I.-L. Prejbeanu, B. Diény, P. Pirro, and B. Hillebrands, Journal of Magnetism and Magnetic Materials \textbf{509}, (2020).

\phantomsection
\label{csl:106}106 A.H. Jaafar, R.J. Gray, E. Verrelli, M. O{\textquotesingle}Neill, S.M. Kelly, and N.T. Kemp, Nanoscale \textbf{9}, 17091 (2017).

\phantomsection
\label{csl:107}107 L.E. Srouji, A. Krishnan, R. Ravichandran, Y. Lee, M. On, X. Xiao, and S.J.B. Yoo, {APL} Photonics \textbf{7}, 051101 (2022).

\phantomsection
\label{csl:108}108 J.C. Gartside, K.D. Stenning, A. Vanstone, H.H. Holder, D.M. Arroo, T. Dion, F. Caravelli, H. Kurebayashi, and W.R. Branford, Nature Nanotechnology \textbf{17}, 460 (2022).

\phantomsection
\label{csl:109}109 D.D. Yaremkevich, A.V. Scherbakov, L. De Clerk, S.M. Kukhtaruk, A. Nadzeyka, R. Campion, A.W. Rushforth, S. Savel’ev, A.G. Balanov, and M. Bayer, Nature Communications \textbf{14}, (2023).

\phantomsection
\label{csl:110}110 K. Vandoorne, P. Mechet, T.V. Vaerenbergh, M. Fiers, G. Morthier, D. Verstraeten, B. Schrauwen, J. Dambre, and P. Bienstman, Nature Communications \textbf{5}, (2014).

\phantomsection
\label{csl:111}111 D.J. Gauthier, E. Bollt, A. Griffith, and W.A.S. Barbosa, Nature Communications \textbf{12}, (2021).

\phantomsection
\label{csl:112}112 S.-in Yi, J.D. Kendall, R.S. Williams, and S. Kumar, Nature Electronics (2022).

\phantomsection
\label{csl:113}113 A. Ac{\'{\i}}n, I. Bloch, H. Buhrman, T. Calarco, C. Eichler, J. Eisert, D. Esteve, N. Gisin, S.J. Glaser, F. Jelezko, S. Kuhr, M. Lewenstein, M.F. Riedel, P.O. Schmidt, R. Thew, A. Wallraff, I. Walmsley, and F.K. Wilhelm, New Journal of Physics \textbf{20}, 080201 (2018).

\phantomsection
\label{csl:114}114 V. Dunjko and H.J. Briegel, Reports on Progress in Physics \textbf{81}, 074001 (2018).

\phantomsection
\label{csl:115}115 P. Mujal, R. Mart{\'{\i}}nez-Pe{\~{n}}a, J. Nokkala, J. Garc{\'{\i}}a-Beni, G.L. Giorgi, M.C. Soriano, and R. Zambrini, Advanced Quantum Technologies \textbf{4}, 2100027 (2021).

\phantomsection
\label{csl:116}116 P. Pfeiffer, I.L. Egusquiza, M.D. Ventra, M. Sanz, and E. Solano, Scientific Reports \textbf{6}, (2016).

\phantomsection
\label{csl:117}117 J. Salmilehto, F. Deppe, M.D. Ventra, M. Sanz, and E. Solano, Scientific Reports \textbf{7}, (2017).

\phantomsection
\label{csl:118}118 M. Spagnolo, J. Morris, S. Piacentini, M. Antesberger, F. Massa, A. Crespi, F. Ceccarelli, R. Osellame, and P. Walther, Nature Photonics \textbf{16}, 318 (2022).

\phantomsection
\label{csl:119}119 K. Friston, Nature Reviews Neuroscience \textbf{11}, 127 (2010).

\phantomsection
\label{csl:120}120 N.B. Janson and C.J. Marsden, Scientific Reports \textbf{7}, (2017).

\phantomsection
\label{csl:121}121 H.-T. Zhang, T.J. Park, A.N.M.N. Islam, D.S.J. Tran, S. Manna, Q. Wang, S. Mondal, H. Yu, S. Banik, S. Cheng, H. Zhou, S. Gamage, S. Mahapatra, Y. Zhu, Y. Abate, N. Jiang, S.K.R.S. Sankaranarayanan, A. Sengupta, C. Teuscher, and S. Ramanathan, Science \textbf{375}, 533 (2022).

\phantomsection
\label{csl:122}122 M.J. Colbrook, V. Antun, and A.C. Hansen, Proceedings of the National Academy of Sciences \textbf{119}, (2022).

\phantomsection
\label{csl:123}123 T. Song, W. Rim, S. Park, Y. Kim, J. Jung, G. Yang, S. Baek, J. Choi, B. Kwon, Y. Lee, S. Kim, G. Kim, H.-S. Won, J.-H. Ku, S.S. Paak, E.S. Jung, S.S. Park, and K. Kim, in \textit{2016 {IEEE} International Solid-State Circuits Conference ({ISSCC})} ({IEEE}, 2016).

\phantomsection
\label{csl:124}124 T. Song, J. Jung, W. Rim, H. Kim, Y. Kim, C. Park, J. Do, S. Park, S. Cho, H. Jung, B. Kwon, H.-S. Choi, J.S. Choi, and J.S. Yoon, in \textit{2018 {IEEE} International Solid - State Circuits Conference - ({ISSCC})} ({IEEE}, 2018).

\phantomsection
\label{csl:125}125 G. Yeap, X. Chen, B.R. Yang, C.P. Lin, F.C. Yang, Y.K. Leung, D.W. Lin, C.P. Chen, K.F. Yu, D.H. Chen, C.Y. Chang, S.S. Lin, H.K. Chen, P. Hung, C.S. Hou, Y.K. Cheng, J. Chang, L. Yuan, C.K. Lin, C.C. Chen, Y.C. Yeo, M.H. Tsai, Y.M. Chen, H.T. Lin, C.O. Chui, K.B. Huang, W. Chang, H.J. Lin, K.W. Chen, R. Chen, S.H. Sun, Q. Fu, H.T. Yang, H.L. Shang, H.T. Chiang, C.C. Yeh, T.L. Lee, C.H. Wang, S.L. Shue, C.W. Wu, R. Lu, W.R. Lin, J. Wu, F. Lai, P.W. Wang, Y.H. Wu, B.Z. Tien, Y.C. Huang, L.C. Lu, J. He, Y. Ku, J. Lin, M. Cao, T.S. Chang, S.M. Jang, H.C. Lin, Y.C. Peng, J.Y. Sheu, and M. Wang, in \textit{2019 {IEEE} International Electron Devices Meeting ({IEDM})} ({IEEE}, 2019).

\phantomsection
\label{csl:126}126 Y. Kim, C. Ong, A.M. Pillai, H. Jagadeesh, G. Baek, I. Rajwani, Z. Guo, and E. Karl, {IEEE} Journal of Solid-State Circuits 1 (2022).

\phantomsection
\label{csl:127}127 F. Dong, X. Si, and M.-F. Chang, in \textit{2022 IEEE 16th International Conference on Solid-State \ Integrated Circuit Technology (ICSICT)} (IEEE, 2022).

\phantomsection
\label{csl:128}128 X. Si, M.-F. Chang, W.-S. Khwa, J.-J. Chen, J.-F. Li, X. Sun, R. Liu, S. Yu, H. Yamauchi, and Q. Li, {IEEE} Transactions on Circuits and Systems I: Regular Papers \textbf{66}, 4172 (2019).

\phantomsection
\label{csl:129}129 Y. Wang, S. Zhang, Y. Li, J. Chen, W. Zhao, and Y. Ha, {IEEE} Transactions on Very Large Scale Integration ({VLSI}) Systems \textbf{31}, 684 (2023).

\phantomsection
\label{csl:130}130 Q. Dong, S. Jeloka, M. Saligane, Y. Kim, M. Kawaminami, A. Harada, S. Miyoshi, M. Yasuda, D. Blaauw, and D. Sylvester, {IEEE} Journal of Solid-State Circuits \textbf{53}, 1006 (2018).

\phantomsection
\label{csl:131}131 B. Pan, G. Wang, H. Zhang, W. Kang, and W. Zhao, IEEE Transactions on Circuits and Systems II: Express Briefs \textbf{69}, (2022).

\phantomsection
\label{csl:132}132 J. Zhang, Z. Wang, and N. Verma, IEEE Journal of Solid-State Circuits \textbf{52}, (2017).

\phantomsection
\label{csl:133}133 A.S. Rekhi, B. Zimmer, N. Nedovic, N. Liu, R. Venkatesan, M. Wang, B. Khailany, W.J. Dally, and C.T. Gray, in \textit{Proceedings of the 56th Annual Design Automation Conference 2019} ({ACM}, 2019).

\phantomsection
\label{csl:134}134 S.K. Gonugondla, C. Sakr, H. Dbouk, and N.R. Shanbhag, in \textit{Proceedings of the 39th International Conference on Computer-Aided Design} ({ACM}, 2020).

\phantomsection
\label{csl:135}135 B. Murmann, {IEEE} Transactions on Very Large Scale Integration ({VLSI}) Systems \textbf{29}, 3 (2021).

\phantomsection
\label{csl:136}136 S. Spetalnick and A. Raychowdhury, {IEEE} Transactions on Circuits and Systems I: Regular Papers \textbf{69}, 1466 (2022).

\phantomsection
\label{csl:137} J. Saikia, S. Yin, S.K. Cherupally, B. Zhang, J. Meng, M. Seok, and J.-S. Seo, in \textit{2021 Design Automation \& Test in Europe Conference \& Exhibition (DATE)} (IEEE, 2021).

\phantomsection
\label{csl:138}138 Y.-D. Chih, P.-H. Lee, H. Fujiwara, Y.-C. Shih, C.-F. Lee, R. Naous, Y.-L. Chen, C.-P. Lo, C.-H. Lu, H. Mori, W.-C. Zhao, D. Sun, M.E. Sinangil, Y.-H. Chen, T.-L. Chou, K. Akarvardar, H.-J. Liao, Y. Wang, M.-F. Chang, and T.-Y.J. Chang, in \textit{2021 {IEEE} International Solid- State Circuits Conference ({ISSCC})} ({IEEE}, 2021).

\phantomsection
\label{csl:139}139 G. Desoli, N. Chawla, T. Boesch, M. Avodhyawasi, H. Rawat, H. Chawla, V.S. Abhijith, P. Zambotti, A. Sharma, C. Cappetta, M. Rossi, A. De Vita, and F. Girardi, in \textit{2023 IEEE International Solid- State Circuits Conference (ISSCC)} (IEEE, 2023).

\phantomsection
\label{csl:140}140 J. Zhang, Z. Wang, and N. Verma, in \textit{2016 {IEEE} Symposium on {VLSI} Circuits ({VLSI}-Circuits)} ({IEEE}, 2016).

\phantomsection
\label{csl:141}141 M. Kang, S.K. Gonugondla, A. Patil, and N.R. Shanbhag, {IEEE} Journal of Solid-State Circuits \textbf{53}, 642 (2018).

\phantomsection
\label{csl:142}142 H. Valavi, P.J. Ramadge, E. Nestler, and N. Verma, in \textit{2018 {IEEE} Symposium on {VLSI} Circuits} ({IEEE}, 2018).

\phantomsection
\label{csl:143}143 H. Jia, H. Valavi, Y. Tang, J. Zhang, and N. Verma, in \textit{2019 {IEEE} Hot Chips 31 Symposium ({HCS})} ({IEEE}, 2019).

\phantomsection
\label{csl:144}144 H. Jia, M. Ozatay, Y. Tang, H. Valavi, R. Pathak, J. Lee, and N. Verma, in \textit{2021 {IEEE} International Solid- State Circuits Conference ({ISSCC})} ({IEEE}, 2021).

\phantomsection
\label{csl:145}145 K. Ueyoshi, I.A. Papistas, P. Houshmand, G.M. Sarda, V. Jain, M. Shi, Q. Zheng, S. Giraldo, P. Vrancx, J. Doevenspeck, D. Bhattacharjee, S. Cosemans, A. Mallik, P. Debacker, D. Verkest, and M. Verhelst, in \textit{2022 {IEEE} International Solid- State Circuits Conference ({ISSCC})} ({IEEE}, 2022).

\phantomsection
\label{csl:146}146 R. Bez, E. Camerlenghi, A. Modelli, and A. Visconti, Proc. {IEEE} \textbf{91}, 489 (2003).

\phantomsection
\label{csl:147}147 C.M. Compagnoni, A. Goda, A.S. Spinelli, P. Feeley, A.L. Lacaita, and A. Visconti, Proc. {IEEE} \textbf{105}, 1609 (2017).

\phantomsection
\label{csl:148}148 T. Pekny, L. Vu, J. Tsai, D. Srinivasan, E. Yu, J. Pabustan, J. Xu, S. Deshmukh, K.-F. Chan, M. Piccardi, K. Xu, G. Wang, K. Shakeri, V. Patel, T. Iwasaki, T. Wang, P. Musunuri, C. Gu, A. Mohammadzadeh, A. Ghalam, V. Moschiano, T. Vali, J. Park, J. Lee, and R. Ghodsi, IEEE Int. Solid-State Circuits Conf. (ISSCC) Dig. Tech. Papers 132 (2022).

\phantomsection
\label{csl:149}149 F. Merrikh-Bayat, X. Guo, M. Klachko, M. Prezioso, K.K. Likharev, and D.B. Strukov, {IEEE} Trans. Neural Netw. Learn. Syst \textbf{29}, 4782 (2018).

\phantomsection
\label{csl:150}150 X. Guo, F. Merrikh-Bayat, M. Prezioso, Y. Chen, B. Nguyen, N. Do, and D.B. Strukov, in \textit{Proc. CICC} (2017), pp. 1–4.

\phantomsection
\label{csl:151}151 M. Bavandpour, S. Sahay, M.R. Mahmoodi, and D.B. Strukov, Neuromorph. Comput. Eng. \textbf{1}, 014001 (2021).

\phantomsection
\label{csl:152}152 P. Wang, F. Xu, B. Wang, B. Gao, H. Wu, H. Qian, and S. Yu, IEEE Trans. VLSI Syst. \textbf{27}, 988 (2019).

\phantomsection
\label{csl:153}153 W. Shim and S. Yu, {IEEE} Electron Device Lett. \textbf{42}, 160 (2021).

\phantomsection
\label{csl:154}154 S.-T. Lee, G. Yeom, H. Yoo, H.-S. Kim, S. Lim, J.-H. Bae, B.-G. Park, and J.-H. Lee, IEEE Trans. Electron Devices \textbf{68}, 3365 (2021).

\phantomsection
\label{csl:155}155 S.-T. Lee and J.-H. Lee, Front. Neurosci. \textbf{14}, 14 (2020).

\phantomsection
\label{csl:156}156 H.-T. Lue, P.-K. Hsu, M.-L. Wei, T.-H. Yeh, P.-Y. Du, W.-C. Chen, K.-C. Wang, and C.-Y. Lu, IEDM Tech. Dig. 915 (2019).

\phantomsection
\label{csl:157}157 H.-N. Yoo, J.-W. Back, N.-H. Kim, D. Kwon, B.-G.- Park, and J.-H. Lee, Proc. IEEE Symp. VLSI Technol. Circuits 304 (2022).

\phantomsection
\label{csl:158}158 G. Malavena, M. Filippi, A.S. Spinelli, and C.M. Compagnoni, IEEE Trans. Electron Devices \textbf{66}, 4727 (2019).

\phantomsection
\label{csl:159}159 G. Malavena, M. Filippi, A.S. Spinelli, and C.M. Compagnoni, IEEE Trans. Electron Devices \textbf{66}, 4733 (2019).

\phantomsection
\label{csl:160}160 C.M. Compagnoni and A.S. Spinelli, {IEEE} Transactions on Electron Devices \textbf{66}, 4504 (2019).

\phantomsection
\label{csl:161}161 G. Malavena, S. Petrò, A.S. Spinelli, and C.M. Compagnoni, Proc. ESSDERC 122 (2019).

\phantomsection
\label{csl:162}162 D.-H. Kwon, K.M. Kim, J.H. Jang, J.M. Jeon, M.H. Lee, G.H. Kim, X.-S. Li, G.-S. Park, B. Lee, S. Han, M. Kim, and C.S. Hwang, Nature Nanotechnology \textbf{5}, 148 (2010).

\phantomsection
\label{csl:163}163 L. Song, Y. Zhuo, X. Qian, H. Li, and Y. Chen, in \textit{2018 {IEEE} International Symposium on High Performance Computer Architecture ({HPCA})} ({IEEE}, 2018).

\phantomsection
\label{csl:164}164 G. Dai, T. Huang, Y. Wang, H. Yang, and J. Wawrzynek, in \textit{Proceedings of the 24th Asia and South Pacific Design Automation Conference} ({ACM}, 2019).

\phantomsection
\label{csl:165}165 S. Wang, Y. Li, D. Wang, W. Zhang, X. Chen, D. Dong, S. Wang, X. Zhang, P. Lin, C. Gallicchio, X. Xu, Q. Liu, K.-T. Cheng, Z. Wang, D. Shang, and M. Liu, Nature Machine Intelligence \textbf{5}, 104 (2023).

\phantomsection
\label{csl:166}166 W. Zhang, S. Wang, Y. Li, X. Xu, D. Dong, N. Jiang, F. Wang, Z. Guo, R. Fang, C. Dou, K. Ni, Z. Wang, D. Shang, and M. Liu, in \textit{2022 {IEEE} Symposium on {VLSI} Technology and Circuits ({VLSI} Technology and Circuits)} ({IEEE}, 2022).

\phantomsection
\label{csl:167}167 F. Alibart, E. Zamanidoost, and D.B. Strukov, Nature Communications \textbf{4}, (2013).

\phantomsection
\label{csl:168}168 F.M. Bayat, M. Prezioso, B. Chakrabarti, H. Nili, I. Kataeva, and D. Strukov, Nature Communications \textbf{9}, (2018).

\phantomsection
\label{csl:169}169 M. Hu, C.E. Graves, C. Li, Y. Li, N. Ge, E. Montgomery, N. Davila, H. Jiang, R.S. Williams, J.J. Yang, Q. Xia, and J.P. Strachan, Advanced Materials \textbf{30}, 1705914 (2018).

\phantomsection
\label{csl:170}170 Q. Duan, Z. Jing, X. Zou, Y. Wang, K. Yang, T. Zhang, S. Wu, R. Huang, and Y. Yang, Nature Communications \textbf{11}, (2020).

\phantomsection
\label{csl:171}171 M. Ishii, U. Shin, K. Hosokawa, M. BrightSky, W. Haensch, S. Kim, S. Lewis, A. Okazaki, J. Okazawa, M. Ito, M. Rasch, W. Kim, and A. Nomura, in \textit{2019 {IEEE} International Electron Devices Meeting ({IEDM})} ({IEEE}, 2019).

\phantomsection
\label{csl:172}172 S. Yu, Z. Li, P.-Y. Chen, H. Wu, B. Gao, D. Wang, W. Wu, and H. Qian, in \textit{2016 {IEEE} International Electron Devices Meeting ({IEDM})} ({IEEE}, 2016).

\phantomsection
\label{csl:173}173 B. Yan, Q. Yang, W.-H. Chen, K.-T. Chang, J.-W. Su, C.-H. Hsu, S.-H. Li, H.-Y. Lee, S.-S. Sheu, M.-S. Ho, Q. Wu, M.-F. Chang, Y. Chen, and H. Li, in \textit{2019 Symposium on {VLSI} Technology} ({IEEE}, 2019).

\phantomsection
\label{csl:174}174 C.-X. Xue, T.-Y. Huang, J.-S. Liu, T.-W. Chang, H.-Y. Kao, J.-H. Wang, T.-W. Liu, S.-Y. Wei, S.-P. Huang, W.-C. Wei, Y.-R. Chen, T.-H. Hsu, Y.-K. Chen, Y.-C. Lo, T.-H. Wen, C.-C. Lo, R.-S. Liu, C.-C. Hsieh, K.-T. Tang, and M.-F. Chang, in \textit{2020 {IEEE} International Solid- State Circuits Conference - ({ISSCC})} ({IEEE}, 2020).

\phantomsection
\label{csl:175}175 Y. Lu, X. Li, L. Yan, T. Zhang, Y. Yang, Z. Song, and R. Huang, in \textit{2020 {IEEE} International Electron Devices Meeting ({IEDM})} ({IEEE}, 2020).

\phantomsection
\label{csl:176}176 V. Joshi, M.L. Gallo, S. Haefeli, I. Boybat, S.R. Nandakumar, C. Piveteau, M. Dazzi, B. Rajendran, A. Sebastian, and E. Eleftheriou, Nature Communications \textbf{11}, (2020).

\phantomsection
\label{csl:177}177 H. Li, W.-C. Chen, A. Levy, C.-H. Wang, H. Wang, P.-H. Chen, W. Wan, W.-S. Khwa, H. Chuang, Y.-D. Chih, M.-F. Chang, H.-S.P. Wong, and P. Raina, {IEEE} Transactions on Electron Devices \textbf{68}, 6637 (2021).

\phantomsection
\label{csl:178}178 P. Yao, H. Wu, B. Gao, S.B. Eryilmaz, X. Huang, W. Zhang, Q. Zhang, N. Deng, L. Shi, H.-S.P. Wong, and H. Qian, Nature Communications \textbf{8}, (2017).

\phantomsection
\label{csl:179}179 P. Yao, H. Wu, B. Gao, J. Tang, Q. Zhang, W. Zhang, J.J. Yang, and H. Qian, Nature \textbf{577}, 641 (2020).

\phantomsection
\label{csl:180}180 S. Ambrogio, P. Narayanan, H. Tsai, R.M. Shelby, I. Boybat, C. di Nolfo, S. Sidler, M. Giordano, M. Bodini, N.C.P. Farinha, B. Killeen, C. Cheng, Y. Jaoudi, and G.W. Burr, Nature \textbf{558}, 60 (2018).

\phantomsection
\label{csl:181}181 G. Karunaratne, M. Schmuck, M.L. Gallo, G. Cherubini, L. Benini, A. Sebastian, and A. Rahimi, Nature Communications \textbf{12}, (2021).

\phantomsection
\label{csl:182}182 P.M. Sheridan, F. Cai, C. Du, W. Ma, Z. Zhang, and W.D. Lu, Nature Nanotechnology \textbf{12}, 784 (2017).

\phantomsection
\label{csl:183}183 W. Wan, R. Kubendran, S.B. Eryilmaz, W. Zhang, Y. Liao, D. Wu, S. Deiss, B. Gao, P. Raina, S. Joshi, H. Wu, G. Cauwenberghs, and H.-S.P. Wong, in \textit{2020 {IEEE} International Solid- State Circuits Conference - ({ISSCC})} ({IEEE}, 2020).

\phantomsection
\label{csl:184}184 S. Wo{\'{z}}niak, A. Pantazi, T. Bohnstingl, and E. Eleftheriou, Nature Machine Intelligence \textbf{2}, 325 (2020).

\phantomsection
\label{csl:185}185 C. Du, F. Cai, M.A. Zidan, W. Ma, S.H. Lee, and W.D. Lu, Nature Communications \textbf{8}, (2017).

\phantomsection
\label{csl:186}186 J. Moon, W. Ma, J.H. Shin, F. Cai, C. Du, S.H. Lee, and W.D. Lu, Nature Electronics \textbf{2}, 480 (2019).

\phantomsection
\label{csl:187}187 G. Milano, G. Pedretti, K. Montano, S. Ricci, S. Hashemkhani, L. Boarino, D. Ielmini, and C. Ricciardi, Nature Materials \textbf{21}, 195 (2021).

\phantomsection
\label{csl:188}188 H. Tsai, S. Ambrogio, C. Mackin, P. Narayanan, R.M. Shelby, K. Rocki, A. Chen, and G.W. Burr, in \textit{2019 Symposium on {VLSI} Technology} ({IEEE}, 2019).

\phantomsection
\label{csl:189}189 G. Karunaratne, M.L. Gallo, G. Cherubini, L. Benini, A. Rahimi, and A. Sebastian, Nature Electronics \textbf{3}, 327 (2020).

\phantomsection
\label{csl:190}190 Z. Liu, J. Tang, B. Gao, P. Yao, X. Li, D. Liu, Y. Zhou, H. Qian, B. Hong, and H. Wu, Nature Communications \textbf{11}, (2020).

\phantomsection
\label{csl:191}191 Y. Zhong, J. Tang, X. Li, B. Gao, H. Qian, and H. Wu, Nature Communications \textbf{12}, (2021).

\phantomsection
\label{csl:192}192 Z. Sun, G. Pedretti, A. Bricalli, and D. Ielmini, Science Advances \textbf{6}, (2020).

\phantomsection
\label{csl:193}193 Y.J. Jeong, J. Lee, J. Moon, J.H. Shin, and W.D. Lu, Nano Letters \textbf{18}, 4447 (2018).

\phantomsection
\label{csl:194}194 S. Choi, J.H. Shin, J. Lee, P. Sheridan, and W.D. Lu, Nano Letters \textbf{17}, 3113 (2017).

\phantomsection
\label{csl:195}195 A. Sebastian, T. Tuma, N. Papandreou, M.L. Gallo, L. Kull, T. Parnell, and E. Eleftheriou, Nature Communications \textbf{8}, (2017).

\phantomsection
\label{csl:196}196 S. Choi, S.H. Tan, Z. Li, Y. Kim, C. Choi, P.-Y. Chen, H. Yeon, S. Yu, and J. Kim, Nature Materials \textbf{17}, 335 (2018).

\phantomsection
\label{csl:197}197 J. Yang, M. Rao, H. Tang, J.-B. Wu, W. Song, M. Zhang, W. Yin, Y. Zhuo, F. Kiani, B. Chen, H. Liu, H.-Y. Chen, R. Midya, F. Ye, H. Jiang, Z. Wang, M. Wu, M. Hu, H. Wang, Q. Xia, G. Ge, and J. Li, (2022).

\phantomsection
\label{csl:198}198 F. Kiani, J. Yin, Z. Wang, J.J. Yang, and Q. Xia, Science Advances \textbf{7}, (2021).

\phantomsection
\label{csl:199}199 T. Dalgaty, N. Castellani, C. Turck, K.-E. Harabi, D. Querlioz, and E. Vianello, Nature Electronics \textbf{4}, 151 (2021).

\phantomsection
\label{csl:200}200 Z. Wang, C. Li, P. Lin, M. Rao, Y. Nie, W. Song, Q. Qiu, Y. Li, P. Yan, J.P. Strachan, N. Ge, N. McDonald, Q. Wu, M. Hu, H. Wu, R.S. Williams, Q. Xia, and J.J. Yang, Nature Machine Intelligence \textbf{1}, 434 (2019).

\phantomsection
\label{csl:201}201 M.L. Gallo, A. Sebastian, R. Mathis, M. Manica, H. Giefers, T. Tuma, C. Bekas, A. Curioni, and E. Eleftheriou, Nature Electronics \textbf{1}, 246 (2018).

\phantomsection
\label{csl:202}202 F. Cai, S. Kumar, T.V. Vaerenbergh, X. Sheng, R. Liu, C. Li, Z. Liu, M. Foltin, S. Yu, Q. Xia, J.J. Yang, R. Beausoleil, W.D. Lu, and J.P. Strachan, Nature Electronics \textbf{3}, 409 (2020).

\phantomsection
\label{csl:203}203 M.R. Mahmoodi, M. Prezioso, and D.B. Strukov, Nature Communications \textbf{10}, (2019).

\phantomsection
\label{csl:204}204 M. Le Gallo and A. Sebastian, Journal of Physics D: Applied Physics \textbf{53}, 213002 (2020).

\phantomsection
\label{csl:205}205 S.G. Sarwat, Materials Science and Technology \textbf{33}, 1890 (2017).

\phantomsection
\label{csl:206}206 A. Sebastian, M. Le Gallo, G.W. Burr, S. Kim, M. BrightSky, and E. Eleftheriou, Journal of Applied Physics \textbf{124}, 111101 (2018).

\phantomsection
\label{csl:207}207 S.R. Ovshinsky and B. Pashmakov, MRS Online Proceedings Library (OPL) \textbf{803}, (2003).

\phantomsection
\label{csl:208}208 M.L. Gallo, R. Khaddam-Aljameh, M. Stanisavljevic, A. Vasilopoulos, B. Kersting, M. Dazzi, G. Karunaratne, M. Braendli, A. Singh, S.M. Mueller, and others, Nature Electronics (2023).

\phantomsection
\label{csl:209}209 S.R. Nandakumar, M. Le Gallo, C. Piveteau, V. Joshi, G. Mariani, I. Boybat, G. Karunaratne, R. Khaddam-Aljameh, U. Egger, A. Petropoulos, and others, Frontiers in Neuroscience \textbf{14}, 406 (2020).

\phantomsection
\label{csl:210}210 A. Sebastian, T. Tuma, N. Papandreou, M. Le Gallo, L. Kull, T. Parnell, and E. Eleftheriou, Nature Communications \textbf{8}, 1115 (2017).

\phantomsection
\label{csl:211}211 G. Karunaratne, M. Hersche, J. Langeneager, G. Cherubini, M. Le Gallo, U. Egger, K. Brew, S. Choi, I. Ok, C. Silvestre, and others, in \textit{ESSCIRC 2022-IEEE 48th European Solid State Circuits Conference (ESSCIRC)} (IEEE, 2022), pp. 105–108.

\phantomsection
\label{csl:212}212 S. Ambrogio, N. Ciocchini, M. Laudato, V. Milo, A. Pirovano, P. Fantini, and D. Ielmini, Frontiers in Neuroscience \textbf{10}, 56 (2016).

\phantomsection
\label{csl:213}213 T. Tuma, M. Le Gallo, A. Sebastian, and E. Eleftheriou, IEEE Electron Device Letters \textbf{37}, 1238 (2016).

\phantomsection
\label{csl:214}214 T. Tuma, A. Pantazi, M. Le Gallo, A. Sebastian, and E. Eleftheriou, Nature Nanotechnology \textbf{11}, 693 (2016).

\phantomsection
\label{csl:215}
215 \href{https://download.intel.com/newsroom/2021/archive/2015-07-28-news-releases-intel-and-micron-produce-breakthrough-memory-technology.pdf}{\textit{https://download.intel.com/newsroom/2021/archive/2015-07-28-news-releases-intel-and-micron-produce-breakthrough-memory-technology.pdf}}.

\phantomsection
\label{csl:216}216 F. Arnaud, P. Ferreira, F. Piazza, A. Gandolfo, P. Zuliani, P. Mattavelli, E. Gomiero, G. Samanni, J. Jasse, C. Jahan, and others, in \textit{International Electron Devices Meeting (IEDM)} (IEEE, 2020), pp. 24–2.

\phantomsection
\label{csl:217}217 M. Lanza, A. Sebastian, W.D. Lu, M. Le Gallo, M.-F. Chang, D. Akinwande, F.M. Puglisi, H.N. Alshareef, M. Liu, and J.B. Roldan, Science \textbf{376}, 9979 (2022).

\phantomsection
\label{csl:218}218 S. Raoux, Annual Review of Materials Research \textbf{39}, 25 (2009).

\phantomsection
\label{csl:219}219 J. Liang, S. Yeh, S.S. Wong, and H.-S.P. Wong, 4th IEEE International Memory Workshop 1 (2012).

\phantomsection
\label{csl:220}220 S. Yoo, H.D. Lee, S. Lee, H. Choi, and T. Kim, IEEE Transactions on Electron Devices \textbf{67}, 1454 (2020).

\phantomsection
\label{csl:221}221 S.H. Lee, M.S. Kim, G.S. Do, S.G. Kim, H.J. Lee, J.S. Sim, N.G. Park, S.B. Hong, Y.H. Jeon, K.S. Choi, and others, in \textit{2010 Symposium on VLSI Technology} (IEEE, 2010), pp. 199–200.

\phantomsection
\label{csl:222}222 T. Mikolajick, M.H. Park, L. Begon-Lours, and S. Slesazeck, Adv Mater 2206042 (2022).

\phantomsection
\label{csl:223}223 H.P. McAdams, R. Acklin, T. Blake, X.-H. Du, J. Eliason, J. Fong, W.F. Kraus, D. Liu, S. Madan, T. Moise, S. Natarajan, N. Qian, Y. Qiu, K.A. Remack, J. Rodriguez, J. Roscher, A. Seshadri, and S.R. Summerfelt, {IEEE} Journal of Solid-State Circuits \textbf{39}, 667 (2004).

\phantomsection
\label{csl:224}224 S. Beyer, S. Dunkel, M. Trentzsch, J. Muller, A. Hellmich, D. Utess, J. Paul, D. Kleimaier, J. Pellerin, S. Muller, J. Ocker, A. Benoist, H. Zhou, M. Mennenga, M. Schuster, F. Tassan, M. Noack, A. Pourkeramati, F. Muller, M. Lederer, T. Ali, R. Hoffmann, T. Kampfe, K. Seidel, H. Mulaosmanovic, E.T. Breyer, T. Mikolajick, and S. Slesazeck, in \textit{2020 {IEEE} International Memory Workshop ({IMW})} ({IEEE}, 2020).

\phantomsection
\label{csl:225}225 E.Y. Tsymbal and H. Kohlstedt, Science \textbf{313}, 181 (2006).

\phantomsection
\label{csl:226}226 K. Asadi, M. Li, P.W.M. Blom, M. Kemerink, and D.M. de Leeuw, Materials Today \textbf{14}, 592 (2011).

\phantomsection
\label{csl:227}227 D. Bondurant, Ferroelectrics \textbf{112}, 273 (1990).

\phantomsection
\label{csl:228}228 T.S. Böscke, J. Müller, D. Bräuhaus, U. Schröder, and U. Böttger, Applied Physics Letters \textbf{99}, 102903 (2011).

\phantomsection
\label{csl:229}229 H. Mulaosmanovic, P.D. Lomenzo, U. Schroeder, S. Slesazeck, T. Mikolajick, and B. Max, in \textit{2021 {IEEE} International Reliability Physics Symposium ({IRPS})} ({IEEE}, 2021).

\phantomsection
\label{csl:230}230 R. Materlik, C. Künneth, and A. Kersch, Journal of Applied Physics \textbf{117}, 134109 (2015).

\phantomsection
\label{csl:231}231 H. Mulaosmanovic, S. Dunkel, M. Trentzsch, S. Beyer, E.T. Breyer, T. Mikolajick, and S. Slesazeck, {IEEE} Transactions on Electron Devices \textbf{67}, 5804 (2020).

\phantomsection
\label{csl:232}232 M. Materano, P.D. Lomenzo, A. Kersch, M.H. Park, T. Mikolajick, and U. Schroeder, Inorganic Chemistry Frontiers \textbf{8}, 2650 (2021).

\phantomsection
\label{csl:233}233 D.R. Islamov, V.A. Gritsenko, T.V. Perevalov, V.A. Pustovarov, O.M. Orlov, A.G. Chernikova, A.M. Markeev, S. Slesazeck, U. Schroeder, T. Mikolajick, and G.Y. Krasnikov, Acta Materialia \textbf{166}, 47 (2019).

\phantomsection
\label{csl:234}234 L. B{\'{e}}gon-Lours, M. Halter, F.M. Puglisi, L. Benatti, D.F. Falcone, Y. Popoff, D.D. Pineda, M. Sousa, and B.J. Offrein, Advanced Electronic Materials \textbf{8}, 2101395 (2022).

\phantomsection
\label{csl:235}235 P. Nukala, M. Ahmadi, Y. Wei, S. de Graaf, E. Stylianidis, T. Chakrabortty, S. Matzen, H.W. Zandbergen, A. Björling, D. Mannix, D. Carbone, B. Kooi, and B. Noheda, Science \textbf{372}, 630 (2021).

\phantomsection
\label{csl:236}236 S.S. Cheema, D. Kwon, N. Shanker, R. dos Reis, S.-L. Hsu, J. Xiao, H. Zhang, R. Wagner, A. Datar, M.R. McCarter, C.R. Serrao, A.K. Yadav, G. Karbasian, C.-H. Hsu, A.J. Tan, L.-C. Wang, V. Thakare, X. Zhang, A. Mehta, E. Karapetrova, R.V. Chopdekar, P. Shafer, E. Arenholz, C. Hu, R. Proksch, R. Ramesh, J. Ciston, and S. Salahuddin, Nature \textbf{580}, 478 (2020).

\phantomsection
\label{csl:237}237 Y. Wei, G. Vats, and B. Noheda, Neuromorphic Computing and Engineering \textbf{2}, 044007 (2022).

\phantomsection
\label{csl:238}238 B. Dieny, I.L. Prejbeanu, K. Garello, P. Gambardella, P. Freitas, R. Lehndorff, W. Raberg, U. Ebels, S.O. Demokritov, J. Akerman, A. Deac, P. Pirro, C. Adelmann, A. Anane, A.V. Chumak, A. Hirohata, S. Mangin, S.O. Valenzuela, M.C. Onba{\c{s}}l{\i}, M. d'Aquino, G. Prenat, G. Finocchio, L. Lopez-Diaz, R. Chantrell, O. Chubykalo-Fesenko, and P. Bortolotti, Nature Electronics \textbf{3}, 446 (2020).

\phantomsection
\label{csl:239}239 D. Apalkov, B. Dieny, and J.M. Slaughter, Proceedings of the {IEEE} \textbf{104}, 1796 (2016).

\phantomsection
\label{csl:240}240 S. Lequeux, J. Sampaio, V. Cros, K. Yakushiji, A. Fukushima, R. Matsumoto, H. Kubota, S. Yuasa, and J. Grollier, Scientific Reports \textbf{6}, (2016).

\phantomsection
\label{csl:241}241 S. Fukami, C. Zhang, S. DuttaGupta, A. Kurenkov, and H. Ohno, Nature Materials \textbf{15}, 535 (2016).

\phantomsection
\label{csl:242}242 J. Wang, H. Sepehri-Amin, H. Tajiri, T. Nakamura, K. Masuda, Y.K. Takahashi, T. Ina, T. Uruga, I. Suzuki, Y. Miura, and K. Hono, Acta Materialia \textbf{166}, 413 (2019).

\phantomsection
\label{csl:243}243 K.M. Song, J.-S. Jeong, B. Pan, X. Zhang, J. Xia, S. Cha, T.-E. Park, K. Kim, S. Finizio, J. Raabe, J. Chang, Y. Zhou, W. Zhao, W. Kang, H. Ju, and S. Woo, Nature Electronics \textbf{3}, 148 (2020).

\phantomsection
\label{csl:244}244 E. Raymenants, A. Vaysset, D. Wan, M. Manfrini, O. Zografos, O. Bultynck, J. Doevenspeck, M. Heyns, I.P. Radu, and T. Devolder, Journal of Applied Physics \textbf{124}, (2018).

\phantomsection
\label{csl:245}245 M. Prezioso, M.R. Mahmoodi, F.M. Bayat, H. Nili, H. Kim, A. Vincent, and D.B. Strukov, Nature Communications \textbf{9}, (2018).

\phantomsection
\label{csl:246}246 F.-X. Liang, I.-T. Wang, and T.-H. Hou, Advanced Intelligent Systems \textbf{3}, 2100007 (2021).

\phantomsection
\label{csl:247}247 J. Torrejon, M. Riou, F.A. Araujo, S. Tsunegi, G. Khalsa, D. Querlioz, P. Bortolotti, V. Cros, K. Yakushiji, A. Fukushima, H. Kubota, S. Yuasa, M.D. Stiles, and J. Grollier, Nature \textbf{547}, 428 (2017).

\phantomsection
\label{csl:248}248 K. Hayakawa, S. Kanai, T. Funatsu, J. Igarashi, B. Jinnai, W.A. Borders, H. Ohno, and S. Fukami, Physical Review Letters \textbf{126}, (2021).

\phantomsection
\label{csl:249}249 W.A. Borders, A.Z. Pervaiz, S. Fukami, K.Y. Camsari, H. Ohno, and S. Datta, Nature \textbf{573}, 390 (2019).

\phantomsection
\label{csl:250}250 M.-H. Wu, M.-C. Hong, C.-C. Chang, P. Sahu, J.-H. Wei, H.-Y. Lee, S.-S. Shcu, and T.-H. Hou, in \textit{2019 Symposium on {VLSI} Technology} ({IEEE}, 2019).

\phantomsection
\label{csl:251}251 M.-H. Wu, I.-T. Wang, M.-C. Hong, K.-M. Chen, Y.-C. Tseng, J.-H. Wei, and T.-H. Hou, Physical Review Applied \textbf{18}, (2022).

\phantomsection
\label{csl:252}252 M. Schott, A. Bernand-Mantel, L. Ranno, S. Pizzini, J. Vogel, H. Béa, C. Baraduc, S. Auffret, G. Gaudin, and D. Givord, Nano Letters \textbf{17}, (2017).

\phantomsection
\label{csl:253}253 C.E. Fillion, J. Fischer, R. Kumar, A. Fassatoui, S. Pizzini, L. Ranno, D. Ourdani, M. Belmeguenai, Y. Roussigné, S.M. Chérif, S. Auffret, I. Joumard, O. Boulle, G. Gaudin, L. Buda-Prejbeanu, C. Baraduc, and H. Béa, Nat Commun \textbf{13}, 5257 (2022).

\phantomsection
\label{csl:254}254 A. Jaiswal, S. Roy, G. Srinivasan, and K. Roy, {IEEE} Transactions on Electron Devices \textbf{64}, 1818 (2017).

\phantomsection
\label{csl:255}255 M. Zahedinejad, H. Fulara, R. Khymyn, A. Houshang, M. Dvornik, S. Fukami, S. Kanai, H. Ohno, and J. Åkerman, Nat Mater \textbf{21}, 81 (2022).

\phantomsection
\label{csl:256}256 X.-G. Zhang and W.H. Butler, Physical Review B \textbf{70}, (2004).

\phantomsection
\label{csl:257}257 A. Ross, N. Leroux, A. De Riz, D. Marković, D. Sanz-Hernández, J. Trastoy, P. Bortolotti, D. Querlioz, L. Martins, L. Benetti, M.S. Claro, P. Anacleto, A. Schulman, T. Taris, J.-B. Begueret, S. Saïghi, A.S. Jenkins, R. Ferreira, A.F. Vincent, F.A. Mizrahi, and J. Grollier, Nature Nanotechnology \textbf{18}, (2023).

\phantomsection
\label{csl:258}258 S. Kanai, K. Hayakawa, H. Ohno, and S. Fukami, Physical Review B \textbf{103}, (2021).

\phantomsection
\label{csl:259}259 M.-H. Wu, M.-S. Huang, Z. Zhu, F.-X. Liang, M.-C. Hong, J. Deng, J.-H. Wei, S.-S. Sheu, C.-I. Wu, G. Liang, and T.-H. Hou, in \textit{2020 IEEE Symposium on VLSI Technology} (IEEE, 2020).

\phantomsection
\label{csl:260}260 J. Choe, in \textit{2023 IEEE International Memory Workshop (IMW)} (IEEE, 2023).

\phantomsection
\label{csl:261}261 R. Khymyn, I. Lisenkov, J. Voorheis, O. Sulymenko, O. Prokopenko, V. Tiberkevich, J. Akerman, and A. Slavin, Sci Rep \textbf{8}, 15727 (2018).

\phantomsection
\label{csl:262}262 F. Trier, P. Noël, J.-V. Kim, J.-P. Attané, L. Vila, and M. Bibes, Nature Reviews Materials \textbf{7}, (2021).

\phantomsection
\label{csl:263}263 S.S.P. Parkin, M. Hayashi, and L. Thomas, Science \textbf{320}, (2008).

\phantomsection
\label{csl:264}264 A. Fert, V. Cros, and J. Sampaio, Nature Nanotechnology \textbf{8}, (2013).

\phantomsection
\label{csl:265}265 M.S. El Hadri, P. Pirro, C.-H. Lambert, S. Petit-Watelot, Y. Quessab, M. Hehn, F. Montaigne, G. Malinowski, and S. Mangin, Physical Review B \textbf{94}, (2016).

\phantomsection
\label{csl:266}266 A. Fernández-Pacheco, R. Streubel, O. Fruchart, R. Hertel, P. Fischer, and R.P. Cowburn, Nat Commun \textbf{8}, 15756 (2017).

\phantomsection
\label{csl:267}267 Y. Shen, N.C. Harris, S. Skirlo, M. Prabhu, T. Baehr-Jones, M. Hochberg, X. Sun, S. Zhao, H. Larochelle, D. Englund, and M. Solja{\v{c}}i{\'{c}}, Nature Photonics \textbf{11}, 441 (2017).

\phantomsection
\label{csl:268}268 S. Bandyopadhyay, A. Sludds, S. Krastanov, R. Hamerly, N. Harris, D. Bunandar, M. Streshinsky, M. Hochberg, and D. Englund, ArXiv:2208.01623 [Cs.ET] (2022).

\phantomsection
\label{csl:269}269 J. Feldmann, N. Youngblood, C.D. Wright, H. Bhaskaran, and W.H.P. Pernice, Nature \textbf{569}, 208 (2019).

\phantomsection
\label{csl:270}270 J. Zhang, S. Dai, Y. Zhao, J. Zhang, and J. Huang, Advanced Intelligent Systems \textbf{2}, 1900136 (2020).

\phantomsection
\label{csl:271} M. Kumar, S. Abbas, and J. Kim, \textit{ACS Applied Materials \& Interfaces} \textbf{10}, 34370 (2018).

\phantomsection
\label{csl:272}272 I. Taghavi, M. Moridsadat, A. Tofini, S. Raza, N.A.F. Jaeger, L. Chrostowski, B.J. Shastri, and S. Shekhar, Nanophotonics \textbf{11}, 3855 (2022).

\phantomsection
\label{csl:273}273 J. Wu, Z.T. Xie, Y. Sha, H.Y. Fu, and Q. Li, Photonics Research \textbf{9}, 1616 (2021).

\phantomsection
\label{csl:274}274 J. Robertson, P. Kirkland, J.A. Alanis, M. Hejda, J. Bueno, G.D. Caterina, and A. Hurtado, Scientific Reports \textbf{12}, (2022).

\phantomsection
\label{csl:275}275 S.T. Ilie, J. Faneca, I. Zeimpekis, T.D. Bucio, K. Grabska, D.W. Hewak, H.M.H. Chong, and F.Y. Gardes, Scientific Reports \textbf{12}, (2022).

\phantomsection
\label{csl:276}276 M. Wang, S. Cai, C. Pan, C. Wang, X. Lian, Y. Zhuo, K. Xu, T. Cao, X. Pan, B. Wang, S.-J. Liang, J.J. Yang, P. Wang, and F. Miao, Nature Electronics \textbf{1}, 130 (2018).

\phantomsection
\label{csl:277}277 Y. Shi, X. Liang, B. Yuan, V. Chen, H. Li, F. Hui, Z. Yu, F. Yuan, E. Pop, H.-S.P. Wong, and M. Lanza, Nature Electronics \textbf{1}, 458 (2018).

\phantomsection
\label{csl:278}278 L. Sun, Y. Zhang, G. Han, G. Hwang, J. Jiang, B. Joo, K. Watanabe, T. Taniguchi, Y.-M. Kim, W.J. Yu, B.-S. Kong, R. Zhao, and H. Yang, Nature Communications \textbf{10}, (2019).

\phantomsection
\label{csl:279}279 F.M. Puglisi, L. Larcher, C. Pan, N. Xiao, Y. Shi, F. Hui, and M. Lanza, in \textit{2016 {IEEE} International Electron Devices Meeting ({IEDM})} ({IEEE}, 2016).

\phantomsection
\label{csl:280}280 X. Yan, J.H. Qian, V.K. Sangwan, and M.C. Hersam, Advanced Materials \textbf{34}, 2108025 (2022).

\phantomsection
\label{csl:281}281 \url{https://www.fujitsu.com/jp/group/fsm/en/products/reram,} \textit{{Website}}.

\phantomsection
\label{csl:282}282 S. Chen, M.R. Mahmoodi, Y. Shi, C. Mahata, B. Yuan, X. Liang, C. Wen, F. Hui, D. Akinwande, D.B. Strukov, and M. Lanza, Nature Electronics \textbf{3}, 638 (2020).

\phantomsection
\label{csl:283}283 K. Lu, X. Li, Q. Sun, X. Pang, J. Chen, T. Minari, X. Liu, and Y. Song, Materials Horizons \textbf{8}, 447 (2021).

\phantomsection
\label{csl:284}284 S. Pazos, X. Xu, T. Guo, K. Zhu, H.N. Alshareef, and M. Lanza, Nature Reviews Materials \textbf{9}, (2024).

\phantomsection
\label{csl:285}285 T. Mikolajick, M.H. Park, L. Begon‐Lours, and S. Slesazeck, Advanced Materials \textbf{35}, (2023).

\phantomsection
\label{csl:286}286 F. Xue, C. Zhang, Y. Ma, Y. Wen, X. He, B. Yu, and X. Zhang, Advanced Materials \textbf{34}, 2201880 (2022).

\phantomsection
\label{csl:287}287 C. Kaspar, B.J. Ravoo, W.G. van der Wiel, S.V. Wegner, and W.H.P. Pernice, Nature \textbf{594}, (2021).

\phantomsection
\label{csl:288}288 A. Chen, in \textit{71st Device Research Conference} ({IEEE}, 2013).

\phantomsection
\label{csl:289}289 Y. Shen, W. Zheng, K. Zhu, Y. Xiao, C. Wen, Y. Liu, X. Jing, and M. Lanza, Advanced Materials \textbf{33}, 2103656 (2021).

\phantomsection
\label{csl:290}290 B. Tang, H. Veluri, Y. Li, Z.G. Yu, M. Waqar, J.F. Leong, M. Sivan, E. Zamburg, Y.-W. Zhang, J. Wang, and A.V.-Y. Thean, Nature Communications \textbf{13}, (2022).

\phantomsection
\label{csl:291}291 X. Xu, T. Guo, H. Kim, M.K. Hota, R.S. Alsaadi, M. Lanza, X. Zhang, and H.N. Alshareef, Advanced Materials \textbf{34}, (2022).

\phantomsection
\label{csl:292}292 T.-A. Chen, C.-P. Chuu, C.-C. Tseng, C.-K. Wen, H.-S.P. Wong, S. Pan, R. Li, T.-A. Chao, W.-C. Chueh, Y. Zhang, Q. Fu, B.I. Yakobson, W.-H. Chang, and L.-J. Li, Nature \textbf{579}, 219 (2020).

\phantomsection
\label{csl:293}293 K.Y. Ma, L. Zhang, S. Jin, Y. Wang, S.I. Yoon, H. Hwang, J. Oh, D.S. Jeong, M. Wang, S. Chatterjee, G. Kim, A.-R. Jang, J. Yang, S. Ryu, H.Y. Jeong, R.S. Ruoff, M. Chhowalla, F. Ding, and H.S. Shin, Nature \textbf{606}, 88 (2022).

\phantomsection
\label{csl:294}294 G. Migliato Marega, H.G. Ji, Z. Wang, G. Pasquale, M. Tripathi, A. Radenovic, and A. Kis, Nature Electronics \textbf{6}, (2023).

\phantomsection
\label{csl:295}295 M. Lanza, R. Waser, D. Ielmini, J.J. Yang, L. Goux, J. Su{\~{n}}e, A.J. Kenyon, A. Mehonic, S. Spiga, V. Rana, S. Wiefels, S. Menzel, I. Valov, M.A. Villena, E. Miranda, X. Jing, F. Campabadal, M.B. Gonzalez, F. Aguirre, F. Palumbo, K. Zhu, J.B. Roldan, F.M. Puglisi, L. Larcher, T.-H. Hou, T. Prodromakis, Y. Yang, P. Huang, T. Wan, Y. Chai, K.L. Pey, N. Raghavan, S. Due{\~{n}}as, T. Wang, Q. Xia, and S. Pazos, {ACS} Nano \textbf{15}, 17214 (2021).

\phantomsection
\label{csl:296}296 M. Lanza, A. Sebastian, W.D. Lu, G.M. Le, M.F. Chang, D. Akinwande, F.M. Puglisi, H.N. Alshareef, M. Liu, and J.B. Roldan, Science \textbf{376}, 9979 (2022).

\phantomsection
\label{csl:297}297 D. Edelstein, M. Rizzolo, D. Sil, A. Dutta, J. DeBrosse, M. Wordeman, A. Arceo, I.C. Chu, J. Demarest, E.R.J. Edwards, E.R. Evarts, J. Fullam, A. Gasasira, G. Hu, M. Iwatake, R. Johnson, V. Katragadda, T. Levin, J. Li, Y. Liu, C. Long, T. Maffitt, S. McDermott, S. Mehta, V. Mehta, D. Metzler, J. Morillo, Y. Nakamura, S. Nguyen, P. Nieves, V. Pai, R. Patlolla, R. Pujari, R. Southwick, T. Standaert, O. van der Straten, H. Wu, C.-C. Yang, D. Houssameddine, J.M. Slaughter, and D.C. Worledge, in \textit{2020 {IEEE} International Electron Devices Meeting ({IEDM})} ({IEEE}, 2020).

\phantomsection
\label{csl:298}298 F. Xiong, E. Yalon, A. Behnam, C.M. Neumann, K.L. Grosse, S. Deshmukh, and E. Pop, in \textit{2016 {IEEE} International Electron Devices Meeting ({IEDM})} ({IEEE}, 2016).

\phantomsection
\label{csl:299}299 M.L. Gallo and A. Sebastian, Journal of Physics D: Applied Physics \textbf{53}, 213002 (2020).

\phantomsection
\label{csl:300}300 R. Dittmann, S. Menzel, and R. Waser, Advances in Physics \textbf{70}, 155 (2021).

\phantomsection
\label{csl:301}301 V.V. Zhirnov, R. Meade, R.K. Cavin, and G. Sandhu, Nanotechnology \textbf{22}, 254027 (2011).

\phantomsection
\label{csl:302}302 E. Chicca and G. Indiveri, Applied Physics Letters \textbf{116}, 120501 (2020).

\phantomsection
\label{csl:303}303 C. Safranski, G. Hu, J.Z. Sun, P. Hashemi, S.L. Brown, L. Buzi, C.P. D{\textquotesingle}Emic, E.R.J. Edwards, E. Galligan, M.G. Gottwald, O. Gunawan, S. Karimeddiny, H. Jung, J. Kim, K. Latzko, P.L. Trouilloud, S. Zare, and D.C. Worledge, in \textit{2022 {IEEE} Symposium on {VLSI} Technology and Circuits ({VLSI} Technology and Circuits)} ({IEEE}, 2022).

\phantomsection
\label{csl:304}304 C. Persch, M.J. Müller, A. Yadav, J. Pries, N. Honn{\'{e}}, P. Kerres, S. Wei, H. Tanaka, P. Fantini, E. Varesi, F. Pellizzer, and M. Wuttig, Nature Communications \textbf{12}, (2021).

\phantomsection
\label{csl:305}305 M. von Witzleben, S. Wiefels, A. Kindsmüller, P. Stasner, F. Berg, F. Cüppers, S. Hoffmann-Eifert, R. Waser, S. Menzel, and U. Böttger, {ACS} Applied Electronic Materials \textbf{3}, 5563 (2021).

\phantomsection
\label{csl:306}306 S. Wiefels, M.V. Witzleben, M. Huttemann, U. Bottger, R. Waser, and S. Menzel, {IEEE} Transactions on Electron Devices \textbf{68}, 1024 (2021).

\phantomsection
\label{csl:307} N. Kopperberg, S. Wiefels, S. Liberda, R. Waser, and S. Menzel, \textit{ACS Applied Materials \& Interfaces} \textbf{13}, 58066 (2021).

\phantomsection
\label{csl:308}308 C. Bengel, J. Mohr, S. Wiefels, A. Singh, A. Gebregiorgis, R. Bishnoi, S. Hamdioui, R. Waser, D. Wouters, and S. Menzel, Neuromorphic Computing and Engineering \textbf{2}, 034001 (2022).

\phantomsection
\label{csl:309}309 K. Schnieders, C. Funck, F. Cüppers, S. Aussen, T. Kempen, A. Sarantopoulos, R. Dittmann, S. Menzel, V. Rana, S. Hoffmann-Eifert, and S. Wiefels, {APL} Materials \textbf{10}, 101114 (2022).

\phantomsection
\label{csl:310}310 S. Slesazeck, V. Havel, E. Breyer, H. Mulaosmanovic, M. Hoffmann, B. Max, S. Duenkel, and T. Mikolajick, in \textit{2019 {IEEE} 11th International Memory Workshop ({IMW})} ({IEEE}, 2019).

\phantomsection
\label{csl:311}311 J.H. Yoon, S.J. Song, I.H. Yoo, J.Y. Seok, K.J. Yoon, D.E. Kwon, T.H. Park, and C.S. Hwang, Advanced Functional Materials \textbf{24}, (2014).

\phantomsection
\label{csl:312}312 M. Lanza, H.-S.P. Wong, E. Pop, D. Ielmini, D. Strukov, B.C. Regan, L. Larcher, M.A. Villena, J.J. Yang, L. Goux, A. Belmonte, Y. Yang, F.M. Puglisi, J. Kang, B. Magyari-Köpe, E. Yalon, A. Kenyon, M. Buckwell, A. Mehonic, A. Shluger, H. Li, T.-H. Hou, B. Hudec, D. Akinwande, R. Ge, S. Ambrogio, J.B. Roldan, E. Miranda, J. Su{\~{n}}e, K.L. Pey, X. Wu, N. Raghavan, E. Wu, W.D. Lu, G. Navarro, W. Zhang, H. Wu, R. Li, A. Holleitner, U. Wurstbauer, M.C. Lemme, M. Liu, S. Long, Q. Liu, H. Lv, A. Padovani, P. Pavan, I. Valov, X. Jing, T. Han, K. Zhu, S. Chen, F. Hui, and Y. Shi, Advanced Electronic Materials \textbf{5}, 1800143 (2018).

\phantomsection
\label{csl:313}313 G. Dearnaley, A.M. Stoneham, and D.V. Morgan, Reports on Progress in Physics \textbf{33}, 1129 (1970).

\phantomsection
\label{csl:314}314 Y. Yang, P. Gao, S. Gaba, T. Chang, X. Pan, and W. Lu, Nature Communications \textbf{3}, (2012).

\phantomsection
\label{csl:315}315 Y. Yang, P. Gao, L. Li, X. Pan, S. Tappertzhofen, S.H. Choi, R. Waser, I. Valov, and W.D. Lu, Nature Communications \textbf{5}, (2014).

\phantomsection
\label{csl:316}316 R. Waser, R. Dittmann, G. Staikov, and K. Szot, Advanced Materials \textbf{21}, 2632 (2009).

\phantomsection
\label{csl:317}317 S. Meister, S.B. Kim, J.J. Cha, H.-S.P. Wong, and Y. Cui, ACS Nano \textbf{5}, (2011).

\phantomsection
\label{csl:318}318 S. Chen, L. Jiang, M. Buckwell, X. Jing, Y. Ji, E. Grustan-Gutierrez, F. Hui, Y. Shi, M. Rommel, A. Paskaleva, G. Benstetter, W.H. Ng, A. Mehonic, A.J. Kenyon, and M. Lanza, Advanced Functional Materials \textbf{28}, 1802266 (2018).

\phantomsection
\label{csl:319}319 M. Buckwell, L. Montesi, S. Hudziak, A. Mehonic, and A.J. Kenyon, Nanoscale \textbf{7}, 18030 (2015).

\phantomsection
\label{csl:320}320 A. Mehonic, M. Buckwell, L. Montesi, M.S. Munde, D. Gao, S. Hudziak, R.J. Chater, S. Fearn, D. McPhail, M. Bosman, A.L. Shluger, and A.J. Kenyon, Advanced Materials \textbf{28}, 7486 (2016).

\phantomsection
\label{csl:321}321 K. Szot, W. Speier, G. Bihlmayer, and R. Waser, Nature Materials \textbf{5}, 312 (2006).

\phantomsection
\label{csl:322}322 H.R.J. Cox, M. Buckwell, W.H. Ng, D.J. Mannion, A. Mehonic, P.R. Shearing, S. Fearn, and A.J. Kenyon, {APL} Materials \textbf{9}, 111109 (2021).

\phantomsection
\label{csl:323}323 I. Valov and T. Tsuruoka, Journal of Physics D: Applied Physics \textbf{51}, 413001 (2018).

\phantomsection
\label{csl:324}324 A. Mehonic, A. Vrajitoarea, S. Cueff, S. Hudziak, H. Howe, C. Labb{\'{e}}, R. Rizk, M. Pepper, and A.J. Kenyon, Scientific Reports \textbf{3}, (2013).

\phantomsection
\label{csl:325}325 S. Stathopoulos, L. Michalas, A. Khiat, A. Serb, and T. Prodromakis, Scientific Reports \textbf{9}, (2019).

\phantomsection
\label{csl:326}326 R. Yuan, Q. Duan, P.J. Tiw, G. Li, Z. Xiao, Z. Jing, K. Yang, C. Liu, C. Ge, R. Huang, and Y. Yang, Nature Communications \textbf{13}, (2022).

\phantomsection
\label{csl:327}327 J. del Valle, P. Salev, F. Tesler, N.M. Vargas, Y. Kalcheim, P. Wang, J. Trastoy, M.-H. Lee, G. Kassabian, J.G. Ram{\'{\i}}rez, M.J. Rozenberg, and I.K. Schuller, Nature \textbf{569}, 388 (2019).

\phantomsection
\label{csl:328}328 S.G. Sarwat, B. Kersting, T. Moraitis, V.P. Jonnalagadda, and A. Sebastian, Nature Nanotechnology \textbf{17}, 507 (2022).

\phantomsection
\label{csl:329}329 G.W. Burr, M.J. BrightSky, A. Sebastian, H.-Y. Cheng, J.-Y. Wu, S. Kim, N.E. Sosa, N. Papandreou, H.-L. Lung, H. Pozidis, E. Eleftheriou, and C.H. Lam, IEEE Journal on Emerging and Selected Topics in Circuits and Systems \textbf{6}, (2016).

\phantomsection
\label{csl:330}330 M. Salinga, B. Kersting, I. Ronneberger, V.P. Jonnalagadda, X.T. Vu, G.M. Le, I. Giannopoulos, O. Cojocaru-Mirédin, R. Mazzarello, and A. Sebastian, Nat Mater \textbf{17}, 681 (2018).

\phantomsection
\label{csl:331}331 F. Rao, K. Ding, Y. Zhou, Y. Zheng, M. Xia, S. Lv, Z. Song, S. Feng, I. Ronneberger, R. Mazzarello, W. Zhang, and E. Ma, Science \textbf{358}, 1423 (2017).

\phantomsection
\label{csl:332}332 S. Bhatti, R. Sbiaa, A. Hirohata, H. Ohno, S. Fukami, and S.N. Piramanayagam, Materials Today \textbf{20}, (2017).

\phantomsection
\label{csl:333}333 L.W. Martin and A.M. Rappe, Nature Reviews Materials \textbf{2}, (2016).

\phantomsection
\label{csl:334}334 J. Hwang, Y. Goh, and S. Jeon, Small \textbf{20}, (2023).

\phantomsection
\label{csl:335}335 K.-H. Kim, I. Karpov, R.H. Olsson, and D. Jariwala, Nature Nanotechnology \textbf{18}, (2023).

\phantomsection
\label{csl:336}336 A. Chanthbouala, V. Garcia, R.O. Cherifi, K. Bouzehouane, S. Fusil, X. Moya, S. Xavier, H. Yamada, C. Deranlot, N.D. Mathur, M. Bibes, A. Barth{\'{e}}l{\'{e}}my, and J. Grollier, Nature Materials \textbf{11}, 860 (2012).

\phantomsection
\label{csl:337}337 M. Onen, N. Emond, B. Wang, D. Zhang, F.M. Ross, J. Li, B. Yildiz, and J.A. del Alamo, Science \textbf{377}, 539 (2022).

\phantomsection
\label{csl:338}338 M.L. Gallo, T. Tuma, F. Zipoli, A. Sebastian, and E. Eleftheriou, in \textit{2016 46th European Solid-State Device Research Conference ({ESSDERC})} ({IEEE}, 2016).

\phantomsection
\label{csl:339}339 A. Fukushima, T. Seki, K. Yakushiji, H. Kubota, H. Imamura, S. Yuasa, and K. Ando, Applied Physics Express \textbf{7}, 083001 (2014).

\phantomsection
\label{csl:340}340 D. Vodenicarevic, N. Locatelli, A. Mizrahi, J.S. Friedman, A.F. Vincent, M. Romera, A. Fukushima, K. Yakushiji, H. Kubota, S. Yuasa, S. Tiwari, J. Grollier, and D. Querlioz, Physical Review Applied \textbf{8}, (2017).

\phantomsection
\label{csl:341}341 W. Qian, X. Li, M.D. Riedel, K. Bazargan, and D.J. Lilja, IEEE Transactions on Computers \textbf{60}, (2011).

\phantomsection
\label{csl:342}342 F. Neugebauer, I. Polian, and J.P. Hayes, in \textit{2017 Euromicro Conference on Digital System Design (DSD)} (IEEE, 2017).

\phantomsection
\label{csl:343}343 K.-E. Harabi, T. Hirtzlin, C. Turck, E. Vianello, R. Laurent, J. Droulez, P. Bessi{\`{e}}re, J.-M. Portal, M. Bocquet, and D. Querlioz, Nature Electronics (2022).

\phantomsection
\label{csl:344}344 A. Sengupta, G. Srinivasan, D. Roy, and K. Roy, in \textit{2018 {IEEE} International Electron Devices Meeting ({IEDM})} ({IEEE}, 2018).

\phantomsection
\label{csl:345}345 K.Y. Camsari, R. Faria, B.M. Sutton, and S. Datta, Physical Review X \textbf{7}, (2017).

\phantomsection
\label{csl:346}346 A. Mizrahi, T. Hirtzlin, A. Fukushima, H. Kubota, S. Yuasa, J. Grollier, and D. Querlioz, Nature Communications \textbf{9}, (2018).

\phantomsection
\label{csl:347}347 A. Sengupta, P. Panda, P. Wijesinghe, Y. Kim, and K. Roy, Scientific Reports \textbf{6}, (2016).

\phantomsection
\label{csl:348}348 N. Mohseni, P.L. McMahon, and T. Byrnes, Nature Reviews Physics \textbf{4}, (2022).

\phantomsection
\label{csl:349}349 A. Grimaldi, K. Selcuk, N.A. Aadit, K. Kobayashi, Q. Cao, S. Chowdhury, G. Finocchio, S. Kanai, H. Ohno, S. Fukami, and K.Y. Camsari, in \textit{2022 International Electron Devices Meeting (IEDM)} (IEEE, 2022).

\phantomsection
\label{csl:350}350 N.A. Aadit, A. Grimaldi, M. Carpentieri, L. Theogarajan, J.M. Martinis, G. Finocchio, and K.Y. Camsari, Nature Electronics \textbf{5}, (2022).

\phantomsection
\label{csl:351}351 C.-Y. Huang, W.C. Shen, Y.-H. Tseng, Y.-C. King, and C.-J. Lin, {IEEE} Electron Device Letters \textbf{33}, 1108 (2012).

\phantomsection
\label{csl:352}352 H. Nili, G.C. Adam, B. Hoskins, M. Prezioso, J. Kim, M.R. Mahmoodi, F.M. Bayat, O. Kavehei, and D.B. Strukov, Nature Electronics \textbf{1}, 197 (2018).

\phantomsection
\label{csl:353}353 A.F. Vincent, J. Larroque, N. Locatelli, N.B. Romdhane, O. Bichler, C. Gamrat, W.S. Zhao, J.-O. Klein, S. Galdin-Retailleau, and D. Querlioz, {IEEE} Transactions on Biomedical Circuits and Systems \textbf{9}, 166 (2015).

\phantomsection
\label{csl:354}354 J.Z. Sun, Physical Review B \textbf{62}, 570 (2000).

\phantomsection
\label{csl:355}355 J. Kaiser, W.A. Borders, K.Y. Camsari, S. Fukami, H. Ohno, and S. Datta, Physical Review Applied \textbf{17}, (2022).

\phantomsection
\label{csl:356}356 A. Sebastian, R. Pendurthi, A. Kozhakhmetov, N. Trainor, J.A. Robinson, J.M. Redwing, and S. Das, Nature Communications \textbf{13}, (2022).

\phantomsection
\label{csl:357}357 S. Liu, T.P. Xiao, J. Kwon, B.J. Debusschere, S. Agarwal, J.A.C. Incorvia, and C.H. Bennett, Frontiers in Nanotechnology \textbf{4}, (2022).

\phantomsection
\label{csl:358}358 D. Bonnet, T. Hirtzlin, A. Majumdar, T. Dalgaty, E. Esmanhotto, V. Meli, N. Castellani, S. Martin, J.-F. Nodin, G. Bourgeois, J.-M. Portal, D. Querlioz, and E. Vianello, Nature Communications \textbf{14}, 7530 (2023).

\phantomsection
\label{csl:359}359 K. Roy, A. Jaiswal, and P. Panda, Nature \textbf{575}, 607 (2019).

\phantomsection
\label{csl:360}360 B. Wickramasinghe, S.S. Chowdhury, A.K. Kosta, W. Ponghiran, and K. Roy, IEEE Transactions on Cognitive and Developmental Systems 1 (2024).

\phantomsection
\label{csl:361}361 A. Vaswani, N. Shazeer, N. Parmar, J. Uszkoreit, L. Jones, A.N. Gomez, {\L. Kaiser, and I. Polosukhin, Advances in Neural Information Processing Systems \textbf{30}, (2017}).

\phantomsection
\label{csl:362}362 T. Mikolov, M. Karafi{\'{a}}t, L. Burget, J. {\v{C}}ernock{\'{y}}, and S. Khudanpur, in \textit{Interspeech 2010} ({ISCA}, 2010).

\phantomsection
\label{csl:363}363 A. Graves, A.-rahman Mohamed, and G. Hinton, in \textit{2013 {IEEE} International Conference on Acoustics Speech and Signal Processing} ({IEEE}, 2013).

\phantomsection
\label{csl:364}364 S. Hochreiter and J. Schmidhuber, Neural Comput \textbf{9}, 1735 (1997).

\phantomsection
\label{csl:365}365in \textit{Proceedings of SSST-8, Eighth Workshop on Syntax, Semantics and Structure in Statistical Translation} (\url{https://arxiv.org/abs/1409.1259,} 2014), pp. 103–111.

\phantomsection
\label{csl:366}366in \textit{NIPS 2014 Workshop on Deep Learning} (\url{https://arxiv.org/abs/1412.3555,} 2014).

\phantomsection
\label{csl:367}367IEEE Signal Processing Magazine \textbf{36(6)}, 51 (2019).

\phantomsection
\label{csl:368}368 S.S. Chowdhury, C. Lee, and K. Roy, Neurocomputing \textbf{464}, 83 (2021).

\phantomsection
\label{csl:369}369 W. Ponghiran and K. Roy, in \textit{Proceedings of the AAAI Conference on Artificial Intelligence} (2022), pp. 8001–8008.

\phantomsection
\label{csl:370}370in \textit{In International Conference on Learning Representations} (\url{https://arxiv.org/abs/2010.11929v2,} 2020).

\phantomsection
\label{csl:371}371 P.U. Diehl, D. Neil, J. Binas, M. Cook, S.-C. Liu, and M. Pfeiffer, in \textit{2015 International Joint Conference on Neural Networks ({IJCNN})} ({IEEE}, 2015).

\phantomsection
\label{csl:372}372 C. Lee, S.S. Sarwar, P. Panda, G. Srinivasan, and K. Roy, Front Neurosci \textbf{14}, 119 (2020).

\phantomsection
\label{csl:373}373 J. Kim, H. Kim, S. Huh, J. Lee, and K. Choi, Neurocomputing \textbf{311}, 373 (2018).

\phantomsection
\label{csl:374}374in \textit{Proceedings of the 56th Annual Design Automation Conference} (\url{https://doi.org/10.1145/3316781.3317822,} 2019), pp. 1–6.

\phantomsection
\label{csl:375}375 I. Garg, S.S. Chowdhury, and K. Roy, in \textit{2021 {IEEE}/{CVF} International Conference on Computer Vision ({ICCV})} ({IEEE}, 2021).

\phantomsection
\label{csl:376}376in \textit{57th ACM/IEEE Design Automation Conference (DAC)} (\url{https://dl.acm.org/doi/10.5555/3437539.3437564,} 2020), pp. 1–6.

\phantomsection
\label{csl:377} B. Han and K. Roy, in \textit{Computer Vision \textendash{} ECCV 2020} (Springer International Publishing, 2020), pp. 388–404.

\phantomsection
\label{csl:378}378 N. Rathi and K. Roy, IEEE Trans Neural Netw Learn Syst \textbf{PP}, (2021).

\phantomsection
\label{csl:379}379 B. Rueckauer, I.A. Lungu, Y. Hu, M. Pfeiffer, and S.C. Liu, Front Neurosci \textbf{11}, 682 (2017).

\phantomsection
\label{csl:380}380 C. Brandli, R. Berner, M. Yang, S.-C. Liu, and T. Delbruck, {IEEE} Journal of Solid-State Circuits \textbf{49}, 2333 (2014).

\phantomsection
\label{csl:381}381 A. Zhu, L. Yuan, K. Chaney, and K. Daniilidis, in \textit{Robotics: Science and Systems {XIV}} (Robotics: Science and Systems Foundation, 2018).

\phantomsection
\label{csl:382}382 A.Z. Zhu, L. Yuan, K. Chaney, and K. Daniilidis, in \textit{2019 {IEEE}/{CVF} Conference on Computer Vision and Pattern Recognition Workshops ({CVPRW})} ({IEEE}, 2019).

\phantomsection
\label{csl:383}383 A.K. Kosta and K. Roy, in \textit{IEEE International Conference on Robotics and Automation (ICRA)} (\url{https://arxiv.org/abs/2209.11741,} 2023), pp. 6021–6027.

\phantomsection
\label{csl:384}384 A. Sengupta, Y. Ye, R. Wang, C. Liu, and K. Roy, Frontiers in Neuroscience \textbf{13}, (2019).

\phantomsection
\label{csl:385}385 J. Kaiser, H. Mostafa, and E. Neftci, Front Neurosci \textbf{14}, 424 (2020).

\phantomsection
\label{csl:386}386 D. Huh and T.J. Sejnowski, in \textit{Advances in Neural Information Processing Systems} (\url{https://arxiv.org/abs/1706.04698,} 2018).

\phantomsection
\label{csl:387}387in \textit{International Conference on Learning Representations} (\url{https://arxiv.org/abs/2005.01807,} 2019).

\phantomsection
\label{csl:388}388in \textit{European Conference on Computer Vision (Pp. 709-726). Cham: Springer Nature Switzerland} (\url{https://arxiv.org/abs/2110.05929,} 2022).

\phantomsection
\label{csl:389}389(2022).

\phantomsection
\label{csl:390}390 K. Suetake, S.I. Ikegawa, R. Saiin, and Y. Sawada, Neural Netw \textbf{159}, 208 (2023).

\phantomsection
\label{csl:391}391 P. O’Connor, E. Gavves, and M. Welling, in \textit{The 22nd International Conference on Artificial Intelligence and Statistics} (PMLR, 2019), pp. 1516–1523.

\phantomsection
\label{csl:392}392 M. Xiao, Q. Meng, Z. Zhang, Y. Wang, and Z. Lin, Advances in Neural Information Processing Systems \textbf{34}, 14516 (2021).

\phantomsection
\label{csl:393}393 N. Rathi, I. Chakraborty, A. Kosta, A. Sengupta, A. Ankit, P. Panda, and K. Roy, {ACM} Computing Surveys \textbf{55}, 1 (2023).

\phantomsection
\label{csl:394}394 B.V. Benjamin, P. Gao, E. McQuinn, S. Choudhary, A.R. Chandrasekaran, J.-M. Bussat, R. Alvarez-Icaza, J.V. Arthur, P.A. Merolla, and K. Boahen, Proceedings of the {IEEE} \textbf{102}, 699 (2014).

\phantomsection
\label{csl:395}395 P.A. Merolla, J.V. Arthur, R. Alvarez-Icaza, A.S. Cassidy, J. Sawada, F. Akopyan, B.L. Jackson, N. Imam, C. Guo, Y. Nakamura, B. Brezzo, I. Vo, S.K. Esser, R. Appuswamy, B. Taba, A. Amir, M.D. Flickner, W.P. Risk, R. Manohar, and D.S. Modha, Science \textbf{345}, 668 (2014).

\phantomsection
\label{csl:396}396 D.G. Elliott, M. Stumm, W.M. Snelgrove, C. Cojocaru, and R. Mckenzie, {IEEE} Design {\&} Test of Computers \textbf{16}, 32 (1999).

\phantomsection
\label{csl:397}397 A. Agrawal, M. Ali, M. Koo, N. Rathi, A. Jaiswal, and K. Roy, {IEEE} Solid-State Circuits Letters \textbf{4}, 137 (2021).

\phantomsection
\label{csl:398}398 I. Chakraborty, A. Jaiswal, A.K. Saha, S.K. Gupta, and K. Roy, Applied Physics Reviews \textbf{7}, 021308 (2020).

\phantomsection
\label{csl:399}399 R. Waser, R. Dittmann, G. Staikov, and K. Szot, Adv Mater \textbf{21}, 2632 (2009).

\phantomsection
\label{csl:400}400 G.W. Burr, M.J. BrightSky, A. Sebastian, H.-Y. Cheng, J.-Y. Wu, S. Kim, N.E. Sosa, N. Papandreou, H.-L. Lung, H. Pozidis, E. Eleftheriou, and C.H. Lam, {IEEE} Journal on Emerging and Selected Topics in Circuits and Systems \textbf{6}, 146 (2016).

\phantomsection
\label{csl:401}401 M. Hosomi, H. Yamagishi, T. Yamamoto, K. Bessho, Y. Higo, K. Yamane, H. Yamada, M. Shoji, H. Hachino, C. Fukumoto, H. Nagao, and H. Kano, in \textit{{IEEE} {InternationalElectron} Devices Meeting 2005. {IEDM} Technical Digest.} ({IEEE}, n.d.).

\phantomsection
\label{csl:402}402 A. Sengupta and K. Roy, Applied Physics Reviews \textbf{4}, 041105 (2017).

\phantomsection
\label{csl:403}403 T. Sharma, C. Wang, A. Agrawal, and K. Roy, in \textit{2021 {IEEE}/{ACM} International Symposium on Low Power Electronics and Design ({ISLPED})} ({IEEE}, 2021).

\phantomsection
\label{csl:404}404 F. Merrikh-Bayat, X. Guo, M. Klachko, M. Prezioso, K.K. Likharev, and D.B. Strukov, IEEE Trans Neural Netw Learn Syst \textbf{29}, 4782 (2018).

\phantomsection
\label{csl:405}405 P. Panda and K. Roy, in \textit{International Joint Conference on Neural Networks (IJCNN)} (\url{https://arxiv.org/abs/1602.01510,} 2016), pp. 299–306.

\phantomsection
\label{csl:406}406 Z. Sun and D. Ielmini, {IEEE} Transactions on Circuits and Systems {II}: Express Briefs \textbf{69}, 3024 (2022).

\phantomsection
\label{csl:407}407 Z. Sun, S. Kvatinsky, X. Si, A. Mehonic, Y. Cai, and R. Huang, Nature Electronics \textbf{6}, (2023).

\phantomsection
\label{csl:408}408 M. Le Gallo, A. Sebastian, R. Mathis, M. Manica, H. Giefers, T. Tuma, C. Bekas, A. Curioni, and E. Eleftheriou, Nature Electronics \textbf{1}, 246 (2018).

\phantomsection
\label{csl:409}409 M. Le Gallo, A. Sebastian, G. Cherubini, H. Giefers, and E. Eleftheriou, IEEE Transactions on Electron Devices \textbf{65}, (2018).

\phantomsection
\label{csl:410}410 E.J. Fuller, S.T. Keene, A. Melianas, Z. Wang, S. Agarwal, Y. Li, Y. Tuchman, C.D. James, M.J. Marinella, J.J. Yang, A. Salleo, and A.A. Talin, Science \textbf{364}, 570 (2019).

\phantomsection
\label{csl:411}411 T.P. Xiao, B. Feinberg, C.H. Bennett, V. Agrawal, P. Saxena, V. Prabhakar, K. Ramkumar, H. Medu, V. Raghavan, R. Chettuvetty, S. Agarwal, and M.J. Marinella, IEEE Transactions on Circuits and Systems I: Regular Papers \textbf{69}, (2022).

\phantomsection
\label{csl:412}412.

\phantomsection
\label{csl:413}413 Y. Luo, S. Wang, P. Zuo, Z. Sun, and R. Huang, {IEEE} Transactions on Circuits and Systems I: Regular Papers 1 (2022).

\phantomsection
\label{csl:414}414 Z. Sun, G. Pedretti, E. Ambrosi, A. Bricalli, W. Wang, and D. Ielmini, Proceedings of the National Academy of Sciences \textbf{116}, 4123 (2019).

\phantomsection
\label{csl:415}415 B. Feinberg, R. Wong, T.P. Xiao, C.H. Bennett, J.N. Rohan, E.G. Boman, M.J. Marinella, S. Agarwal, and E. Ipek, in \textit{2021 {IEEE} International Symposium on High-Performance Computer Architecture ({HPCA})} ({IEEE}, 2021).

\phantomsection
\label{csl:416}416 P. Mannocci, E. Melacarne, A. Pezzoli, G. Pedretti, C. Villa, F. Sancandi, U. Spagnolini, and D. Ielmini, in \textit{2023 International Electron Devices Meeting (IEDM)} (IEEE, 2023).

\phantomsection
\label{csl:417}417 Q. Zeng, J. Liu, M. Jiang, J. Lan, Y. Gong, Z. Wang, Y. Li, C. Li, J. Ignowski, and K. Huang, IEEE Internet of Things Journal \textbf{11}, (2024).

\phantomsection
\label{csl:418}418 S. Wang, Y. Luo, P. Zuo, L. Pan, Y. Li, and Z. Sun, Science Advances \textbf{9}, (2023).

\phantomsection
\label{csl:419}419 Z. Chen, Y. Zhou, H. Xu, Y. Fu, Y. Li, Y. He, X.-S. Miao, P. Mannocci, and D. Ielmini, in \textit{2023 International Electron Devices Meeting (IEDM)} (IEEE, 2023).

\phantomsection
\label{csl:420}420 J. Woo and S. Yu, IEEE Nanotechnology Magazine \textbf{12}, (2018).

\phantomsection
\label{csl:421}421 S. Achour, R. Sarpeshkar, and M.C. Rinard, {ACM} {SIGPLAN} Notices \textbf{51}, 177 (2016).

\phantomsection
\label{csl:422}422 J. Zhao, S. Huang, O. Yousuf, Y. Gao, B.D. Hoskins, and G.C. Adam, Frontiers in Neuroscience \textbf{15}, (2021).

\phantomsection
\label{csl:423}423 W. Yi, F. Perner, M.S. Qureshi, H. Abdalla, M.D. Pickett, J.J. Yang, M.-X.M. Zhang, G. Medeiros-Ribeiro, and R.S. Williams, Applied Physics A \textbf{102}, (2011).

\phantomsection
\label{csl:424}424.

\phantomsection
\label{csl:425}425 B. Feinberg, U.K.R. Vengalam, N. Whitehair, S. Wang, and E. Ipek, in \textit{2018 ACM/IEEE 45th Annual International Symposium on Computer Architecture (ISCA)} (IEEE, 2018).

\phantomsection
\label{csl:426}426 P. Zuo, Z. Sun, and R. Huang, IEEE Transactions on Circuits and Systems II: Express Briefs \textbf{70}, (2023).

\phantomsection
\label{csl:427}427 P. Mannocci, E. Giannone, and D. Ielmini, in \textit{2023 IEEE International Conference on Metrology for EXtended Reality, Artificial Intelligence and Neural Engineering (MetroXRAINE)} (IEEE, 2023).

\phantomsection
\label{csl:428}428 P. Mannocci and D. Ielmini, IEEE Journal on Exploratory Solid-State Computational Devices and Circuits \textbf{9}, (2023).

\phantomsection
\label{csl:429}429 S. Shapero, A.S. Charles, C.J. Rozell, and P. Hasler, {IEEE} Journal on Emerging and Selected Topics in Circuits and Systems \textbf{2}, 530 (2012).

\phantomsection
\label{csl:430}430 S. Rehman, M.F. Khan, H.-D. Kim, and S. Kim, Nature Communications \textbf{13}, (2022).

\phantomsection
\label{csl:431}431 C.E. Graves, C. Li, G. Pedretti, and J.P. Strachan, in \textit{Memristor Computing Systems} (Springer International Publishing, 2022), pp. 105–139.

\phantomsection
\label{csl:432}432 C. Li, C.E. Graves, X. Sheng, D. Miller, M. Foltin, G. Pedretti, and J.P. Strachan, Nature Communications \textbf{11}, (2020).

\phantomsection
\label{csl:433}433 X.S. Hu, M. Niemier, A. Kazemi, A.F. Laguna, K. Ni, R. Rajaei, M.M. Sharifi, and X. Yin, in \textit{2021 {IEEE} International Electron Devices Meeting ({IEDM})} ({IEEE}, 2021).

\phantomsection
\label{csl:434}434 L. Grinsztajn, E. Oyallon, and G. Varoquaux, NeurIPS 2022 Datasets and Benchmarks Track (2022).

\phantomsection
\label{csl:435}435 Z. Xie, W. Dong, J. Liu, H. Liu, and D. Li, in \textit{Proceedings of the Sixteenth European Conference on Computer Systems} ({ACM}, 2021).

\phantomsection
\label{csl:436}436 G. Pedretti, C.E. Graves, S. Serebryakov, R. Mao, X. Sheng, M. Foltin, C. Li, and J.P. Strachan, Nature Communications \textbf{12}, (2021).

\phantomsection
\label{csl:437}437 A. Shafiee, A. Nag, N. Muralimanohar, R. Balasubramonian, J.P. Strachan, M. Hu, R.S. Williams, and V. Srikumar, in \textit{2016 {ACM}/{IEEE} 43rd Annual International Symposium on Computer Architecture ({ISCA})} ({IEEE}, 2016).

\phantomsection
\label{csl:438}438 G. Pedretti, J. Moon, P. Bruel, S. Serebryakov, R.M. Roth, L. Buonanno, A. Gajjar, T. Ziegler, C. Xu, M. Foltin, P. Faraboschi, J. Ignowski, and C.E. Graves, ArXiv (2023).

\phantomsection
\label{csl:439}439 K. Pagiamtzis and A. Sheikholeslami, {IEEE} Journal of Solid-State Circuits \textbf{41}, 712 (2006).

\phantomsection
\label{csl:440}440 G. Pedretti, C.E. Graves, T.V. Vaerenbergh, S. Serebryakov, M. Foltin, X. Sheng, R. Mao, C. Li, and J.P. Strachan, Advanced Electronic Materials \textbf{8}, 2101198 (2022).

\phantomsection
\label{csl:441}441 H. Farzaneh, J.P.C. de Lima, M. Li, A.A. Khan, X.S. Hu, and J. Castrillon, ArXiv (2023).

\phantomsection
\label{csl:442}442 A. Bremler-Barr, D. Hay, D. Hendler, and R.M. Roth, {IEEE}/{ACM} Transactions on Networking \textbf{18}, 1665 (2010).

\phantomsection
\label{csl:443}443 A. Lucas, Frontiers in Physics \textbf{2}, (2014).

\phantomsection
\label{csl:444}444 J.J. Hopfield and D.W. Tank, Biological Cybernetics \textbf{52}, 141 (1985).

\phantomsection
\label{csl:445}445 E. Farhi, J. Goldstone, and S. Gutmann, (2014).

\phantomsection
\label{csl:446}446 M. Ercsey-Ravasz and Z. Toroczkai, Nature Physics \textbf{7}, 966 (2011).

\phantomsection
\label{csl:447}447 B. Moln{\'{a}}r, F. Moln{\'{a}}r, M. Varga, Z. Toroczkai, and M. Ercsey-Ravasz, Nature Communications \textbf{9}, (2018).

\phantomsection
\label{csl:448}448 T.J. Sejnowski, in \textit{{AIP} Conference Proceedings} ({AIP}, 1986).

\phantomsection
\label{csl:449}449 R. Hamerly, T. Inagaki, P.L. McMahon, D. Venturelli, A. Marandi, T. Onodera, E. Ng, C. Langrock, K. Inaba, T. Honjo, and others, Feedback \textbf{1}, 2 (2018).

\phantomsection
\label{csl:450}450 T.-J. Yang and V. Sze, in \textit{2019 IEEE International Electron Devices Meeting (IEDM)} (IEEE, 2019), pp. 22–1.

\phantomsection
\label{csl:451}451 J. Deguchi, D. Miyashita, A. Maki, S. Sasaki, K. Nakata, and F. Tachibana, in \textit{2019 IEEE International Electron Devices Meeting (IEDM)} (IEEE, 2019), pp. 22–4.

\phantomsection
\label{csl:452}452 H. Jia, M. Ozatay, Y. Tang, H. Valavi, R. Pathak, J. Lee, and N. Verma, {IEEE} Journal of Solid-State Circuits \textbf{57}, 198 (2022).

\phantomsection
\label{csl:453}453 L. Fick, S. Skrzyniarz, M. Parikh, M.B. Henry, and D. Fick, in \textit{2022 {IEEE} International Solid- State Circuits Conference ({ISSCC})} ({IEEE}, 2022).

\phantomsection
\label{csl:454}454 P. Narayanan, S. Ambrogio, A. Okazaki, K. Hosokawa, H. Tsai, A. Nomura, T. Yasuda, C. Mackin, S.C. Lewis, A. Friz, M. Ishii, Y. Kohda, H. Mori, K. Spoon, R. Khaddam-Aljameh, N. Saulnier, M. Bergendahl, J. Demarest, K.W. Brew, V. Chan, S. Choi, I. Ok, I. Ahsan, F.L. Lie, W. Haensch, V. Narayanan, and G.W. Burr, {IEEE} Transactions on Electron Devices \textbf{68}, 6629 (2021).

\phantomsection
\label{csl:455}455 M.L. Gallo, R. Khaddam-Aljameh, M. Stanisavljevic, A. Vasilopoulos, B. Kersting, M. Dazzi, G. Karunaratne, M. Brändli, A. Singh, S.M. Müller, J. Büchel, X. Timoneda, V. Joshi, M.J. Rasch, U. Egger, A. Garofalo, A. Petropoulos, T. Antonakopoulos, K. Brew, S. Choi, I. Ok, T. Philip, V. Chan, C. Silvestre, I. Ahsan, N. Saulnier, V. Narayanan, P.A. Francese, E. Eleftheriou, and A. Sebastian, Nature Electronics (2023).

\phantomsection
\label{csl:456}456 S. Kim, M. Ishii, S. Lewis, T. Perri, M. BrightSky, W. Kim, R. Jordan, G.W. Burr, N. Sosa, A. Ray, J.-P. Han, C. Miller, K. Hosokawa, and C. Lam, in \textit{2015 {IEEE} International Electron Devices Meeting ({IEDM})} ({IEEE}, 2015).

\phantomsection
\label{csl:457}457 S. Kim, S. Kim, S. Um, S. Kim, K. Kim, and H.-J. Yoo, in \textit{2022 {IEEE} Symposium on {VLSI} Technology and Circuits ({VLSI} Technology and Circuits)} ({IEEE}, 2022).

\phantomsection
\label{csl:458}458 M. Anderson, B. Chen, S. Chen, S. Deng, J. Fix, M. Gschwind, A. Kalaiah, C. Kim, J. Lee, J. Liang, and et al, ArXiv Preprint ArXiv:2107.04140 (2021).

\phantomsection
\label{csl:459}459 R. Khaddam-Aljameh, M. Stanisavljevic, J.F. Mas, G. Karunaratne, M. Brandli, F. Liu, A. Singh, S.M. Muller, U. Egger, A. Petropoulos, T. Antonakopoulos, K. Brew, S. Choi, I. Ok, F.L. Lie, N. Saulnier, V. Chan, I. Ahsan, V. Narayanan, S.R. Nandakumar, M.L. Gallo, P.A. Francese, A. Sebastian, and E. Eleftheriou, {IEEE} Journal of Solid-State Circuits \textbf{57}, 1027 (2022).

\phantomsection
\label{csl:460}460 S. Jain, H. Tsai, C.-T. Chen, R. Muralidhar, I. Boybat, M.M. Frank, S. Wozniak, M. Stanisavljevic, P. Adusumilli, P. Narayanan, K. Hosokawa, M. Ishii, A. Kumar, V. Narayanan, and G.W. Burr, {IEEE} Transactions on Very Large Scale Integration ({VLSI}) Systems \textbf{31}, 114 (2023).

\phantomsection
\label{csl:461}461 A.I. Khan, A. Daus, R. Islam, K.M. Neilson, H.R. Lee, H.P. Wong, and E. Pop, Science \textbf{373}, 1243 (2021).

\phantomsection
\label{csl:462}462 B. Prasad, S. Parkin, T. Prodromakis, C.-B. Eom, J. Sort, and J.L. MacManus-Driscoll, APL Materials \textbf{10}, (2022).

\phantomsection
\label{csl:463}463 G.W. Burr, A. Sebastian, E. Vianello, R. Waser, and S. Parkin, APL Materials \textbf{8}, (2020).

\phantomsection
\label{csl:464}464 T. Venkatesan and S. Williams, Applied Physics Reviews \textbf{9}, (2022).

\phantomsection
\label{csl:465}465 T. Chen, J. van Gelder, B. van de Ven, S.V. Amitonov, B. de Wilde, H.-C. Ruiz Euler, H. Broersma, P.A. Bobbert, F.A. Zwanenburg, and W.G. van der Wiel, Nature \textbf{577}, (2020).

\phantomsection
\label{csl:466}466 R.A. John, Y. Demirağ, Y. Shynkarenko, Y. Berezovska, N. Ohannessian, M. Payvand, P. Zeng, M.I. Bodnarchuk, F. Krumeich, G. Kara, I. Shorubalko, M.V. Nair, G.A. Cooke, T. Lippert, G. Indiveri, and M.V. Kovalenko, Nature Communications \textbf{13}, (2022).

\phantomsection
\label{csl:467}467 N. Yantara, S.E. Ng, D. Sharma, B. Zhou, P.S.V. Sun, H.M. Chua, N.F. Jamaludin, A. Basu, and N. Mathews, Advanced Materials \textbf{36}, (2023).

\phantomsection
\label{csl:468}468 E.J. Fuller, S.T. Keene, A. Melianas, Z. Wang, S. Agarwal, Y. Li, Y. Tuchman, C.D. James, M.J. Marinella, J.J. Yang, A. Salleo, and A.A. Talin, Science \textbf{364}, (2019).

\phantomsection
\label{csl:469}469 Y. van de Burgt, E. Lubberman, E.J. Fuller, S.T. Keene, G.C. Faria, S. Agarwal, M.J. Marinella, A. Alec Talin, and A. Salleo, Nature Materials \textbf{16}, (2017).

\phantomsection
\label{csl:470}470 S. Goswami, R. Pramanick, A. Patra, S.P. Rath, M. Foltin, A. Ariando, D. Thompson, T. Venkatesan, S. Goswami, and R.S. Williams, Nature \textbf{597}, (2021).

\phantomsection
\label{csl:471}471 S. Goswami, A.J. Matula, S.P. Rath, S. Hedström, S. Saha, M. Annamalai, D. Sengupta, A. Patra, S. Ghosh, H. Jani, S. Sarkar, M.R. Motapothula, C.A. Nijhuis, J. Martin, S. Goswami, V.S. Batista, and T. Venkatesan, Nature Materials \textbf{16}, (2017).

\phantomsection
\label{csl:472}472 M. Lanza, G. Molas, and I. Naveh, Nature Electronics \textbf{6}, (2023).


@unpublished{vlsi,
  howpublished = {https://www.esann.org/sites/default/files/proceedings/legacy/es2012-44.pdf},
  url = {https://www.esann.org/sites/default/files/proceedings/legacy/es2012-44.pdf},
  title = {{Towards biologically realistic multi-compartment neuron model emulation in analog VLSI}},
  author = {Sebastian Millner, Andreas Hartel, Johannes Schemmel and Karlheinz Meier},
}
\end{document}